\documentstyle [12pt,dina4,german,epsfig]{book}

\germanTeX
\begin{document}

\def\ep{e^{+}}
\def\emm{e^{-}}
\def\epm{e^{\pm}}
\def\pp{p}
\def\cms{\hbox{\scriptsize cms}}
\def\IN#1{{$\left[#1\right]$}}
\def\GeV{\mbox{GeV}}
\def\MeV{\mbox{MeV}}
\def\eV{\mbox{eV}}
\def\MHz{\mbox{MHz}}
\def\min{\mbox{min}}
\def\sec{\mbox{s}}
\def\meter{\mbox{m}}
\def\cm{\mbox{cm}}
\def\mm{\mbox{mm}}
\def\tesla{\mbox{T}}
\def\muamp{\mbox{$\mu$A}}
\def\pb{\mbox{pb}}
\def\acos{\mbox{acos}}
\def\myxi{x_{i/p}}
\def\vv#1{\hbox{\bf #1}}
\def\prot{p}
\def\ele{e}
\def\ycut{y_{\mbox{\scriptsize cut}}}
\def\mb#1{{\mbox{\scriptsize #1}}}
\def\tjet{\theta_{\mb{jet}}}
\def\tquark{\theta_{\mb{Quark}}}
\def\had{\hbox{\scriptsize had}}

\thispagestyle{empty}

\newcommand{\PInst}{I.}       
\newcommand{\Pdat}{Februar 1996}       
\newcommand{\Pnum}{96/31}            
\newcommand{\Name}{Hadig}             
\newcommand{\Vname}{Thomas}         
\newcommand{\Ptit}{                 
       Untersuchungen zur \\
       Jet--Parton--Korrelation \\
       in der tief-inelastischen Streuung 
}

\begin{figure}[ht]
\vspace*{-1.5cm} 
{
\huge
\parbox[b]{6.9cm}{
\vbox to 22mm { \vfill
 \leavevmode\kern-0.6cm
     \epsfysize=2.2cm
     \epsffile{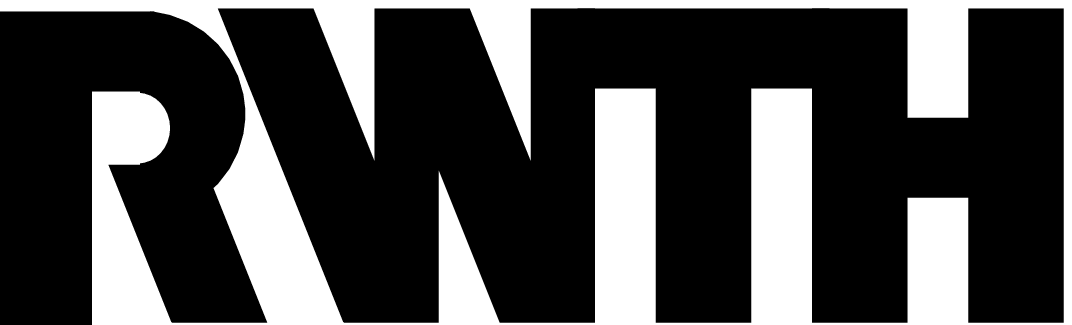} \vfill
}
}         
\hfill
\leavevmode\kern-0.4cm
\parbox[b]{3.4cm}{
\raisebox{0.2cm}{
\vbox to 22mm {
{\normalsize \sf  
  \begin{flushleft}
RHEINISCH\\     
WESTF"ALISCHE\\ 
TECHNISCHE\\   
HOCHSCHULE\\    
AACHEN
  \end{flushleft}
}
}
}
}
  \hfill
\parbox[b]{4.0cm}{
\vbox to 33mm {
{\footnotesize 
  \begin{flushright}
   \raisebox{2.6mm}{\Large \sf hep-ex/9709011}\\
   \raisebox{1.3mm}{\Large \sf PITHA \Pnum}\\
    \rule[0.2cm]{4.0cm}{1.0mm}\\
   {\Large \sf \Pdat }\\
  \end{flushright}
}
}
}

\vbox to 22.0cm{

\begin{center}
   \vfill \bigskip  \bigskip  \bigskip    \bigskip   \bigskip   \bigskip
   {\huge \sf \Ptit}
   \vfill
      \bigskip   \bigskip   
{\Large \sf
                          \Vname~\Name}\\
 \bigskip   \bigskip   \bigskip   \bigskip   \bigskip
   \bigskip   \bigskip   \bigskip   \bigskip
   \bigskip   \bigskip   \bigskip   \bigskip
   \bigskip   
\end{center}

\vfill

\begin{center}
{   \large \sf
   \PInst Physikalisches Institut der Technischen Hochschule Aachen \\
}
\end{center}

\vspace*{-0.7cm}

 \leavevmode\kern-0.6cm
\rule{15.8cm}{0.2mm}


\begin{center}
\huge  \sf
  PHYSIKALISCHE INSTITUTE \\
     RWTH AACHEN \\
  52056 AACHEN, GERMANY
\end{center}

}
}
\end{figure}

\clearpage

\thispagestyle{empty}

{\bf Abstract}

The jet parton correlation has been studied for different cut scenarios and 
various jet algorithms.
These migrations are getting important for measuring the strong coupling 
constant $\alpha_s$ and the
NLO gluon density $g(\xi,Q^2)$ from jet rates in DIS at HERA. 

\title{      
       Untersuchungen zur \\
       Jet--Parton--Korrelation \\
       in der tief-inelastischen Streuung }
\author{
\vspace{1 cm} \\
        von 
\vspace{1. cm} \\
        Thomas Hadig \\
\vspace{1. cm} \\
        Diplomarbeit in Physik 
\vspace{1. cm} \\
        vorgelegt der
\vspace{1. cm} \\
        Mathematisch-Naturwissenschaftlichen Fakult"at der \\
        Rheinisch-Westf"alischen Technischen Hochschule Aachen 
\vspace{1. cm} \\
        im Februar 1996
\vspace{1. cm} \\
        angefertigt im \\
        I\@.\ Physikalischen Institut}
\date{}        
\maketitle

\tableofcontents

\chapter{Einleitung}
\label {kapeinl}

Ein wesentlicher Charakterzug der Menschen ist die Neugierde. Immer
schon haben wir uns nicht mit der blo"sen Existenz eines Ph"anomens
zufriedengegeben, sondern haben versucht die Ursache daf"ur zu
ergr"unden.

Die "alteste Frage ist die nach dem Woher der Welt und ihrer
Zusammensetzung. Eine entscheidende Idee hatte hierbei schon der
griechische Philosoph Demokrit, f"ur den die Welt aus elementaren,
unteilbaren Bausteinen (\grqq atomos\grqq ) aufgebaut war.

\begin{figure}[tbp]
\begin{center}
\epsfig{file=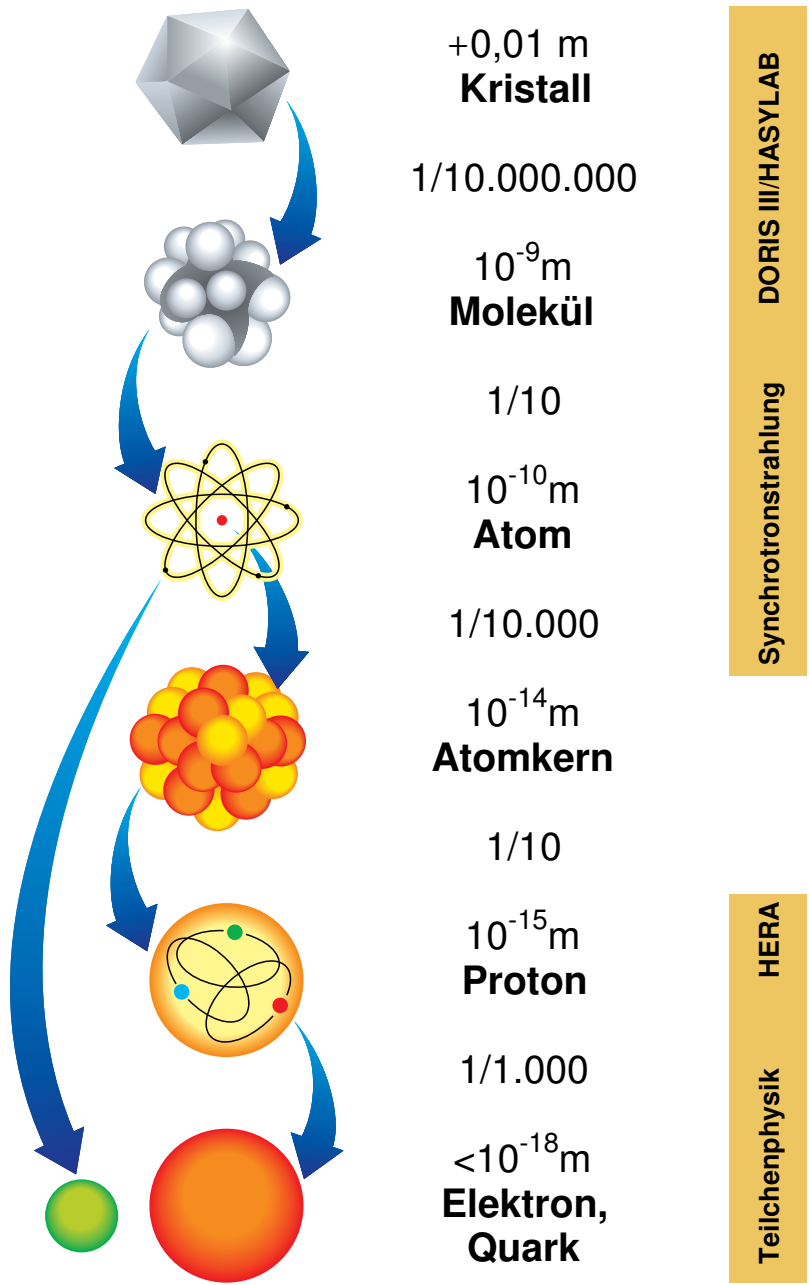,width=0.7\hsize}
\end{center}
\caption[Vergleich der Gr"o"senordnungen]{{\bf Vergleich der
Gr"o"senordnungen verschiedener Strukturen}{\it\ (gedruckt mit freundlicher
Genehmigung der DESY PR Abteilung)}}
\label{abbsizes}
\end{figure}

Heute ist diese Frage immer noch nicht beantwortet, auch wenn die
moderne Naturwissenschaft Schritt f"ur Schritt zu immer kleineren
Strukturen vordringt (siehe Abbildung \ref{abbsizes}).

Eine Voraussetzung daf"ur war die Entwicklung des Modellbegriffes. Unter
einem Modell verstehen wir in der Physik eine Theorie, die einen
bestimmten Teil der Natur beschreibt, ohne da"s Messungen den
Vorhersagen des Modells widersprechen. Aus den meisten Theorien k"onnen
wir zus"atzliche Aussagen "uber bisher nicht beobachtbare Bereiche und
neue Ph"anomene treffen. Die Aufgabe des Experimentalphysikers ist es,
dieses zu "uberpr"ufen. Stimmen die Messungen nicht mit der Theorie
"uberein, so sind die bestehenden Modelle in diesem Bereich nicht mehr
g"ultig und die Theoretiker sind dazu aufgerufen, diese zu erweitern
oder neue Modelle zu entwickeln. Dieses gegenseitige Wechselspiel ist
sicher ein wesentlicher Grund f"ur den Erfolg der heutigen
Naturwissenschaft.

Auch wenn die heute so h"aufig beschrieene Anwendbarkeit der
Forschungsergebnisse bei der Grundlagenforschung nicht direkt erkennbar
ist, so befriedigt sie dennoch die Neugier der Menschen und leistet
einen Beitrag zum technischen und philosophischen Fortschritt. Als
Beispiele seien hier nur die Relativit"atstheorie oder das
Heisenbergsche Unsch"arfeprinzip erw"ahnt, die das Weltbild und die
Philosophie dieses Jahrhunderts wie kaum eine andere Entwicklung gepr"agt
haben.

\def\lefam#1#2{\displaystyle \begin{array}{c} {#1} \\ {#2} \end{array}}
\def\lefamk#1#2{$\displaystyle \left( \lefam{#1}{#2}\right)_L$}

\def\av{\displaystyle{A\over B}}
\def\vb{\vphantom{$\av$}} 
\def\v3{\vphantom{$\displaystyle{\av\over A}$}}
\def\h1{\hphantom{test}}

\begin{table}[tb]
\begin{center}
\begin{tabular}{@{\extracolsep{\fill}} l ccc}
$\lefam{\mb{Leptonen}}{\mb{(Spin ${1\over 2}$)}}$ & \lefamk{{\nu_e}}{{e^-}} & 
\lefamk{{\nu_{\mu} }}{{\mu^-}} & \lefamk{{\nu_{\tau} }}{{\tau^-}} \\
\vb& $e^-_R$ & $\mu^-_R$ & $\tau^-_R$ \\
\hline
\v3$\lefam{\mb{Quarks}}{\mb{(Spin ${1\over 2}$)}}$ & \lefamk{u}{d} & 
\lefamk{c}{s} & \lefamk{t}{b} \\
\vb& $u_R, d_R$ & $c_R, s_R$ & $t_R, b_R$ \\
\end{tabular}\\
\end{center}

\begin{center}
\begin{tabular}{@{\extracolsep{\fill}}| l| c|}
\hline
\multicolumn{2}{|c|}{\vb Austauschteilchen (Spin 1)} \\
\hline
\vb elektromagnetisch\h1 & $\gamma$ \\
\vb schwach           & $Z^0, W^{\pm}$ \\
\vb stark             & \h1$g_i, i=1,\dots,8$\h1 \\
\hline
\hline
\multicolumn{2}{|c|}{\vb Skalare (Spin 0)} \\
\hline
\vb Higgsfeld & $h$ \\
\hline
\end{tabular}\\
\end{center}
\caption[Elementarteilchen im Standardmodell]
{{\bf Elementarteilchen im Standardmodell}}
\label{tabstdmod}
\end{table}

Das zur Zeit g"ultige Standardmodell der Elementarteilchenphysik ist in
der Lage alle heute bekannten Effekte der Hochenergiephysik durch drei
Quark- und drei Lepton-Familien zu beschreiben, deren Mitglieder "uber
Austauschteilchen miteinander wechselwirken. Die Partikel sind in
Tabelle \ref{tabstdmod} aufgef"uhrt. Es wird zwischen links- (L) und
rechtsh"andigen (R) Teilchen unterschieden, da Neutrinos nur linksh"andig
in der Natur vorkommen. Die Quarks sind alle bereits experimentell
beobachtet worden, wobei die Existenz des Top-Quarks $t$ nach ersten
Hinweisen 1994 nun experimentell best"atigt wurde \cite{Fermi:95-022-E},
\cite{Fermi:95-028-E}.
Bei den Leptonen ist nur das Tau-Neutrino $\nu_{\tau}$ bisher
unbeobachtet.

Die ebenfalls in Tabelle \ref{tabstdmod} gezeigten Austauschteilchen
"ubertragen die Wechselwirkung zwischen den Teilchen. Diese Partikel
sind alle experimentell best"atigt. Ein wichtiges Ziel in der
Hochenergiephysik ist, die verschiedenen Wechselwirkungen durch den
gleichen Mechanismus mit einer einzigen Kopplungskonstante zu
beschreiben. Diese sogenannte {\it Vereinheitlichung} ist bei der
elektromagnetischen und der schwachen Kraft bereits gelungen. Um dies
durch die Modelle der Gro"sen Vereinheitlichung (GUT, {\it grand unified
theory}) ebenfalls f"ur die starke Kraft zu erreichen, ist eine genaue
Kenntnis der starken Kopplungskonstante wesentlich. Die Vereinigung
der Kopplungskonstanten wird bei einer heute nicht erreichbaren
Energie erwartet.

Ein Problem des Standardmodells ist die Masse der Teilchen zu
erkl"aren. Der Higgs-Mechanismus erreicht dies durch ein zus"atzliches
Teilchen, das Higgs-Boson $h.$ Es besteht die Hoffnung, da"s die Existenz
dieses Teilchens bei den geplanten Beschleunigern LEP 200 oder LHC
nachgewiesen werden kann.

Das Standardmodell ist sicher nicht die letzte Theorie. Neben der
gro"sen vereinheitlichten Theorie ist die hohe Anzahl an
Eingabeparametern unbefriedigend. Ein Zusammenhang zwischen diesen
Parametern wird in neueren Modellen wie der Supersymmetrie und der
String-Theorie gesucht. Gleichzeitig stellen diese neue Vorhersagen auf,
so da"s sich in den n"achsten Jahren die M"oglichkeit weiterer
Experimente und Messungen ergeben wird.

Es bleiben somit noch viele Fragen offen, deren Beantwortung wir jedoch
mit den heute bestehenden und zuk"unftigen Beschleunigern sowie durch
engere Kooperation mit anderen Bereichen physikalischer Forschung
(z.B.\ der Kosmologie) immer n"aherkommen werden. So werden wir
hoffentlich irgendwann die Frage nach dem Woher und dem Woraus der Welt
beantworten k"onnen.

\chapter {Experimentelle Grundlagen}
\label{kapdetector}

\section {Teilchenbeschleuniger}

In der Teilchenphysik werden Beschleuniger zur Bestimmung der
Eigenschaften von Elementarteilchen eingesetzt.

Zwei Teilchenreaktionen erm"oglichen es, Daten zu gewinnen, Streuungen
und Annihilationen.

Zum einen kann man "uber {\bf Streuungen} durch Beschu"s des zu
untersuchenden Teilchens dessen innere Struktur vermessen. Die
erreichbare Aufl"osung $\Delta x$ wird vom Viererimpuls"ubertrag
$\sqrt{Q^2}$ des Austauschteilchens "uber die Beziehung
\begin{eqnarray}
\Delta x &=& {\hbar \over \sqrt{Q^2}} \label{eqnaufl}
\end{eqnarray}
bestimmt. Der Viererimpuls"ubertrag ist durch die zur
Verf"ugung stehende Schwerpunktsenergie begrenzt.

Zum anderen wird durch {\bf Annihilation} von Teilchen und Antiteilchen
eine definierte Schwerpunktsenergie erreicht, mit der neue Teilchen
durch Resonanzen im Wirkungsquerschnitt nachgewiesen und untersucht
werden k"onnen bzw.\ deren Masse und Lebensdauer bestimmt wird. Dazu
mu"s jedoch die Sto"senergie in der Gr"o"senordnung der Masse des zu
erzeugenden Teilchens sein.

Beim Bau eines Beschleunigers, mit dem Teilchenreaktionen mit hoher
Genauigkeit vermessen werden sollen, m"ussen verschiedene
Designparameter beachtet werden. Von gro"ser Bedeutung sind hier die
Teilchensorte der Sto"spartner, die maximale Energie und die
Luminosit"at.

Eine m"oglichst hohe {\bf Schwerpunktsenergie} erreicht man, indem zwei
bewegte Partikel aneinander streuen bzw.\ annihilieren. Aus diesem
Grunde sind sogenannte Fixed-Target-Experimente, d.h.\ Beschleuniger bei
denen ein beschleunigtes Teilchen auf ein feststehendes, ruhendes Ziel
geschossen wird, nur noch selten anzutreffen. Die Berechnung der
Schwerpunktsenergie erfolgt unter Vernachl"assigung der Teilchenmassen
bewegter Teilchen durch\footnote{Formel f"ur Collider gilt nur f"ur
Kollisionen von Teilchen aus entgegengesetzten Richtungen.} 
\begin{eqnarray}
E_{\mb{Schwerpunkt}}^2 &=& (E_1 + E_2)^2 - (\vec p_1 + \vec p_2)^2\\
E_{\mb{Collider}} &=&  \sqrt{4 E_1 E_2} \\
E_{\mb{Fixed-Target}} &=& \sqrt{2 E_1 m_2}
\end{eqnarray}

Die verbreitetste Bauart von Beschleunigern ist die eines Ringes, in dem
die beiden Sto"spartner gegenl"aufig in einem Kreis umlaufen und nach
und nach auf die Maximalenergie beschleunigt werden. Die Nachteile
dieses Verfahrens liegen in der n"otigen Synchronisation der
Beschleunigungsvorg"ange und dem Energieverlust durch
Synchrotronstrahlung aufgrund der gekr"ummten Bahn. F"ur die Zukunft
sind jedoch auch einige Experimente geplant, bei denen die Sto"spartner
durch zwei Linearbeschleuniger, die die Teilchen in einer geraden R"ohre
beschleunigen, aufeinander geschossen werden. Dazu mu"s jedoch die
St"arke der beschleunigenden elektromagnetischen Felder vergr"o"sert
werden, da sonst die Beschleunigungsstrecken zu lang sind. Auch hier ist
die Synchronisation der beiden Teilchenstrahlen kompliziert. Ein
weiterer Nachteil ist, da"s sich die Teilchenstrahlen nur einmal kreuzen
und danach das Beschleunigersystem verlassen.

Neben der Schwerpunktsenergie ist die {\bf spezifische Luminosit"at
{$L$}} einer der grundlegenden Parameter. Zusammen mit der
Maschinenlaufzeit $t$ l"a"st sich die integrierte Luminosit"at $${\cal
L} = \displaystyle\int_t L(t) dt$$ berechnen. Die zu erwartetende
Ereigniszahl $N$ f"ur einen Proze"s mit gegebenen Wirkungsquerschnitt
$\sigma$ ergibt sich aus $$N = {\cal L} \cdot \sigma.$$

Eine hohe Luminosit"at erreicht man durch kleine Strahlquerschnitte und
hohe Teilchenstr"ome.

\begin{table}[tbp]
\begin{center}
\begin{minipage}{\hsize}
\begin{tabular*}{\hsize}{@{\extracolsep{\fill}}|l|l|c|c|c|}
\hline
& & & maximale & spezifische \\
Name & Ort & Teilchensorte & Strahlenergie & Luminosit"at \\ 
& & & \IN{\GeV} & \IN{10^{30} \cm^{-2}\sec^{-1}}\\
\hline
PETRA\footnote{Datennahme 1986 beendet} & Hamburg &
 $\ep\emm$ & 23.4 & 24 bei $17.5\GeV$ \\
PEP\footnote{Datennahme 1990 beendet} & Stanford &
 $\ep\emm$ & 15 & 60 \\
TRISTAN & Tsukuba & $\ep\emm$ & 32 & 37 \\
SLC & Stanford & $\ep\emm$ & 50 & 0.35 \\
LEP & Genf & $\ep\emm$ & 55 & 11 \\
LEP 200 & Genf & $\ep\emm$ & 70\footnote{stufenweise Erh"ohung der Energie
von LEP auf eine Schwerpunktenergie von max. $193\;\GeV$} & 11 \\
\hline
$Sp\bar pS$ & Genf & $\pp\pb$ & 315 & 6 \\
Tevatron I & Chicago & $\pp\pb$ & 1000 & 10 \\
\hline
HERA & Hamburg & $\epm\pp$ & 30/820 & 16 \\
\hline
LHC\footnote{geplant, Fertigstellung voraussichtlich 2004} & 
CERN & $\pp\pp$ & 7000 & 10000\\
\hline
\end{tabular*}
\end{minipage}
\end{center}
\caption[Liste der Beschleuniger]{{\bf Liste der Beschleuniger~:}
{\it\ Daten, au"ser LEP200, entnommen \cite{PDG:94} }}
\label{tabbeschl}
\end{table}


Alle Colliderexperimente sind in eine der drei folgenden Kategorien nach
Art der {\bf Sto"spartner} einzuordnen. Die Vorteile bei der
Beschleunigung von Elektronen entstehen aufgrund der geringen Masse.
Schon bei einer Energie von $37\;\MeV$ erreichen die Elektronen eine
Geschwindigkeit von 99.99\% der Lichtgeschwindigkeit. Dadurch m"ussen
bei der Beschleunigung im Ring nur kleine zeitliche Synchronisationen
ber"ucksichtigt werden. Die Probleme bei der Verwendung von Elektronen
bestehen in den hohen Energieverlusten pro Umlauf durch
Synchrotronstrahlung, die die maximale Strahlenergie begrenzen oder den
Durchmesser des Ringes ansteigen lassen. Bei den schwereren Protonen
hingegen ist ein Geschwindigkeitsbruchteil von 99.99\% der
Lichtgeschwindigkeit erst bei einer Teilchenenergie von $66.5\;\GeV$
erreicht, dadurch wird der Beschleunigungsvorgang komplizierter. Die
erreichbaren Energien liegen h"oher, wobei ber"ucksichtigt werden mu"s,
da"s bei tief-inelastischen St"o"sen nur ein Bruchteil $x_B$ des
Protonimpules im harten Proze"s zur Verf"ugung steht.

Bei {\it $\ep\emm$ Beschleunigern} werden Elektronen mit Positronen mit 
entgegengesetzt gleichem Impuls zur Kollision gebracht. Die dabei
freiwerdende Energie kann zur Erzeugung von Teilchen verwendet werden,
die die Quantenzahlen des Vakuums haben. Die Vorteile liegen in der
einfachen Reaktionskinematik, da das Laborsystem gleichzeitig
Schwerpunktsystem ist und bei Annihilation eine immer konstante,
durch die Strahlenergie gegebene Schwerpunktsenergie vorliegt.

Bei {\it $\pp\bar p$ Maschinen} streuen Protonen an deren Antiteilchen. Der
Vorteil dieser Beschleuniger liegt in der h"oheren Energie der
Strahlteilchen. Die Nachteile entstehen durch die innere Substruktur der
Protonen, so da"s nur ein Teil des Protonimpules in der Reaktion zur
Verf"ugung steht und die Reste der Protonen zus"atzliche Energie im
Detektor hinterlassen. Dadurch entstehen hohe Jetmultiplizit"aten. Ein
weiteres Problem entsteht dadurch, da"s die Strahlteilchen durch die starke
Wechselwirkung im Anfangszustand zus"atzliche Hintergrundereignisse
hervorrufen k"onnen. F"ur $\pp\pp$ Collider, wie z.B. der
inzwischen genehmigt LHC, gelten die gleichen Vor- und Nachteile,
hinzu kommt jedoch, da"s eine Annihilation der Protonen oder deren
Konstituenten hier nicht m"oglich ist.

Schlie"slich existiert noch die M"oglichkeit Partikel unterschiedlicher
Teilchensorte kollidieren zu lassen. Bei {\it $\epm\pp$ Beschleunigern}
werden hochenergetische Protonen auf Elektronen oder Positronen
geschossen. Die Vorteile liegen in der inneren Substruktur nur eines der
Strahlteilchen, des Protons. Diese kann durch Beobachtung des gestreuten
Elektrons und des hadronischen Endzustandes vermessen werden. Da der
Energie"ubertrag von Ereignis zu Ereignis unterschiedlich ist, k"onnen
die Untersuchungen mit verschiedenen Schwerpunktsenergien ohne
Ver"anderung der Strahlenergien durchgef"uhrt werden. Die Nachteile
entstehen durch die Verschiedenheit von Labor- und Schwerpunktsystem.

In der Tabelle \ref{tabbeschl} sind die wichtigsten Beschleuniger
aufgef"uhrt.

\section{HERA}

Der Beschleuniger HERA ({\bf H}adron {\bf E}lektron {\bf R}ing {\bf
A}nlage) ist der einzige $\epm\pp$ Beschleuniger und wurde nach
siebenj"ahriger Bauzeit 1991 in Betrieb genommen.

\begin{figure}[tb]
\epsfig{file=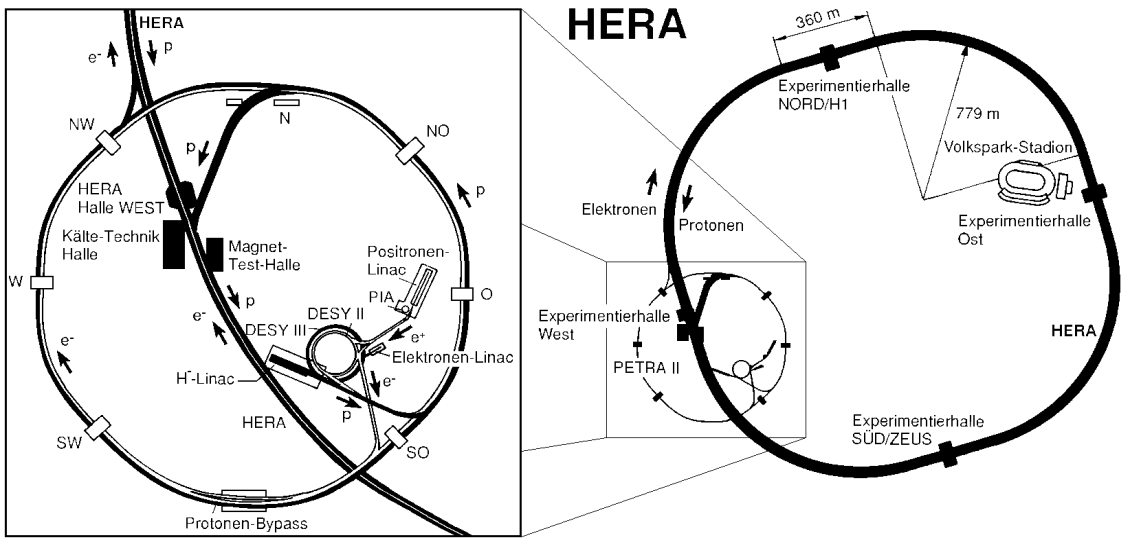,width=\hsize}
\caption[HERA mit Vorbeschleunigern]{{\bf Der HERA Speicherring und
die Vorbeschleuniger}}
\label{abbhera}
\end{figure}

Die Beschleuniger, die auf dem DESY-Gel"ande ( {\bf D}eutsches {\bf
E}lektronen {\bf Sy}n"-chro"-tron) bereits vorhanden waren, werden
hierbei als Vorbeschleuniger benutzt. Eine "Ubersicht "uber die
einzelnen Ger"ate gibt die Abbildung {\ref{abbhera}}.

Der Protonenring ist mit supraleitenden Magneten versehen. Bei der
Injektion in HERA sind die Protonen bereits auf $40\;\GeV$
vorbeschleunigt. Die entgegengesetzt umlaufenden Elektronen\footnote{In
diesem Ring k"onnen sowohl Elektronen als auch deren Antiteilchen, die
Positronen, beschleunigt werden. Nachdem bis Mitte 1994 Elektronen
verwendet wurden, werden nun Positronen als Sto"spartner der Protonen
benutzt. Im folgenden wird trotzdem meist ausschlie"slich der Begriff
Elektronen verwendet, da die Betrachtungen hier unabh"angig von der Art
der Strahlleptonen sind. Unterschiede ergeben sich aber z.B.\ in der
Berechnung der Wirkungsquerschnitte elektroschwacher Prozesse.} haben
eine Einschu"senergie in HERA von $14\;\GeV.$ Dort werden herk"ommliche
Magnete verwendet, da die Energie und Masse der Elektronen wesentlich
niedriger ist und so kleinere Magnetfelder ben"otigt werden.

Die Strahlteilchen kollidieren in den Wechselwirkungspunkten unter einem
Winkel von $180^{\circ}$. Die Protonen und Elektronen werden zu
sogenannten {\it Bunches} zusammengefa"st, d.h.\ zu Paketen aus
Teilchen, die aufgrund der Beschleunigung durch elektrische
Wechselfelder in Strahlrichtung eine begrenzte Ausdehnung haben.

Es existieren vier Hallen, in denen Experimente aufgebaut werden
k"onnen. Neben den beiden Vielzweckdetektoren H1 und ZEUS existiert seit
1995 mit HERMES ein drittes Experiment, das "uber ein Gastarget die
Streuung an polarisierten Elektronen bzw.\ Positronen untersucht. Zur
Zeit ist ein viertes Experiment HERA-B in Planung.

Die technischen Daten von HERA sind in Tabelle \ref{tabhera}
zusammengefa"st.

\begin{table}[tb]
\begin{minipage}{\hsize}
\begin{tabular}{|l|ll|ll|}
\hline
& \multicolumn{2}{|c|}{Designparameter} & \multicolumn{2}{|c|}{1994} \\
& $\pp$-Ring & $\epm$-Ring & $\pp$-Ring & $\epm$-Ring \\
\hline
Maximalenergie \IN{\GeV} & 820 & 30 & 820 & 27.55 \\
Anzahl Teilchenpakete & 210 & 210 & 153+17\footnote{Die zweite Zahl gibt
die Anzahl der Pilotbunche an, das sind die Strahlpakete, die nicht auf
ein Paket des anderen Strahls treffen und so zur Messung des Strahl--Wand und
Strahl--Gas Untergrundes dienen.} & 153+15 \\
Strahlstrom/Paket \IN{\muamp} & 760 & 290 & 241 & 101 \\
\hline
Einschu"senergie in HERA \IN{\GeV} & 40 & 14 & & \\
magnetisches Ablenkfeld \IN{\tesla} & 4.65 & 0.165 & & \\
Temperatur der Magnete & 4.4 K & 294 K & & \\
Strahlungsverlust pro Umlauf & $125\MeV$ & $6\eV$ & & \\
\hline
Schwerpunktenergie \IN{\GeV} & \multicolumn{2}{|c|}{314} & \multicolumn{2}{|c|}{300.6} \\
spez. Luminosit"at \IN{10^{30}\cm^{-2}\sec^{-1}} & \multicolumn{2}{|c|}{15} & 
\multicolumn{2}{|c|}{1.4\footnote{Der Wert von 1994 entspricht der bei H1
gemessenen durchschnittlichen spezifischen Luminosit"at. Der h"ochste Wert war
$4.3 \cdot 10^{30}\cm^{-2}\sec^{-1}$}} \\
\hline
Umfang \IN{\meter} & \multicolumn{4}{|c|}{6336}\\
Kollisionsrate \IN{\MHz} & \multicolumn{4}{|c|}{10.4}\\
\hline
\end{tabular}
\end{minipage}
\label{tabhera}
\caption[Einige technische Parameter von HERA]
{{\bf Einige technische Parameter von HERA :}
{\it\ Die Designwerte wurden bei der Datennahme 1994 teilweise noch nicht
erreicht.}}
\end{table}

\section {Der H1 Detektor}

Der H1 Detektor ist einer der beiden Vielzweck $4\pi$ Detektoren am HERA
Speicherring. Seine Aufgaben sind die Detektierung und Bestimmung von
Reaktionsprodukten bei St"o"sen von Protonen und Elektronen.

Dazu ist der nahezu komplette Raumbereich um den Wechselwirkungspunkt
mit Spurkammern und Kalorimetern ausgestattet. Spurkammern dienen zur
Bestimmung der Bahn geladener Teilchen. An der Kr"ummung dieser Bahnen
in einem externen Magnetfeld l"a"st sich das Ladungsvorzeichen und in
Verbindung mit der St"arke der Ionisation die Art des Teilchens
bestimmen. Kalorimeter dienen zur Energiemessung, indem sie die
Menge der Ionisationsladungen in den einzelnen Zellen bestimmen.

Aus diesen Informationen kann man die Flugbahnen der Reaktionsprodukte
und die Vertexposition, d.i.\ der Ort der Kollision, rekonstruieren.
Au"serdem sind noch Informationen "uber die Zeitpunkte der Ereignisse
hilfreich.

Da die Protonen eine h"ohere Strahlenergie haben und bei
tief-inelastischen Streuungen nicht intakt bleiben, entspricht das
Laborsystem nicht dem Schwerpunktsys"-tem der Reaktion. Aus diesem Grunde
sind die Anforderungen an die Me"sger"ate in Proton- und in
Elektronrichtung unterschiedlich und der Detektor selber nicht
symmetrisch aufgebaut.

\begin{figure}[p]
\epsfig{file=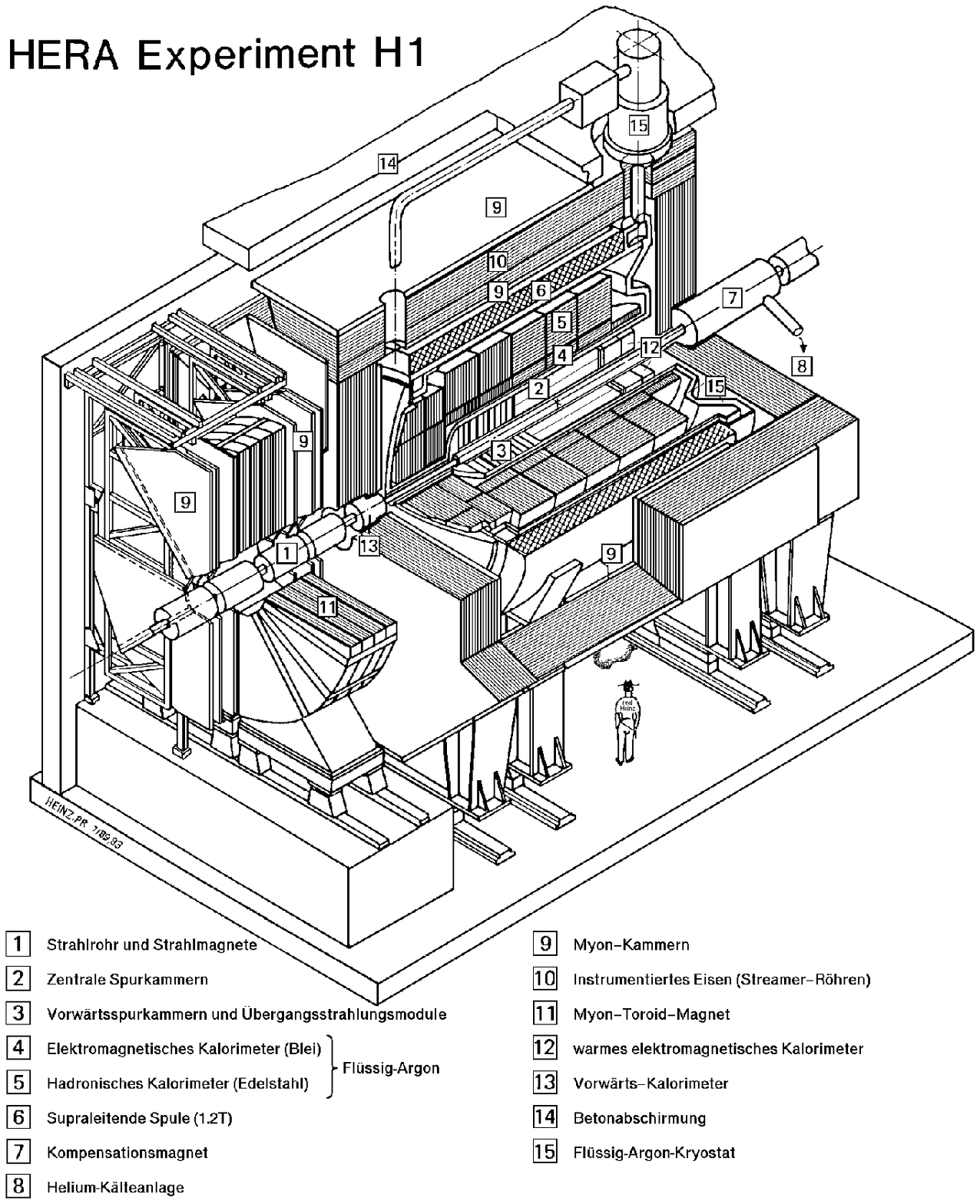}
\caption[Der H1 Detektor]{{\bf Der H1 Detektor}}
\label{abbh1}
\end{figure}

Die Abbildung {\ref {abbh1}} zeigt den H1 Detektor in einer
Ri"szeichnung. Das H1 Koordinatensystem ist rechth"andig und liegt so,
da"s die $z$-Achse entlang des Strahlrohres in Richtung der Protonen
weist. Die $x$-Achse zeigt zum Speicherringmittelpunkt und die $y$-Achse
somit nach oben. Der Nullpunkt liegt im nominalen Wechselwirkungspunkt.
Meist erfolgt die Beschreibung jedoch in Polarkoordinaten, wobei
$\theta$ den Polarwinkel zur $z$-Achse darstellt. Die ungestreuten
Protonen laufen unter dem Winkel $\theta = 0^{\circ}$ aus. Der
Azimutalwinkel $\varphi$ beschreibt die Lage in der $x-y$-Ebene.

Die f"ur diese Arbeit wichtigsten Komponenten werden hier nun n"aher
erl"autert, die Zahlen beziehen sich dabei auf Abbildung {\ref{abbh1}}
\cite{DESY:93-103}~:

\begin{description}
\item[Spule (6)~:] Die supraleitende Spule erzeugt ein Magnetfeld, das eine
Kr"ummung der Flugbahnen geladener Teilchen bewirkt. Dadurch kann das
Ladungvorzeichen und der Impuls eines Teilchen bestimmt werden. Die
Spule ist $5.16\;\meter$ lang und hat einen Radius von $2.75\;\meter.$
Bei den Spurkammern in der Mitte des Detektors erzeugt sie ein nahezu
homogenes Magnetfeld von $1.2\;\tesla.$
\item[Fl"ussig-Argon-Kalorimeter (LAr, 4 und 5)~:] Das
mit fl"ussigem Argon gef"ullte Kalorimeter deckt den Winkelbereich von
vier bis 153 Grad ab. Durch seine Zweiteilung in einen hadronischen
"au"seren Teil und ein elektromagnetisches inneres Kalorimeter ist es in
der Lage, nicht nur die Energie der Teilchen zu messen, sondern auf
Grund des Verh"altnisses der Ladungsdepositionen in den beiden
Komponenten einen Hinweis auf die Art des Teilchens zu geben. Dies ist
f"ur die Berechnung der Energie wesentlich, da der Energieverlust pro
Wegl"ange bei Hadronen und Leptonen unterschiedlich ist. Die
charakterischen Gr"o"sen zu den beiden Bereichen sind in Tabelle
\ref{tablar} zusammengefa"st. Die Energieaufl"osung ist dabei "uber
die Formel $$ {\sigma_E \over E/1\;\GeV} = \sqrt { {A^2 \over E} + {B^2
\over E^2} + C^2 }$$ mit den Anteilen durch die Samplingfrequenz (A),
das elektronische Rauschen (B) und die Kalibrationsunsicherheit (C)
gegeben.

\begin{table}[tb]
\begin{center}
\begin{minipage}{14cm}
\begin{tabular*}{14cm}{@{\extracolsep{\fill}}|l|c|c|}
\hline
& hadronisch & el.mag. \\
\hline
Absorbermaterial & Stahl & Blei \\
Dicke der Absorberplatten \IN{\mm} &  19 & 2.4 \\
Dicke der fl"ussigen Argonschichten \IN{\mm} & $2 \cdot 2.4$ & 2.35 \\
Tiefe in charakteristischen Einheiten & 
4.5 - 8\footnote{in hadronischen Wechselwirkungsl"angen inklusive
elektromagnetischem Kalorimeter} & 20 - 30\footnote{in Stahlungsl"angen} \\
\hline
A \IN{10^{-2} \sqrt{\GeV}} &  $46.1 \pm 0.7$ & $11.16 \pm 0.05 $ \\
B \IN{\MeV} & $730 \pm 30$ & $152 \pm 4$ \\
C \IN{10^{-2}} & $2.6 \pm 0.2$ & $0.64 \pm 0.07$ \\
bei Energie \IN{\GeV} & 3.7 - 170 & 3.7 - 80 \\
\hline
\end{tabular*}
\end{minipage}
\end{center}
\caption[Daten des Fl"ussig-Argon-Kalorimeters]
{{\bf Daten des Fl"ussig-Argon-Kalorimeters.}{\it\ Daten aus
\cite{DESY:93-078}} }
\label{tablar}
\end{table}

\item[R"uckw"artiges elektromagnetisches Kalorimeter (BEMC, 12)~:]
Das BEMC\linebreak[4] ist in R"uckw"artsrichtung montiert und "uberdeckt den
Winkelbereich von $151^{\circ}$ bis $177^{\circ}.$ Da man dort nur bei
sehr kleinen Protonimpulsbruchteilen $x_B$ hadronische Aktivit"at
erwartet und das gestreute Elektron bei geringen
Viererimpuls"ubertr"agen $(Q^2 < 100 \GeV^2)$ genau in diesen Bereich
f"allt, ist das BEMC nur als elektromagnetisches Kalorimeter ausgelegt.
Es besteht aus $2.5\;\mm$ dicken Bleiabsorberplatten und $4\;\mm$
starken Szintillatorplatten, deren Licht "uber Photodioden ausgelesen
wird. Die Energieaufl"osung f"ur Elektronen betr"agt $$\left. {\sigma_E
\over E} \right|_{\epm}= { 10 \% \over \sqrt {E/\GeV}},$$
f"ur Hadronen jedoch nur $$\left. {\sigma_E \over E}
\right|_{\hbox{\scriptsize had}} = { 80 \% \over \sqrt {E/\GeV}}.$$

\item[Instrumentiertes Eisen (Iron, 10)~:]
Zur R"uckf"uhrung des magnetischen Flusses befindet sich au"serhalb der
Spule ein Eisenjoch. Zwischen den zehn Eisenplatten der Dicke $75\mm$
sind Streamerrohrkammern eingeschoben. Diese sind mit Pads best"uckt,
mit denen eine Energiemessung der nicht in den Kalorimetern absorbierten
Teilchen m"oglich ist.
\end{description}

Die anderen Detektorkomponenten werden bei dieser Analyse nicht direkt
benutzt. Die Informationen sind jedoch in der Ereignisvorselektion meist
"uber Veto- und Triggersignale bzw.\ in der Rekonstruktion als Zeit-
oder Spurmessungen verwendet worden. N"ahere Informationen zum Detektor
oder zum Beschleuniger kann in der angegebenen Literatur gefunden werden
(\cite{DESY:93-103},
\cite{Itterbeck:93},\cite{Koehler:95},\cite{Rosenbauer:95}).

\chapter{Theoretische Grundlagen}
\label {kap_theorie}

\section {Tief-inelastische Lepton-Nukleon-Streuung}

Protonen haben im Gegensatz zu Elektronen eine innere Struktur,
weshalb man zwischen elastischer und inelastischer Streuung
unterscheiden mu"s.

Bei der Lepton-Nukleon-Streuung wird zwischen {\bf elastischer}
Streuung, d.h.\ das Proton bleibt intakt, und {\bf inelastischer}
unterschieden, d.h.\ das Proton zerbricht und die Streuung findet an
einem Parton des Protons statt.
         
Bei den in dieser Arbeit untersuchten {\bf tief-inelastischen St"o"sen}
wird ein virtuelles Photon ausgetauscht, d.h.\ ein Photon dessen
Viererimpulsquadrat $\vv{q}^2$ negativ ist.

Dieser Proze"s kann mit $e p \rightarrow e X$ bezeichnet werden, wobei
$X$ den gesamten hadronischen Endzustand beschreibt.

\subsection{Kinematik}
\label{kapkine}

"Ublicherweise wird in der Hochenergiephysik ein Teilchen durch
Vierervektoren beschrieben. Ein wichtiger Vektor ist der
Impulsvierervektor, der definiert ist durch~:
$$\vv{p} = \left( \begin{array}{c} E \\ 
\vec{p} \end{array} \right)=
\left( \begin{array}{c} E \\ p_x \\ p_y \\ p_z \end{array} \right) $$

Das Skalarprodukt eines Impulsvierervektors mit sich selbst ergibt die
invariante Masse (mit Einsteinscher Summenkonvention) aus~:
\begin{eqnarray} 
\vv{p}^2 = \vv{p} \cdot \vv{p} &=& 
\vv{p}_{\mu} \vv{p}^{\mu} = g^{\mu\nu} \vv{p}_{\mu} \vv{p}_{\nu} \nonumber\\
&=& E^2 - p^2 = (E+p)\cdot(E-p) = m^2.
\end{eqnarray}

Im folgenden werden die Ruhemassen vernachl"assigt, da diese bei HERA
f"ur die einlaufenden Teilchen klein im Vergleich zu den jeweiligen
Gesamtenergien sind. F"ur die Impulse der Strahlteilchen ergibt sich
dann
$$\vv{P} = \left( 
\begin{array}{c} E_p \\ \vec P \end{array} \right) = \left(
\begin{array}{c} 820\;\GeV \\ 0 \\ 0 \\ 820\;\GeV \end{array} \right) $$
f"ur die Protonen und f"ur die Elektronen
$$\vv{k} = \left( 
\begin{array}{c} E_e \\ \vec k \end{array} \right) = \left(
\begin{array}{c} 27.55\;\GeV \\ 0 \\ 0 \\ -27.55\;\GeV \end{array} \right) $$

\begin{figure}[tb]
\epsfig{file=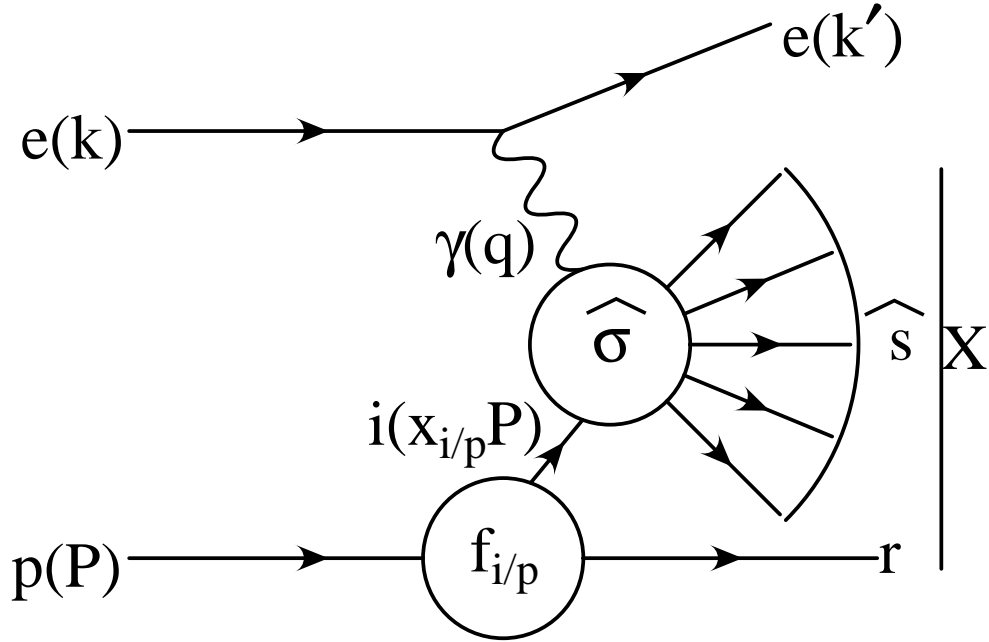,width=\hsize}
\caption[Kinematik bei der Elektron Proton Streuung]
{{\bf Schematische Darstellung einer Elektron Proton Streuung~:}
{\it\ Dargestellt werden hier die Bezeichnungen der wichtigsten Vektoren und
Variablen.}}
\label{abbdis}
\end{figure}

In Abbildung \ref{abbdis} wird die tief-inelastische Streuung
verdeutlicht.

Das unter dem Winkel $\theta_e$ gestreute Elektron hat den Vierervektor
$\vv{k}^{\prime}.$ Das Austauschteilchen $\gamma$ oder $Z^0$ tr"agt den
Impuls $\vv{q} =\vv{k} - \vv{k}^{\prime}.$

In den n"achsten Abschnitten betrachten wir die Mandelstam-Variablen
$s,t$ und $u$ f"ur diesen Proze"s.

\begin{eqnarray}
s_{ep} = s &=& ( \vv{P} + \vv{k} )^2 \\
  &=& m_p^2 + 2 \vv{P} \cdot \vv{k} + m_e^2 \nonumber \\
  &\approx& 4 E_p E_e
\end{eqnarray}

Dies entspricht dem Quadrat der Schwerpunktsenergie $E_{\cms}.$ Diese
ist aufgrund der festen Strahlenergien konstant $\sqrt{s} = 300.6\;\GeV.$

F"ur den Viererimpuls"ubertrag gilt
\begin{eqnarray}
t = \vv{q}^2 &=& (\vv{k}-\vv{k}^{\prime})^2.
\end{eqnarray}
Da $t = \vv{q}^2$ jedoch negativ ist, wird meist
$$Q^2 = -\vv{q}^2 > 0$$
benutzt.

Betrachten wir nur den harten Proze"s, der in der Abbildung {\ref{abbdis}}
durch ein $\hat \sigma$ gekennzeichnet ist, so erhalten wir nun als
einlaufende Teilchen das Photon $\vv{q}$ und ein Parton $i,$ d.h.\ ein
Teil des Protons, mit dem Impuls $\myxi \vv{P}.$

Dann ergibt sich die Mandelstam-Variable $\hat s$ zu
\begin{eqnarray}
s_{\gamma i} = \hat s &=& (\vv{q} + \myxi \vv{P})^2.
\label{eqnshat}
\end{eqnarray}

Bei der Bestimmung $\hat s$ mu"s der Protonrest $r$ von den Produkten
des harten Streuprozesses getrennt werden. Rechnen wir mit dem gesamten
hadronischen Endzustand $X,$ so erhalten wir
\begin{eqnarray}
s_{\gamma p} = W^2 &=& (\vv{q} + \vv{P})^2. \nonumber\\
&\approx& -Q^2 + 2 \vv{P} \cdot \vv{q}
\end{eqnarray}

F"uhren wir die beiden Bjorken Skalenvariabeln
\begin{eqnarray}
x_B \equiv x &=& {Q^2 \over 2 \vv{P} \cdot \vv{q}}\label{eqnxb}
\end{eqnarray}
und
\begin{eqnarray}
y_B \equiv y &=& {\vv{P} \cdot \vv{q} \over \vv{P} \cdot \vv{k}}\label{eqnyb}
\end{eqnarray}
ein, so ergeben sich die folgenden Beziehungen
\begin{eqnarray}
Q^2 &=& x y s \label{eqnqxys}\\
W^2 &\approx& Q^2 {1-x \over x} = y s - Q^2 \label{eqnw2}\\
\myxi &=& x (1 + {\hat s \over Q^2}) \label{eqnxishat}
\end{eqnarray}

Ist $\hat s$ Null oder wesentlich kleiner als $Q^2,$ so entspricht
$\myxi \approx x$ und $x$ kann als der Impulsbruchteil des Protons
betrachtet werden, den das Parton tr"agt.

Analog ist $y$ der Anteil der Elektronenenergie, der am harten Streuproze"s
teilnimmt.

Es gilt f"ur beide Skalenvariablen $x,y \in [0,1].$

\subsection{Berechnung der kinematischen Gr"o"sen}
\label{kapberechn}

Die Reaktionskinematik kann unter Ber"ucksichtigung der Symmetrie
bez"uglich des Azimutalwinkels $\varphi$ nach Formel \ref{eqnqxys} bei
konstanter Schwerpunktsenergie $s$ "uber zwei Gr"o"sen beschrieben
werden. Die Messung dieser Gr"o"sen kann dabei "uber die Daten des
gestreuten Elektrons und den hadronischen Endzustand erfolgen. Nun
werden einige Methoden vorgestellt.

Benutzt man nur das gestreute Elektronen $e$ zur Bestimmung der Daten,
erh"alt man die Gr"o"sen $Q^2$ und $y$ "uber
\begin{eqnarray}
y_e &=& 1 - {E_e^{\prime} - (k_z^{\prime})_e \over 2 E_e} \\
Q^2_e &=& 2 E_e E_e^{\prime} (1 + \cos \theta_e).
\end{eqnarray}

Die Berechnung mit den Kalorimeterinformationen "uber den hadronischen
Endzustand (had) erfolgt durch
\begin{eqnarray}
y_{\had} &=& {E_{\had} -
(p_z)_{\had} \over 2 E_e} \\
Q^2_{\had} &=& {p^2_{T\had} \over
1 - y_{\had}}.
\end{eqnarray}

Die Doppelwinkelmethode DA benutzt nur die Winkel des gestreuten
Elektrons $\theta_e$ und den Winkel der Impulssumme "uber den
hadronischen Endzustand $\theta_{\had}$ und ist somit von der
Energiekalibration weitgehend unabh"angig.
\begin{eqnarray}
y_{\hbox{\scriptsize DA}} &=& {
\displaystyle \tan {\theta_{\had}\over 2} \over
\displaystyle \tan {\theta_{\had}\over 2} +
\tan {\theta_e \over 2}} \\
Q^2_{\hbox{\scriptsize DA}} &=& 4 E_e^2 {
\displaystyle \cot {\theta_e\over 2} \over
\displaystyle \tan {\theta_{\had}\over 2} +
\tan {\theta_e \over 2}}
\end{eqnarray}

Andere Methoden k"onnen in der angegebenen Literatur 
 (\cite{DESY:94-231}, \cite{Rosenbauer:95})
nachgelesen werden.

\def\Punkt{{ \hbox{\scriptsize Punkt} }}

\subsection {Quark Parton Modell}

Schreiben wir den Wirkungsquerschnitt der tief-inelastischen Streuung in
Abh"angigkeit der Lorentz Tensoren so ergibt sich
\begin{eqnarray}
d\sigma \sim |{\cal M}|^2 &=& {e^4 \over Q^4}
L^e_{\mu\nu} W^{\mu\nu}\label{eqndisxsec}
\end{eqnarray}
mit dem Lorentz Tensor f"ur punktf"ormige Teilchen gemittelt "uber alle
Spineinstellungen \cite{Halzen:84}
\begin{eqnarray}
L_e^{\mu\nu} &=& {1\over 2} \sum_{e\;\hbox{\scriptsize spins}}
[ \bar u(\vv{k}) \gamma^{\mu} u(\vv{k}^{\prime})] 
[ \bar u(\vv{k}) \gamma^{\nu} u(\vv{k}^{\prime})]^{*} 
\label{eqnlept}
\end{eqnarray}

W"are das Proton ebenfalls elementar, d.i.\ punktf"ormig, dann k"onnten
wir f"ur das Proton einen analogen Lorentz Tensor angeben. Da wir in
der tief-inelastischen Streuung diese Annahme jedoch nicht machen
k"onnen, setzen wir den allgemeinsten symmetrischen Lorentz Tensor an
\cite{Halzen:84}
\begin{eqnarray}
W^{\mu\nu} &=& - W_1 g^{\mu\nu} + {W_2 \over M^2} P^{\mu} P^{\nu} +
{W_4 \over M^2} q^{\mu} q^{\nu} +
{W_5 \over M^2} (P^{\mu} q^{\nu} + q^{\mu} P^{\nu}).\label{eqnwuv1}
\end{eqnarray}

Eine Beziehung zwischen den $W_i$ Funktionen ergibt sich durch Betrachtung
der Stromerhaltung am Protonvertex.
\begin{eqnarray}
W_5 &=& {\vv{P}\cdot \vv{q}\over Q^2} W_2 \\
W_4 &=& \left( {\vv{P} \cdot \vv{q} \over Q^2} \right)^2 W_2 -
{M^2 \over Q^2} W_1
\end{eqnarray}

Daraus folgt f"ur \ref{eqnwuv1}
\begin{eqnarray}
W^{\mu\nu} &=& - \left( g^{\mu\nu} + {q^{\mu} q^{\nu} \over Q^2} \right) W_1
+ \left. \left( \vv{P} + {\vv{P}\cdot\vv{q}\over Q^2}\vv{q}\right)^2
\right|^{\mu\nu}
{W_2 \over M^2} \label{eqnwuv}
\end{eqnarray}

Aus \ref{eqndisxsec} ergibt sich mit \ref{eqnlept} und \ref{eqnwuv}
\begin{eqnarray}
L^{e\mu\nu}W_{\mu\nu} &=& 4 W_1 (\vv{k} \cdot \vv{k}^{\prime}) +
{2 W_2 \over M^2}
\left[ 2 (\vv{P} \cdot \vv{k}) (\vv{P} \cdot \vv{k}^{\prime})
- M^2 (\vv{k} \cdot \vv{k}^{\prime}) \right]
\end{eqnarray}

F"ur die Lorentz Skalenvariablen $W_i$ gilt bei punktf"ormigen Teilchen,
d.i.\ wenn \ref{eqnwuv} in \ref{eqnlept} "ubergeht,
\begin{eqnarray}
2 M W_1^{\Punkt} = 2 F_1^{\Punkt}(\omega) &=& {Q^2 \over 2 M \nu}
\delta\left( 1 - {Q^2\over 2 M \nu}\right) \nonumber\\
&=& {1\over\omega} \delta \left( 1 - {1\over\omega }\right) \nonumber\\
\nu W_2^{\Punkt} = F_2^{\Punkt}(\omega) &=&
\delta\left( 1 - {Q^2\over 2 M \nu}\right) \nonumber\\
&=& \delta \left(1-{1\over\omega}\right) \label{eqnwpoint}
\end{eqnarray}

Bei kleinem $Q^2$ gilt, d.h.\ wenn wir das Proton als Teilchen mit Formfaktor
$G^2(Q^2)$ ausfassen~:
\begin{eqnarray}
2 M W_1^{\hbox{\scriptsize elastisch}}
&=& {Q^2 \over 2 M\nu} G^2(Q^2)
\;\delta\left( 1 - {Q^2\over 2 M \nu}\right) \nonumber\\
\nu W_2^{\hbox{\scriptsize elastisch}}
&=& G^2(Q^2)
\;\delta\left( 1 - {Q^2\over 2 M \nu}\right) \label{eqnwela}
\end{eqnarray}

Ein wesentlicher Unterschied ist, da"s bei punktf"ormigen Teilchen der
Wirkungsquerschnitt nur vom Verh"altnis $\omega = {2M \nu\over Q^2},$ bei
strukturbehafteten jedoch sowohl von $Q^2$ als auch von $\nu$ abh"angt.

Zerbricht nun das Proton in punktf"ormige Partonen, so k"onnen wir die
Lepton Parton Wechselwirkung durch einen punktf"ormigen Lorentz Tensor
beschreiben, wenn wir die Masse und den Impuls durch den
jeweiligen Bruchteil $x$ ersetzen.

Dadurch k"onnen wir den Proze"s zweiteilen. Zum einen ben"otigen wir
Partondichtefunktionen $f_{i/p}$ (particle density function, PDF), die die
Wahrscheinlichkeit ein Parton des Typs $i$ im Proton anzutreffen in
Abh"angigkeit der kinematischen Variablen beschreiben. Zum anderen bleibt
dann noch die Streuung des Leptons am Parton "ubrig, welche, wie wir oben
gesehen haben, einfach zu beschreiben ist (siehe auch Abbildung
\ref{abbdis}).

F"ur die Partondichtefunktionen mu"s die Normierungsbedingung
\begin{eqnarray}
\sum_i \int dx xf_{i/p}(x) &=&1 \label{eqnnormpdf}
\end{eqnarray}
gelten, da das ganze Proton aus Partonen aufgebaut ist.

Mit der oben definierten Variable $\omega$ gilt dann aus \ref{eqnwpoint}
\begin{eqnarray}
F_1^i(\omega)= x F_1^p &=& mW_1^{\Punkt}
={1\over 2 x\omega} \delta\left( 1 - {1\over x\omega}\right) \nonumber\\
F_2^i(\omega) &=& \delta\left( 1 - {1\over x\omega}\right)
= x \delta\left( x - {1\over \omega}\right)
\end{eqnarray}
und daraus bei Summation "uber alle Partonen eines Protons
\begin{eqnarray}
F_2^p(x) &=& \sum_i \int dx e_i^2 f_{i/p}(x) F_2^i(\omega)\nonumber\\
&=& \sum_i e_i^2 x f_{i/p}(x) \nonumber\\
F_1^p(x) &=& {1\over 2x}F_2(x)
\end{eqnarray}
$e_i$ ist die Ladung des Partons $i$ in Einheiten der Elementarladung.

Die Funktionen $F_i$ sind ausschlie"slich von einer Variablen abh"angig,
d.h.\ bei festem $x$ ist $F_i$ konstant bei ver"andertem $Q^2.$ Dieses
Verhalten wird Bjorken Skalenverhalten (Bjorken scaling) genannt.

Der Aufbau des Protons aus nicht strukturbehafteten Teilchen kann dadurch
nachgewiesen werden.

\subsection {Quantenchromodynamik}

Bevor wir uns die Partondichtefunktionen genauer ansehen, haben wir noch
ein Problem zu l"osen~: Wenn das Proton aus mehreren Partonen besteht,
wieso sind diese Partonen nicht einzeln, sondern nur in einem Teilchen
gebunden vorzufinden.

Direkt damit gekoppelt ist das Problem, da"s Teilchen existieren, die
aus drei identischen Quarks mit gleichen Quantenzahlen aufgebaut sind
(z.b. $|\Delta^{++}> = |uuu>|\uparrow\uparrow\uparrow>).$ Dies ist
jedoch f"ur Fermionen nach dem Pauli-Prinzip verboten \cite{Berger:92}.

Das Problem l"a"st sich durch die Einf"uhrung einer neuen Quantenzahl
l"osen. Diese wird in Analogie zu den drei Grundfarben Farbladung
genannt, d.h.\ es existieren rote, gr"une und blaue Quarks bzw.\
anti-rote, anti-gr"une und anti-blaue Antiquarks.

Das Proton ist, wie jedes andere Teilchen, das wir direkt beobachten
k"onnen, farblos und somit entweder eine Kombination aus drei
unterschiedlich farbigen Quarks (Baryonen) bzw.\ Antiquarks
(Anti-Baryonen) oder ein Zusammenschlu"s aus einem Quark und einem
Antiquark mit der zugeh"origen Antifarbe (Mesonen).

Dieses Verhalten bedingt die Existenz einer neuen Wechselwirkung, der
starken Kraft, und der entsprechenden Austauschteilchen. Im Gegensatz zu
der elektro"-mag"-netischen Wechselwirkung, die nur ein neutrales
Austauschteilchen hat, existieren hier acht linear unabh"angige
Farb--Antifarb--Kombinationen und somit acht Austauschteilchen, die wir
Gluonen nennen.

Sehen wir uns nun die Verteilung von $\myxi$ an, so k"onnen wir
verschiedene Szenarien unterscheiden.

\begin{figure}[tbp]
\epsfig{file=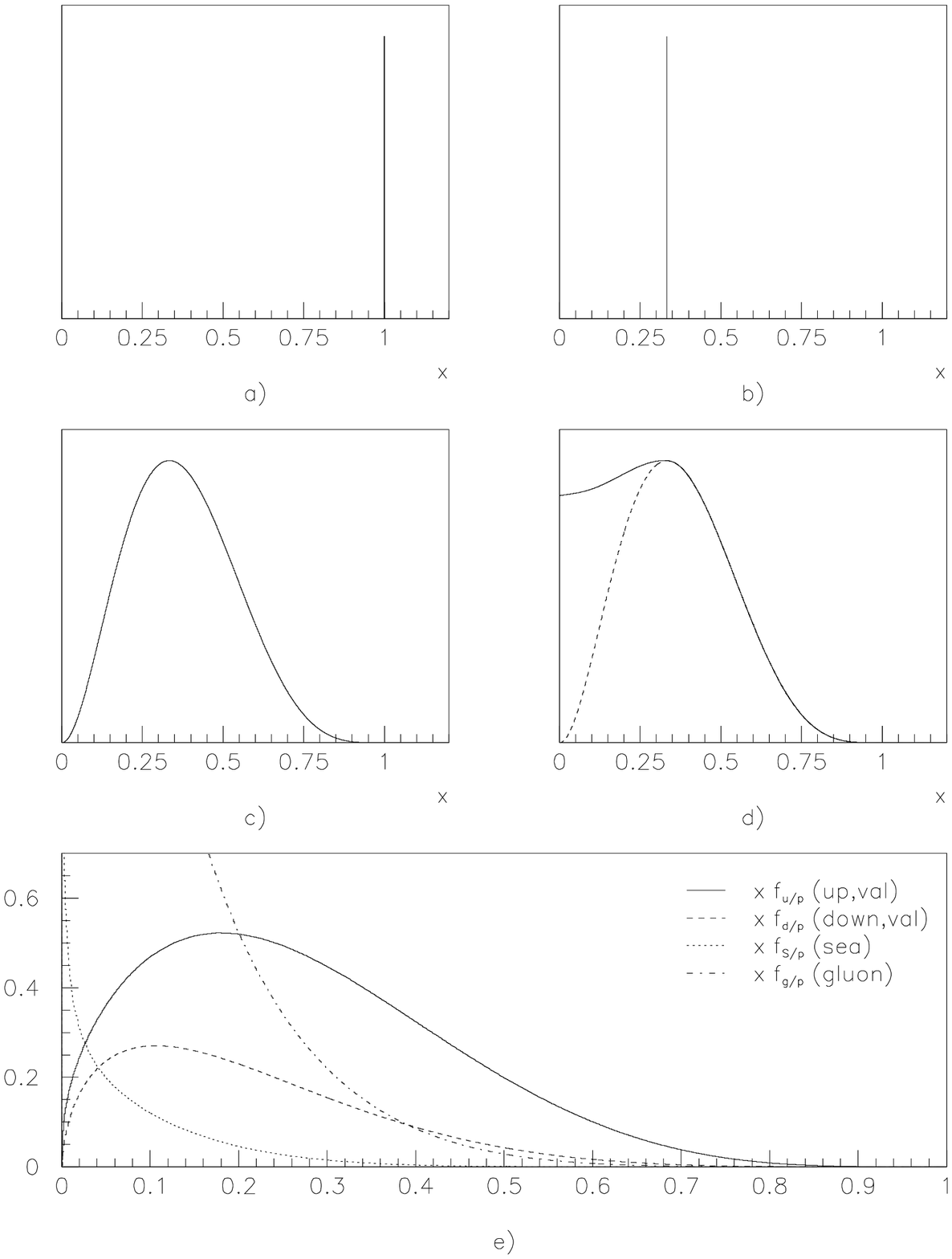,width=\hsize}
\caption[$x$ Verteilung]{{\bf $\myxi$ Verteilung f"ur unterschiedliche
Protonstrukturannahmen~:}
{\it\ a)~Proton ist punktf"ormig, b)~Proton besteht aus drei Quarks,
c)~zus"atzliche Wechselwirkung zwischen Quarks, d)~zus"atzliche
Seequarks, e)~Parametrisierung der Messungen nach MRS(H) bei $\mu_f =
10\;\GeV$ \cite{PDF:606};
siehe Erl"auterungen im Text oder \cite{Halzen:84} }}
\label{abbxbdist}
\end{figure}

Ist das Proton ein punktf"ormiges, nicht strukturbehaftetes Teilchen, so
nimmt an einer Streuung immer der gesammte Impuls $\vv{P}$ teil, d.h.\
$\myxi \equiv 1$ und es ergibt sich Bild \ref{abbxbdist}a.

Besteht das Proton hingegen aus Partonen, die sich in Richtung der
Protonen bewegen und nicht miteinander wechselwirken, so haben diese
Partonen die gleiche Geschwindigkeit und somit --- unter
Vernachl"assigung des Massenunterschiedes --- den gleichen Impuls. F"ur
unsere Verteilung gilt dann $\myxi \equiv {1\over n}.$ Bei einem Aufbau
aus drei Quarks ergibt sich somit Bild \ref{abbxbdist}b.

Wie wir jedoch bereits gesehen haben, sind bisher keine einzelnen Quarks
beobachtet worden. Die Wahrscheinlichkeit ist also gro"s, da"s es eine
Partonwechselwirkung gibt und damit Impuls zwischen diesen "ubertragen
wird. Die $x_B$-Verteilung hat in diesem Fall weiterhin ein Maximum bei
${1\over 3},$ sie ist jedoch verschmiert (Abbildung \ref{abbxbdist}c).

In der Quantenelektrodynamik gibt es die M"oglichkeit der
Vakuumpolarisation, d.h.\ ein hochenergetisches Photon kann ein
Teilchen--Antiteilchen--Paar erzeugen. Dies ist bei Gluonen ebenfalls
m"oglich, wir erwarten zu den drei Quarks eine der Bremsstrahlung
"ahnliche Quark-See-Verteilung. Dadurch erh"oht sich die
Dichteverteilung bei kleinen Impulsbruchteilen (Abbildung
\ref{abbxbdist}d).

Der genaue Verlauf der Dichtefunktionen ist --- wie bereits erw"ahnt ---
von der kinematischen Region abh"angig und in machen Phasenraumbereichen
noch nicht gemessen worden. In Abbildung \ref{abbxbdist}e ist eine
Parametrisierung der Dichtefunktionen, wie sie sich aus den Messungen
ergibt, gezeigt. Wir erkennen, da"s die Verteilungen ungef"ahr unserem
Modell d entspricht, dies jedoch ein noch zu einfacher Ansatz ist.

Integriert man "uber alle $x$ und summiert "uber alle Valenz- und
Seequarks, so ergibt sich beim Proton jedoch nicht 1, wie Gleichung
\ref{eqnnormpdf} erwarten l"a"st. Es m"ussen zus"atzlich noch die 
Gluonen ber"ucksichtigt werden. Da die Gluonen in Seequarks zerfallen,
ist deren Spektrum "ahnlich der Seequarkverteilung (siehe auch Abbildung
\ref{abbxbdist}e). Da die Gluonen jedoch nicht elektrisch geladen sind, 
koppeln sie nur stark und sind durch Streuung an Elektronen nicht direkt
zu messen.

Als Einf"uhrung in die Theorie der Quantenchromodynamik kann
z.B.\ \cite{Berger:92}, \cite{Halzen:84} oder \cite{Nachtmann:92}
empfohlen werden. 

\subsection {Die laufende Kopplungskonstante $\alpha_s$}
\label{kapstoer}

In der Quantenelektrodynamik (QED) hat sich die St"orungstheorie
sinnvoll zur Berechnung der Wirkungsquerschnitte erwiesen.

Mit ihrer Hilfe k"onnen wir diese in Terme in unterschiedlicher Ordnung
in der Kopplungskonstanten $\alpha$ zerlegen~:
\begin{eqnarray}
\sigma &=& A_0 \alpha^0 + A_1 \alpha^1 + A_2 \alpha^2 + \dots +
A_i \alpha^i + \dots \label{eqnstoerqed}
\end{eqnarray}

Die Grundvoraussetzung ist jedoch, da"s die Kopplungskonstante
wesentlich kleiner als eins ist und wir somit die Terme in h"oherer
Ordnung vernachl"assigen k"onnen.

Die verschiedenen Terme $A_i$ k"onnen dann mit Hilfe sogenannter
Feynman-Gra"-phen veranschaulicht und mit den zugeh"origen Regeln
berechnet werden. N"aheres hierzu findet sich in der angegebenen
Literatur (\cite{Halzen:84}, \cite{Itzykson:85}).

Bei den heute m"oglichen genauen Messungen in der QED stellen wir fest,
da"s die Kopplungskonstante $\alpha = {e^2\over 4\pi}$ nicht konstant
ist, sondern bei hohen Energien und damit nach \ref{eqnaufl} bei
kleineren Abst"anden gr"o"ser wird. So steigt der Wert von $1\over 137$
bei Niedrigenergieexperimenten auf $1\over127$ bei LEP, d.h.\ Energien
in der Gr"o"se der Z-Boson-Masse $M_Z.$ Dies kann durch
Vakuumpolarisation erkl"art werden. Um die auftretenden Divergenzen zu
umgehen wird anstelle der Ladung $e_0$ die Ladung $e$
\begin{eqnarray}
\parbox{1.2cm}{\hfill\epsfig{file=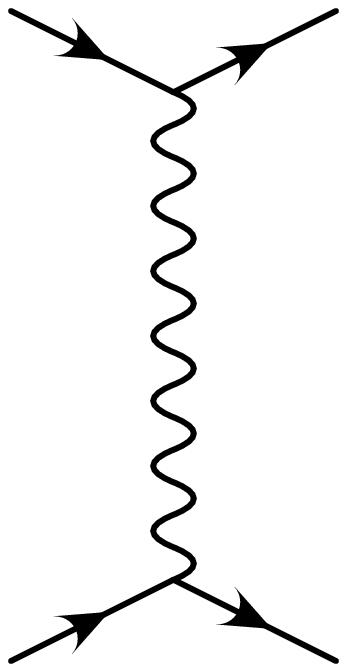,width=1cm}\hfill } 
& \longrightarrow &
\parbox{1.2cm}{\hfill\epsfig{file=eps/qed.ps,width=1cm}\hfill }
\left[ 1 -
\parbox{0.7cm}{\hfill\epsfig{file=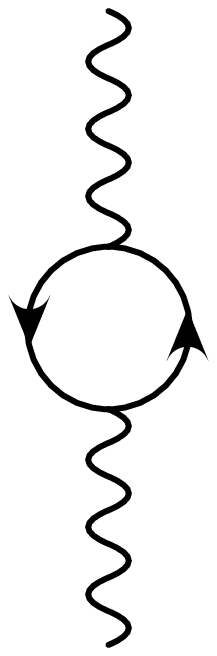,width=0.5cm}\hfill }
+ 
\parbox{0.6cm}{\hfill\epsfig{file=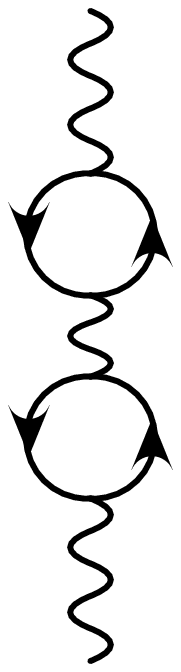,width=0.4cm}\hfill }
- \dots\right]
\end{eqnarray}
verwendet. Dieser Vorgang wird {\it Renormalisierung} genannt.

Der Wert der Kopplungskonstante "andert sich von $\alpha\equiv\alpha(0)$
auf \cite{Halzen:84}
\begin{eqnarray}
\alpha(Q^2) &=& {\alpha(\mu^2) \over  
\displaystyle { 1- {\alpha(\mu^2)\over 3\pi } 
                \log\left( {Q^2 \over \mu^2} \right)}}
\end{eqnarray}

Wenden wir nun das gleiche Prinzip auf die Quantenchromodynamik (QCD) an.
Schreiben wir den Wirkungsquerschnitt in Potenzen der starken
Kopplungskonstante $\alpha_s,$ so wird aus Gleichung \ref{eqnstoerqed}
\begin{eqnarray}
\sigma &=& a_0 \alpha_s^0 + a_1 \alpha_s^1 + a_2 \alpha_s^2 + \dots +
a_i \alpha_s^i + \dots \label{eqnstoerqcd}
\end{eqnarray}

Auch hier gibt es die Vakuumpolarisation, hinzu kommt jedoch noch die
Kopplung von Austauschteilchen aneinander, die in der QED wegen der
Ladungsneutralit"at der Photonen nicht auftritt. Somit renormalisieren
wir die starke Kopplungskonstante durch
\begin{eqnarray}
\parbox{1.2cm}{\hfill\epsfig{file=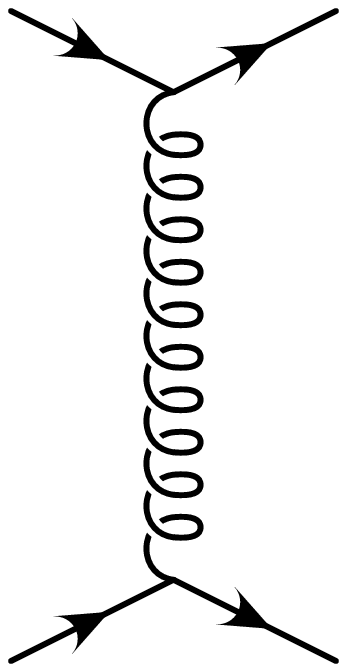,width=1cm}\hfill }
&\longrightarrow &
\parbox{1.2cm}{\hfill\epsfig{file=eps/qcd.ps,width=1cm}\hfill }
\left[ 1 -
\parbox{0.7cm}{\hfill\epsfig{file=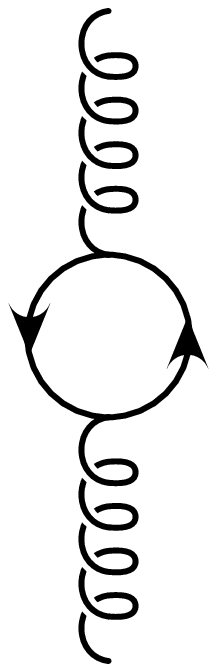,width=0.5cm}\hfill }
-
\parbox{1.2cm}{\hfill\epsfig{file=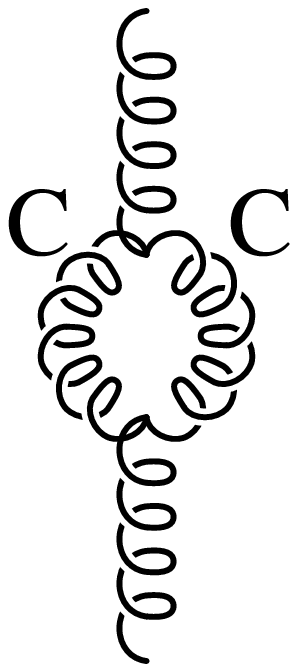,width=1.cm}\hfill }
-
\parbox{1.2cm}{\hfill\epsfig{file=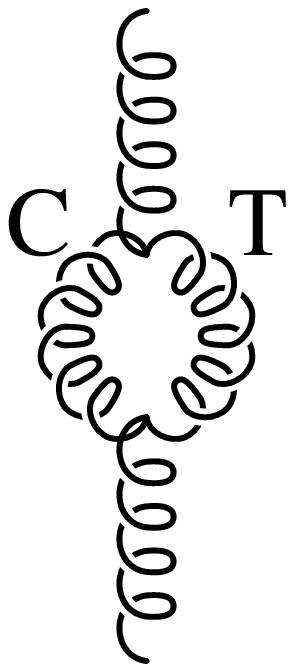,width=1.cm}\hfill }
+\dots\right]
\end{eqnarray}
T steht hier f"ur transversale Polarisation, C f"ur Coulomb Quarks,
d.i.\ longitudinale bzw.\ skalare Polarisation (\cite{Halzen:84}).

In Formeln ausgedr"uckt mit der Anzahl Quarkarten $n_f$ gilt dann
\begin{eqnarray}
\alpha_s(Q^2) &=&
{\alpha_s(\mu^2) \over 
\displaystyle { 1 + {\alpha_s(\mu^2)\over 12\pi}
   (33 - 2 n_f) \log \left( {Q^2 \over \mu^2} \right)}} \\
&=& {12 \pi \over 
\displaystyle {(33 - 2 n_f) \log \left( {Q^2 \over \Lambda^2} \right)}}
\end{eqnarray}
wobei in der letzten Gleichung der freie QCD-Parameter
\begin{eqnarray}
\Lambda^2 &=&
\mu^2 exp \left( {-12\pi \over (33-2n_f)\alpha_s(\mu^2)} \right)
\end{eqnarray}
eingesetzt wurde.

Wir erkennen hier, da"s f"ur die sechs uns bekannten Quarks der
logarithmische Term im Gegensatz zur QED ein positives Vorzeichen hat.
Dies bewirkt, da"s die starke Kopplungskonstante bei kleineren Energien
immer gr"o"ser wird und wir somit nur farblose Teilchen sehen. Dieser
Umstand wird {\it color confinement} genannt.

\subsection {Prozesse in f"uhrender und n"achst-zu-f"uhrender Ordnung}
\label{kapkorrek}

Im folgenden werden wir uns die Feynman-Graphen f"ur einige der harten
Streuprozesse ansehen.

\begin{figure}[tbp]
\begin{center}
\epsfig{file=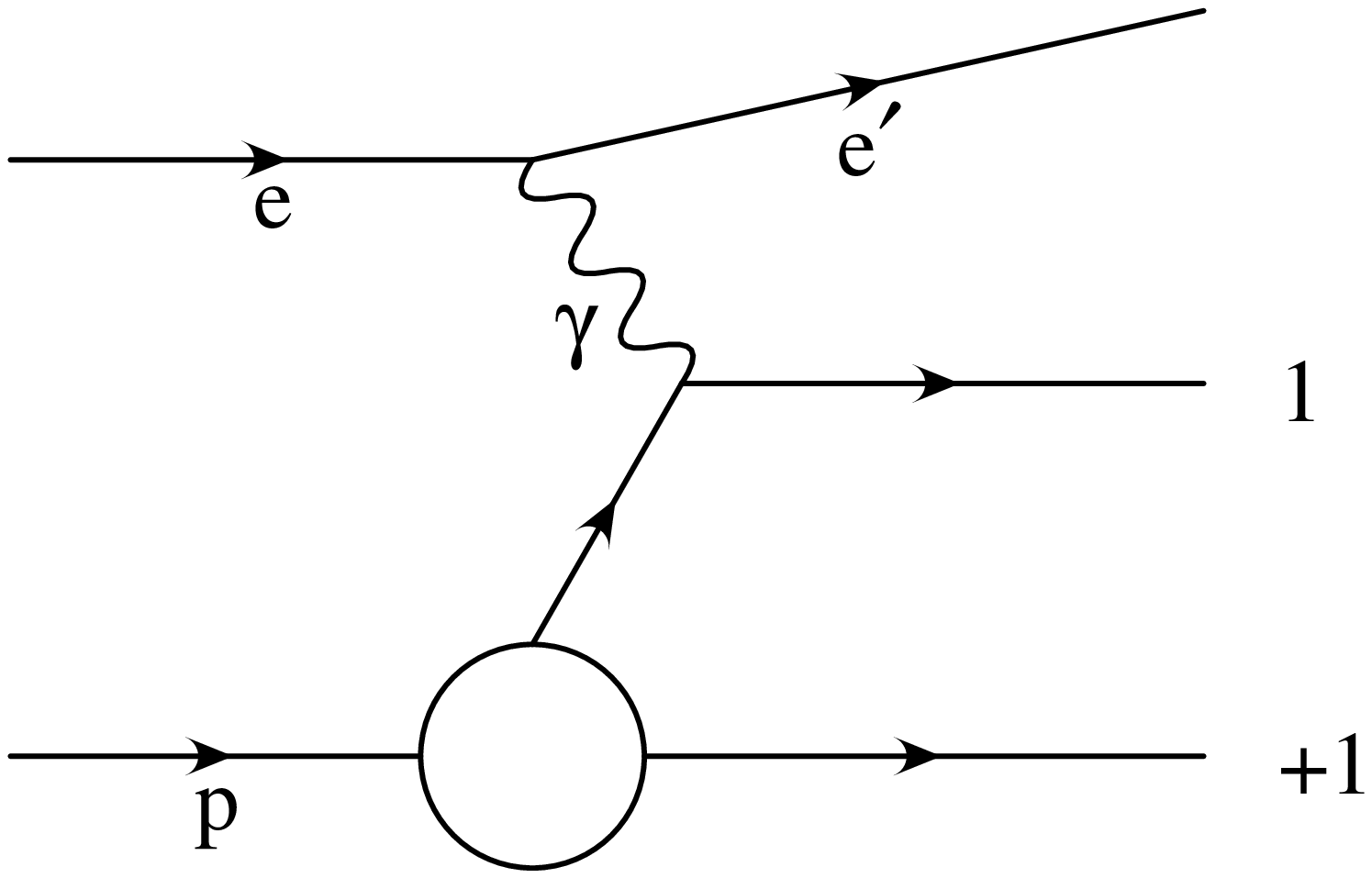,width=.5\hsize}
\end{center}
\caption[1+1 Ereignis in f"uhrender Ordnung]
{{\bf 1+1 Ereignis in f"uhrender Ordnung} }
\label{abbfirlo}
\end{figure}

Der einfachste Graph ist in Abbildung \ref{abbfirlo} gezeigt.
Die Streuung enth"alt keine starke Wechselwirkung, somit geh"ort sie
in die Klasse der ${\cal O}(\alpha_s^0)$ Prozesse.
Da Glu"-onen elektrisch neutral sind, kann dieser Proze"s nur mit
Quarks und Antiquarks als Partonen auftreten. Wir erhalten hier ein Parton
zus"atzlich zum Protonrest, deshalb wird dieser Proze"s als 1+1 Ereignis
bezeichnet.

\begin{figure}[tbp]
\epsfig{file=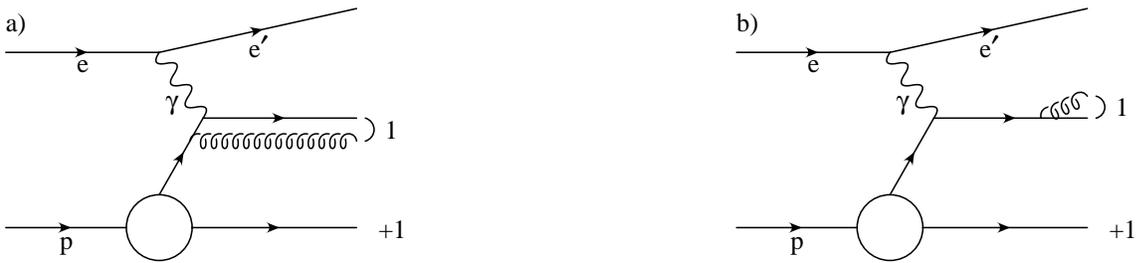,width=\hsize}
\caption[1+1 Ereignis in n"achst-zu-f"uhrender Ordnung]
{{\bf 1+1 Ereignis in n"achst zu f"uhrender Ordnung~:} {\it a) Initial state 
Abstrahlung, b) Final state Abstrahlung} }
\label{abbfirnlo}
\end{figure}

F"ugen wir in diesen Graphen eine starke Wechselwirkung ein, so existiert
die M"oglichkeit ein Gluon vom Quark abzustrahlen. Dies kann vor
der Wechselwirkung mit dem Photon geschehen, dann wird der Proze"s
{ \it initial state radiation} (\ref{abbfirnlo}a) genannt. Erfolgt die
Abstrahlung nach dem Sto"s, so haben wir eine { \it final state radiation}
(\ref{abbfirnlo}b).

Ist das abgestrahlte Gluon weich, d.h.\ das Gluon hat eine geringe Energie,
oder Quark und Gluon sind kollinear, d.h.\ die Impulsrichtung ist gleich,
dann werden wir diesen Proze"s als 1+1 Ereignis klassifizieren.
Da jedoch der Wirkungsquerschnitt f"ur diesen Proze"s ${\cal O}(\alpha_s)$
ist, ist dies ein Beitrag zur n"achst-zur-f"uhrenden Ordnung (NLO,
Next-to-Leading Order).

\begin{figure}[tbp]
\epsfig{file=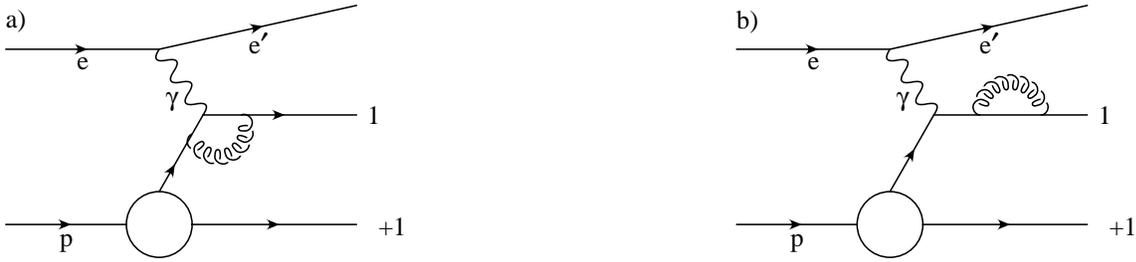,width=\hsize}
\caption[1+1 Ereignis, virtuelle Korrekturen]
{{\bf virtuelle Korrekturen zu 1+1 Ereignis}}
\label{abbfirvir}
\end{figure}

Einen weiteren Anteil erhalten wir durch virtuelle Korrekturen,
d.h.\ Prozesse, die ein internes virtuelles Teilchen austauschen
(Abbildung \ref{abbfirvir}). Der Wirkungsquerschnitt dieser Prozesse
selber ist zwar ${\cal O}(\alpha_s^2),$ jedoch l"a"st sich der
Endzustand nicht vom Proze"s in f"uhrender Ordnung unterscheiden. Durch
die deshalb notwendige Addition der Matrixelemente entsteht bei der
Berechnung des Wirkungsquerschnittes ein Interferenzterm in 
${\cal O}(\alpha_s).$
\begin{eqnarray}
\sigma &\sim&
\left|
\parbox{2cm}{\hfill\epsfig{file=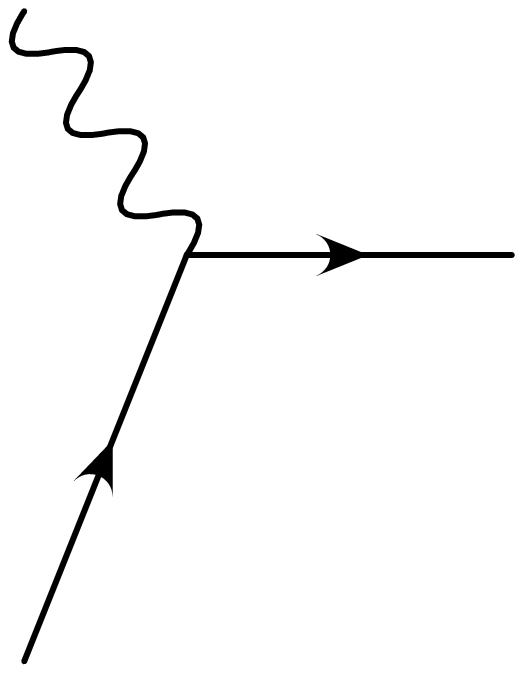,width=1.7cm}\hfill}
+
\parbox{2cm}{\hfill\epsfig{file=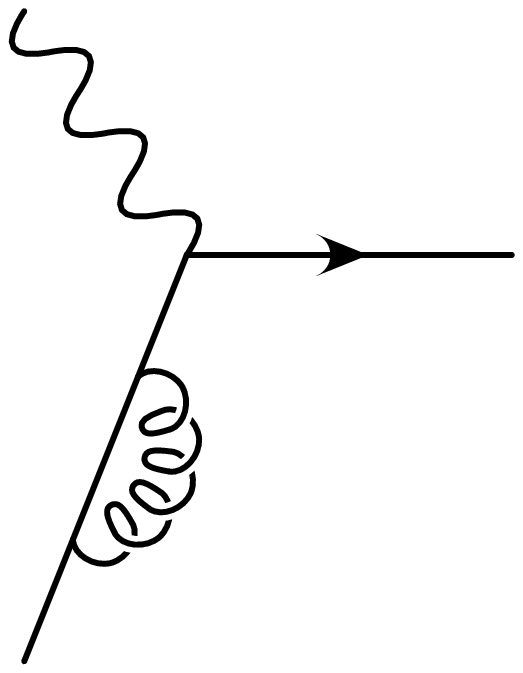,width=1.7cm}\hfill}
+       
\parbox{2cm}{\hfill\epsfig{file=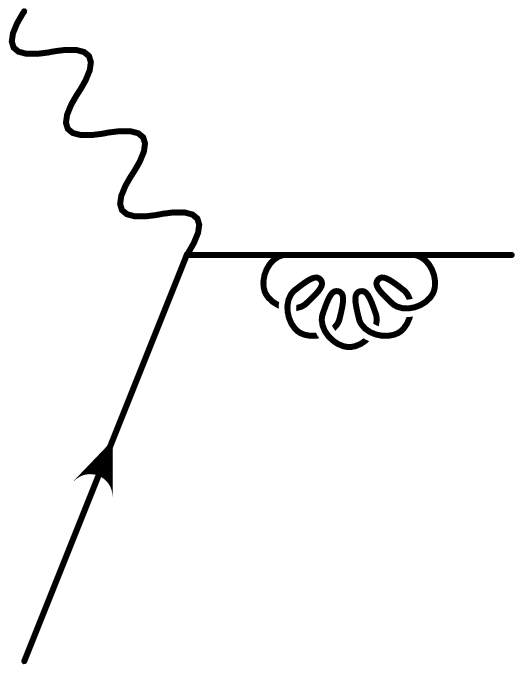,width=1.7cm}\hfill}
+       
\parbox{2cm}{\hfill\epsfig{file=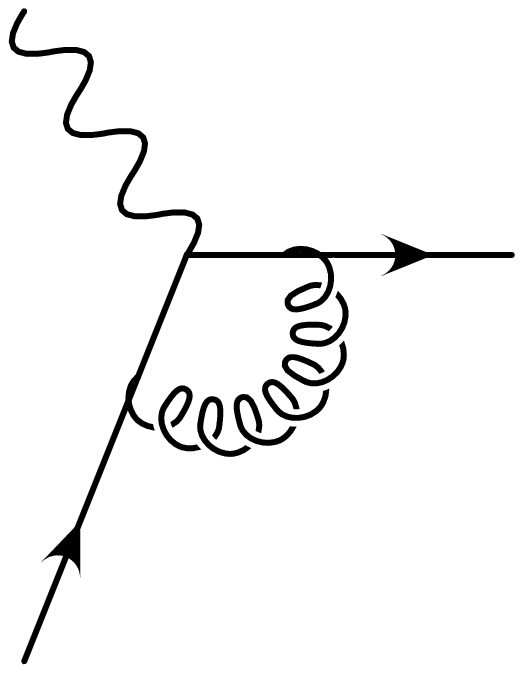,width=1.7cm}\hfill}
+\dots
\right|^2
\end{eqnarray}

Anzumerken ist hier noch, da"s beim Proze"s \ref{abbfirvir}a beim Sto"s
zwischen Parton und Photon nicht der komplette Partonimpulsbruchteil $x$
zur Verf"ugung steht, da das virtuelle Gluon ein Teil des Partonimpulses
erh"alt.

\begin{figure}[tbp]
\epsfig{file=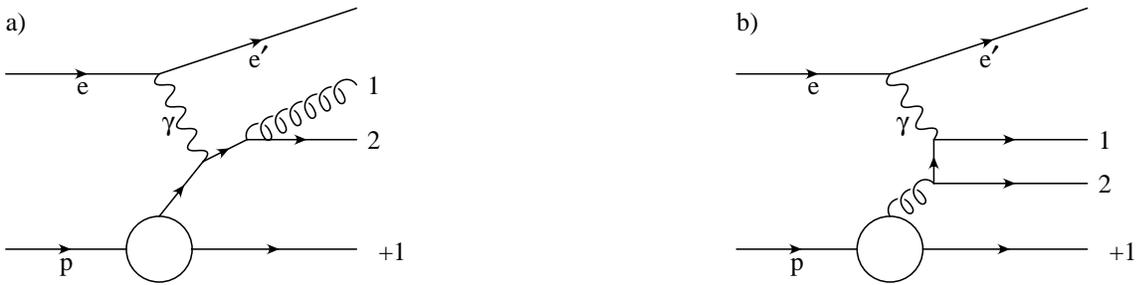,width=\hsize}
\caption[2+1 Ereignis in f"uhrender Ordnung]
{{\bf 2+1 Ereignis in f"uhrender Ordnung~:}{\it a) QCD-Compton, b) Boson 
Gluon Fusion}}
\label{abbseclo}
\end{figure}

Gehen wir nun zur n"achsten Proze"sklasse "uber. Wird entgegen unserer
bisherigen Annahme ein hartes Gluon beim NLO 1+1 Ereignis (Abbildung
\ref{abbfirnlo}) abgestrahlt, so erhalten wir zwei harte Partonen zum
Protonrest. Dieser Proze"s wird in Analogie zur Compton Streuung in der
Quantenelektrodynamik QCD-Compton Proze"s genannt (Abbildung
\ref{abbseclo}a). Der Graph tr"agt somit zur f"uhrenden Ordnung der 2+1
Ereignisse bei und ist ${\cal O}(\alpha_s).$ Die Entscheidung, wann
ein Proze"s zur 2+1 oder zur 1+1 Ereignisklasse geh"ort, mu"s durch
einen Abschneideparameter getroffen werden. Diese Unterscheidung
ist auch deshalb wichtig, da der Wirkungsquerschnitt f"ur die
Abstrahlung kollinearer bzw.\ weicher Partonen divergiert. Eine
genauere Betrachung wird bei der Behandlung des analogen Problems
in Monte-Carlo-Generatoren in Kapitel \ref{kapsimul} durchgef"uhrt.

Bei der sogenannten Boson-Gluon-Fusion (Abbildung \ref{abbseclo}b) wird
ein Gluon vom Proton abgestrahlt. Das Gluon zerf"allt dann in ein
Quark--Antiquark--Paar. Die Streuung des Photons kann entweder am
Quark oder am Antiquark erfolgen. Auch dieser Proze"s geh"ort zur
Klasse der LO 2+1 Ereignisse.

\begin{figure}[tbp]
\epsfig{file=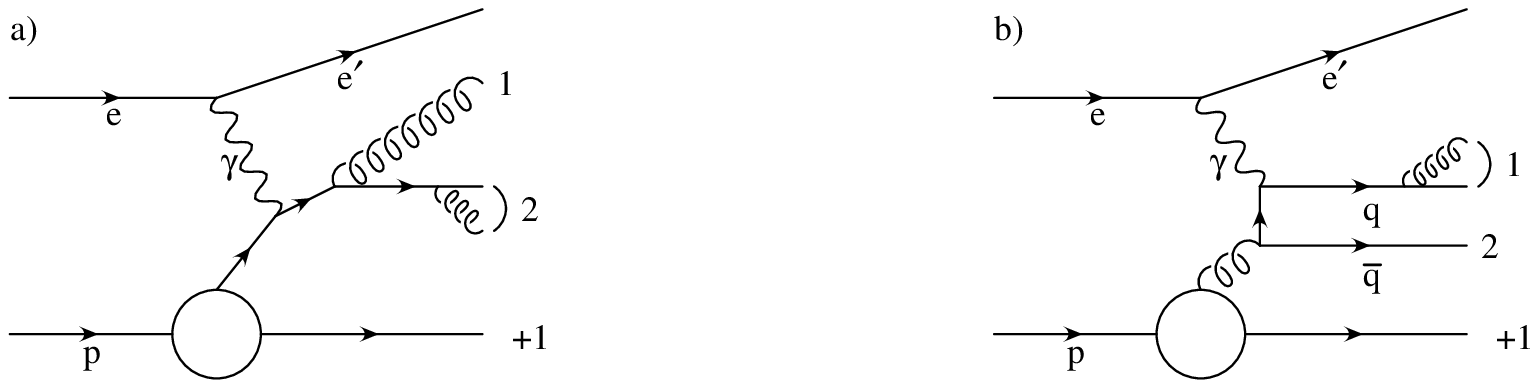,width=\hsize}
\caption[2+1 Ereignis in n"achst-zu-f"uhrender Ordnung]
{{\bf 2+1 Ereignis in n"achst-zu-f"uhrender Ordnung} }
\label{abbsecnlo}
\end{figure}

Die NLO 2+1 Prozesse entstehen analog den 1+1 Ereignissen durch
Abstrahlung weicher Gluonen oder durch Interferenzterme bei den
virtuellen Korrekturen (Abbildung \ref{abbsecnlo}). Dies sind
Beitr"age zur St"orungsrechnung in ${\cal O}(\alpha_s^2).$

\section {Monte-Carlo-Simulationen}
\label{kapsimul}

Um sich ein Bild von den zu erwartenden Detektordaten zu machen, ist es
notwendig Monte-Carlo-Simulationen einzusetzen.

\begin{figure}[tb]
\epsfig{file=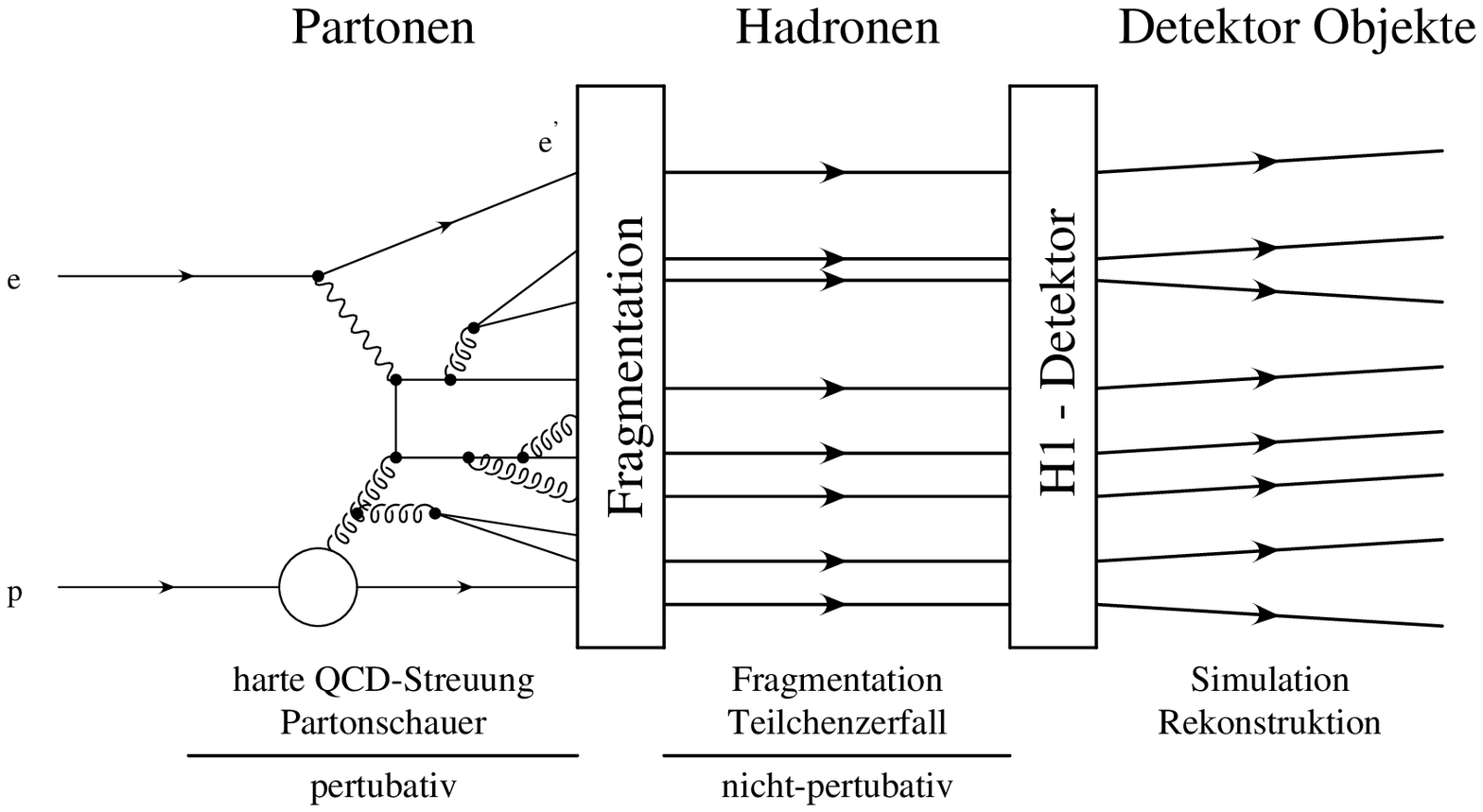,width=\hsize}
\caption["Ubersicht "uber die Vorg"ange bei Monte-Carlo-Generationen] 
{{\bf "Ubersicht "uber die Vorg"ange bei Monte-Carlo-Generationen}} 
\label{abbmonte}
\end{figure}

Die Simulation von Ereignissen geschieht hierbei in mehreren Schritten.
Einen allgemeinen "Uberblick "uber die unterschiedlichen Stufen eines
Monte-Carlo-Generators gibt Abbildung \ref{abbmonte}. Begonnen wird mit
dem harten Streuproze"s, dessen Art und Kinematik den theoretischen
Modellen entnommen wird. Nachdem zus"atzliche Abstrahlungen von Partonen
ber"ucksichtigt worden sind, wird die Hadronisierung durchgef"uhrt,
d.h.\ die farbigen Quarks werden zu farblosen Hadronen zusammengefa"st.
Der Zerfall kurzlebiger Teilchen mu"s dann ebenso beachtet werden wie
Detektoreffekte, die durch unsensitve Detektorregionen oder die
unterschiedlichen Detektoreffizienzen entstehen.

Im folgenden werden die Schritte kurz erl"autert. Genaueres ist der
Literatur "uber Monte-Carlo-Generatoren bzw.\ den Programmanleitungen zu
entnehmen (\cite{Seymour:95}, \cite{Hampel:93}, \cite{Lepto}, \cite{Herwig},
\cite{Ariadne}, \cite{Django}).

\subsection {Beschreibung auf Parton-Niveau}

Wir haben bereits im vorigen Abschnitt die einzelnen, f"ur uns
interessanten Prozesse betrachtet. Anhand dieser Wirkungsquerschnitte
wird dann bei den Simulationen ein Ereignis gew"urfelt.

Leider gibt es keinen Ereignisgenerator, der Ereignisse auf der Basis
von NLO Berechnungen erzeugen kann. Alle vorhandenen Programme benutzen
Matrixelemente in f"uhrender Ordnung (LO) f"ur die Erzeugung von 2+1
Ereignissen.

Um dennoch die Korrekturen ber"ucksichtigen zu k"onnen, die durch
Gluonabstrahlung in h"oheren Ordnungen entstehen, werden als Ausweg
QCD-inspirierte Partonschauer (PS, {\it parton shower}) verwendet. Diese
Partonschauer m"ussen von den Abstrahlungen in h"oheren Ordnungen der QCD
unterschieden werden, da die Schauer unabh"angig vom harten Subproze"s
angewandt werden. Es treten somit keine Interferenzen oder virtuelle
Korrekturen auf. Partonschauer haben sich zusammen mit der im n"achsten
Abschnitt erl"auterten Hadronisierung als n"utzliches Hilfsmittel bei der
Beschreibung der Daten erwiesen.

Wir m"ussen hierbei zwischen zwei F"allen unterscheiden~: Zum einen kann
die Gluonabstrahlung vor der Streuung erfolgen {\it (initial state
parton shower),} zum anderen danach {\it (final state parton shower).}

Bei den final state Partonschauern treten zwei Divergenzen auf, d.h.\ an
diesen Stellen im Phasenraum wird die Abstrahlwahrscheinlichkeit eins
bzw.\ der Wirkungsquerschnitt f"ur diesen Proze"s unendlich. Eine
Divergenz erhalten wir bei Kollinearit"at, d.h.\ wenn der Impuls des
abgestrahlten Gluons in die gleiche Richtung zeigt wie der des
abstrahlenden Partons. Dieses Problem k"onnen wir durch einen
Abschneideparameter beseitigen, der z.B.\ eine minimale invariante Masse
des Gluon Parton Paares verlangt. Der andere Fall tritt bei weicher
Abstrahlung auf, d.i.\ wenn das Gluon eine sehr geringe Energie erh"alt.
Auch dieses Problem l"a"st sich software-technisch umgehen, wobei es
jedoch mehrere Vorgehensweisen gibt, deren Erkl"arung hier zu weit
f"uhren w"urde.

Bei initial state Partonschauer ist die Situation noch wesentlich
komplizierter. Das gr"o"ste zus"atzliche Hindernis ist die Ver"anderung
der Energie des Partons durch die Abstrahlung und damit der Kinematik
des harten Prozesses. Dies k"onnen wir durch R"uckw"artsentwicklung
umgehen, d.h.\ es wird nicht die Energie des Partons nach Abstrahlung
verringert, sondern die vor der Abstrahlung vergr"o"sert.

Es ergeben sich weitere Probleme, wenn wir mehrfache Gluonabstrahlung
ber"ucksichtigen. Die verschiedenen Modelle haben unterschiedliche
Algorithmen, die z.B.\ die Reihenfolge, in der die Gluonen entstehen,
und den Abschneidepunkt, wann dieses Verfahren abzubrechen ist,
betreffen.

Das Ariadne Monte-Carlo-Programm benutzt mit dem {\it Colour Dipole
Model} ein v"ollig anderes Schema, das Partonschauer komplett umgeht.
Die einzelnen Farb--Antifarb--Paare werden als Dipole verstanden, die
analog der Elektrodynamik durch Polarisation neue Dipole hervorrufen.

\subsection {Die Hadronisierung}

Bei der Hadronisierung werden aus den farbigen Partonen farblose
Hadronen gebildet. Es ist bisher noch nicht gelungen diesen Vorgang
theoretisch zu beschreiben, da hier der pertubative Bereich verlassen
wird und die St"orungsrechnung somit nicht mehr funktioniert.

Die beiden meistbenutzten Modelle sind die Cluster Fragmentation und die
String Fragmentation.

Beim Cluster Modell werden jeweils zwei im Phasenraum benachbarte Partonen
entsprechend ihren Farbladungen zu einem sogenannten Cluster kombiniert,
der dann entsprechend seiner Energie, Ladung und der vorhandenen Quarkarten
in Hadronen zerf"allt.

Das String Modell hingegen benutzt ein sehr anschauliches Analogon.
Es zieht zwischen der Farbladung und der
zugeh"origen Antifarbladung eine \grqq Saite\grqq,\ die analog den
Feldlinien in der Elektrodynamik Energie enth"alt. Aufgrund des
color confinements ist die Energie hier jedoch proportional zur L"ange der
Feldlinien. Wird die Saite zu lang und damit die Spannung zu gro"s, so rei"st
diese und es entsteht ein neues Quark Antiquark Paar. Die Saiten, die
am Ende des Verfahrens vorhanden sind, bilden dann die Hadronen.

Die Probleme in beiden Modellen liegen in der Erzeugung von Baryonen, da
diese nicht aus Quark Antiquark Paaren, sondern aus jeweils drei Quarks 
bestehen.  Leider versagen die Modelle hier und nur durch einige Tricks und 
eine gro"se Anzahl freier Parameter ist es im String Modell gelungen, eine 
Beschreibung der existierenden Daten zu erreichen.

Im abschlie"senden Schritt auf Hadronniveau m"ussen die Teilchenbahnen
nun im Magnetfeld berechnet und der Zerfall kurzlebiger Komponenten in
Betracht gezogen werden. Dies ist durch die Kenntnisse der
Zerfallszeiten und der verschiedenen Kan"ale jedoch verglichen mit den
vorhergehenden Schritten einfach und durch die Theorie beschrieben.

\subsection {Beschreibung auf Detektor-Niveau}

Im letzten Schritt werden die Detektoreffekte simuliert. Die danach
vorhandenen Daten entsprechen denen gemessener Ereignisse. Wir haben nur
noch solche Werte, die z.B.\ der Ladungsverteilung in den Kalorimetern
oder den Treffern in den Spurkammern entsprechen.

Diese Stufe ist unabh"angig von den Monte-Carlo-Generatoren und
wird von GE"-ANT \cite{Geant:321} mit speziellen H1 Anpassungen
durchgef"uhrt \cite{H1sim}.

Die Daten lassen sich nun mit den wirklichen Me"sergebnissen des
Detektors vergleichen. Um dies einfacher zu gestalten werden im
abschlie"senden Rekonstruktionsschritt Teilchenidentifikationen,
Energieumrechnungen, Klassifizierungen und andere Berechnungen
durchgef"uhrt \cite{H1rec}.


\chapter{Messungen mit Jet-Raten}
\label {kap_jets}

\section {Jets}
\label{kapjets}

Wir werden nun Messungen durchf"uhren und dann mit den dabei gewonnenen
Detektordaten die Theorie testen. Dazu ist es notwendig eine Verbindung
zwischen dem Detektorniveau und dem Partonniveau zu suchen.

\begin{figure}[tbp]                
\begin{center}
\epsfig{file=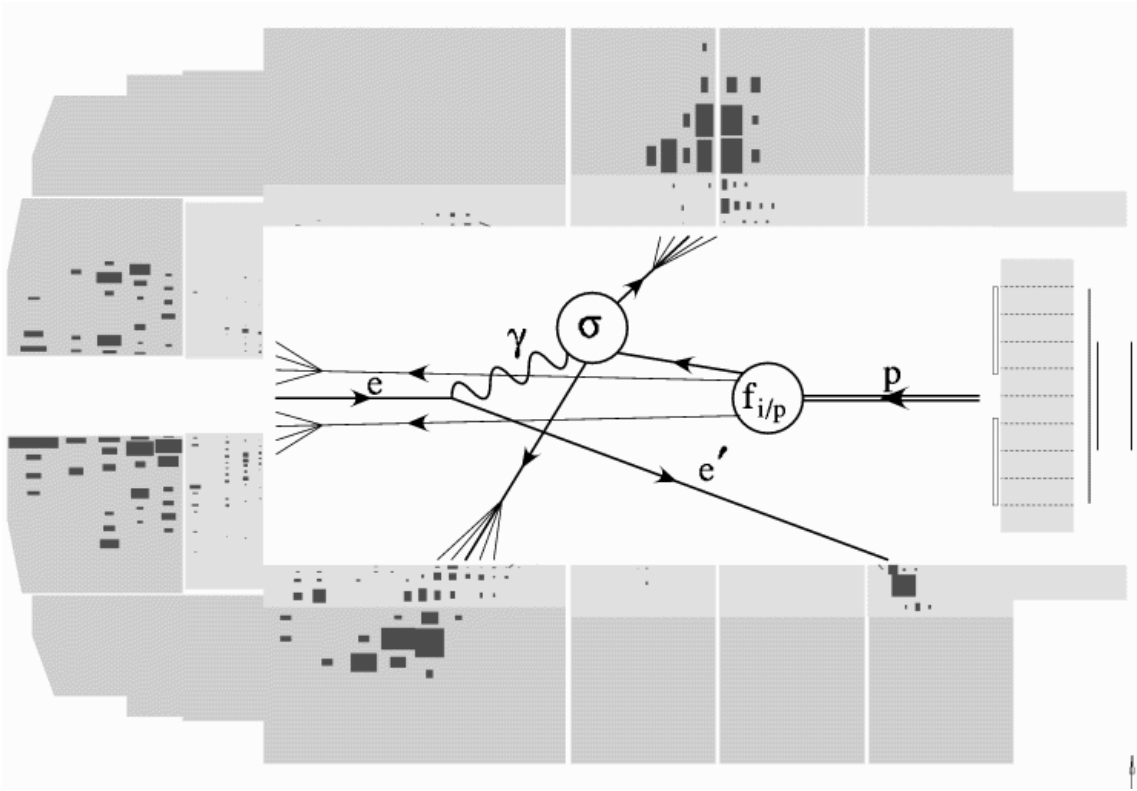,width=\hsize}
\end{center}
\caption[Graphische Darstellung der unterschiedlichen Niveaus
]{{\bf Graphische Darstellung der unterschiedlichen Niveaus~:}{\it
Zu sehen sind im "au"seren Bereich die beiden zentralen Kalorimeter 
(LAr, BEMC) mit der Energieverteilung in den Clustern (dunkle Punkte,
Detektorlevel). Im inneren ist ein Graph eines Prozesses gezeichnet, der 
diese Messung hervorgerufen haben k"onnte. Die Strahlteilchen laufen von 
den Seiten in den Detektor ein und streuen aneinander. Es ergeben sich der 
Protonrest und zwei harte Partonen (Partonlevel, nach Partonschauern). 
Diese bilden dann Hadronen, die durch die Auff"acherung am Ende der
Partonlinien verdeutlicht werden (Hadronlevel).}}
\label{abbniveau}
\end{figure}

Einen "Uberblick "uber die einzelnen Niveaus gibt Abbildung \ref{abbniveau}
(siehe auch Kapitel \ref{kapsimul}).

Wenn wir erst den Zwischenschritt "uber das Hadronniveau machen, so k"onnen
wir erkennen, da"s aus einigen wenigen Partonen eine wesentlich gr"o"sere
Anzahl Hadronen entstanden ist. Ein einfaches Modell, das eine Lokale Parton
Hadron Dualit"at (LPHD, {\it local parton hadron duality}) vorhersagt, ist
dabei erstaunlich erfolgreich. Die Grundidee ist hierbei, da"s die Anzahl
der Hadronen in einem Phasenraumbereich proportional zur Anzahl der Partonen 
ist \cite{Seymour:95}. Die Aussagekraft des Modells ist zwar stark begrenzt, 
aber es zeigt uns, da"s wir mehrere Hadronen zu einem Gebilde, das wir 
{\bf Jet} nennen, zusammenfassen m"ussen.

Auf Detektorniveau tritt das gleiche Ph"anomen auf, da die Hadronen
auf ihrem Weg durch den Detektor an unterschiedlichen Stellen Energie
deponieren und nicht immer sofort gestoppt werden. Aufgrund der begrenzten
Raumaufl"osung der Detektorzellen k"onnen wir jedoch die genaue
Anzahl der Ladungsdepositionen kaum erfassen. Es ist deshalb sinnvoller
Bereiche mit hohem Ladungsinhalt zu sogenannten {\bf Clustern} 
zusammenzufassen.

\begin{figure}[tbp]
\epsfig{file=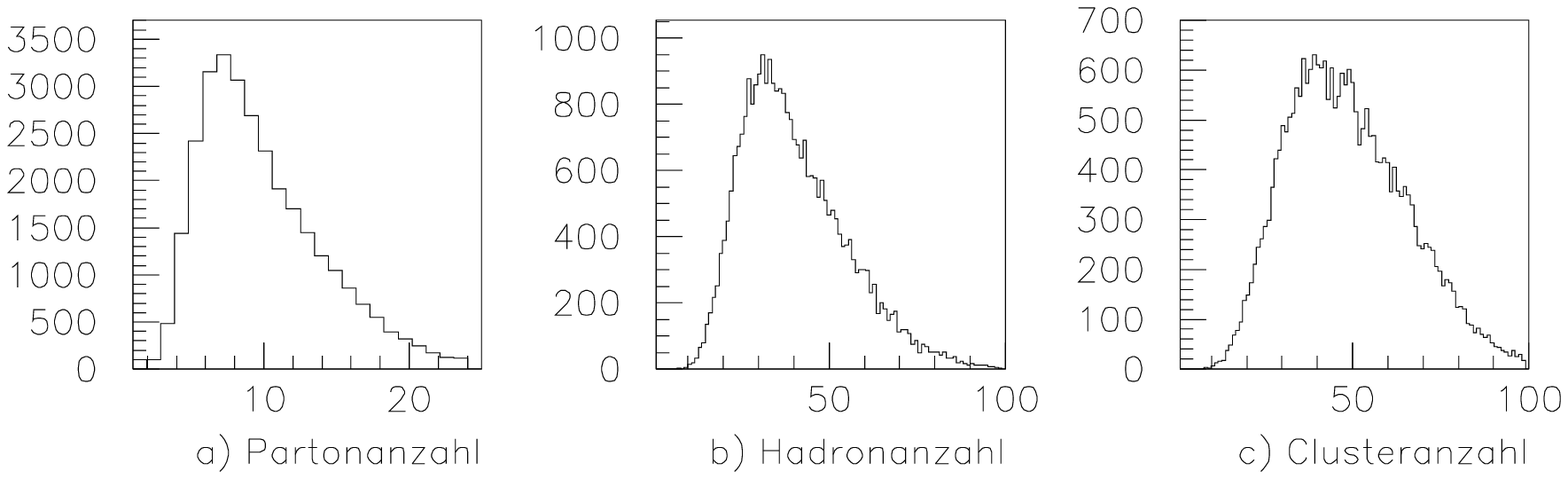,width=\hsize}
\caption[Anzahl Partikel pro Ereignis auf unterschiedlichen Niveaus]{{\bf
Anzahl Partikel pro Ereignis auf den folgenden Niveaus~:}{\it\ a)~Parton,
b) Hadron, c) Cluster. Dargestellt sind alle Ereignisse des
MEHQ Datensatzes (siehe Tabelle \ref{tabfiles}).}}
\label{abbnaphc}
\end{figure}

Einen "Uberblick "uber die Anzahl Partonen, Hadronen und Cluster pro Ereignis
gibt Abbildung \ref{abbnaphc}.

Wir m"ussen somit einen Algorithmus entwerfen, der Hadronen oder Cluster
zu Jets zusammenfa"st. Sind zwei Partonen eng benachbart, so kann es
vorkommen, da"s beide Partonen auf Hadronniveau nur einen Jet ergeben.
Deshalb ist es sinnvoll den gleichen Algorithmus auch auf die Partonen
anzuwenden und diese Fehlinterpretationen im Wirkungsquerschnitt zu
ber"ucksichtigen. Dies haben wir im Kapitel \ref{kapkorrek} jedoch schon
getan. Der Begriff {\it Objekt} beschreibt im folgenden somit Partonen,
Hadronen bzw.\ Cluster in Abh"angigkeit des betrachteten Niveaus. 

Einige Jetalgorithmen werden nun vorgestellt. Zur Berechnung der NLO
Wirkungsquerschnitte existieren nur die beiden Programme
PROJET\cite{Projet} und DIS"-JET\cite{Disjet}. Da in diesen nur der JADE
Algorithmus eingebaut ist, sind wir gezwungen diesen zu verwenden. F"ur
die Zukunft sind verschiedene neue Programme angek"undigt bzw.\ befinden
sich zur Zeit in der Testphase, die mehrere verschiedene Algorithmen
enthalten. Daher beschr"anken wir uns hier nicht auf den JADE
Algorithmus.

\subsection {Cone Algorithmus}

Der Cone Algorithmus ist sicher der intuitivste, da er r"aumlich benachbarte
Partikel zu einem Jet zusammenfa"st. Die Abstandsberechnung geschieht hierbei
"uber den Azimutalwinkel $\varphi$ und die Pseudorapidit"at $\eta,$ die mit
dem Polarwinkel "uber
\begin{eqnarray}
\eta & \equiv & - \ln \left(\tan \left({\theta \over 2}\right)\right)
\end{eqnarray}
zusammenh"angt. Der Algorithmus startet mit einer Anzahl Jet Kandidaten,
die Partikeln mit einer transversalen Energie gr"o"ser als das einstellbare
Minimum entsprechen. Dann werden alle Partikel hinzugef"ugt, die in einem
Radius
\begin{eqnarray}
R &=& \sqrt{\delta\eta^2 + \delta\varphi^2} < R_{\mbox{\scriptsize max}}
\end{eqnarray}
um den Kandidaten mit dem h"ochsten transversalen Impuls liegen. Dies
wird dann der Reihe nach mit allen Kandidaten wiederholt. Am Ende werden
alle Kandidaten, die eine bestimmte transversale Mindestenergie
besitzen, als Jet aufgefa"st.

\subsection {JADE und verwandte Algorithmen}
\label{kapjade}

Bei den Clusteralgorithmen wird nicht der r"aumliche Abstand als Ma"s
verwendet, sondern die invariante Masse zweier Partikel. Meist wird die
invariante Masse jedoch durch eine Skala geteilt, um unabh"angig von
der Energie des stattfindenden Prozesses zu werden. Bei $\ep\emm$
Collidern wird die Schwerpunktenergie $E^2_{CM}=Q^2$ verwendet, bei HERA
verwenden wir die invariante Masse des hadronischen Endzustandes $W^2.$
Eine Begr"undung hierzu kann in \cite{Nisius:94} Kapitel 5.1 gefunden
werden. Es ergibt sich dann als Schnittgr"o"se
\begin{eqnarray}
y_{ij} &\equiv & {m_{ij}^2 \over W^2}.
\end{eqnarray}

Der Algorithmus besteht aus drei Schritten, die bis zur Erf"ullung der
Abbruchbedingung wiederholt werden.

\begin{enumerate}
\item F"ur alle m"oglichen Objektpaare $(i,j)$ wird die invariante Masse
$m_{ij}$ berechnet. Es wird der niedrigste Wert $m_{kl}$ bestimmt.
\item Ist $\displaystyle {m_{kl}\over W^2} > \ycut,$ so wird das Verfahren
abgebrochen und die "ubrig gebliebenen Objekte entsprechen den Jets.
Ansonsten wird mit der Berechnung fortgefahren.
\item Die Teilchen $k$ und $l,$ werden zusammengefa"st und das
Summenteilchen wird an ihrer Stelle in die Liste aufgenommen. Der Algorithmus
f"angt nun wieder mit Schritt 1, der Berechnung der invarianten Massen, von
vorne an.
\end{enumerate}

F"ur die Rekombination zweier Teilchen gibt es keine vorgeschriebene
Regel, da in der Theorie mit masselosen Objekten gerechnet wird. In der
Praxis haben die Partikel jedoch eine nicht verschwindende Masse.
Zus"atzlich ergibt sich bei Addition der Vierervektoren eine Masse, die
nicht der Summe der Massen der Ausgangsteilchen entspricht. Die
unterschiedlichen Ideen haben zu verschiedenen Algorithmen gef"uhrt, die
in Tabelle \ref{tabalgo} aufgelistet sind.

\begin{table}[tb]
\begin{center}
\begin{tabular}{@{\extracolsep{\fill}}|c|c|c|c|}
\hline
Schema & Berechnung von $m_{ij}^2$ & Rekombination & Anmerkungen \\
\hline
JADE & $2 E_i E_j (1- \cos(\theta_{ij}))$ & $\vv{p}_k = \vv{p}_i + \vv{p}_j$
& Massen werden bei $m_{ij}^2$ \\
&&& Berechnung vernachl"assigt \\
\hline
E & $(\vv{p}_i + \vv{p}_j)^2$ & $\vv{p}_k = \vv{p}_i + \vv{p}_j$
& lorentz-invariant \\
\hline
E0 & $(\vv{p}_i + \vv{p}_j)^2$ & $E_k = (E_i + E_j)$ & ${\vec p}$ ist nicht
erhalten \\
&& $\vec p_k = {E_k \over |\vec p_i + \vec p_k|} (\vec p_i + \vec p_j)$ &\\
\hline
P & $(\vv{p}_i + \vv{p}_j)^2$ & $\vec{p}_k = \vec{p}_i + \vec{p}_j$ & $E$
nicht erhalten \\
&& $E_k = | \vec p_k |$ &\\
\hline
\end{tabular}
\end{center}
\caption[Rekombinationsschemata von JADE und verwandten Algorithmen
]{{\bf Rekombinationsschemata von JADE und verwandten Algorithmen}}
\label{tabalgo}
\end{table}

Im Gegensatz zum Cone Algorithmus wird hier keine minimale transversale
Energie f"ur die Jets verlangt. Somit sollte ein Jet, der durch den
Protonrest {\it (remnant)} entsteht, gefunden werden. Dies geschieht
auch auf Parton- und Hadronniveau. Auf dem Detektorniveau hingegen
verl"a"st der gr"o"ste Teil des Remnants den Detektor durch das
Strahlrohr. Aus diesem Grunde werden die Cluster um ein Pseudoteilchen
erweitert, dessen transversaler Impuls zu Null gesetzt wird. 
\begin{eqnarray}
\vec p_{\mbox {\scriptsize Ps.,z}} &=& \vec P_z + \vec k_z -
\sum_{\mbox{\scriptsize Cluster}} \vec p_{i,z} \\
\vec p_x = \vec p_y &=& 0 \nonumber\\
E &=& |\vec p|,\nonumber
\end{eqnarray}
wobei die Summe "uber alle Cluster einschlie"slich derer des gestreuten
Elektron l"auft. Zus"atzlich ist die Kenntnis "uber den gesamten
hadronischen Endzustand, inklusive Protonrestes, zur $W^2$ Berechnung
erforderlich. Weiteres zum Pseudoteilchenansatz ist in
\cite{Graudenz:91} und \cite{Nisius:94} zu finden, der JADE Algorithmus
ist in \cite{JADE:86} n"aher beschrieben.

\subsection {$k_T$ Algorithmus}
\label{kapkt}

Der $k_T$ oder Durham Algorithmus (\cite{Webber:92}, \cite{H1-12:93-336}) 
geh"ort auch in die Klasse der Clusteralgorithmen.       

\begin{figure}[tbp]
\begin{center}
\epsfig{file=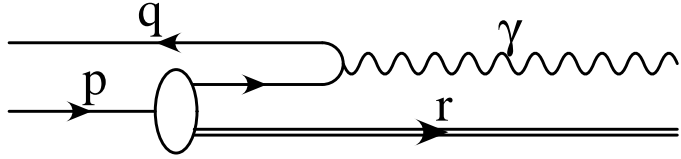,width=0.6\hsize}
\end{center}
\caption[Schematische Darstellung der Kinematik im Breitsystem]{{\bf
Schematische Darstellung der Kinematik im Breitsystem}}
\label{abbbreit}
\end{figure}

Der Algorithmus benutzt die Partikel nicht im Labor- sondern im
Breitsystem, in dem das an der Streuung teilnehmende Parton durch das
Photon wie an einer Wand zur"uckgeworfen wird (siehe auch Abbildung
\ref{abbbreit}).

Wir k"onnen uns das Verfahren zweigeteilt vorstellen. Im ersten Schritt
wird der Protonrest von den "ubrigen sogenannten Macro Jets separiert und
diese Macro Jets werden dann im abschlie"senden Schritt aufgetrennt.

Der erste Schritt kann dabei folgenderma"sen beschrieben werden~:
\begin{enumerate}
\item Zuerst wird eine Skala $E_t$ definiert, die die harte Streuung 
umschreibt.
$$Q^2 \ge E_t^2 \ge \Lambda^2$$
\item Dann wird der Schnittwert f"ur alle Objekte $i$ relativ zum 
einlaufenden Proton 
$$y_{ip} = {2 (1-\cos(\theta_{ip})) \over E_t^2}\;E_i^2$$
und f"ur alle Objektpaare $(i,j)$
$$y_{ij} = {2 (1-\cos(\theta_{ij})) \over E_t^2}\;\min (E_i^2,E_j^2)$$
bestimmt.
\item Es wird das Minimum aller $\left\{y_{ip},y_{ij}\right\}$ berechnet und
in Abh"angigkeit des Ergebnisses einer der folgenden Schritte eingeleitet~:
\begin{itemize}
\item Sind alle Schnittwerte gr"o"ser 1, so wird das Verfahren abgebrochen
und mit dem zweiten Teil des Algorithmus begonnen.
\item Ist ein $y_{ip}$ kleiner als alle anderen Schnittwerte, so wird
das Objekt $i$ in den Beamjet aufgenommen, d.h. f"ur die weitere Berechnung
verworfen.
\item Ist ein $y_{ij}$ kleiner als alle anderen Schnittwerte, so werden
die beiden Objekte $i$ und $j$ zu einem neuen Partikel zusammengefa"st.
\end{itemize}                                                          
\item Die Schritte 2 und 3 werden solange wiederholt, bis die Abbruchbedingung
$\min(\{y_{ip},y_{ij}\})\ge 1$ erf"ullt ist. Die "ubrigbleibenden Partikel
werden als Macro Jets bezeichnet.
\end{enumerate}

Im zweiten Teil des Verfahrens werden nun nur noch die nicht zum Beamjet
geh"orenden Objekte verwendet, d.h.\ die im ersten Teil entstandenen
Macro Jets werden wieder in die Ursprungscluster zerlegt.

\begin{enumerate}
\item Wir definieren nun einen Schnittwert 
$$\ycut = {Q_0^2 \over E_t^2} < 1.$$
\item F"ur alle Okjektpaare $(i,j)$ wird nun analog zum ersten Teil $y_{ij}$
berechnet.
\item Ist das Minimum aller $y_{ij}$ kleiner als unser Aufl"osungsparameter
$\ycut,$ so wird das Verfahren fortgesetzt, ansonsten ist der Jetalgorithmus
hier beendet und die verbliebenen Partikel ergeben die gesuchten Jets.
\item Die beiden Objekte, die das minimale $y_{ij}$ bilden, werden
rekombiniert. Die verschiedenen Schemata sind analog denen des JADE 
Algorithmus definiert. Danach wird das Verfahren ab Schritt 2 wiederholt.
\end{enumerate}


\section {Die Messung der starken Kopplungskonstante}
\label{kapasmess}

Wenden wir uns nun der Messung eines der entscheidenden QCD-Parameter
zu. Die Kopplung der starken Wechselwirkung wird durch die laufende
Kopplungskonstante $\alpha_s$ beschrieben. Wie wir schon in Kapitel
\ref{kapstoer} gesehen haben, k"onnen wir den Wirkungsquerschnitt durch
die St"orungsrechnung in Potenzen der Kopplungskonstante beschreiben.

Tun wir dies nun bis zur n"achst-zur-f"uhrenden Ordnung in Abh"angigkeit
der Jetanzahlen, so erhalten wir~:
\begin{eqnarray}
d\sigma_{1+1}(Q^2, \alpha_s, \ycut) &=& a_{20}(Q^2)
+ \alpha_s(Q^2,\Lambda^2) a_{21}(Q^2,\ycut) \\
d\sigma_{2+1}(Q^2, \alpha_s, \ycut) &=&
\alpha_s(Q^2,\Lambda^2) a_{31}(Q^2,\ycut)
+ \alpha_s(Q^2,\Lambda^2)^2 a_{32}(Q^2,\ycut)
\end{eqnarray}

Hierbei lassen sich die $a_{ij}$ aus den Wirkungsquerschnitten
berechnen. Sie beschreiben Ereignisse mit $i$ Jets, d.h.\ $(i-1)$ harten
Jets und dem Protonrest, und sind Beitr"age in Ordnung $\alpha_s^j.$

Da wir die Wirkungsquerschnitte nicht direkt messen, sondern nur die
Anzahl an $n+1$ Jetereignissen bestimmen k"onnen, ist es sinnvoll mit
der Jetrate zu rechnen. Au"serdem erreichen wir dann Unabh"angigkeit von
Fehlern in der Luminosit"ats- und Effizienzenbestimmung.

Die 2+1 Jetrate ist definiert durch~:
\begin{eqnarray}
R_{2+1} &=& {N_{2+1} \over N_{\mbox{\scriptsize tot}}} = {N_{2+1} \over
\tilde N_{2+1} + N_{1+1}} \\
&=& {\alpha_s(Q^2,\Lambda^2) a_{31}(Q^2,\ycut) + \alpha_s^2(Q^2,\Lambda^2)
a_{32}(Q^2,\ycut)  \over  a_{20}(Q^2) + \alpha_s(Q^2,\Lambda^2)
[a_{21}(Q^2,\ycut)+\tilde a_{31}(Q^2,\ycut)]}
\label{eqnrate}
\end{eqnarray}

$\tilde N_{2+1}$ und $\tilde a_{31}$ tr"agt der M"oglichkeit Rechnung,
da"s manche 2+1 Ereignisse zwar zum totalen Wirkungsquerschnitt
gerechnet werden, aufgrund von Phasenraumschnitten, z.B.\ im Jetwinkel,
jedoch nicht zur 2+1 Jetanzahl.

Wir k"onnen diese Gleichung nun nach $\alpha_s$ aufl"osen und erhalten
dann
\begin{eqnarray}
\alpha_s(Q^2) &=& {\cal F}[a_{ij}(Q^2,\ycut),R_{2+1}(Q^2,\ycut)].
\end{eqnarray}

Damit k"onnen wir durch Messung der Jetrate und Berechnung der
Koeffizienten $a_{ij}$ bei verschiedenen $Q^2,$  die St"arke der starken
Kopplung und deren Laufen mit $Q^2$ bestimmen.

Einsichtig ist, da"s wir die Faktoren $a_{ij}$ nur von der Theorie,
d.h.\ auf Partonniveau, berechnen k"onnen. Daher m"ussen wir daf"ur sorgen,
da"s unsere auf Detektorniveau gemessene Jetrate der auf Partonniveau
entspricht. Ansonsten m"ussen wir einen Korrekturfaktor
\begin{eqnarray}
R &=& {R_{2+1,\mbox{\scriptsize parton}} \over
       R_{2+1,\mbox{\scriptsize detektor}}}
\end{eqnarray}
einf"uhren.

Mit Hilfe eines Fitprogrammes l"a"st sich somit nun unser Me"swert bestimmen.
Dies wurde zum ersten Mal von der H1 Collaboration durchgef"uhrt
(\cite{Nisius:94}) und ist inzwischen auch von anderen Experimenten
erfolgreich angewandt worden \cite{DESY:95-182}.

\section {Die Messung der Gluondichtefunktion}
\label{kapgluon}

Nahezu analog gestaltet sich die Berechnung der Gluondichtefunktion 
$f_{g/p}(x).$

Hier beruht die Messung auf der Aufteilung des Wirkungsquerschnittes in
einen quark-induzierten und einen gluon-induzierten Teil.
\begin{eqnarray}
\sigma &=& \int d\myxi \left[ f_{q/p}(\myxi,\mu^2) \sigma_q(\myxi,\mu^2)
+ f_{g/p}(\myxi,\mu^2) \sigma_g(\myxi,\mu^2) \right] \label{eqngluon}
\end{eqnarray}

Um die Messung direkt auszuf"uhren ist es notwendig, QCD--Compton
Ereignisse (Abbildung \ref{abbseclo}a) von Boson-Gluon-Fusion
Ereignissen (Abbildung \ref{abbseclo}b) zu trennen, da wir so die
Integrale "uber die beiden Terme einzeln messen k"onnen. Um dann die
Dichtefunktion zu bestimmen, m"ussen wir die einzelnen
Wirkungsquerschnitte berechnen und eine Parametrisierung der
Quarkdichtefunktionen einsetzen. Die Dichtefunktionen der Quarks sind
wesentlich genauer bekannt als die der Glu"-onen.

Auch hier kann man durch Teilen der Gleichung \ref{eqngluon} durch die
Effizienzen und die Luminosit"at zu unseren Me"sgr"o"sen, den
Jetanzahlen gelangen. F"ur dieses Verfahren ist also eine verl"a"sliche
Jeterkennung und die richtige Behandlung der
Partonjet--Detektorjet--Korrelation wichtig.

Bei einer NLO Berechnung k"onnen wir die oben beschriebene Aufteilung
nicht mehr vornehmen. Dies ist an Abbildung \ref{abbsecnlo}b zu
erkennen. Wir k"onnen den Proze"s als Boson--Gluon--Proze"s rechnen,
wenn wir, wie in dem Feynman--Graphen angedeutet, das Gluon zum Quark
fassen, oder wir k"onnen das Ereignis zu den QCD--Compton Ereignissen
z"ahlen, wenn das aus dem Paarbildungsproze"s entstandene Antiquark
kollinear zum Protonrest ist und wir es somit zum Remnant packen.

Die einzige M"oglichkeit ist also, das komplette Integral zu fitten. Um
den Fit durchzuf"uhren, m"ussen wir jedoch die Wirkungsquerschnitte f"ur
jeden Fitschritt neu berechnen. Dies ist mit der heutigen
Rechnertechnologie nicht m"oglich.

Ein Ausweg bietet die Methode der Mellin Transformation. Gehen wir durch
eine solche Transformation in den Momentenraum "uber, so l"a"st sich
dort das Moment des Wirkungsquerschnittes als Produkt der Momente der
Partondichtefunktionen $f_{i/p}$ und der Momente der
Einzelwirkungsquerschnitte $\sigma_i$ f"ur den aktuellen kinematischen
Bereich berechnen. F"uhren wir nun den Fit nur in bestimmten
vorgegebenen Phasenraumintervallen durch, so m"ussen die Momente der
Einzelwirkungsquerschnitte an den St"utzstellen nur einmal berechnet
werden. Da dies der zeitaufwendige Teil ist, die Transformation der
Dichtefunktionen und die R"ucktransformation der Momente aber schnell
erfolgen kann, haben wir hiermit eine M"oglichkeit gefunden den Fit in
n"achst-zu-f"uhrender Ordnung durchzuf"uhren \cite{DESY:95-107}.

\chapter{Ergebnisse der Untersuchungen}
\label {kap_unter}

\section {Datens"atze}

Wir wollen nun die Korrelation zwischen den Jets auf Parton- und denen
auf Detektorniveau genauer untersuchen. Dazu ben"otigen wir die schon
angesprochenen Monte-Carlo-Simulationen. In dieser Arbeit verwenden wir
Datens"atze, die mit den Programmen Lepto 6.1 \cite{Lepto}, Herwig
5.8 \cite{Herwig} und Ariadne 4.3 \cite{Ariadne} erzeugt wurden. Meist
wird eine Unterteilung in einen niedrigen und einen hohen $Q^2$ Bereich
vorgenommen, da das Elektron dann ins r"uckw"artige Kalorimeter $(Q^2 <
100\;\GeV^2)$ bzw.\ ins Fl"ussig--Argon--Kalorimeter $(Q^2 >
100\;\GeV^2)$ gestreut wird. Die einzelnen Datens"atze sind in Tabelle
\ref{tabfiles}
aufgef"uhrt.

\begin{table}[tb]
\begin{center}
\begin{minipage}{\hsize}
\begin{tabular}{@{\extracolsep{\fill}}|c|c|c|c|}
\hline                         
Generator & Anzahl Ereignisse & Simulationsversion     & K"urzel       \\
Datei     & Format            & Rekonstruktionsversion & $Q^2$ Bereich \\
\hline
\hline
Lepto 6.1 & 29998 & 3.06/06 & MEPSHQ \\ 
MEPS3621 & DST\footnote{DST bedeutet Data Summary Tape} & 
  6.00/20 & $Q^2 > 90 \GeV^2$ \\
\hline
Lepto 6.1 & 99031 & 3.06/06 & MEPSLQ \\
MEPS3620 & DST & 6.00/20 & $Q^2 > 9 \GeV^2$ \\
\hline
Ariadne & 22715 & 3.05/04 & ARHQ \\
MEAR3652 & V2\footnote{Version 2 der Files, H1 Standardformat} &
  5.02/18 & $Q^2 > 100 \GeV^2$ \\
\hline
Ariadne & 87248 & 3.05/03 & ARLQ \\
MEAR3641 & DST & 6.00/12 & $Q^2 > 6 \GeV^2$ \\
\hline
Herwig 5.8 & 34998 & 3.06/27 & HEHQ \\
MEPS3604 & DST & 6.01/08 & $Q^2 > 90 \GeV^2$ \\
\hline                    
Herwig 5.8 & 63039 & 3.06/14 & HELQ \\
MEPS3602 & DST & 6.01/05 & $Q^2 > 6 \GeV^2$ \\
\hline                    
\hline                    
Generator & Anzahl Ereignisse & Format & K"urzel       \\
\hline                    
Daten & 21651 & DST6 & DAHQ \\
\hline                    
Daten & 162892 & DST6 & DALQ \\
\hline
\end{tabular}
\end{minipage}
\end{center}
\caption[Liste der verwendeten Datens"atze]{{\bf
Liste der verwendeten Datens"atze}}
\label{tabfiles}
\end{table}

Zum Vergleich mit den Daten wurden die vorselektierten\footnote{Die
Vorselektion wurde von der H1-ELAN Gruppe vorgenommen, eine genauere
Selektion auf DIS Ereignisse von der Jet-Gruppe.} Me"swerte aus der
Datennahme des Jahres 1994 verwendet. Um eine Verf"alschung der Daten
gegen"uber den zur Verf"ugung stehenden Monte-Carlos zu verhindern,
wurden nur Positron Runs mit nominellem Vertex verwendet. Es gibt sich
somit eine zur Verf"ugung stehende integrierte Luminosit"at von ${\cal
L} = 2.74\;\pb^{-1}.$ Die Daten sind ebenfalls in die beiden $Q^2$
Bereiche aufgeteilt. Genaueres ist der Tabelle
\ref{tabfiles} zu entnehmen.

\section {Standardschnitte}

Die in diesem Kapitel erl"auterten Schnitte werden ben"otigt, um zum einen
m"oglichst viele Nicht--DIS Ereignisse zu verwerfen und zum anderen
spezielle Eigenarten der Hard- und Software zu ber"ucksichtigen. 

\vfill
\subsection {Kinematische Variablen}
\label{kapkinvar}

Wie in Kapitel \ref{kapberechn} erl"autert, ben"otigen wir f"ur die
Berechnung der Reaktionskinematik Kenntnis "uber nur zwei unabh"angige
Variablen. Wir wollen nun die verschiedenen Berechnungsmethoden genauer
betrachten.

\begin{figure}[p]                
\begin{center}
\epsfig{file=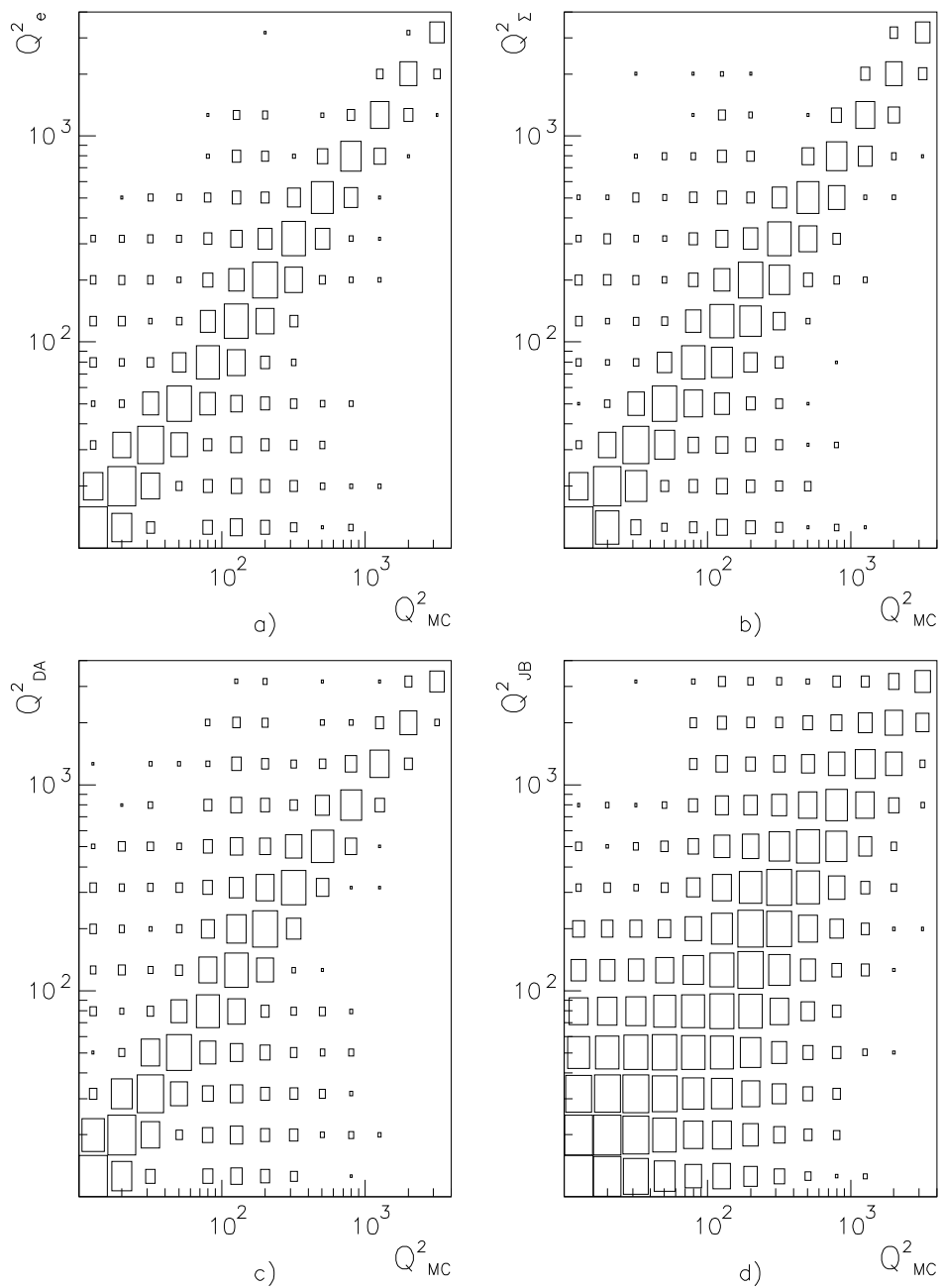,width=\hsize}
\end{center}
\caption[$Q^2$ Rekonstruktion
]{{\bf $Q^2$ Rekonstruktion mit unterschiedlichen Methoden~:
}{\it a)~Elektronmethode, b)~Sigma ($\Sigma$),
c)~Doppelwinkelmethode (DA), d)~Jacquet-Blondel (JB)}}
\label{abbq2qual}
\end{figure}

In Abbildung \ref{abbq2qual} ist die Qualit"at der $Q^2$ Bestimmung f"ur
vier Methoden gezeigt. Auf der Abszisse ist das vom Generator zur
Erzeugung des Ereignisses benutzte negative Quadrat des
Viererimpuls"ubertrages $Q^2_{MC}$ aufgetragen, auf der Ordinate jeweils
der auf Detektorniveau rekonstruierte Wert. Verwendet wurden die
Elektron- $(Q^2_{e}),$ die Doppelwinkel- $(Q^2_{DA}),$ die Sigma-
$(Q^2_{\Sigma})$ \cite{DESY:94-231} und die Jacquet-Blondel-Methode
$(Q^2_{JB}).$ Gezeigt sind die Daten der Lepto-Monte-Carlos (MEPSHQ,
MEPSLQ). Die Gr"o"se der Boxen in diesen Diagrammen gibt jeweils die
Anzahl der Ereignisse in diesem Bereich an. Der Ma"sstab ist hierbei
logarithmisch, so da"s auch einzelne Eintr"age noch eine kleine Box
erzeugen. Insgesamt erkennen wir, da"s die Rekonstruktion mit allen
Methoden gut funktioniert. Die Jacquet-Blondel Methode ist von der
Auswertung des hadronischen Endzustandes besonders abh"angig. Dadurch
haben Detektorverluste und die hadronische Energiekalibration einen
gro"sen Einflu"s. Deshalb sind hier auch die gr"o"sten Abweichungen
erkennbar. Die Sigma Methode ist der Jacquet-Blondel Methode sehr
"ahnlich, durch Verwendung der Gr"o"se $E-p_z$ des totalen Ereignisses
im Gegensatz zur zweifachen Strahlelektronenergie ist die Abh"angigkeit
von der Bestimmung des hadronischen Endzustandes geringer. Au"serdem
ber"ucksichtigt sie damit Photonabstrahlungen am einlaufenden
Elektronast. Diese Abstrahlungen sind in unseren Datens"atzen nicht
vorhanden und die Sigma Methode kann ihren Vorteil nicht voll
ausspielen. Die Elektronmethode, die vollkommen unabh"angig vom
hadronischen Endzustand ist, ist "uber den gesamten kinematischen
Bereich gesehen die geeigneteste.

\begin{figure}[p]                
\begin{center}
\epsfig{file=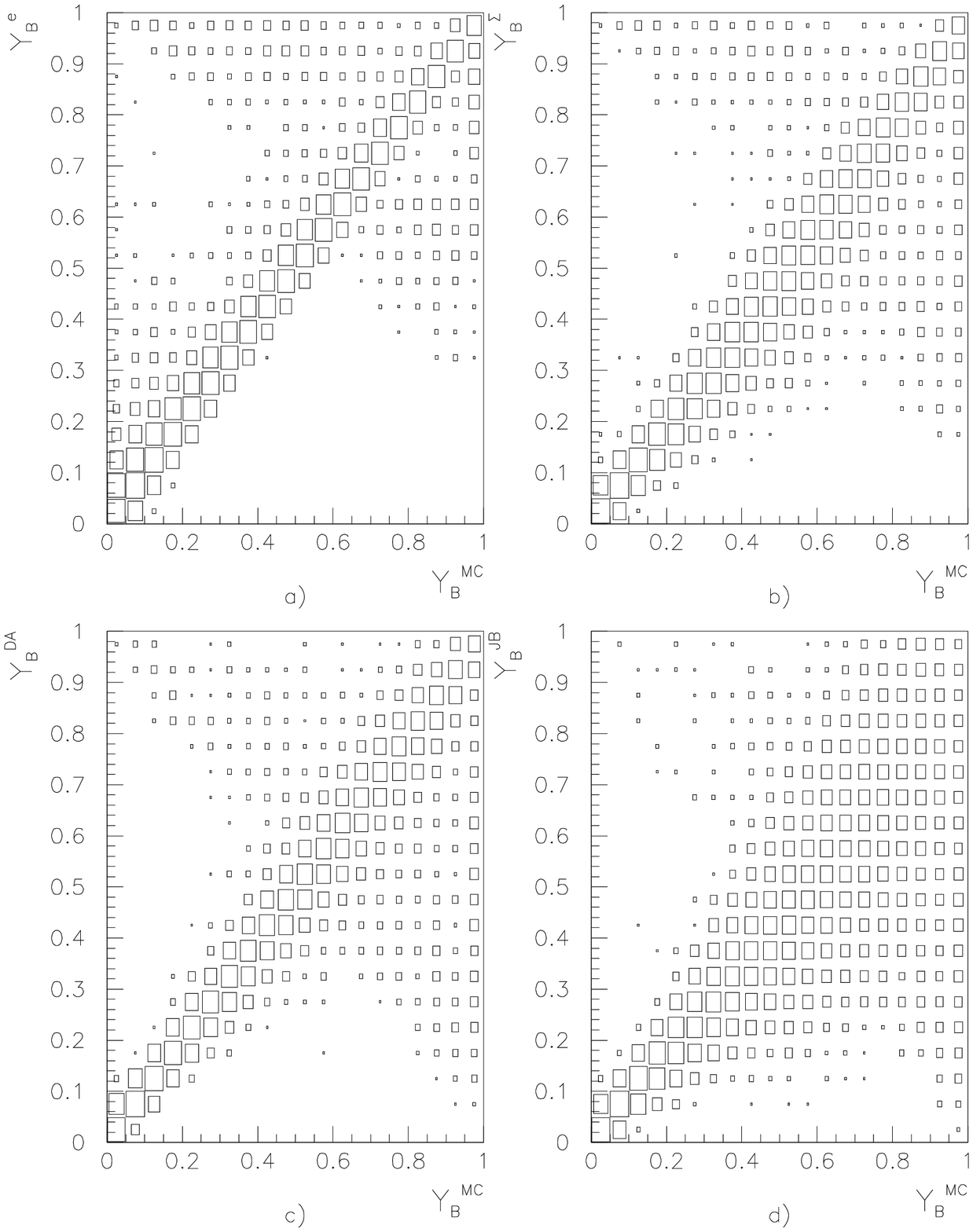,width=\hsize}
\end{center}
\caption[$y_B$ Rekonstruktion bei $Q^2>100\;\GeV$
]{{\bf $y_B$ Rekonstruktion bei $Q^2>100\;\GeV$ mit unterschiedlichen
Methoden~: }{\it a)~Elektronmethode, b)~Sigma ($\Sigma$),
c)~Doppelwinkelmethode (DA), d)~Jacquet-Blondel (JB)}}
\label{abbybhqqual}
\end{figure}

\begin{figure}[p]                
\begin{center}
\epsfig{file=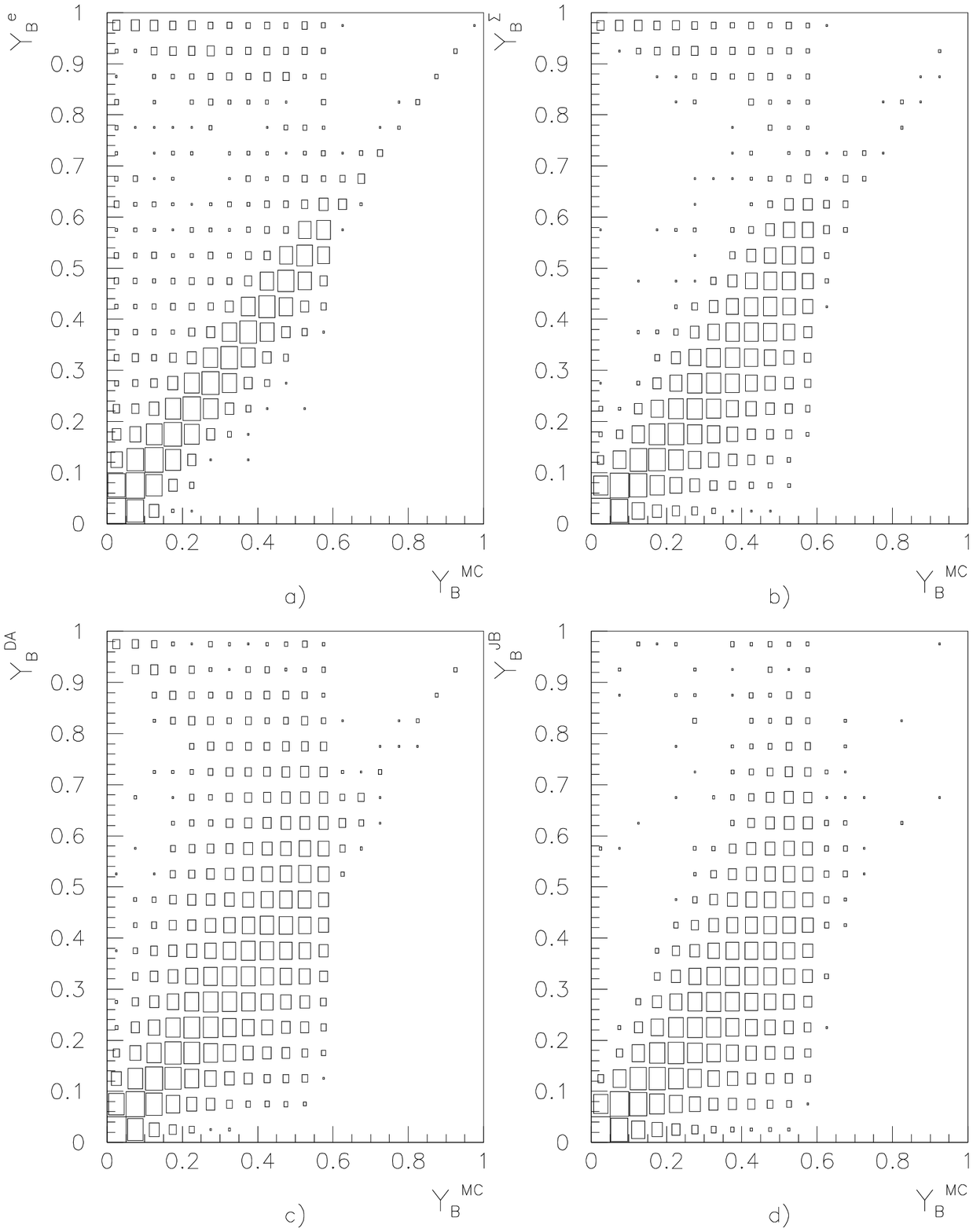,width=\hsize}
\end{center}
\caption[$y_B$ Rekonstruktion bei $Q^2<100\;\GeV$
]{{\bf $y_B$ Rekonstruktion bei $Q^2 < 100\;\GeV$ mit unterschiedlichen 
Methoden~:
}{\it a)~Elektronmethode, b)~Sigma ($\Sigma$),
c)~Doppelwinkelmethode (DA), d)~Jacquet-Blondel (JB)}}
\label{abbyblqqual}
\end{figure}

Die Bestimmung der zweiten, von $Q^2$ unabh"angigen Variable $y_B$ kann
ebenfalls "uber die obengenannten Methoden erfolgen. Die entsprechenden
Verteilungen sind f"ur die hohen $Q^2$ in Abbildung \ref{abbybhqqual},
f"ur die niedrigen in Abbildung \ref{abbyblqqual} gezeigt.

Bei hohen $Q^2$ zeigt die $y_B$ Verteilung ebenfalls eine gute
"Ubereinstimmung bei allen Methoden. Bei $y_B$ gr"o"ser $0.7$ werden
jedoch auch die Abweichungen gr"o"ser. In diesem Bereich ist der
Untergrund durch Photoproduktionsereignisse gro"s, da das gestreute
Elektron den Detektor unbeobachtet durch das Strahlrohr verl"a"st und
niederenergetische Hadronen ein Elektron mit hohem $y_B$ vert"auschen
k"onnen. Deshalb betrachten wir im weiteren nur Ereignisse mit $y_B <
0.7.$ Vergleichen wir die unterschiedlichen Methoden, so ist auch hier
die Elektronmethode die genaueste. Nur in Bereich $y_B < 0.2$ erhalten
wir mit der Doppelwinkelmethode die h"ochste Genauigkeit.
 
F"ur $Q^2$ kleiner $100\;\GeV^2$ f"allt auf, da"s kaum Eintr"age mit $y_B
> 0.6$ vorhanden sind. Dies ist durch den Zusammenhang zwischen $y_B$,
der Energie des gestreuten Elektrons $E_e^{\prime}$ und dessen Winkel
$\theta_e$ durch die Gleichung \ref{eqnyb} erkl"arbar. Der Rand des BEMC
bei $151^{\circ}$ und die minimale Clusterenergie des Elektronkandidaten
von $10\;\GeV$ ergeben dann eine Absch"atzung f"ur den Grenzwert.

\begin{eqnarray}
y_{B,\mbox{\scriptsize max}} &=& \left. {\vv{P} \cdot 
(\vv{k} - \vv{k}^{\prime}) \over \vv{P} \cdot \vv{k}} \right|_{\mbox{max}}= 
\left. {2 E_e - E_e^{\prime} (1 - \cos(\theta_e)) \over 2 E_e}
\right|_{\mbox{max}}\nonumber \\ 
&=& 1 - {E_{e,\mbox{\scriptsize min}}^{\prime} 
(1-\cos(\theta_{e,\mbox{\scriptsize min}})) \over 2 E_e} = 
1-{10 (1-\cos (151^{\circ})) \over 2 \cdot 27.5} = 0.66
\end{eqnarray}

Die geringe Ereignisanzahl mit $y_B^MC > 0.6$ entsteht ausnahmslos durch
Ereignisse mit generiertem $Q^2 > 100\;\GeV,$ die durch das Fehlen eines
oberen $Q^2$ Schnittes im MEPSLQ Monte-Carlo entstehen. Diese Elektronen
fallen deshalb nicht in den Bereich des BEMC und die oben angegebene
Schranke gilt nicht. Eine Aufteilung und Untersuchung der gestreuten
Elektronen in Abh"angigkeit der Detektorkomponente wird in Kapitel
\ref{kapelfin} vorgenommen. Ansonsten erkennen wir auch hier ein Verhalten
analog dem bei hohen $Q^2$ Werten.

Zusammenfassend k"onnen wir also aus der Rekonstruktion der kinematischen
Gr"o"sen schlie"sen~:

\begin{itemize}
\item Die Berechnung von $Q^2$ erfolgt im gesamten Phasenraum am besten
durch die Elektronmethode.
\item Die Berechnung von $y_B$ erfolgt f"ur $y_B^e > 0.2$ mit der 
Elektronmethode, an"-sons"-ten mit der Doppelwinkelmethode.
\item Ein Schnitt auf $y_B^e < 0.7$ ist aufgrund von Problemen bei der
Rekonstruktion und aufgrund des Photoproduktionsuntergrundes sinnvoll.
\end{itemize}

\subsection {Kinematische Schnitte}
\label{kapkinschn}

Ein weiterer Schnitt ergibt sich aus den Strahlungskorrekturen. Diese sind
bei kleinem $y_B$ bzw.\ bei kleinem $W^2$ besonders gro"s.

\begin{figure}[tb]                
\begin{center}
\epsfig{file=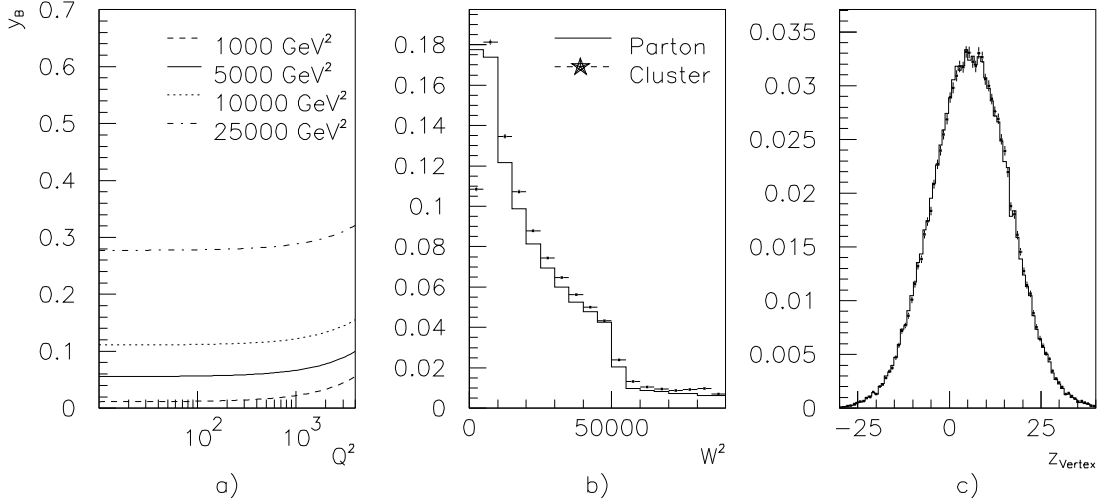,width=\hsize}
\end{center}
\caption[ISO-$W^2$ Linien, $W^2$ und $z_{\mbox{\scriptsize Vertex}}$ Verteilung
]{{\bf a) ISO-$W^2$ Linien, b) $W^2$ Verteilung und c) 
$z_{\mbox{\scriptsize Vertex}}$ Verteilung.}{\ \it Die Verteilungen
b) und c) sind auf gleiche Fl"achen normiert.}}
\label{abbw2}
\end{figure}

Nach Gleichung \ref{eqnw2} gilt $W^2 = y s - Q^2.$ Dies ist in Abbildung
\ref{abbw2}a verdeutlicht. Wir erkennen, da"s ein Schnitt von $W^2 >
5000\;\GeV^2$ ungef"ahr einem $y_B > 0.05$ Schnitt entspricht. Dies gilt
nahezu unabh"angig von $Q^2.$ Die $W^2$ Verteilung f"ur alle MEPS
Ereignisse ist in Abbildung \ref{abbw2}b gezeigt.

Wie wir in Kapitel \ref{kapasmess} gesehen haben, sind unsere Jetraten
unabh"angig von den Effizienzen und der Luminosit"at. Dies gilt
nat"urlich nur dann, wenn wir keinen Effizienzunterschied in den
einzelnen Ereignissen ber"ucksichtigen m"ussen. Es ist sinnvoll den
Bereich des prim"aren Vertex einzuschr"anken, da so bei jedem Ereignis
die Detektorkomponenten unter den gleichen Winkeln erscheinen und wir
damit f"ur alle Ereignisse gleiche Bedingungen erhalten. Die
einlaufenden Teilchenstrahlen sind in der $x-y$-Ebene stark geb"undelt.
Daher ist nur ein Schnitt in Strahlrichtung sinnvoll. Die
$z_{\mb{Vertex}}$ Verteilung f"ur die beiden Lepto-Monte-Carlo sind in
Abbildung \ref{abbw2}c gezeigt. Der nominale Wechselwirkungspunkt liegt
bei $z=4\;\cm.$ Einen Schnitt von $-21\;\cm < z_{\mbox{\scriptsize
Vertex}} < 29\;\cm$ erf"ullen 2\% der Ereignisse nicht.

Wir f"uhren somit folgende kinematischen Schnitte ein~:               
\begin{itemize}
\item $W^2 = y_B^{DA}\;s - Q^2_e> 5000\;\GeV^2$
\item $-21\;\cm < z_{\mbox{\scriptsize Vertex}} < 29\;\cm$
\end{itemize}
 
\subsection {Methoden zur Identifizierung des gestreuten Elektrons}
\label{kapelfin}

Wir haben gesehen, wie wichtig es f"ur die kinematischen Berechnungen
bei Verwendung der Elektronmethode ist, das gestreute Elektron richtig
im Detektor zu identifizieren.

Da das Elektron in zwei verschiedene Detektorkomponenten gestreut werden
kann, benutzen wir f"ur jede eine eigene Routine, sogenannte {\it
Elektronfinder.} Im Fl"ussig--Argon--Ka"-lo"-ri"-me"-ter wird die
Routine QFSELH, im r"uckw"artigen elektromag"-netischen Kalorimeter die
Routine QFSELM verwendet. Beide sind in der offiziellen H1
Funktionsbibliothek H1PHAN\cite{H1phan} enthalten und wurden mit
Ausnahme des $y_B$ Schnittes mit den voreingestellten Steuerkarten
benutzt. Der $y_B$ Schnitt wurde nicht in der QFSELH Routine
durchgef"uhrt, sondern erst nach der Suche der Elektronkandidaten.

Wurden keine Elektronkandidaten gefunden, so wird das Ereignis auf
Detektorniveau verworfen. Ergibt die Suche in nur einer Routine einen
Kandidaten, wird dieser f"ur die weitere Betrachtung verwendet. Finden
jedoch beide Routinen einen Kandidaten, so wird der Kandidat im BEMC als
gestreutes Elektron identifiziert, wenn dessen Energie gr"o"ser als
$10\;\GeV$ ist. Ansonsten wird f"ur die nachfolgenden Schnitte der LAr
Kandidat verwendet. Dadurch verwerfen wir Elektronkandidaten, die durch
Hadronen im BEMC vorget"auscht werden.

Nachdem somit ein Kandidat festgelegt ist, werden in Abh"angigkeit von der
registrierenden Detektorkomponente die in den folgenden Abschnitten 
erl"auterten Bedingungen "uberpr"uft.

Auf Partonniveau werden die Schnitte anhand des $Q^2$ Bereiches ausgew"ahlt,
d.h.\ f"ur $Q^2 < 100\;\GeV$ werden die BEMC Schnitte, ansonsten die
LAr  Schnitte durchgef"uhrt.

\subsubsection{Elektronkandidaten im BEMC} 

F"ur Elektronkandidaten im BEMC werden einige zus"atzliche Schnitte
eingef"uhrt.

Im Bereich kleiner Viererimpuls"ubertr"age ist eine Kontamination des
Datensatzes durch Photoproduktionsereignisse zu erwarten. Diese
Ereignisse verwerfen wir durch ein minimales $Q^2$ von $10\;\GeV^2$ und
eine minimale Energie des Kandidaten von $14\;\GeV.$

Ebenso sollte durch eine Winkel"uberpr"ufung sichergestellt werden, da"s
das gestreute Elektron auch wirklich in den Bereich des BEMC f"allt.

Da im BEMC in der Mitte aufgrund des Strahlrohres ein quadratischer
Bereich unbest"uckt ist, m"ussen wir Kandidaten aus diesen Bereich durch
einen Schnitt auf die $x$ und $y$ Koordinate des
Elektronclusterschwerpunktes ({\it center of gravity}) aus unserem
Datensatz entfernen, damit das gestreute Elektron vollst"andig im
Detektor enthalten ist.

Ein weiterer Schnitt ergibt sich aus der Tatsache, da"s au"ser dem
Kalorimeter noch eine R"uck"-w"ar"-ti"-ge Proportional Kammer ({BPC, \it
backward proportional chamber}) vorhanden ist. Hier fordern wir einen
Treffer im Umkreis von $15\;\cm$ um unseren Kandidaten.
                                   
Zusammengefa"st ergibt sich somit~:

\begin{itemize}
\item $10\;\GeV^2 < Q^2 < 100\;\GeV^2$
\item $160^{\circ} < \theta_e < 172.5^{\circ}$
\item $E_e > 14\;\GeV$
\item $|x_{CG}|> 14\;\cm$ oder $|y_{CG}| > 14\;\cm$
\item $d_{BPC} < 15\;\cm$ 
\end{itemize}
 

\subsubsection{Elektronkandidaten im LAr}

Auch hier machen wir Schnitte in $Q^2$ und im Elektronwinkel $\theta_e$.

Da unter kleineren Winkeln auch vermehrt Hadronen zu finden sind,
m"ussen wir hier die Energieverteilung in den beiden unterschiedlichen
Detektorteilen getrennt betrachten. Im hadronischen Kalorimeter sollte
nur wenig Energie vorhanden sein, da Elektronen im vorgelagerten
elektromagnetischen Teil ihre Energie komplett abgeben. Im
elektromagnetischen Teil selber sollte eine relativ kleine, im Gegensatz
zu hadronischen Jets kompaktere Ladungsverteilung zu erkennen sein.

Zus"atzlich wird noch ein Schnitt in der Gr"o"se $E-p_z$ f"ur das
gesamte Ereignis durchgef"uhrt. Ein Wert in H"ohe der zweifachen
Strahlelektronenergie stellt sicher, da"s der hadronische Endzustand
vollst"andig erkannt wurde. Dies ist nicht der Fall, wenn der
Elektronkandidat durch ein Hadron vorget"auscht wurde und das wirkliche
gestreute Elektron undetektiert bleibt. N"aheres zur Rekonstruktion des
hadronischen Endzustandes wird im n"achsten Abschnitt erl"autert.

Zusammengefa"st ergeben sich f"ur LAr Kalorimeter Elektronen die
Schnitte~:

\begin{itemize}
\item $Q^2 > 100\;\GeV^2$
\item $10^{\circ} < \theta_e < 148^{\circ}$
\item $E_{EMC} < 1.2\;\GeV$ in einem Konus $15\;\cm \le r \le 30\;\cm$ 
im elektromagnetischen Kalorimeter um den Elektronkandidaten.
\item $E_{HAD} < 0.5\;\GeV$ in einem Konus $r \le 30\;\cm$ im hadronischen 
Kalorimeter um den Elektronkandidaten.
\item $38\;\GeV < \sum E-p_z < 70\;\GeV$
\end{itemize}


\subsection {Rekonstruktionsqualit"at des hadronischen Endzustandes}
\label{kaprekon}

Wir m"ussen zun"achst die Partikel bestimmen, die wir in unserem
Jetalgorithmus zu Jets zusammenfassen wollen. Dies sind auf Partonniveau
die vom Generator erzeugten Partonen und auf Detektorniveau die vom
Rekonstruktionsprogramm H1"-REC\cite{H1rec} bestimmten hadronischen
Cluster mit dem f"ur die Algorithmen ben"otigten Pseudoteilchen (siehe
hierzu Kapitel
\ref{kapjade}).

Die richtige Partikelidentifizierung k"onnen wir "uberpr"ufen,
indem wir uns die Energie- und Impulsverteilungen ansehen.

\begin{figure}[tbp]                
\begin{center}
\epsfig{file=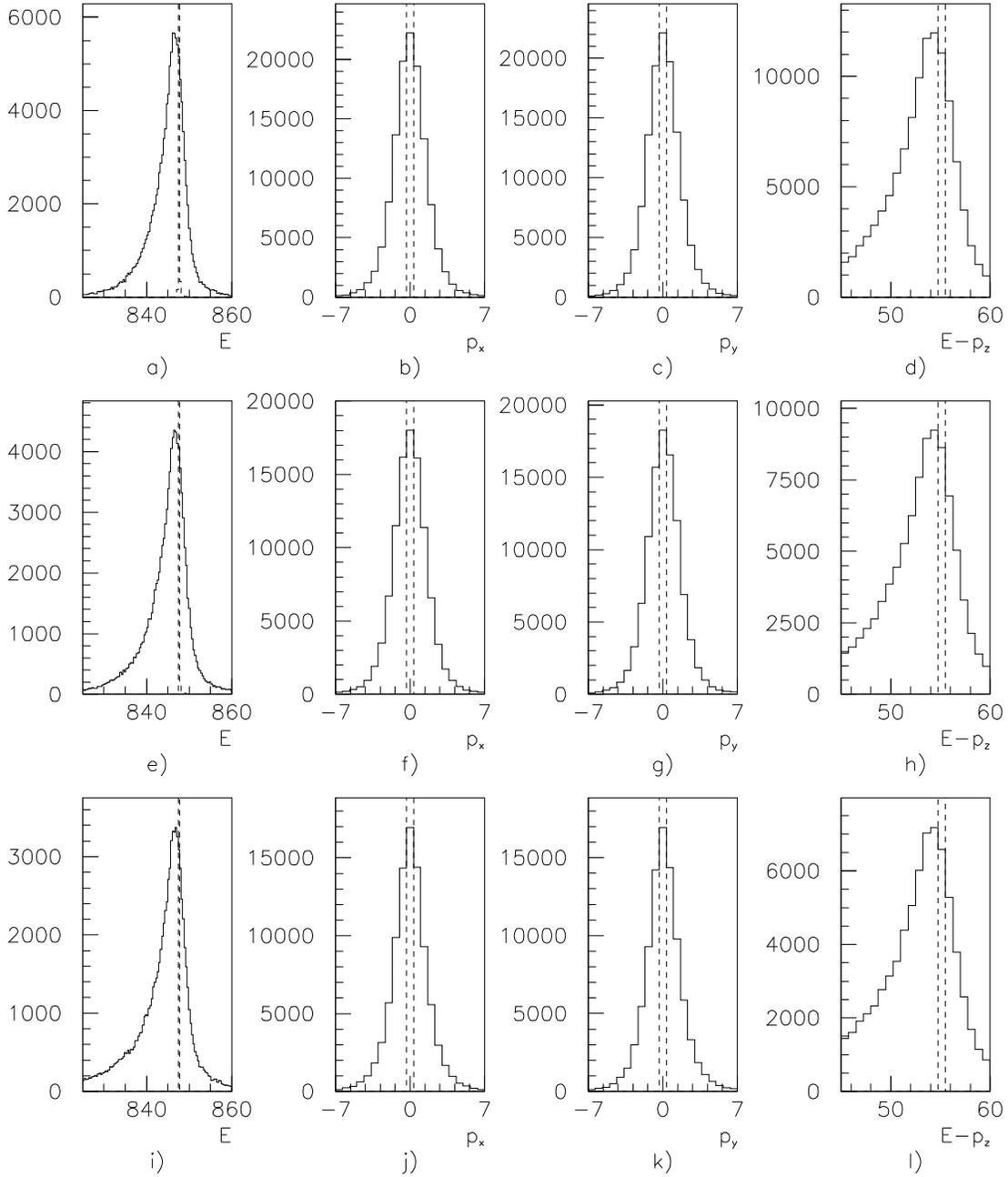,width=\hsize}
\end{center}
\caption[Verteilung der Energie- und Impulssummen]{{\bf
Verteilung der Energie- und Impulssummen.}{\it\ Dargestellt
f"ur Lepto (a-d), Ariadne (e-h) und Herwig (i-l) mit Detektor- (durchgezogen)
und Partonniveau (gestrichelt).}}
\label{abbsum}
\end{figure}

F"ur die Energie erwarten wir die Summe der Energien der Strahlteilchen
\begin{eqnarray}
E_{\mb{tot}} &=& 820.0\;\GeV + 27.5\;\GeV = 847.5\;\GeV.
\end{eqnarray}
Die Verteilungen f"ur die Ereignisse, die mit den drei
Monte-Carlo-Generatoren erzeugt wurden, sehen wir in Abbildung
\ref{abbsum}a,e und i.

Wir erkennen, da"s auf Partonniveau alle Ereignisse bei der erwarteten
Energie auftreten. Bei den Clustern ist die Verteilung wesentlich
breiter, der Peak ist jedoch auch hier --- im Rahmen der Ungenauigkeit
durch die Verluste --- an der erwarteten Stelle. Die Form der
Verteilung ist zum Teil durch die globale Kalibration der Kalorimeter zu
erkl"aren. Untersuchungen haben Unterschiede in der Empfindlichkeit der
Detektoroktanten und der Detektorscheiben von einigen Prozent ergeben
\cite{H1-12:95-466}. Der gr"o"ste Teil wird aber durch Teilchen verursacht, 
die durch totes Material aufgehalten worden sind oder den Detektor
unbeobachtet verlassen haben. Durch die Einf"uhrung des Pseudoteilchens
ist die Aussagekraft der hadronischen Gesamtenergieverteilung jedoch
eingeschr"ankt.

Die Qualit"at der Rekonstruktion des Protonrestes durch das
Pseudoteilchen l"a"st sich an der Verteilung der zur Strahlachse
transversalen Impulse erkennen. Da beide einlaufenden Strahlteilchen
keine transversalen Impulse besitzen, erwarten wir auch f"ur die
Impulssumme "uber alle Endzustandspartikel eine Ausrichtung entlang der
Strahlachse. Die Verteilungen im Partonniveau best"atigen dies
(Abbildung \ref{abbsum}b, c, f, g, j, k). Auf Detektorniveau ist
ebenfalls ein Maximum bei Null zu sehen, jedoch ist auch hier die
Verteilung breiter. Die Werte liegen im Rahmen der Me"sungenauigkeit. Da
das Pseudoteilchen keinen transversalen Impuls erh"alt, ist die
Rekonstruktion des hadronischen Endzustandes somit gelungen und der
Pseudoteilchenansatz gerechtfertigt.

Die Impulssumme in Richtung der Strahlachse stimmt auf Detektorniveau
durch die Definition des Pseudoteilchens mit dem Erwartungswert
"uberein. Dies ist hier deshalb nicht gezeigt.

In Abbildung \ref{abbsum}d, h und l ist die Verteilung der Summe von
$E-p_z$ gezeigt. Aus den obengenannten folgt, da"s wir hier eine Kopie
der Energieverteilung sehen, jetzt jedoch um den Erwartungswert $2 E_e.$

\begin{figure}[tbp]                
\begin{center}
\epsfig{file=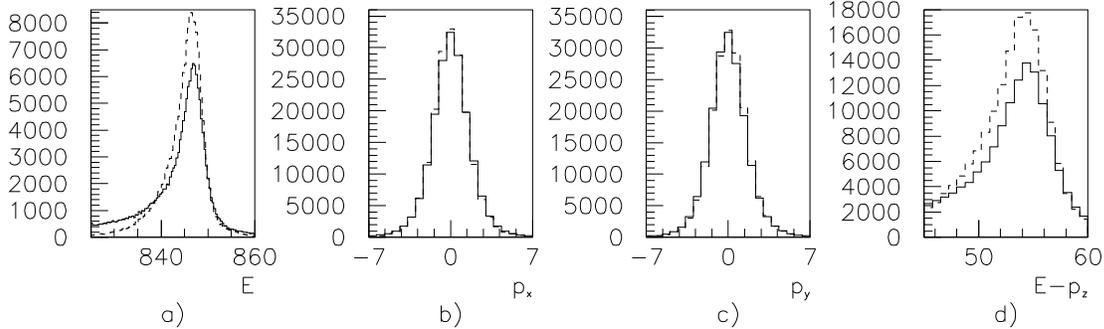,width=\hsize}
\end{center}
\caption[
Verteilung der Energie- und Impulssummen]{{\bf
Verteilung der Energie- und Impulssummen.}{\it\ Dargestellt
f"ur Daten (durchgezogen) und Lepto-Monte-Carlo (gestrichelt), nur 
Detektorniveau.}}
\label{abbsumb}
\end{figure}

In Abbildung \ref{abbsumb} sind die gleichen Verteilungen f"ur die Daten
gezeigt. Die "Ubereinstimmung zwischen den Daten und dem
Monte-Carlo-Detektorniveau ist gut, die Datenverteilung der Energie ist
jedoch etwas breiter.
                                      
Aufgrund der Verteilung ist es sinnvoll Schnitte einzuf"uhren, die
Ereignisse mit zu gro"sen Abweichungen verwerfen. Wir werden deshalb in
dieser Arbeit folgenden Grenzen verwenden~:

\begin{itemize}
\item Die Energiesumme mu"s auf Partonniveau besser als $1\%,$ auf
Detektorniveau besser als $4\%$ mit dem Erwartungswert "ubereinstimmen.
\item Die transversalen Impulse $p_x$ und $p_y$ m"ussen jeweils kleiner
als $1\;\GeV$ auf Parton- und $5\;\GeV$ auf Detektorniveau sein.
\end{itemize}

\subsection {$Q^2$ Einteilung}

F"ur die Messung der starken Kopplungskonstanten $\alpha_s$ in
verschiedenen $Q^2$ Bereichen, m"ussen wir eine Einteilung in sogenannte
{\it Bins} vornehmen.

Bei der ersten Messung der starken Kopplungskonstante $\alpha_s$ mit den Daten
von 1993 wurde eine F"unfteilung in drei niedrige und zwei hohe $Q^2$ Bins
vorgenommen \cite{Nisius:94},\cite{NisQCD:94},\cite{Nisius:95}. Mit der
h"oheren Statistik aus dem Jahr 1994 ist eine Einteilung in acht Bins, f"unf
niedrige und drei hohe, m"oglich \cite{NisBrue:95},\cite{Eisele:95}.

Die Einteilung wurde so gew"ahlt, da"s die Anzahl der 2+1 Ereignisse
in den Bins bei den Daten ungef"ahr gleich gro"s ist,
gleichzeitig der Bereich aber nicht zu weit wird. Diese Wahl ist sinnvoll,
da der statistische Fehler der Jetrate fast ausschlie"slich durch den
Fehler der 2+1 Jetanzahl bestimmt wird. Ich behalte diese Einteilung
deshalb bei. Die genauen Angaben "uber die einzelnen Bins sind in
Tabelle \ref{tabbins} aufgef"uhrt.

\begin{table}[tb]
\begin{center}
\begin{tabular}{@{\extracolsep{\fill}}|c|c|c|r|r|r|r|}
\hline                         
Bin & Bereich & mittlerer Wert & \multicolumn{2}{|c|}{Anzahl Ereignisse} &
\multicolumn{2}{|c|}{Anzahl 2+1 Ereignisse}\\
  & \IN{\GeV^2} & \IN{\GeV^2} & \multicolumn{2}{|c|}{nach Stdschnitten} &
\multicolumn{2}{|c|}{nach Stdschnitten}\\
\hline
1 & 10 --- 14 & 11.9 & \hspace{0.3cm} 14810 & 14951 & 
\hspace{1.2cm}744 & 854 \\
2 & 14 --- 18 & 15.9 & 10788 & 10893 & 582 & 651 \\
3 & 18 --- 25 & 21.4 & 10239 & 10329 & 650 & 674 \\
4 & 25 --- 40 & 32.1 &  9451 & 9578 & 746 & 701 \\
5 & 40 --- 100 & 66.6 & 6604 & 6547 & 628 & 604 \\
\hline
6 & 100 --- 300 & 186.6 & 7917 & 7749 & 1142 & 961  \\
7 & 300 --- 700 & 479.1 & 5047 & 4660 & 1049 & 896  \\
8 & 700 --- 4000 & 2011.7 & 2236 & 1848 & 521 & 396 \\
\hline
\end{tabular}
\end{center}
\caption[Einteilung in $Q^2$ Bins]{{\bf
Einteilung in $Q^2$ Bins.}{\ \it Neben den Grenzen ist das mittlere $Q^2$
im jeweiligen Bereich eingetragen. Die Einteilung wurde so gew"ahlt,
da"s die 2+1 Jetanzahlen in allen Bins ungef"ahr gleich gro"s ist.
Mehr hierzu im Text. Bei den Anzahlen bezieht sich die erste Zahl jeweils
auf das Parton-, die zweite auf das Detektorniveau.}}
\label{tabbins}
\end{table}

Die Anzahl der Ereignisse mit 3 oder mehr harten Jets liegt auf beiden
Niveaus summiert "uber alle Bins unter 3\%\footnote{Weniger als 200
Ereignisse mit mehr als zwei harten Jets pro Niveau.}. Der Einflu"s
dieser Ereignisse kann somit vernachl"assigt werden.


\section {Fehlerbestimmung}

Der statistische Fehler der Jetanzahlen $N$ ist gegeben
durch die Wurzel
\begin{eqnarray}
\Delta N &=& \sqrt{N}.
\end{eqnarray}

Die Jetraten berechnen sich "ahnlich Effizienzen durch
\begin{eqnarray}
R_{n+1} &=& {N_{2+1} \over N_{\mb{tot}}}.
\end{eqnarray}
Wir m"ussen daher einen Effizienzenfehler benutzen, da die beiden Anzahlen
korreliert sind~:
\begin{eqnarray}
\Delta \epsilon &=& \sqrt{{\epsilon (1-\epsilon)\over N}}
\end{eqnarray}
oder in Jetanzahlen geschrieben
\begin{eqnarray}
\Delta R_{2+1} &=& \sqrt{{N_{2+1}\over N_{\mb{tot}}^2}
\left( 1-{N_{2+1}\over N_{\mb{tot}} } \right)} \nonumber\\
&=& \sqrt{ {R_{2+1} \left( 1-R_{2+1}\right) \over N_{\mb{tot}} }}
\end{eqnarray}

Den Fehler f"ur den Korrekturfaktor
\begin{eqnarray}
R &=& {R_{2+1}^{\mb{Parton}} \over R_{2+1}^{\mb{Detektor}} }
\end{eqnarray}
hingegen berechnen wir durch lineare Fehlerfortpflanzung.
Der Fehler f"ur einen Quotienten zweier Gr"o"sen ist dann
\begin{eqnarray}
R &=& {R_1 \over R_2} \\
\Delta R &=& {\Delta R_1 R_2 + \Delta R_2 R_1 \over R_2^2}
\label{eqnfehkor}
\end{eqnarray}

Auch hier haben wir zwar eine Korrelation, da die gleichen Ereignisse
verwendet wurden, um die Raten auf den unterschiedlichen Niveaus zu
bestimmen. Durch Benutzung verschiedener Monte-Carlos mit gleichen
Einstellungen f"ur Parton- und Detektorniveau w"are es m"ogliche diese
Korrelation vollst"andig auszuschalten. Wir k"onnen sie jedoch bei den
hohen Ereigniszahlen vernachl"assigen.

\section {Motivation zur Untersuchung der Jet--Parton--Korrelation}
\markright{\ref{kapmotiv} MOTIVATION}
\label{kapmotiv}

Betrachten wir die Korrelation der $(n+1)$ Jet Ereignisanzahlen zwischen
Parton- und Detektorniveau, so ergeben sich Matrizen der Art
\begin{center}
\begin{tabular}{@{\extracolsep{\fill}}c|c||c|c}
\multicolumn{2}{c||}{Bin} & \multicolumn{2}{c}{Parton} \\
\multicolumn{2}{c||}{1} & 1+1 & 2+1 \\
\hline
\hline
Clus- & 1+1 & $a_{11}$ & $a_{21}$ \\
\cline{2-4}
ter & 2+1 & $a_{12}$ & $a_{22}$ \\
\end{tabular}
\end{center}

Die Hauptdiagonalelemente $a_{11}$ und $a_{22}$ beschreiben die auf
beiden Niveaus gleich klassifizierten Ereignisse und die
Nebendiagonalelemente $a_{21}$ und $a_{12}$ die Migrationen, d.i.\ der
Fall, in dem durch Fehl-Erkennung ein Ereignis beim "Ubergang von
Parton- auf Detektorniveau die Jetklasse wechselt. Diese Elemente
sollten m"oglichst niedrige Werte annehmen.

Benutzen wir nur die Standardschnitte, so ergeben sich die in Tabelle
\ref{tabstdmigmeps} verzeichneten Migrationen f"ur die MEPS Datens"atze
in den verschiedenen $Q^2$ Bins. Es wurde hier der JADE Algorithmus
verwendet\footnote{Da der JADE Algorithmus mit dem Aufl"osungsparameter
$\ycut = 0.02$ unsere Standardwahl ist, wird im folgenden der
Algorithmus nur bei Verwendung eines anderen Algorithmuses explizit
angegeben.}.

\begin{table}[tbp]
\begin{center}
\begin{tabular}{@{\extracolsep{\fill}}c|c||c|c}
\multicolumn{2}{c||}{Bin} & \multicolumn{2}{c}{Parton} \\
\multicolumn{2}{c||}{1} & 1+1 & 2+1 \\
\hline
\hline
Clus- & 1+1 & 12458 & 326 \\
\cline{2-4}
ter & 2+1 & 434 & 343 \\
\end{tabular}
\hfill
\begin{tabular}{@{\extracolsep{\fill}}c|c||c|c}
\multicolumn{2}{c||}{Bin} & \multicolumn{2}{c}{Parton} \\
\multicolumn{2}{c||}{2} & 1+1 & 2+1 \\
\hline
\hline
Clus- & 1+1 & 9419 & 247 \\
\cline{2-4}
ter & 2+1 & 336 & 267 \\
\end{tabular}
\hfill
\begin{tabular}{@{\extracolsep{\fill}}c|c||c|c}
\multicolumn{2}{c||}{Bin} & \multicolumn{2}{c}{Parton} \\
\multicolumn{2}{c||}{3} & 1+1 & 2+1 \\
\hline
\hline
Clus- & 1+1 & 8970 & 291 \\
\cline{2-4}
ter & 2+1 & 336 & 287 \\
\end{tabular}
\end{center}
\vspace {0.25cm}
\begin{center}
\begin{tabular}{@{\extracolsep{\fill}}c|c||c|c}
\multicolumn{2}{c||}{Bin} & \multicolumn{2}{c}{Parton} \\
\multicolumn{2}{c||}{4} & 1+1 & 2+1 \\
\hline
\hline
Clus- & 1+1 & 8108 & 345 \\
\cline{2-4}
ter & 2+1 & 305 & 345 \\
\end{tabular}
\hfill
\begin{tabular}{@{\extracolsep{\fill}}c|c||c|c}
\multicolumn{2}{c||}{Bin} & \multicolumn{2}{c}{Parton} \\
\multicolumn{2}{c||}{5} & 1+1 & 2+1 \\
\hline
\hline
Clus- & 1+1 & 5464 & 256 \\
\cline{2-4}
ter & 2+1 & 259 & 299 \\
\end{tabular}
\hfill
\begin{tabular}{@{\extracolsep{\fill}}c|c||c|c}
\multicolumn{2}{c||}{Bin} & \multicolumn{2}{c}{Parton} \\
\multicolumn{2}{c||}{6} & 1+1 & 2+1 \\
\hline
\hline
Clus- & 1+1 & 5934 & 448 \\
\cline{2-4}
ter & 2+1 & 274 & 586 \\
\end{tabular}
\end{center}
\vspace {0.25cm}
\begin{center}
\begin{tabular}{@{\extracolsep{\fill}}c|c||c|c}
\multicolumn{2}{c||}{Bin} & \multicolumn{2}{c}{Parton} \\
\multicolumn{2}{c||}{7} & 1+1 & 2+1 \\
\hline
\hline
Clus- & 1+1 & 3324 & 341 \\
\cline{2-4}
ter & 2+1 & 234 & 605 \\
\end{tabular}
\hfill
\begin{tabular}{@{\extracolsep{\fill}}c|c||c|c}
\multicolumn{2}{c||}{Bin} & \multicolumn{2}{c}{Parton} \\
\multicolumn{2}{c||}{8} & 1+1 & 2+1 \\
\hline
\hline
Clus- & 1+1 & 1287 & 154 \\
\cline{2-4}
ter & 2+1 & 99 & 286 \\
\end{tabular}
\hfill
\hphantom{
\begin{tabular}{@{\extracolsep{\fill}}c|c||c|c}
\multicolumn{2}{c||}{} & \multicolumn{2}{c}{Parton} \\
\multicolumn{2}{c||}{} & 1+1 & 2+1 \\
\hline
\hline
Clus- & 1+1 & & \\
\cline{2-4}
ter & 2+1 & & \\
\end{tabular}
}
\end{center}
\vspace {0.25cm}
\begin{center}
\begin{tabular}{@{\extracolsep{\fill}}|c||c|c|c|c|c|c|c|c|}
\hline
Bin & 1 & 2 & 3 & 4 & 5 & 6 & 7 & 8 \\
\hline
Rate Parton&  4.9 &  5.0 &  5.8 &  7.6 &  8.8 & 14.3 & 21.0 & 24.1\\
in \% & $\pm 0.2$&$\pm 0.3$&$\pm 0.3$&$\pm 0.3$&$\pm 0.4$&$\pm 0.5$&$\pm 0.7$&$\pm 1.1$\\
\hline
Rate Detektor&  5.7 &  5.9 &  6.3 &  7.1 &  8.9 & 11.9 & 18.6 & 21.1\\
in \%&$\pm 0.2$&$\pm 0.3$&$\pm 0.3$&$\pm 0.3$&$\pm 0.4$&$\pm 0.4$&$\pm 0.6$&$\pm 1.0$\\
\hline
Korrektur- & 0.86 & 0.85 & 0.93 & 1.06 & 0.99 & 1.20 & 1.13 & 1.14\\
faktor &$\pm0.07$&$\pm0.08$&$\pm0.08$&$\pm0.08$&$\pm0.09$&$\pm0.08$&$\pm0.07$&$\pm0.10$\\
\hline
Reinheit in \%& 44.1 & 44.3 & 46.1 & 53.1 & 53.6 & 68.1 & 72.1 & 74.3\\
\hline
\end{tabular}
\end{center}
\caption[Migrationen und Korrekturfaktoren bei Lepto, Standardschnitte]{{\bf
Migrationen und Korrekturfaktoren bei Lepto mit Standardschnitten,
in $Q^2$ Bins}}
\label{tabstdmigmeps}
\end{table}

Aus den Migrationen, beschrieben durch die Elemente 
$a_{\mbox{\scriptsize parton,cluster}}$  lassen sich die
Raten einfach berechnen
\begin{eqnarray}
R_{2+1}^{\mbox{\scriptsize Detektor}} 
&=& {a_{22}+a_{12}\over \sum_{i,j} a_{ij}} \\
R_{2+1}^{\mbox{\scriptsize Parton}} 
&=& {a_{22}+a_{21}\over \sum_{i,j} a_{ij}}
\end{eqnarray}

Neben den Raten ist in der Tabelle \ref{tabstdmigmeps} auch der
Korrekturfaktor 
\begin{eqnarray}
R &=& {a_{22} + a_{21} \over a_{22} + a_{12}}
\end{eqnarray}

berechnet worden. Hier ist jedoch anzumerken, da"s bei
den Migrationstabellen nur solche Ereignisse verwendet werden konnten,
die auf beiden Niveaus die Standardschnitte erf"ullen. Im sechsten Bin
haben wir z.B.\ insgesamt 7242 Ereignisse, obwohl in diesem Bin 7749
Ereignisse die Standardschnitte auf Detektorniveau erf"ullen und sogar
7917 die auf Partonniveau (siehe Tabelle \ref{tabbins}).

F"ur unsere sp"ateren Betrachtungen benutzen wir die Raten, die
sich aus den kompletten S"atzen ergeben. Diese k"onnen sich somit
leicht von den hier angegebenen unterscheiden.

Bei den Raten f"allt uns auf, da"s diese mit anwachsendem $Q^2$ ebenfalls
steigen. Wir erwarten f"ur die starke Kopplungskonstante $\alpha_s$
jedoch genau einen umgekehrten Verlauf. Die $Q^2$ Abh"angigkeit der
Vorfaktoren in Gleichung \ref{eqnrate} ist jedoch st"arker, so da"s der
Abfall der starken Kopplungskonstanten mit $Q^2$ "uberkompensiert wird.

Die Korrekturfaktoren liegen bei eins und zeigen keine gro"sen Probleme
bei der Jet--Parton--Korrelation an.

Sehen wir uns die Migrationen genauer an, so f"allt jedoch auf, da"s die
2+1 Jetklassen auf den beiden Niveaus nicht gut "ubereinstimmen. In den
unteren Bins liegt die Reinheit der 2+1 Jetereignisse auf Detektorniveau
\begin{eqnarray} 
\rho &=& {N_{2+1}^{\mb{Parton und Detektor}} \over N_{2+1}^{\mb{Detektor}}}
\nonumber\\
&=& {a_{22} \over a_{22} + a_{12}}
\end{eqnarray}
bei nur 45 bis 50\%. In den hohen Bins ist sie mit 65 bis 75\% zwar
besser, aber sicherlich nicht zufriedenstellend.

Zur Bek"ampfung von Partonschauern wurde schon bei der Analyse der im
Jahr 1993 genommenen Daten ein Jetwinkelschnitt eingef"uhrt
\begin{eqnarray} 
10^{\circ} < &\theta_{\mb{jet}}& < 145^{\circ}.
\end{eqnarray}

Eine genauere Untersuchung dieses Schnittes wird in Kapitel
\ref{kaputhejet} durchgef"uhrt. In Tabelle \ref{tabt10migmeps} sind
die Migrationen f"ur diesen Schnitt notiert.

\begin{table}[tbp]
\begin{center}
\begin{tabular}{@{\extracolsep{\fill}}c|c||c|c}
\multicolumn{2}{c||}{Bin} & \multicolumn{2}{c}{Parton} \\
\multicolumn{2}{c||}{1} & 1+1 & 2+1 \\
\hline
\hline
Clus- & 1+1 & 6830 & 219 \\
\cline{2-4}
ter & 2+1 & 136 & 147 \\
\end{tabular}
\hfill
\begin{tabular}{@{\extracolsep{\fill}}c|c||c|c}
\multicolumn{2}{c||}{Bin} & \multicolumn{2}{c}{Parton} \\
\multicolumn{2}{c||}{2} & 1+1 & 2+1 \\
\hline
\hline
Clus- & 1+1 & 6500 & 135 \\
\cline{2-4}
ter & 2+1 & 71 & 97 \\
\end{tabular}
\hfill
\begin{tabular}{@{\extracolsep{\fill}}c|c||c|c}
\multicolumn{2}{c||}{Bin} & \multicolumn{2}{c}{Parton} \\
\multicolumn{2}{c||}{3} & 1+1 & 2+1 \\
\hline
\hline
Clus- & 1+1 & 6712 & 182 \\
\cline{2-4}
ter & 2+1 & 95 & 109 \\
\end{tabular}
\end{center}
\vspace {0.25cm}
\begin{center}
\begin{tabular}{@{\extracolsep{\fill}}c|c||c|c}
\multicolumn{2}{c||}{Bin} & \multicolumn{2}{c}{Parton} \\
\multicolumn{2}{c||}{4} & 1+1 & 2+1 \\
\hline
\hline
Clus- & 1+1 & 6707 & 183 \\
\cline{2-4}
ter & 2+1 & 91 & 122 \\
\end{tabular}
\hfill
\begin{tabular}{@{\extracolsep{\fill}}c|c||c|c}
\multicolumn{2}{c||}{Bin} & \multicolumn{2}{c}{Parton} \\
\multicolumn{2}{c||}{5} & 1+1 & 2+1 \\
\hline
\hline
Clus- & 1+1 & 5333 & 138 \\
\cline{2-4}
ter & 2+1 & 67 & 114 \\
\end{tabular}
\hfill
\begin{tabular}{@{\extracolsep{\fill}}c|c||c|c}
\multicolumn{2}{c||}{Bin} & \multicolumn{2}{c}{Parton} \\
\multicolumn{2}{c||}{6} & 1+1 & 2+1 \\
\hline
\hline
Clus- & 1+1 & 5197 & 323 \\
\cline{2-4}
ter & 2+1 & 141 & 274 \\
\end{tabular}
\end{center}
\vspace {0.25cm}
\begin{center}
\begin{tabular}{@{\extracolsep{\fill}}c|c||c|c}
\multicolumn{2}{c||}{Bin} & \multicolumn{2}{c}{Parton} \\
\multicolumn{2}{c||}{7} & 1+1 & 2+1 \\
\hline
\hline
Clus- & 1+1 & 3804 & 246 \\
\cline{2-4}
ter & 2+1 & 109 & 249 \\
\end{tabular}
\hfill
\begin{tabular}{@{\extracolsep{\fill}}c|c||c|c}
\multicolumn{2}{c||}{Bin} & \multicolumn{2}{c}{Parton} \\
\multicolumn{2}{c||}{8} & 1+1 & 2+1 \\
\hline
\hline
Clus- & 1+1 & 1506 & 105 \\
\cline{2-4}
ter & 2+1 & 48 & 141 \\
\end{tabular}
\hfill
\hphantom{
\begin{tabular}{@{\extracolsep{\fill}}c|c||c|c}
\multicolumn{2}{c||}{} & \multicolumn{2}{c}{Parton} \\
\multicolumn{2}{c||}{} & 1+1 & 2+1 \\
\hline
\hline
Clus- & 1+1 & & \\
\cline{2-4}
ter & 2+1 & & \\
\end{tabular}
}
\end{center}
\vspace {0.25cm}
\begin{center}
\begin{tabular}{@{\extracolsep{\fill}}|c||c|c|c|c|c|c|c|c|}
\hline
Bin & 1 & 2 & 3 & 4 & 5 & 6 & 7 & 8 \\
\hline
Rate Parton&  5.0 &  3.4 &  4.1 &  4.3 &  4.5 & 10.1 & 11.2 & 13.7\\
in \% & $\pm 0.3$&$\pm 0.3$&$\pm 0.3$&$\pm 0.3$&$\pm 0.3$&$\pm 0.4$&$\pm 0.5$&$\pm 0.9$\\
\hline
Rate Detektor&  3.9 &  2.5 &  2.9 &  3.0 &  3.2 &  7.0 &  8.1 & 10.5\\
in \%&$\pm 0.3$&$\pm 0.2$&$\pm 0.2$&$\pm 0.3$&$\pm 0.3$&$\pm 0.4$&$\pm 0.5$&$\pm 0.8$\\
\hline
Korrektur- & 1.29 & 1.38 & 1.43 & 1.43 & 1.39 & 1.44 & 1.38 & 1.30\\
faktor &$\pm0.15$&$\pm0.20$&$\pm0.19$&$\pm0.18$&$\pm0.19$&$\pm0.13$&$\pm0.13$&$\pm0.17$\\
\hline
Reinheit in \%& 51.9 & 57.7 & 53.4 & 57.3 & 63.0 & 66.0 & 69.6 & 74.6\\
\hline
\end{tabular}
\end{center}
\caption[Migrationen und Korrekturfaktoren bei Lepto, $\tjet$]{{\bf
Migrationen und Korrekturfaktoren bei Lepto mit Standardschnitten und
zus"atzlichem $10^{\circ} < \tjet < 145^{\circ}$ Schnitt,
in $Q^2$ Bins}}
\label{tabt10migmeps}
\end{table}

Wir erkennen hier die gleiche Systematik. F"ur die $Q^2$ Abh"angigkeit
gilt auch hier das obengesagte. Der Korrekturfaktor liegt nun jedoch mit
durchschnittlich 1.4 weiter von 1 entfernt als ohne $\theta_{\mb{jet}}$
Schnitt. Um dies zu erkl"aren, vergleichen wir die Migrationen n"aher.

Betrachten wir zuerst die niedrigen $Q^2$ Bins. Durch den zus"atzlichen
Schnitt hat sich die Anzahl der Ereignisse, die in beiden Niveaus zwei
harte Jets haben (2+1 Jet,$\;a_{22}$) auf ca.\ 40\% reduziert. Die
Ereignisse $a_{12},$ die auf Partonniveau einen harten Jet haben, auf
Detektorniveau jedoch zwei, sind zu "uber 70\% unterdr"uckt worden. Im
Gegensatz dazu werden nur etwa 42\% der Ereignisse, die auf
Detektorniveau einen harten Jet, auf Partonniveau jedoch zwei haben,
verworfen. Die Konsequenzen abgesehen vom vergr"o"serten Korrekturfaktor
sind kleinere Raten auf beiden Niveaus und bei kleinen $Q^2$ eine leicht
verbesserte Reinheit unserer Detektorprobe.

Dieser Schnitt deutet auf Unterschiede in der Parton- und der
Detektor-Jet"-win"-kel"-ver"-tei"-lung hin. Hierauf werden wir in
Kapitel \ref{kaputhejet} zur"uckkommen.

Bei den hohen $Q^2$ Bins ist die Unterdr"uckung um 55\% f"ur $a_{22},$
51\% f"ur $a_{12}$ und 29\% f"ur $a_{21}$ nicht ganz so extrem, aber
trotzdem sehr unterschiedlich. Insbesondere ist der Effekt auf die auf
beiden Niveaus gleich erkannten Ereignisse gro"s, was die h"oheren
relativen Fehler bewirkt.

Um die St"arke der Migrationen zu charakterisieren gen"ugt also nicht
nur der Korrekturfaktor, sondern die Reinheit unserer Jetklassifizierung
ist ebenfalls wichtig.

Unser Problem bei der Jetanalyse ist deutlich geworden. Wir suchen einen
oder mehrere Schnitte, die die Migrationen, d.h.\ die Elemente $a_{12}$
und $a_{21}$ der Migrationsmatrix unterdr"ucken. Au"serdem sollte die
Reduktion m"oglichst auf beide Elemente gleich stark ausfallen, da sonst
der Korrekturfaktor zu gro"s\footnote{Unter einem gro"sen
Korrekturfaktor verstehen wir einen Wert, der eine gro"se Abweichung vom
idealen Wert 1 bedeutet. Es ist dabei nebens"achlich, ob die Abweichung
nach unten, d.i.\ ein kleiner Korrekturfaktorwert, oder nach oben, d.i.\
ein gro"ser Korrekturfaktorwert, ausf"allt.} wird. W"unschenswert w"are
zus"atzlich, da"s die auf beiden Niveaus gleicherkannten Ereignisse,
insbesondere die, die beide Male als 2+1 Jetereignisse klassifiziert
wurden, nicht zu sehr reduziert werden. Dies w"urde sonst den
statistischen Fehler stark erh"ohen.

\section {Verschiedene Monte-Carlo-Generatoren}

Bevor wir nun weitere Schnitte testen, ist es sinnvoll die
Abh"angigkeit vom Monte-Carlo-Generator und
dem Jetalgorithmus zu betrachten. Ersteres untersuchen wir in diesem
Abschnitt, letzteres im n"achsten.

Wir haben neben unserem Lepto-MEPS-Monte-Carlo noch Simulationen von
Ereignissen, die mit Ariadne bzw.\ mit Herwig erzeugt wurden. Die
genauen Angaben zu diesen Datens"atzen sind in Tabelle \ref{tabfiles}
aufgef"uhrt.

Die Migrationen sind in Tabelle \ref{tabstdmigar} f"ur Ariadne und in
Tabelle \ref{tabstdmighe} f"ur Herwig mit der gleichen $Q^2$ Einteilung
(siehe Tabelle \ref{tabbins}) unter Verwendung der Standardschnitte
aufgef"uhrt.

\begin{table}[tbp]
\begin{center}
\begin{tabular}{@{\extracolsep{\fill}}c|c||c|c}
\multicolumn{2}{c||}{Bin} & \multicolumn{2}{c}{Parton} \\
\multicolumn{2}{c||}{1} & 1+1 & 2+1 \\
\hline
\hline
Clus- & 1+1 & 5728 & 454 \\
\cline{2-4}
ter & 2+1 & 305 & 296 \\
\end{tabular}
\hfill
\begin{tabular}{@{\extracolsep{\fill}}c|c||c|c}
\multicolumn{2}{c||}{Bin} & \multicolumn{2}{c}{Parton} \\
\multicolumn{2}{c||}{2} & 1+1 & 2+1 \\
\hline
\hline
Clus- & 1+1 & 4497 & 400 \\
\cline{2-4}
ter & 2+1 & 243 & 262 \\
\end{tabular}
\hfill
\begin{tabular}{@{\extracolsep{\fill}}c|c||c|c}
\multicolumn{2}{c||}{Bin} & \multicolumn{2}{c}{Parton} \\
\multicolumn{2}{c||}{3} & 1+1 & 2+1 \\
\hline
\hline
Clus- & 1+1 & 4363 & 391 \\
\cline{2-4}
ter & 2+1 & 227 & 277 \\
\end{tabular}
\end{center}
\vspace {0.25cm}
\begin{center}
\begin{tabular}{@{\extracolsep{\fill}}c|c||c|c}
\multicolumn{2}{c||}{Bin} & \multicolumn{2}{c}{Parton} \\
\multicolumn{2}{c||}{4} & 1+1 & 2+1 \\
\hline
\hline
Clus- & 1+1 & 3997 & 417 \\
\cline{2-4}
ter & 2+1 & 203 & 282 \\
\end{tabular}
\hfill
\begin{tabular}{@{\extracolsep{\fill}}c|c||c|c}
\multicolumn{2}{c||}{Bin} & \multicolumn{2}{c}{Parton} \\
\multicolumn{2}{c||}{5} & 1+1 & 2+1 \\
\hline
\hline
Clus- & 1+1 & 2552 & 301 \\
\cline{2-4}
ter & 2+1 & 154 & 229 \\
\end{tabular}
\hfill
\begin{tabular}{@{\extracolsep{\fill}}c|c||c|c}
\multicolumn{2}{c||}{Bin} & \multicolumn{2}{c}{Parton} \\
\multicolumn{2}{c||}{6} & 1+1 & 2+1 \\
\hline
\hline
Clus- & 1+1 & 2875 & 258 \\
\cline{2-4}
ter & 2+1 & 137 & 263 \\
\end{tabular}
\end{center}
\vspace {0.25cm}
\begin{center}
\begin{tabular}{@{\extracolsep{\fill}}c|c||c|c}
\multicolumn{2}{c||}{Bin} & \multicolumn{2}{c}{Parton} \\
\multicolumn{2}{c||}{7} & 1+1 & 2+1 \\
\hline
\hline
Clus- & 1+1 & 1663 & 169 \\
\cline{2-4}
ter & 2+1 & 138 & 214 \\
\end{tabular}
\hfill
\begin{tabular}{@{\extracolsep{\fill}}c|c||c|c}
\multicolumn{2}{c||}{Bin} & \multicolumn{2}{c}{Parton} \\
\multicolumn{2}{c||}{8} & 1+1 & 2+1 \\
\hline
\hline
Clus- & 1+1 & 673 & 77 \\
\cline{2-4}
ter & 2+1 & 75 & 86 \\
\end{tabular}
\hfill
\hphantom{
\begin{tabular}{@{\extracolsep{\fill}}c|c||c|c}
\multicolumn{2}{c||}{} & \multicolumn{2}{c}{Parton} \\
\multicolumn{2}{c||}{} & 1+1 & 2+1 \\
\hline
\hline
Clus- & 1+1 & & \\
\cline{2-4}
ter & 2+1 & & \\
\end{tabular}
}
\end{center}
\vspace {0.25cm}
\begin{center}
\begin{tabular}{@{\extracolsep{\fill}}|c||c|c|c|c|c|c|c|c|}
\hline
Bin & 1 & 2 & 3 & 4 & 5 & 6 & 7 & 8 \\
\hline
Rate Parton& 11.1 & 12.3 & 12.7 & 14.3 & 16.4 & 14.7 & 17.5 & 17.9\\
in \% & $\pm 0.4$&$\pm 0.5$&$\pm 0.5$&$\pm 0.5$&$\pm 0.7$&$\pm 0.6$&$\pm 0.9$&$\pm 1.3$\\
\hline
Rate Detektor&  8.9 &  9.3 &  9.6 &  9.9 & 11.8 & 11.3 & 16.1 & 17.7\\
in \%&$\pm 0.4$&$\pm 0.4$&$\pm 0.5$&$\pm 0.5$&$\pm 0.6$&$\pm 0.6$&$\pm 0.8$&$\pm 1.3$\\
\hline
Korrektur- & 1.25 & 1.31 & 1.33 & 1.44 & 1.38 & 1.30 & 1.09 & 1.01\\
faktor &$\pm0.10$&$\pm0.11$&$\pm0.11$&$\pm0.12$&$\pm0.13$&$\pm0.12$&$\pm0.11$&$\pm0.15$\\
\hline
Reinheit in \%& 49.3 & 51.9 & 55.0 & 58.1 & 59.8 & 65.8 & 60.8 & 53.4\\
\hline
\end{tabular}
\end{center}
\caption[Migrationen und Korrekturfaktoren bei Ariadne, Standardschnitte]{{\bf
Migrationen und Korrekturfaktoren bei Ariadne mit Standardschnitten,
in $Q^2$ Bins}}
\label{tabstdmigar}
\end{table}

\begin{table}[tbp]
\begin{center}
\begin{tabular}{@{\extracolsep{\fill}}c|c||c|c}
\multicolumn{2}{c||}{Bin} & \multicolumn{2}{c}{Parton} \\
\multicolumn{2}{c||}{1} & 1+1 & 2+1 \\
\hline
\hline
Clus- & 1+1 & 4589 & 56 \\
\cline{2-4}
ter & 2+1 & 129 & 43 \\
\end{tabular}
\hfill
\begin{tabular}{@{\extracolsep{\fill}}c|c||c|c}
\multicolumn{2}{c||}{Bin} & \multicolumn{2}{c}{Parton} \\
\multicolumn{2}{c||}{2} & 1+1 & 2+1 \\
\hline
\hline
Clus- & 1+1 & 3426 & 94 \\
\cline{2-4}
ter & 2+1 & 116 & 74 \\
\end{tabular}
\hfill
\begin{tabular}{@{\extracolsep{\fill}}c|c||c|c}
\multicolumn{2}{c||}{Bin} & \multicolumn{2}{c}{Parton} \\
\multicolumn{2}{c||}{3} & 1+1 & 2+1 \\
\hline
\hline
Clus- & 1+1 & 3167 & 108 \\
\cline{2-4}
ter & 2+1 & 122 & 78 \\
\end{tabular}
\end{center}
\vspace {0.25cm}
\begin{center}
\begin{tabular}{@{\extracolsep{\fill}}c|c||c|c}
\multicolumn{2}{c||}{Bin} & \multicolumn{2}{c}{Parton} \\
\multicolumn{2}{c||}{4} & 1+1 & 2+1 \\
\hline
\hline
Clus- & 1+1 & 2950 & 90 \\
\cline{2-4}
ter & 2+1 & 135 & 109 \\
\end{tabular}
\hfill
\begin{tabular}{@{\extracolsep{\fill}}c|c||c|c}
\multicolumn{2}{c||}{Bin} & \multicolumn{2}{c}{Parton} \\
\multicolumn{2}{c||}{5} & 1+1 & 2+1 \\
\hline
\hline
Clus- & 1+1 & 1879 & 91 \\
\cline{2-4}
ter & 2+1 & 94 & 93 \\
\end{tabular}
\hfill
\begin{tabular}{@{\extracolsep{\fill}}c|c||c|c}
\multicolumn{2}{c||}{Bin} & \multicolumn{2}{c}{Parton} \\
\multicolumn{2}{c||}{6} & 1+1 & 2+1 \\
\hline
\hline
Clus- & 1+1 & 6223 & 304 \\
\cline{2-4}
ter & 2+1 & 217 & 491 \\
\end{tabular}
\end{center}
\vspace {0.25cm}
\begin{center}
\begin{tabular}{@{\extracolsep{\fill}}c|c||c|c}
\multicolumn{2}{c||}{Bin} & \multicolumn{2}{c}{Parton} \\
\multicolumn{2}{c||}{7} & 1+1 & 2+1 \\
\hline
\hline
Clus- & 1+1 & 3570 & 266 \\
\cline{2-4}
ter & 2+1 & 189 & 522 \\
\end{tabular}
\hfill
\begin{tabular}{@{\extracolsep{\fill}}c|c||c|c}
\multicolumn{2}{c||}{Bin} & \multicolumn{2}{c}{Parton} \\
\multicolumn{2}{c||}{8} & 1+1 & 2+1 \\
\hline
\hline
Clus- & 1+1 & 1237 & 116 \\
\cline{2-4}
ter & 2+1 & 86 & 288 \\
\end{tabular}
\hfill
\hphantom{
\begin{tabular}{@{\extracolsep{\fill}}c|c||c|c}
\multicolumn{2}{c||}{} & \multicolumn{2}{c}{Parton} \\
\multicolumn{2}{c||}{} & 1+1 & 2+1 \\
\hline
\hline
Clus- & 1+1 & & \\
\cline{2-4}
ter & 2+1 & & \\
\end{tabular}
}
\end{center}
\vspace {0.25cm}
\begin{center}
\begin{tabular}{@{\extracolsep{\fill}}|c||c|c|c|c|c|c|c|c|}
\hline
Bin & 1 & 2 & 3 & 4 & 5 & 6 & 7 & 8 \\
\hline
Rate Parton&  2.1 &  4.5 &  5.4 &  6.1 &  8.5 & 11.0 & 17.3 & 23.4\\
in \% & $\pm 0.3$&$\pm 0.4$&$\pm 0.4$&$\pm 0.5$&$\pm 0.7$&$\pm 0.4$&$\pm 0.6$&$\pm 1.1$\\
\hline
Rate Detektor&  3.6 &  5.1 &  5.8 &  7.4 &  8.7 &  9.8 & 15.6 & 21.7\\
in \%&$\pm 0.3$&$\pm 0.4$&$\pm 0.4$&$\pm 0.5$&$\pm 0.7$&$\pm 0.4$&$\pm 0.6$&$\pm 1.0$\\
\hline
Korrektur- & 0.58 & 0.88 & 0.93 & 0.82 & 0.98 & 1.12 & 1.11 & 1.08\\
faktor &$\pm0.11$&$\pm0.13$&$\pm0.14$&$\pm0.11$&$\pm0.14$&$\pm0.08$&$\pm0.08$&$\pm0.10$\\
\hline
Reinheit in \%& 25.0 & 38.9 & 39.0 & 44.7 & 49.7 & 69.4 & 73.4 & 77.0\\
\hline
\end{tabular}
\end{center}
\caption[Migrationen und Korrekturfaktoren bei Herwig, Standardschnitte]{{\bf
Migrationen und Korrekturfaktoren bei Herwig mit Standardschnitten,
in $Q^2$ Bins}}
\label{tabstdmighe}
\end{table}

Vergleichen wir die drei Tabellen, stellen wir fest, da"s die $Q^2$
Abh"angigkeit der Raten bei den drei Monte-Carlos unterschiedlich ist.
Bei dem mit Herwig generierten Datensatz ist sie am gr"o"sten und beim
Ariadne Datensatz am geringsten.

Damit sind nat"urlich auch die Raten unterschiedlich. Die Abweichungen sind
bei den niedrigen $Q^2$ Bins jedoch etwas gr"o"ser. 

Bei den Korrekturfaktoren wird die $Q^2$ Abh"angigkeit und die
Abh"angigkeit vom benutzten Monte-Carlo besonders deutlich. Im ersten
Bin reichen die Werte von 0.58 bis 1.25, im zweiten immerhin noch von
0.85 bis 1.31. In den letzten beiden Bins liegen die Werte zwischen
1.01 und 1.14 und stimmen im Rahmen der Fehler "uberein. Dies ist
bereits ein deutlicher Hinweis darauf, da"s die Probleme bei den
Jet--Parton--Korrelationen in den niedrigen $Q^2$ Bins wesentlich
gr"o"ser sind.

Zusammenfassend k"onnen wir also sagen, da"s unsere Berechnungen
abh"angig vom benutzen Monte-Carlo-Generator sind. Wir werden deshalb in
Kapitel \ref{kapschnitte} verschiedene Verteilungen untersuchen, um
Probleme bei unseren simulierten Ereignissen im Vergleich mit den Daten
zu finden und die entsprechenden Phasenraumbereiche dann ausschlie"sen
zu k"onnen.

\section {Verschiedene Jetalgorithmen}

Wie bereits im letzten Kapitel angesprochen, m"ussen wir auch die
Abh"angigkeit der Werte vom verwendeten Jetalgorithmus n"aher
betrachten. Dazu wurden auf den beiden Niveaus der Lepto Simulationen
der JADE und der Cone Algorithmus angewandt. Aus Kapitel \ref{kapjade}
wissen wir, da"s bei der Berechnung der invarianten Masse im JADE Schema
die Teilchenmassen vernachl"assigt werden. Dies bedeutet jedoch, da"s
der Jetalgorithmus insgesamt nicht lorentz-invariant ist. Um diesen
Effekt n"aher zu betrachten, wurde der JADE Algorithmus mit JADE Schema
in drei unterschiedlichen Bezugssystemen angewendet. Dazu wurde neben
dem Laborsystem, das unsere Standard"-einstellung ist, das hadronische
Schwerpunktsystem und das Breitsystem (siehe Kapitel \ref{kapkt})
verwendet. Aufgrund des hohen Energieunterschiedes von Strahlproton und
-elektron k"onnen jedoch Lorentz Transformationen mit extremen Werten in
den Rotationsmatrizen auftreten \footnote{In diesen F"allen ist $\beta =
{v\over c} \approx 1.$}. Die Lorentz-Transformationen sind deshalb nicht
unproblematisch und spezielle Tests mu"sten eingebaut werden, um
nichtphysikalische Werte zu verhindern. Die einzelnen Partikel sind vor
der Berechnung der Jets in das entsprechende System transformiert worden
und nach der Anwendung des JADE Algorithmus wurden die Jets ins
Laborsystem zur"ucktransformiert, um einen Vergleich der drei Versionen
zu erm"oglichen.

Als weitere Variante wurde der JADE Algorithmus mit einem ver"anderten
Schnittwert von $\ycut = 0.03$ benutzt, um die Gr"o"senordnung der
Abh"angigkeit vom Auf"-l"o"-sungs"-pa"-ra"-me"-ter zu bestimmen (siehe
Kapitel \ref{kapasmess}).

Zum Schlu"s wurde auch der JADE-Algorithmus unter Verwendung der
Massenberechnung und Rekombination durch das E Schema eingesetzt, das
den Algorithmus lorentz-invariant macht. Fr"uhere Untersuchungen der
JADE Varianten haben jedoch schon gezeigt, da"s gerade dieser
Algorithmus gro"se Korrekturfaktoren bewirkt (\cite{Nisius:94} Kapitel
10, Abbildung 10.19 und 10.20).

Eine Aufstellung der verwendeten Algorithmen ist in Tabelle
\ref{tabualgo} wiedergegeben.

\begin{table}[tb]
\begin{center}
\begin{minipage}{12.5cm}
\begin{tabular}{@{\extracolsep{\fill}}|c|c|c|c|}
\hline                                          
Abk"urzung & Algorithmus & Lorentzsystem & Schnittparameter \\
\hline                                                        
JL & JADE & Labor & 0.02 \\
JC & JADE & hadr.Schwerpunkt & 0.02 \\
JB & JADE & Breit & 0.02 \\
3L & JADE & Labor & 0.03 \\
CL & Cone\footnote{Mit den Parametern $E_{t,\mb{min}} = 1.5\;\GeV$ und
$E_{t,\mb{sum,min}} = 7.0\;\GeV$} & Labor & 0.7 \\
EL & E Schema & Labor & 0.02 \\
\hline
\end{tabular}
\end{minipage}
\end{center}
\caption[Aufstellung der verwendeten Jetalgorithmen
]{{\bf Aufstellung der in den Abbildungen \ref{abbmejetstd4} bis
\ref{abbmejetstd8} benutzten Jetalgorithmen.}{\it\ Genaueres zu den
einzelnen Algorithmen ist in Kapitel \ref{kapjets} zu finden.}}
\label{tabualgo}
\end{table}

\begin{figure}[tbp]                
\begin{center}
\epsfig{file=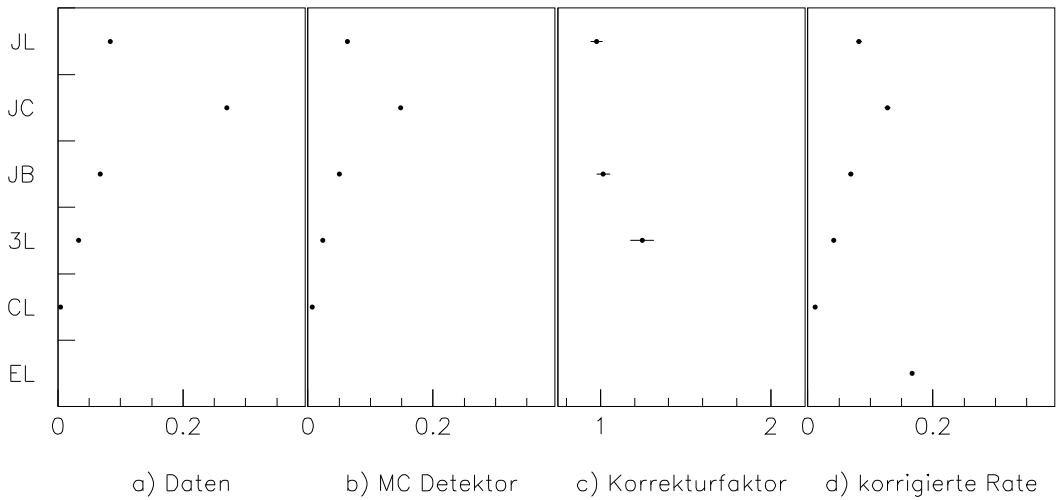,width=\hsize}
\end{center}
\caption[Vergleich verschiedener Jetalgorithmen, Bin 1-4]{{\bf
Vergleich verschiedener Jetalgorithmen, Bin 1-4.}{\it\ Dargestellt sind
auf der Ordinate die Jetalgorithmen JADE im Laborsystem (JL), JADE im
hadronischen Schwerpunktsystem (JC), JADE im Breitsystem (JB), JADE mit
$\ycut = 0.03$ (3L), Cone Algorithmus (CL) und der JADE Algorithmus mit
E-Schema (EL) (siehe auch Tabelle \ref{tabualgo}). In den vier Graphen
sind die Raten $R_{2+1}$ f"ur die Daten (a) und das
Monte-Carlo-Detektorniveau (b) gezeigt. Der Korrekturfaktor ist in c
dargestellt und die sich ergebende korrigierte Rate in d.}}
\label{abbmejetstd4}
\end{figure}

\begin{figure}[tbp]                
\begin{center}
\epsfig{file=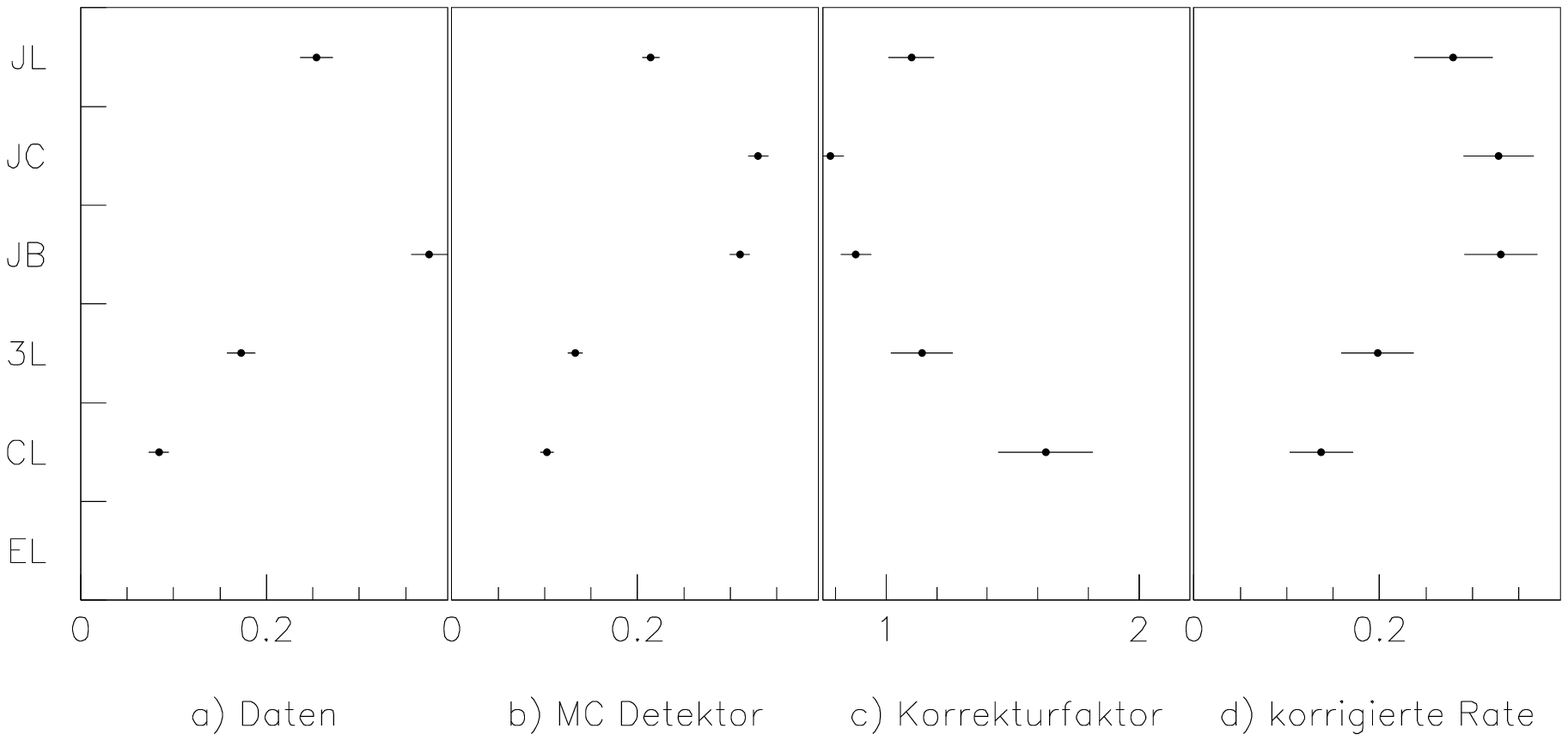,width=\hsize}
\end{center}
\caption[Vergleich verschiedener Jetalgorithmen, Bin 8]{{\bf
Vergleich verschiedener Jetalgorithmen, Bin 8.}{\it\ F"ur n"ahere
Erl"auterungen siehe Abbildung \ref{abbmejetstd4} und Tabelle
\ref{tabualgo} }}
\label{abbmejetstd8}
\end{figure}

In den Abbildungen \ref{abbmejetstd4} bis \ref{abbmejetstd8} sind die
entsprechenden Werte graphisch f"ur verschiedene $Q^2$ Bins dargestellt.
Die Ereignisse aus den ersten vier Bins wurden dabei zu einer Abbildung
zusammengefa"st. Zum einfacheren Vergleich sind die Graphen in einer
festen Skalierung gezeigt. Die Werte f"ur alle acht Bins sind in den
Tabellen \ref{tabjetalgo1} bis \ref{tabjetalgo3} zusammengefa"st.

\begin{figure}[tbp]                
\begin{center}
\epsfig{file=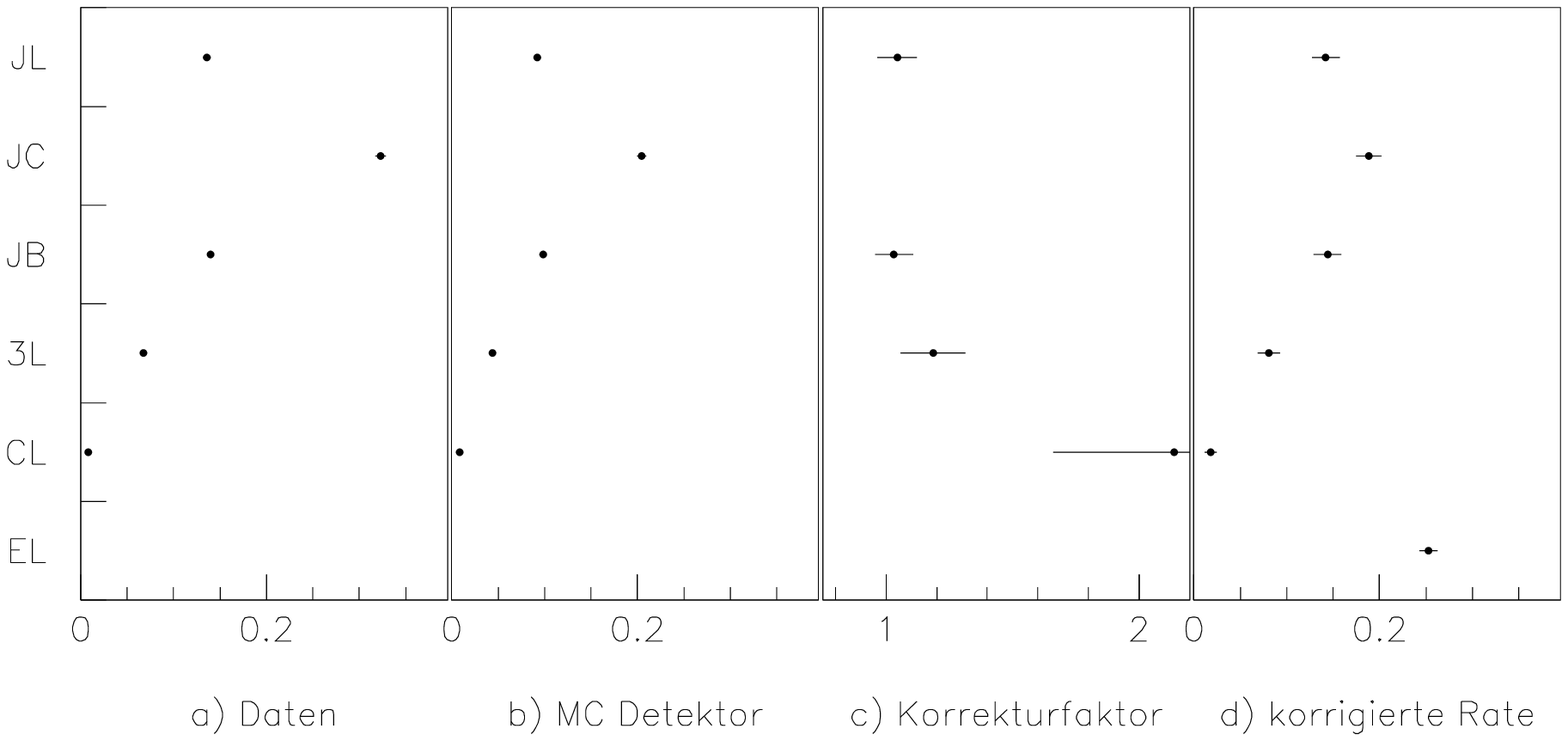,width=\hsize}
\end{center}
\caption[Vergleich verschiedener Jetalgorithmen, Bin 5]{{\bf
Vergleich verschiedener Jetalgorithmen, Bin 5.}{\it\ F"ur n"ahere
Erl"auterungen siehe Abbildung \ref{abbmejetstd4} und Tabelle
\ref{tabualgo} }}
\label{abbmejetstd5}
\end{figure}

\begin{figure}[tbp]                
\begin{center}
\epsfig{file=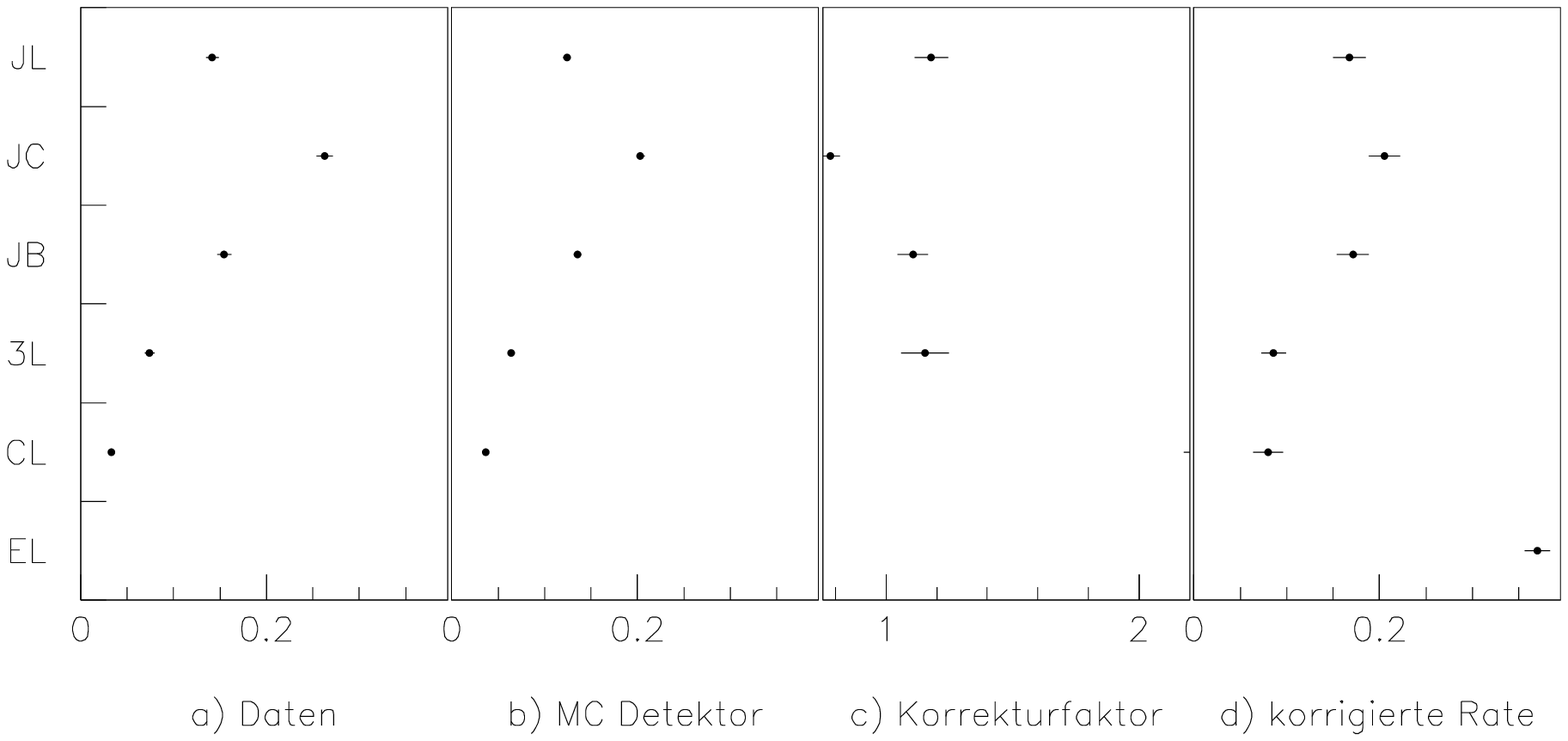,width=\hsize}
\end{center}
\caption[Vergleich verschiedener Jetalgorithmen, Bin 6]{{\bf
Vergleich verschiedener Jetalgorithmen, Bin 6.}{\it\ F"ur n"ahere
Erl"auterungen siehe Abbildung \ref{abbmejetstd4} und Tabelle
\ref{tabualgo} }}
\label{abbmejetstd6}
\end{figure}

\begin{figure}[tbp]                
\begin{center}
\epsfig{file=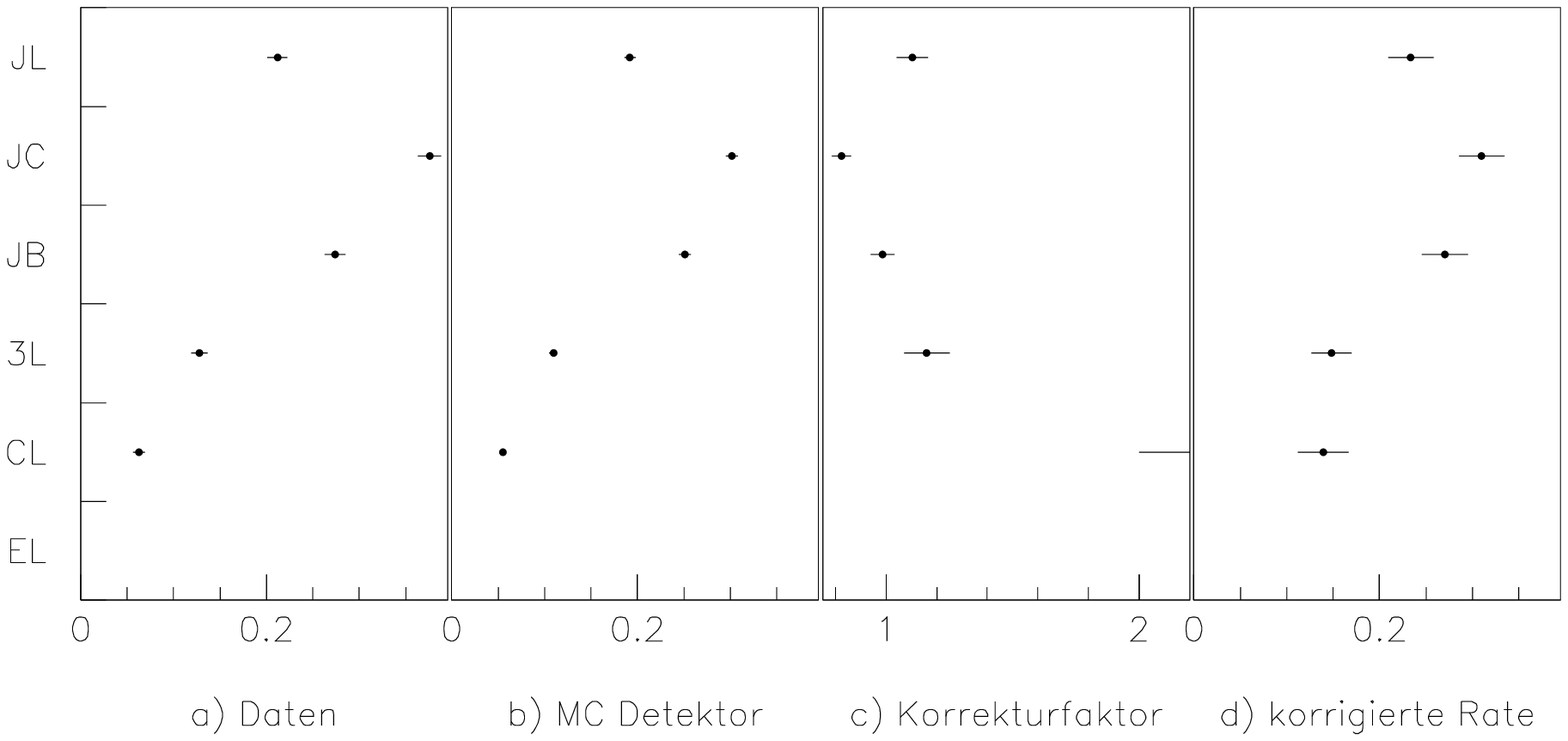,width=\hsize}
\end{center}
\caption[Vergleich verschiedener Jetalgorithmen, Bin 7]{{\bf
Vergleich verschiedener Jetalgorithmen, Bin 7.}{\it\ F"ur n"ahere
Erl"auterungen siehe Abbildung \ref{abbmejetstd4} und Tabelle
\ref{tabualgo} }}
\label{abbmejetstd7}
\end{figure}

\begin{table}[tbp]
\begin{center}
\begin{tabular}{@{\extracolsep{\fill}}|c||c|c|c|c|c|c|c|c|}
\hline
& 1 & 2 & 3 & 4 & 5 & 6 & 7 & 8\\
\hline
\hline
\multicolumn{9}{|c|}{Datenrate $R_{2+1,\mb{Daten}}$ in \%}\\
\hline
\hline
JL &  6.9 &  8.2 &  8.8 & 10.2 & 13.6 & 14.2 & 21.2 & 25.4\\
   & $\pm 0.7$ & $\pm 0.8$ & $\pm 0.8$ & $\pm 0.8$ & $\pm 0.9$ & $\pm 1.2$ & $\pm 1.6$ & $\pm 2.3$\\
\hline
JC & 25.4 & 26.9 & 27.8 & 28.4 & 32.3 & 26.3 & 37.6 & 42.1\\
   & $\pm 0.9$ & $\pm 0.9$ & $\pm 0.9$ & $\pm 0.9$ & $\pm 1.0$ & $\pm 1.4$ & $\pm 1.7$ & $\pm 2.5$\\
\hline
JB &  4.7 &  6.5 &  7.1 &  9.6 & 14.0 & 15.5 & 27.4 & 37.5\\
   & $\pm 0.7$ & $\pm 0.7$ & $\pm 0.7$ & $\pm 0.8$ & $\pm 0.9$ & $\pm 1.2$ & $\pm 1.6$ & $\pm 2.4$\\
\hline
3L &  2.5 &  3.3 &  3.4 &  4.5 &  6.8 &  7.4 & 12.8 & 17.3\\
   & $\pm 0.6$ & $\pm 0.7$ & $\pm 0.7$ & $\pm 0.7$ & $\pm 0.8$ & $\pm 1.0$ & $\pm 1.4$ & $\pm 2.0$\\
\hline
CL &  0.3 &  0.4 &  0.4 &  0.6 &  0.8 &  3.3 &  6.3 &  8.4\\
   & $\pm 0.5$ & $\pm 0.6$ & $\pm 0.6$ & $\pm 0.6$ & $\pm 0.6$ & $\pm 0.9$ & $\pm 1.1$ & $\pm 1.6$\\
\hline
EL & 67.6 & 69.8 & 69.8 & 71.3 & 73.1 & 70.3 & 71.3 & 68.6\\
   & $\pm 0.9$ & $\pm 0.9$ & $\pm 0.9$ & $\pm 0.9$ & $\pm 1.0$ & $\pm 1.4$ & $\pm 1.7$ & $\pm 2.4$\\
\hline
\end{tabular}
\end{center}
\caption[Zusammenfassung des Vergleichs verschiedener Jetalgorithmen]{{\bf
Zusammenfassung des Vergleichs verschiedener Jetalgorithmen}}
\label{tabjetalgo1}
\end{table}

\begin{table}[tbp]
\begin{center}
\begin{tabular}{@{\extracolsep{\fill}}|c||c|c|c|c|c|c|c|c|}
\hline
& 1 & 2 & 3 & 4 & 5 & 6 & 7 & 8\\
\hline
\hline
\multicolumn{9}{|c|}{Monte-Carlo-Detektorrate $R_{2+1,\mb{Cluster}}$ in \%}\\
\hline
\hline
JL &  5.7 &  6.0 &  6.5 &  7.3 &  9.2 & 12.4 & 19.2 & 21.5\\
   & $\pm 0.7$ & $\pm 0.7$ & $\pm 0.7$ & $\pm 0.8$ & $\pm 0.9$ & $\pm 0.9$ & $\pm 1.1$ & $\pm 1.5$\\
\hline
JC & 13.2 & 14.5 & 15.5 & 17.0 & 20.4 & 20.3 & 30.2 & 33.0\\
   & $\pm 0.8$ & $\pm 0.8$ & $\pm 0.9$ & $\pm 0.9$ & $\pm 1.0$ & $\pm 1.0$ & $\pm 1.2$ & $\pm 1.6$\\
\hline
JB &  3.9 &  4.5 &  5.4 &  7.0 &  9.8 & 13.6 & 25.1 & 31.0\\
   & $\pm 0.7$ & $\pm 0.7$ & $\pm 0.7$ & $\pm 0.8$ & $\pm 0.9$ & $\pm 0.9$ & $\pm 1.1$ & $\pm 1.6$\\
\hline
3L &  2.0 &  2.2 &  2.5 &  3.2 &  4.4 &  6.4 & 10.9 & 13.3\\
   & $\pm 0.6$ & $\pm 0.6$ & $\pm 0.7$ & $\pm 0.7$ & $\pm 0.8$ & $\pm 0.8$ & $\pm 1.0$ & $\pm 1.3$\\
\hline
CL &  0.7 &  0.4 &  0.6 &  0.9 &  0.8 &  3.6 &  5.5 & 10.2\\
   & $\pm 0.6$ & $\pm 0.6$ & $\pm 0.6$ & $\pm 0.6$ & $\pm 0.6$ & $\pm 0.7$ & $\pm 0.8$ & $\pm 1.2$\\
\hline
EL & 60.8 & 64.3 & 63.8 & 64.8 & 66.5 & 63.5 & 68.8 & 65.7\\
   & $\pm 0.9$ & $\pm 1.0$ & $\pm 1.0$ & $\pm 1.0$ & $\pm 1.1$ & $\pm 1.0$ & $\pm 1.2$ & $\pm 1.6$\\
\hline
\hline
\multicolumn{9}{|c|}{Monte-Carlo-Partonrate $R_{2+1,\mb{Parton}}$ in \%}\\
\hline
\hline
JL &  5.2 &  5.4 &  6.6 &  8.1 &  9.6 & 14.6 & 21.2 & 23.6\\
   & $\pm 0.7$ & $\pm 0.7$ & $\pm 0.7$ & $\pm 0.8$ & $\pm 0.8$ & $\pm 0.9$ & $\pm 1.1$ & $\pm 1.4$\\
\hline
JC &  5.7 &  6.5 &  7.4 &  9.2 & 11.9 & 15.9 & 24.8 & 25.7\\
   & $\pm 0.7$ & $\pm 0.7$ & $\pm 0.7$ & $\pm 0.8$ & $\pm 0.9$ & $\pm 0.9$ & $\pm 1.1$ & $\pm 1.4$\\
\hline
JB &  3.9 &  4.5 &  5.7 &  7.1 & 10.1 & 15.0 & 24.8 & 27.3\\
   & $\pm 0.7$ & $\pm 0.7$ & $\pm 0.7$ & $\pm 0.8$ & $\pm 0.9$ & $\pm 0.9$ & $\pm 1.1$ & $\pm 1.4$\\
\hline
3L &  2.3 &  2.7 &  3.1 &  4.1 &  5.2 &  7.4 & 12.7 & 15.2\\
   & $\pm 0.6$ & $\pm 0.6$ & $\pm 0.7$ & $\pm 0.7$ & $\pm 0.8$ & $\pm 0.8$ & $\pm 1.0$ & $\pm 1.2$\\
\hline
CL &  1.9 &  1.4 &  1.6 &  2.1 &  1.8 &  8.7 & 12.2 & 16.7\\
   & $\pm 0.6$ & $\pm 0.6$ & $\pm 0.6$ & $\pm 0.6$ & $\pm 0.7$ & $\pm 0.8$ & $\pm 0.9$ & $\pm 1.3$\\
\hline
EL & 13.1 & 13.8 & 15.7 & 19.3 & 23.0 & 33.4 & 43.7 & 45.1\\
   & $\pm 0.8$ & $\pm 0.8$ & $\pm 0.8$ & $\pm 0.9$ & $\pm 1.0$ & $\pm 1.0$ & $\pm 1.2$ & $\pm 1.5$\\
\hline
\end{tabular}
\end{center}
\caption[Zusammenfassung des Vergleichs verschiedener Jetalgorithmen]{{\bf
Zusammenfassung des Vergleichs verschiedener Jetalgorithmen}}
\label{tabjetalgo2}
\end{table}

\begin{table}[tbp]
\begin{center}
\begin{tabular}{@{\extracolsep{\fill}}|c||c|c|c|c|c|c|c|c|}
\hline
& 1 & 2 & 3 & 4 & 5 & 6 & 7 & 8\\
\hline
\hline
\multicolumn{9}{|c|}{Korrekturfaktor $R$}\\
\hline
\hline
JL & 0.91 & 0.90 & 1.01 & 1.11 & 1.04 & 1.18 & 1.10 & 1.10\\
   & $\pm0.07$ & $\pm0.07$ & $\pm0.08$ & $\pm0.08$ & $\pm0.08$ & $\pm0.07$ & $\pm0.07$ & $\pm0.09$\\
\hline
JC & 0.43 & 0.45 & 0.48 & 0.54 & 0.58 & 0.78 & 0.82 & 0.78\\
   & $\pm0.03$ & $\pm0.03$ & $\pm0.03$ & $\pm0.03$ & $\pm0.04$ & $\pm0.04$ & $\pm0.04$ & $\pm0.06$\\
\hline
JB & 0.99 & 1.00 & 1.05 & 1.02 & 1.03 & 1.11 & 0.99 & 0.88\\
   & $\pm0.08$ & $\pm0.09$ & $\pm0.09$ & $\pm0.08$ & $\pm0.08$ & $\pm0.06$ & $\pm0.05$ & $\pm0.07$\\
\hline
3L & 1.19 & 1.23 & 1.26 & 1.30 & 1.19 & 1.16 & 1.16 & 1.14\\
   & $\pm0.13$ & $\pm0.15$ & $\pm0.15$ & $\pm0.14$ & $\pm0.13$ & $\pm0.10$ & $\pm0.09$ & $\pm0.13$\\
\hline
CL & 2.72 & 3.17 & 2.59 & 2.30 & 2.14 & 2.40 & 2.21 & 1.63\\
   & $\pm0.43$ & $\pm0.71$ & $\pm0.53$ & $\pm0.40$ & $\pm0.48$ & $\pm0.23$ & $\pm0.22$ & $\pm0.19$\\
\hline
EL & 0.22 & 0.22 & 0.25 & 0.30 & 0.35 & 0.53 & 0.64 & 0.69\\
   & $\pm0.01$ & $\pm0.01$ & $\pm0.01$ & $\pm0.01$ & $\pm0.02$ & $\pm0.02$ & $\pm0.02$ & $\pm0.03$\\
\hline
\hline
\multicolumn{9}{|c|}{korrigierte Rate $R_{2+1,\mb{korr.}}$ in \%}\\
\hline
\hline
JL &  6.2 &  7.4 &  8.9 & 11.3 & 14.2 & 16.7 & 23.4 & 28.0\\
   & $\pm 1.1$ & $\pm 1.3$ & $\pm 1.4$ & $\pm 1.6$ & $\pm 2.0$ & $\pm 2.3$ & $\pm 3.0$ & $\pm 4.7$\\
\hline
JC & 11.0 & 12.1 & 13.2 & 15.3 & 18.9 & 20.5 & 31.0 & 32.8\\
   & $\pm 1.2$ & $\pm 1.4$ & $\pm 1.4$ & $\pm 1.6$ & $\pm 1.9$ & $\pm 2.2$ & $\pm 2.9$ & $\pm 4.3$\\
\hline
JB &  4.7 &  6.5 &  7.5 &  9.7 & 14.4 & 17.1 & 27.0 & 33.1\\
   & $\pm 1.0$ & $\pm 1.3$ & $\pm 1.4$ & $\pm 1.5$ & $\pm 2.0$ & $\pm 2.2$ & $\pm 3.0$ & $\pm 4.5$\\
\hline
3L &  3.0 &  4.0 &  4.3 &  5.8 &  8.1 &  8.6 & 14.8 & 19.8\\
   & $\pm 1.0$ & $\pm 1.2$ & $\pm 1.2$ & $\pm 1.4$ & $\pm 1.7$ & $\pm 1.8$ & $\pm 2.6$ & $\pm 4.4$\\
\hline
CL &  0.9 &  1.3 &  1.0 &  1.4 &  1.8 &  8.0 & 13.9 & 13.7\\
   & $\pm 0.8$ & $\pm 1.0$ & $\pm 0.8$ & $\pm 0.9$ & $\pm 1.1$ & $\pm 2.1$ & $\pm 3.2$ & $\pm 3.9$\\
\hline
EL & 14.6 & 15.0 & 17.1 & 21.2 & 25.3 & 37.0 & 45.3 & 47.2\\
   & $\pm 1.0$ & $\pm 1.0$ & $\pm 1.1$ & $\pm 1.2$ & $\pm 1.4$ & $\pm 1.9$ & $\pm 2.4$ & $\pm 3.6$\\
\hline
\end{tabular}
\end{center}
\caption[Zusammenfassung des Vergleichs verschiedener Jetalgorithmen]{{\bf
Zusammenfassung des Vergleichs verschiedener Jetalgorithmen}}
\label{tabjetalgo3}
\end{table}

Die Abbildungen erm"oglichen sowohl den Vergleich zwischen Daten und
Monte-Carlo-Detektorniveau als auch die Bewertung der Korrekturfaktoren
$R$ und den Vergleich der korrigierten Raten
\begin{eqnarray}
R_{2+1}^{\mb{korr}} &=& R \cdot R_{2+1}^{\mb{Daten}} \\
&=& {R_{2+1}^{\mb{Parton}}\over R_{2+1}^{\mb{Detektor}} } \cdot
R_{2+1}^{\mb{Daten}}\nonumber
\end{eqnarray}
Zum letzten Punkt ist anzumerken, da"s wir nur f"ur den JADE Algorithmus
mit Schnittwert $\ycut= 0.02$ in den verschiedenen Systemen die gleiche
korrigierte Rate erwarten. Bei Verwendung eines anderen
Schnittparameters oder anderen Algorithmus ver"andern sich die von der
Theorie vorhergesagten Koeffizienten zur Berechnung der starken
Kopplungskonstante (siehe Kapitel \ref{kapasmess}). Da das von uns
verwendete Programm zur Berechnung der theoretischen Werte PROJET nur
den JADE Algorithmus verwendet, ist auf einen Vergleich der Partonraten
mit den theoretischen Werten hier verzichtet worden.

Vergleichen wir zuerst die unterschiedlichen Systeme. Uns f"allt auf,
da"s sich bei der Datenrate gro"se Unterschiede ergeben. Obwohl bei
allen Versionen die Datenrate mit $Q^2$ steigt, ist die St"arke der
Abh"angigkeit unterschiedlich. Vergleichen wir dies mit der Rate auf
Monte-Carlo-Detektorniveau, so ergibt sich ein "ahnliches Bild. Die
Raten selber sind jedoch um etwa 2\% kleiner als bei den Daten. Auf
Partonniveau sieht das Bild schon wesentlich besser aus. Die Werte sind
nun wesentlich enger zusammen und die $Q^2$ Abh"angigkeit ist ebenfalls
sehr "ahnlich. Aus den letzten beiden Punkten ergeben sich zwangsl"aufig
sehr unterschiedliche Korrekturfaktoren. W"ahrend die Korrekturfaktoren
f"ur das Labor- und das Breitsystem zwischen 0.88 und 1.18 im Rahmen
bleiben, sind sie f"ur den JADE Algorithmus im hadronischen
Schwerpunktsystem mit Werten zwischen 0.43 und 0.82 unakzeptabel.

Betrachten wir nun den JADE Algorithmus mit dem gr"o"seren Schnittwert, so
ist die Rate auf allen drei gemessenen Niveaus im Einklang mit unserer
Erwartung kleiner als bei der Standardeinstellung. Die Reduktion
betr"agt in allen F"allen ungef"ahr 50\%, wobei das Verh"altnis in den
niedrigeren Bins auf Detektorlevel eher kleiner, auf Partonlevel etwas
h"oher liegt. F"ur den Korrekturfaktor ergibt sich ein recht
gleichm"a"siger Verlauf mit Werten zwischen 1.14 und 1.30. F"ur die
korrigierte Rate gilt somit ebenfalls eine Reduktion um die H"alfte
gegen"uber den Standardeinstellungen.

Der Cone Algorithmus hat sehr niedrige Raten. Die Beschreibung der Daten
durch das Monte-Carlo-Detektorniveau ist recht gut, aber es zeigen sich
gro"se Abweichungen zwischen den beiden Monte-Carlo-Niveaus. Dies
bedingt auch die unakzeptablen Korrekturfaktoren mit Werten zwischen
1.63 und 3.17. Die sich daraus ergebende korrigierte Rate zeigt ein
merkw"urdiges $Q^2$ Verhalten. W"ahrend die Rate in den unteren f"unf
Bins nahezu konstant ist, tritt ein Anstieg zwischen Bin f"unf und acht
um einen Faktor sieben bis acht auf.

Beim E Schema ergeben sich sehr gro"se Raten auf Daten und
Detektorniveau. Die "Ubereinstimmung hier ist gut. Die Raten auf
Partonniveau sind um ein Vielfaches kleiner und die $Q^2$ Abh"angigkeit
ist hier wesentlich st"arker. Somit ergeben sich unakzeptable, $Q^2$
abh"angige Korrekturfaktoren zwischen 0.22 und 0.69.

Zusammenfassend ergeben sich starke Unterschiede bei Verwendung anderer
Algorithmen oder anderer Einstellparameter. Der Cone Algorithmus und das
E Schema zeigen merkw"urdige $Q^2$ Abh"angigkeiten insbesondere beim
Korrekturfaktor. Eine endg"ul"-ti"-ge Bewertung dieser Algorithmen kann
aber erst nach einem Vergleich mit den Theorieraten erfolgen. Dies ist
zur Zeit noch nicht m"oglich. F"ur die vier JADE Versionen zeigen sich
zu gro"se Abweichungen insbesondere bei Verwendung des Algorithmus im
hadronischen Schwerpunktsystem. Die korrigierte Rate liefert jedoch
akzeptable Werte.
  
\section {Untersuchungen auf dem Hadronniveau}

Der Korrekturfaktor beschreibt die Raten"anderung, die durch mehrere
Effekte eintreten kann. Neben der Hadronisierung und Fragmentation
k"onnen auch Detektoreffekte eine Rolle spielen. Um den Anteil dieser
beiden Schritte an den Migrationen zu untersuchen, spalten wir unsere
Betrachtung auf. Zur Untersuchung der durch die Hadronisierung
verursachten Migrationen sind in Tabelle \ref{tabstdmigmepshad} die
Migrationen vom Parton- zum Hadronniveau verzeichnet. Die Migrationen
durch Detektoreffekte oder Fehl-Erkennung des gestreuten Elektrons sind
in Tabelle \ref{tabstdmigmepshade} beim "Ubergang von Hadron- auf
Detektorniveau notiert. Bei diesem Schritt ist auch zu ber"ucksichtigen,
da"s die Akzeptanz auf Parton- und Hadronniveau im gesamten Raum eins
ist, w"ahrend wir auf Detektorniveau Akzeptanzl"ucken haben. Daher
benutzen wir nur auf Detektorniveau ein Pseudoteilchen als N"aherung
f"ur den Protonrest.

\begin{table}[tbp]
\begin{center}
\begin{tabular}{@{\extracolsep{\fill}}c|c||c|c}
\multicolumn{2}{c||}{Bin} & \multicolumn{2}{c}{Parton} \\
\multicolumn{2}{c||}{1} & 1+1 & 2+1 \\
\hline
\hline
Ha- & 1+1 & 14825 & 327 \\
\cline{2-4}
dron & 2+1 & 433 & 508 \\
\end{tabular}
\hfill
\begin{tabular}{@{\extracolsep{\fill}}c|c||c|c}
\multicolumn{2}{c||}{Bin} & \multicolumn{2}{c}{Parton} \\
\multicolumn{2}{c||}{2} & 1+1 & 2+1 \\
\hline
\hline
Ha- & 1+1 & 10839 & 251 \\
\cline{2-4}
dron & 2+1 & 325 & 385 \\
\end{tabular}
\hfill
\begin{tabular}{@{\extracolsep{\fill}}c|c||c|c}
\multicolumn{2}{c||}{Bin} & \multicolumn{2}{c}{Parton} \\
\multicolumn{2}{c||}{3} & 1+1 & 2+1 \\
\hline
\hline
Ha- & 1+1 & 10095 & 291 \\
\cline{2-4}
dron & 2+1 & 351 & 442 \\
\end{tabular}
\end{center}
\vspace {0.25cm}
\begin{center}
\begin{tabular}{@{\extracolsep{\fill}}c|c||c|c}
\multicolumn{2}{c||}{Bin} & \multicolumn{2}{c}{Parton} \\
\multicolumn{2}{c||}{4} & 1+1 & 2+1 \\
\hline
\hline
Ha- & 1+1 & 9151 & 306 \\
\cline{2-4}
dron & 2+1 & 301 & 528 \\
\end{tabular}
\hfill
\begin{tabular}{@{\extracolsep{\fill}}c|c||c|c}
\multicolumn{2}{c||}{Bin} & \multicolumn{2}{c}{Parton} \\
\multicolumn{2}{c||}{5} & 1+1 & 2+1 \\
\hline
\hline
Ha- & 1+1 & 6262 & 251 \\
\cline{2-4}
dron & 2+1 & 230 & 440 \\
\end{tabular}
\hfill
\begin{tabular}{@{\extracolsep{\fill}}c|c||c|c}
\multicolumn{2}{c||}{Bin} & \multicolumn{2}{c}{Parton} \\
\multicolumn{2}{c||}{6} & 1+1 & 2+1 \\
\hline
\hline
Ha- & 1+1 & 7067 & 423 \\
\cline{2-4}
dron & 2+1 & 363 & 849 \\
\end{tabular}
\end{center}
\vspace {0.25cm}
\begin{center}
\begin{tabular}{@{\extracolsep{\fill}}c|c||c|c}
\multicolumn{2}{c||}{Bin} & \multicolumn{2}{c}{Parton} \\
\multicolumn{2}{c||}{7} & 1+1 & 2+1 \\
\hline
\hline
Ha- & 1+1 & 3990 & 349 \\
\cline{2-4}
dron & 2+1 & 301 & 805 \\
\end{tabular}
\hfill
\begin{tabular}{@{\extracolsep{\fill}}c|c||c|c}
\multicolumn{2}{c||}{Bin} & \multicolumn{2}{c}{Parton} \\
\multicolumn{2}{c||}{8} & 1+1 & 2+1 \\
\hline
\hline
Ha- & 1+1 & 1675 & 128 \\
\cline{2-4}
dron & 2+1 & 128 & 429 \\
\end{tabular}
\hfill
\hphantom{
\begin{tabular}{@{\extracolsep{\fill}}c|c||c|c}
\multicolumn{2}{c||}{} & \multicolumn{2}{c}{Parton} \\
\multicolumn{2}{c||}{} & 1+1 & 2+1 \\
\hline
\hline
Ha- & 1+1 & & \\
\cline{2-4}
dron & 2+1 & & \\
\end{tabular}
}
\end{center}
\vspace {0.25cm}
\begin{center}
\begin{tabular}{@{\extracolsep{\fill}}|c||c|c|c|c|c|c|c|c|}
\hline
Bin & 1 & 2 & 3 & 4 & 5 & 6 & 7 & 8 \\
\hline
Rate Parton&  5.2 &  5.4 &  6.6 &  8.1 &  9.6 & 14.6 & 21.2 & 23.6\\
in \% & $\pm 0.2$&$\pm 0.3$&$\pm 0.3$&$\pm 0.3$&$\pm 0.4$&$\pm 0.4$&$\pm 0.6$&$\pm 0.9$\\
\hline
Rate Hadron &  5.8 &  6.0 &  7.1 &  8.1 &  9.3 & 13.9 & 20.3 & 23.6\\
in \%&$\pm 0.2$&$\pm 0.3$&$\pm 0.3$&$\pm 0.3$&$\pm 0.4$&$\pm 0.4$&$\pm 0.6$&$\pm 0.9$\\
\hline
Korrektur- & 0.89 & 0.90 & 0.92 & 1.01 & 1.03 & 1.05 & 1.04 & 1.00\\
faktor &$\pm0.06$&$\pm0.07$&$\pm0.07$&$\pm0.07$&$\pm0.08$&$\pm0.06$&$\pm0.06$&$\pm0.08$\\
\hline
Reinheit in \%& 54.0 & 54.2 & 55.7 & 63.7 & 65.7 & 70.0 & 72.8 & 77.0\\
\hline
\end{tabular}
\end{center}
\caption[Migrationen und Korrekturfaktoren bei Lepto, Standardschnitte]{{\bf
Migrationen und Korrekturfaktoren bei Lepto mit Standardschnitten, Vergleich
von Parton- und Hadronniveau in $Q^2$ Bins}}
\label{tabstdmigmepshad}
\end{table}

\begin{table}[tbp]
\begin{center}
\begin{tabular}{@{\extracolsep{\fill}}c|c||c|c}
\multicolumn{2}{c||}{Bin} & \multicolumn{2}{c}{Hadron} \\
\multicolumn{2}{c||}{1} & 1+1 & 2+1 \\
\hline
\hline
Clus- & 1+1 & 12395 & 389 \\
\cline{2-4}
ter & 2+1 & 419 & 358 \\
\end{tabular}
\hfill
\begin{tabular}{@{\extracolsep{\fill}}c|c||c|c}
\multicolumn{2}{c||}{Bin} & \multicolumn{2}{c}{Hadron} \\
\multicolumn{2}{c||}{2} & 1+1 & 2+1 \\
\hline
\hline
Clus- & 1+1 & 9358 & 308 \\
\cline{2-4}
ter & 2+1 & 324 & 279 \\
\end{tabular}
\hfill
\begin{tabular}{@{\extracolsep{\fill}}c|c||c|c}
\multicolumn{2}{c||}{Bin} & \multicolumn{2}{c}{Hadron} \\
\multicolumn{2}{c||}{3} & 1+1 & 2+1 \\
\hline
\hline
Clus- & 1+1 & 8934 & 327 \\
\cline{2-4}
ter & 2+1 & 298 & 325 \\
\end{tabular}
\end{center}
\vspace {0.25cm}
\begin{center}
\begin{tabular}{@{\extracolsep{\fill}}c|c||c|c}
\multicolumn{2}{c||}{Bin} & \multicolumn{2}{c}{Hadron} \\
\multicolumn{2}{c||}{4} & 1+1 & 2+1 \\
\hline
\hline
Clus- & 1+1 & 8128 & 325 \\
\cline{2-4}
ter & 2+1 & 282 & 368 \\
\end{tabular}
\hfill
\begin{tabular}{@{\extracolsep{\fill}}c|c||c|c}
\multicolumn{2}{c||}{Bin} & \multicolumn{2}{c}{Hadron} \\
\multicolumn{2}{c||}{5} & 1+1 & 2+1 \\
\hline
\hline
Clus- & 1+1 & 5483 & 237 \\
\cline{2-4}
ter & 2+1 & 242 & 316 \\
\end{tabular}
\hfill
\begin{tabular}{@{\extracolsep{\fill}}c|c||c|c}
\multicolumn{2}{c||}{Bin} & \multicolumn{2}{c}{Hadron} \\
\multicolumn{2}{c||}{6} & 1+1 & 2+1 \\
\hline
\hline
Clus- & 1+1 & 5968 & 414 \\
\cline{2-4}
ter & 2+1 & 270 & 590 \\
\end{tabular}
\end{center}
\vspace {0.25cm}
\begin{center}
\begin{tabular}{@{\extracolsep{\fill}}c|c||c|c}
\multicolumn{2}{c||}{Bin} & \multicolumn{2}{c}{Hadron} \\
\multicolumn{2}{c||}{7} & 1+1 & 2+1 \\
\hline
\hline
Clus- & 1+1 & 3368 & 297 \\
\cline{2-4}
ter & 2+1 & 223 & 616 \\
\end{tabular}
\hfill
\begin{tabular}{@{\extracolsep{\fill}}c|c||c|c}
\multicolumn{2}{c||}{Bin} & \multicolumn{2}{c}{Hadron} \\
\multicolumn{2}{c||}{8} & 1+1 & 2+1 \\
\hline
\hline
Clus- & 1+1 & 1304 & 137 \\
\cline{2-4}
ter & 2+1 & 90 & 295 \\
\end{tabular}
\hfill
\hphantom{
\begin{tabular}{@{\extracolsep{\fill}}c|c||c|c}
\multicolumn{2}{c||}{} & \multicolumn{2}{c}{Hadron} \\
\multicolumn{2}{c||}{} & 1+1 & 2+1 \\
\hline
\hline
Clus- & 1+1 & & \\
\cline{2-4}
ter & 2+1 & & \\
\end{tabular}
}
\end{center}
\vspace {0.25cm}
\begin{center}
\begin{tabular}{@{\extracolsep{\fill}}|c||c|c|c|c|c|c|c|c|}
\hline
Bin & 1 & 2 & 3 & 4 & 5 & 6 & 7 & 8 \\
\hline
Rate Hadron&  5.5 &  5.7 &  6.6 &  7.6 &  8.8 & 13.9 & 20.3 & 23.7\\
in \% & $\pm 0.2$&$\pm 0.3$&$\pm 0.3$&$\pm 0.3$&$\pm 0.4$&$\pm 0.5$&$\pm 0.6$&$\pm 1.0$\\
\hline
Rate Detektor&  5.7 &  5.9 &  6.3 &  7.1 &  8.9 & 11.9 & 18.6 & 21.1\\
in \%&$\pm 0.2$&$\pm 0.3$&$\pm 0.3$&$\pm 0.3$&$\pm 0.4$&$\pm 0.4$&$\pm 0.6$&$\pm 1.0$\\
\hline
Korrektur- & 0.96 & 0.97 & 1.05 & 1.07 & 0.99 & 1.17 & 1.09 & 1.12\\
faktor &$\pm0.07$&$\pm0.08$&$\pm0.09$&$\pm0.08$&$\pm0.09$&$\pm0.08$&$\pm0.07$&$\pm0.10$\\
\hline
Reinheit in \%& 46.1 & 46.3 & 52.2 & 56.6 & 56.6 & 68.6 & 73.4 & 76.6\\
\hline
\end{tabular}
\end{center}
\caption[Migrationen und Korrekturfaktoren bei Lepto, Standardschnitte]{{\bf
Migrationen und Korrekturfaktoren bei Lepto mit Standardschnitten, Vergleich
von Hadron- und Detektorniveau in $Q^2$ Bins}}
\label{tabstdmigmepshade}
\end{table}

Vergleichen wir diese beiden Tabellen mit den Gesamtmigrationen in
Tabelle \ref{tabstdmigmeps}, so stellen wir fest, da"s in beiden
Teilschritten die Reinheit in den hohen $Q^2$ Bins in etwa gleich ist.
Gegen"uber den Parton--Cluster--Migrationen hat sich die Reinheit nur
leicht verbessert. In den niedrigen $Q^2$ Bins ist die Reinheit beim
"Ubergang zwischen Parton- und Hadronniveau besser, ansonsten gilt das
gleiche wie f"ur die hohen Bins. Der Korrekturfaktor l"a"st ebenfalls
nur den Schlu"s zu, da"s die Probleme in beiden Bereichen gleich stark
sind.

Zum Vergleich der Raten ziehen wir die Partonrate von Tabelle
\ref{tabstdmigmeps} und die Hadron- und Detektorrate von Tabelle
\ref{tabstdmigmepshade} heran, da zur Berechnung der Tabelle
\ref{tabstdmigmepshad} auch die Ereignisse verwendet wurden, die auf
Detektorniveau die Standardschnitte nicht erf"ullen. Die Hadronrate liegt
in f"unf Bins zwischen den beiden anderen Raten, in zwei Bins
entsprechen sie den Partonraten und in nur einem Bin ist die Rate h"oher
als die auf Parton- und Detektorniveau. Die Ursache der Migrationen
scheint somit ein Effekt zu sein, der sowohl beim "Ubergang vom Parton-
zum Hadronniveau alsauch bei Ber"ucksichtigung der Detektoreffekte
auftritt. Um dies zu beweisen, sind jedoch weitere Studien und ein
besseres Verst"andnis der Hadronisierung und der Detektoreffekte n"otig.

Um den Einflu"s der Hadronanzahl zu untersuchen, habe ich die Ereignisse
in Bins entsprechend ihrer Hadron"-anzahl geteilt. Die sich daraus
ergebenden Migrationen sind in Tabelle \ref{tabhadmultimeps}
verzeichnet. Wir erkennen, da"s die Reinheit mit der Hadronanzahl stark
ansteigt. Unsere Probleme scheinen also nicht durch die von Niveau zu
Niveau steigende Partikelanzahl ausgel"ost zu werden. Der
Korrekturfaktor ist im Rahmen der Fehler in allen Bereichen gleich. Der
Grund f"ur das Ansteigen der Raten mit der Multiplizit"at ist unbekannt
und deutet die Wichtigkeit von weiteren Untersuchungen der
Hadronisierungseffekte an. Diese w"urden jedoch den Rahmen der
Diplomarbeit sprengen.

\begin{table}[tbp]
\begin{center}
\begin{tabular}{@{\extracolsep{\fill}}c|c||c|c}
\multicolumn{2}{c||}{Bin} & \multicolumn{2}{c}{Parton} \\
\multicolumn{2}{c||}{1} & 1+1 & 2+1 \\
\hline
\hline
Ha- & 1+1 & 1890 & 37 \\
\cline{2-4}
dron & 2+1 & 39 & 19 \\
\end{tabular}
\hfill
\begin{tabular}{@{\extracolsep{\fill}}c|c||c|c}
\multicolumn{2}{c||}{Bin} & \multicolumn{2}{c}{Parton} \\
\multicolumn{2}{c||}{2} & 1+1 & 2+1 \\
\hline
\hline
Ha- & 1+1 & 8191 & 443 \\
\cline{2-4}
dron & 2+1 & 409 & 942 \\
\end{tabular}
\hfill
\begin{tabular}{@{\extracolsep{\fill}}c|c||c|c}
\multicolumn{2}{c||}{Bin} & \multicolumn{2}{c}{Parton} \\
\multicolumn{2}{c||}{3} & 1+1 & 2+1 \\
\hline
\hline
Ha- & 1+1 & 2164 & 382 \\
\cline{2-4}
dron & 2+1 & 318 & 1037 \\
\end{tabular}
\end{center}
\vspace {0.25cm}
\begin{center}
\begin{tabular}{@{\extracolsep{\fill}}|c||c|c|c|}
\hline
Hadronanzahl im Ereignis & kleiner $25$ & 25 bis 50 & gr"o"ser $>50$ \\
\hline
Rate Parton&  2.8 & 13.9 & 36.4 \\
in \% & $\pm 0.4$&$\pm 0.4$&$\pm 0.8$\\
\hline
Rate Hadron &  2.9 & 13.5 & 34.7 \\
in \%&$\pm 0.4$&$\pm 0.4$&$\pm 0.8$\\
\hline
Korrektur- & 0.97 & 1.03 & 1.05 \\
faktor &$\pm0.26$&$\pm0.06$&$\pm0.05$\\
\hline
Reinheit in \%& 32.8 & 69.7 & 76.5 \\
\hline
\end{tabular}
\end{center}
\caption[Migrationen und Korrekturfaktoren bei Lepto in Abh"angigkeit von der Hadronmultiplzit"at]{{\bf
Migrationen und Korrekturfaktoren bei Lepto mit Standardschnitten,
in Bins bez"uglich der Hadronmultiplizit"at}}
\label{tabhadmultimeps}
\end{table}

\vfill
\section {Einf"uhrung zus"atzlicher Schnitte}
\label{kapschnitte}

\subsection{Jetwinkel $\tjet$}
\label{kaputhejet}

\begin{figure}[tbp]                
\begin{center}
\epsfig{file=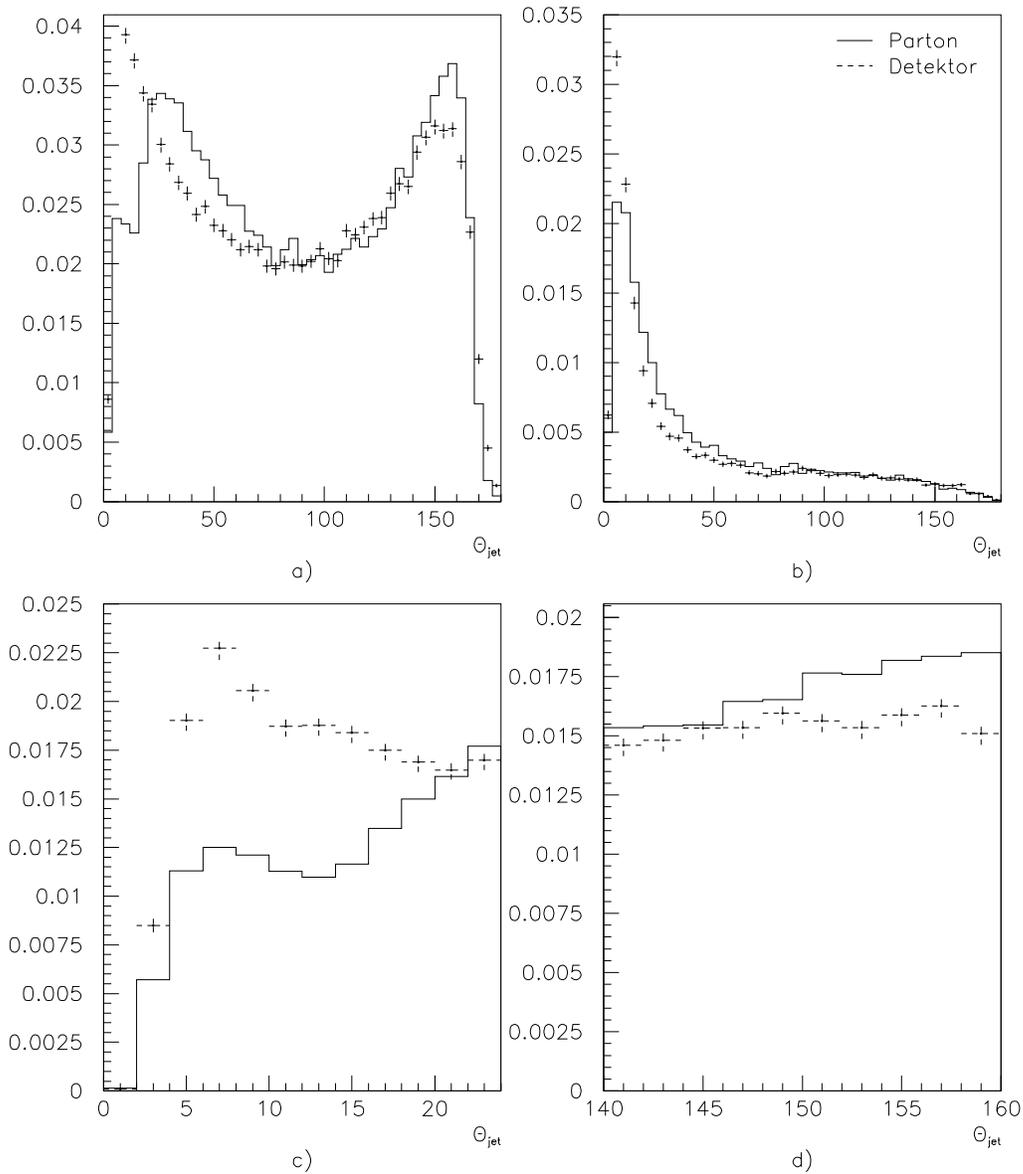,width=0.9\hsize}
\end{center}
\caption[Vergleich der normierten $\theta_{\mb{jet}}$ Verteilungen]{{\bf
Vergleich der $\theta_{\mb{jet}}$ Verteilungen auf Parton- und
Detektorniveau }{\it\ a)~alle Ereignisse, b)~2+1 Ereignisse, c) und d)
sind Vergr"o"serungen von a. Alle Plots sind mit Standardschnitten
berechnet worden und auf gleiche Fl"achen normiert. Dargestellt sind
Parton- (durchgezogen) und Detektorniveau (gestrichelt).}}
\label{abbthejet}
\end{figure}

Aufgrund der Hinweise von Kapitel \ref{kapmotiv} schauen wir uns nun
die Winkelverteilung der harten Jets an. In Abbildung \ref{abbthejet}a
sind die Verteilungen f"ur das Parton- und das Detektorniveau f"ur alle
Ereignisse und in Bild b die inklusiven Verteilungen f"ur die
2+1 Ereignisse alleine gezeigt. Beide Verteilungen sind auf gleiche
Fl"achen normiert. 

Wir erkennen schon hier, da"s im Vorw"arts- und im R"uckw"artsbereich 
die Beschreibungen nicht gut "ubereinstimmen. Deshalb sind die beiden
Bereiche f"ur alle Ereignisse in c und d nochmal vergr"o"sert dargestellt. 
Dazu wurde die gleiche Normierung wie in Bild a benutzt. 

Die Beschreibung der Bereiche kleiner 70 Grad und gr"o"ser 150 Grad
zeigt Abweichungen zwischen den Monte-Carlo-Niveaus. Im Bereich
dazwischen stimmt die Verteilung der 2+1 Ereignisse gut und bei allen
Ereignissen unter Ber"ucksichtigung der dann leicht ge"anderten
Normierung einigerma"sen "uberein. Es scheint jedoch eine Verschiebung
von gro"sen Winkeln auf Partonniveau zu kleineren auf Detektorniveau
vorzuliegen.

Betrachten wir jedoch zuerst den Randbereich bei hohen Winkeln
(Abbildung \ref{abbthejet}d). Das LAr Kalorimeter endet bei 153 Grad und
der auf Hadronen empfindlichste Bereich bereits etwas eher. Dies hat
jedoch auch schon Auswirkungen bei leicht kleineren Winkeln, da dann der
Jet einseitig abgeschnitten wird und sich der Jetwinkel leicht zu
kleineren Werten verschiebt. Die Abweichungen hier sind also
verst"andlich und es empfiehlt sich ein Schnitt auf Jetwinkel kleiner
als 150 Grad.

\begin{figure}[tbp]                
\begin{center}
\epsfig{file=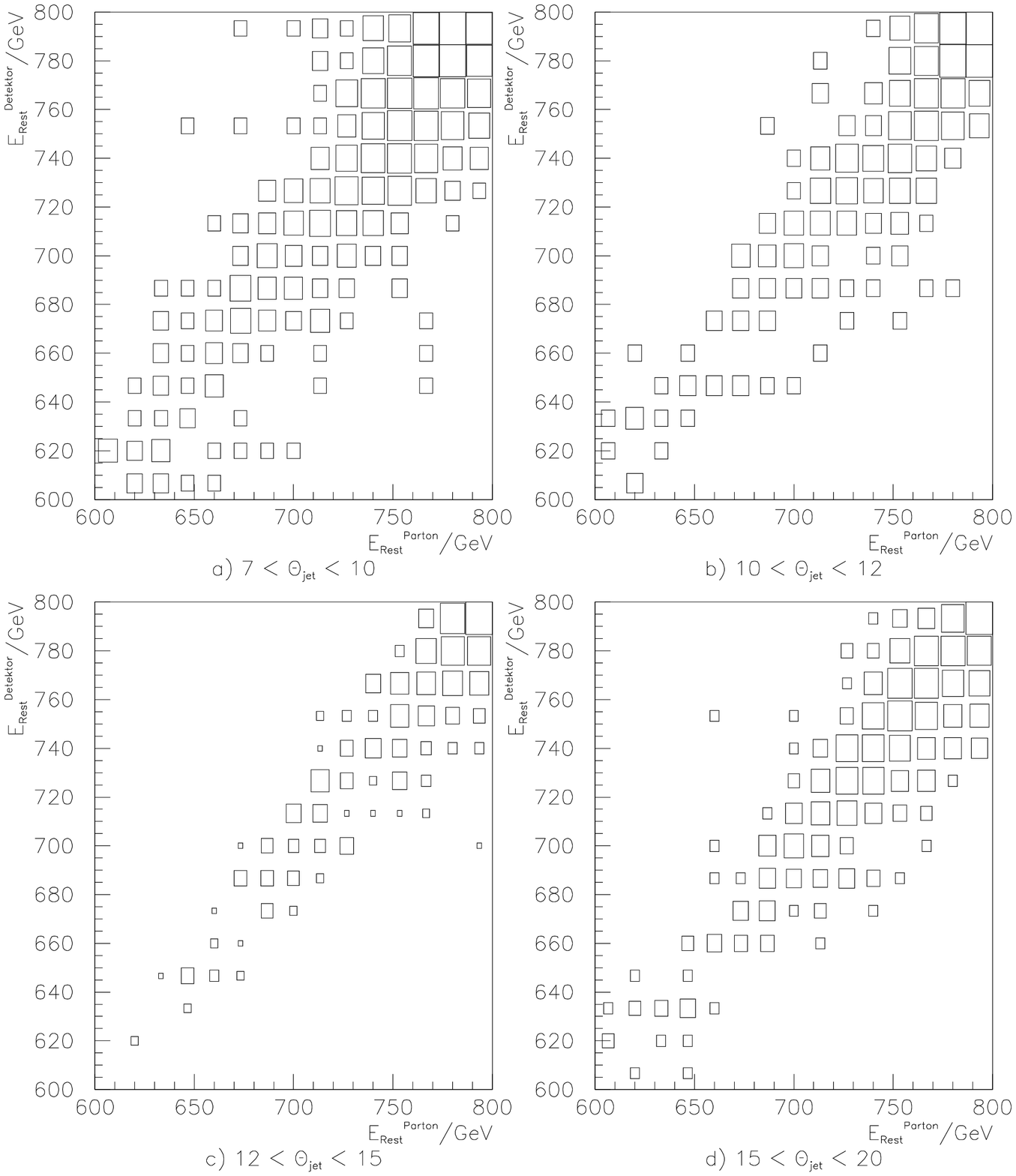,width=\hsize}
\end{center}
\caption[Vergleich der Protonrestenergien]{{\bf
Vergleich der Protonrestenergien f"ur Ereignisse mit Jets im Bereich
kleiner $\tjet.$}
{\it\ a) $7^{\circ} < \theta_{\mb{jet}} < 10^{\circ}$,
b) $10^{\circ} < \theta_{\mb{jet}} < 12^{\circ}$,
c) $12^{\circ} < \theta_{\mb{jet}} < 15^{\circ}$,
d) $15^{\circ} < \theta_{\mb{jet}} < 20^{\circ}$. Die Gr"o"se der Boxen
ist logarithmisch.}}
\label{abbthejetremn}
\end{figure}

Im Bereich kleiner 20 Grad liegt die Situation etwas anders. Hier
"ubersteigt die Detektorverteilung die auf Partonniveau. Der
Detektoranfang liegt bei vier Grad und dort ist auch ein entsprechender
Abfall auf beiden Niveaus zu erkennen. Ein anderer Effekt erh"oht jedoch
die Anzahl Jets unter kleinen Winkeln. Der Protonrest verl"a"st den
Detektor ungesehen durch das Strahlrohr. Dieser Rest kann durch
Abstrahlung oder Hadronisierung in mehrere Hadronen zerfallen, die
"ahnlich den harten Jets einen gewissen Bereich im Detektor "uberdecken.
Diese Hadronen fallen dann unter kleinen Winkeln in das Kalorimeter und
hinterlassen dort ihre Energie. Wird ein harter Jet unter kleinen
Winkeln gestreut, k"onnen sich diese "uberlagern und es ergeben sich
Verf"alschungen in der Energie und dem Winkel der Jets. Dies ist in
Abbildung \ref{abbthejetremn} zu erkennen. Dort ist die Energie des
Protonrestes auf Partonniveau gegen die Energie des Jets aufgetragen,
der aus dem Pseudoteilchen und eventuell zus"atzlichen, rekonstruierten
Clustern gebildet wurde. Die Gr"o"se der Boxen ist wieder logarithmisch.
In den Teilbildern wurden jeweils unterschiedliche Jetwinkel Bereiche
f"ur den unteren harten Jet ausgew"ahlt.
 
Im Bild links oben sind die Ereignisse gezeigt, deren unterster harter
Jet im Bereich zwischen sieben Grad und zehn Grad liegt. Wir erkennen
hier eine sehr breite Verteilung, die auf eine mangelnde Rekonstruktion
schlie"sen l"a"st. In der Abbildung daneben ist der Bereich zwischen
zehn und zw"olf Grad gezeigt. Hier zeigt sich eine nur leicht
verbesserte Korrelation. Erst in den beiden unteren Graphen ist eine
zufriedenstellende Korrelation erreicht. Die Bereiche sind hier zw"olf
bis 15 und 15 bis 20 Grad. Dies legt einen Winkelschnitt von 15 Grad
nahe.

Bei Anwendung des Jetwinkelschnittes werden solche Ereignisse zur
totalen Anzahl gez"ahlt, die einen Winkel in dem angegebenen Bereich
haben, zur 2+1 Jetanzahl jedoch nur dann, wenn beide harten Jets die
Bedingung erf"ullen. Dadurch kann jedoch die Situation eintreten, da"s
in einem Ereignis zwar zwei harte Jets gefunden werden, es aber als 1+1
Ereignis gehandhabt wird, da nur ein Jet den Winkelschnitt erf"ullt.
Dieses Vorgehen ist aber dem kompletten Verwerfen solcher Ereignisse
vorzuziehen, da auch Ereignisse, deren zweiter Jet z.B.\ durch
Detektorakzeptanzen erst garnicht detektiert wurde, beim totalen
Wirkungsquerschnitt ber"ucksichtigt werden.

Die Abh"angigkeit der Rate von der Gr"o"se des unteren Schnittes ist in
den Tabellen \ref{tabtjcuta} und \ref{tabtjcut} f"ur die oberen $Q^2$
Bins gezeigt. Wir erkennen ein Abnehmen der Raten und ein Ansteigen der
Korrekturfaktoren.

\begin{table}[tbp]
\begin{center}
\begin{tabular}{@{\extracolsep{\fill}}|c|c|c|c|c|c|c|}
\hline
\multicolumn{7}{|c|}{Bin 6}\\
\hline
$\tjet$ \IN{^{\circ}}&0&2&4&6&8&10\\
\hline
Parton- & 14.6 & 14.6 & 14.1 & 13.0 & 11.6 & 10.0\\
rate \IN{\%} & $\pm 0.4$ & $\pm 0.4$ & $\pm 0.4$ & $\pm 0.4$ & $\pm 0.3$ & $\pm 0.3$\\
\hline
Detektor- & 12.4 & 12.4 & 11.9 & 10.4 &  8.5 &  6.7\\
rate \IN{\%} & $\pm 0.4$ & $\pm 0.4$ & $\pm 0.4$ & $\pm 0.4$ & $\pm 0.4$ & $\pm 0.3$\\
\hline
Korrektur- & 1.19 & 1.19 & 1.19 & 1.26 & 1.38 & 1.51\\
faktor & $\pm0.07$ & $\pm0.07$ & $\pm0.07$ & $\pm0.08$ & $\pm0.09$ & $\pm0.11$\\
\hline
$\tjet$ \IN{^{\circ}}&12&14&16&18&20&\\
\hline
Parton- &  8.9 &  7.6 &  6.5 &  5.6 &  4.9 &\\
rate \IN{\%} & $\pm 0.3$ & $\pm 0.3$ & $\pm 0.3$ & $\pm 0.2$ & $\pm 0.2$ &\\
\hline
Detektor- &  5.5 &  4.5 &  3.8 &  3.0 &  2.7&\\
rate \IN{\%} & $\pm 0.3$ & $\pm 0.3$ & $\pm 0.3$ & $\pm 0.2$ & $\pm 0.2$ &\\
\hline
Korrektur- & 1.63 & 1.72 & 1.75 & 1.89 & 1.86 &\\
faktor & $\pm0.13$ & $\pm0.16$ & $\pm0.17$ & $\pm0.21$ & $\pm0.22$ &\\
\hline
\hline
\multicolumn{7}{|c|}{Bin 7}\\
\hline
$\tjet$ \IN{^{\circ}}&0&2&4&6&8&10\\
\hline
Parton- & 21.2 & 21.2 & 20.1 & 17.4 & 14.5 & 12.3\\
rate \IN{\%} & $\pm 0.6$ & $\pm 0.6$ & $\pm 0.5$ & $\pm 0.5$ & $\pm 0.5$ & $\pm 0.4$\\
\hline
Detektor- & 19.1 & 19.1 & 17.8 & 14.2 & 10.9 &  8.5\\
rate \IN{\%} & $\pm 0.6$ & $\pm 0.6$ & $\pm 0.6$ & $\pm 0.6$ & $\pm 0.5$ & $\pm 0.5$\\
\hline
Korrektur- & 1.11 & 1.11 & 1.13 & 1.23 & 1.34 & 1.45\\
faktor & $\pm0.06$ & $\pm0.06$ & $\pm0.07$ & $\pm0.08$ & $\pm0.10$ & $\pm0.12$\\
\hline
$\tjet$ \IN{^{\circ}}&12&14&16&18&20&\\
\hline
Parton- & 10.2 &  8.6 &  7.2 &  6.0 &  5.2 &\\
rate \IN{\%} & $\pm 0.4$ & $\pm 0.4$ & $\pm 0.4$ & $\pm 0.3$ & $\pm 0.3$ &\\
\hline
Detektor- &  6.6 &  5.3 &  4.3 &  3.8 &  3.1&\\
rate \IN{\%} & $\pm 0.4$ & $\pm 0.4$ & $\pm 0.4$ & $\pm 0.3$ & $\pm 0.3$ &\\
\hline
Korrektur- & 1.56 & 1.65 & 1.68 & 1.60 & 1.70 &\\
faktor & $\pm0.15$ & $\pm0.18$ & $\pm0.20$ & $\pm0.21$ & $\pm0.25$ &\\
\hline
\end{tabular}
\end{center}
\caption[Raten- und Korrekturfaktorbetrachtung bei $\tjet$ Schnitt]{{\bf
Raten- und Korrekturfaktorbetrachtung bei $\tjet$ Schnitt,
in $Q^2$ Bins}}
\label{tabtjcuta}
\end{table}

\begin{table}[tbp]
\begin{center}
\begin{tabular}{@{\extracolsep{\fill}}|c|c|c|c|c|c|c|}
\hline
\multicolumn{7}{|c|}{Bin 8}\\
\hline
$\tjet$ \IN{^{\circ}}&0&2&4&6&8&10\\
\hline
Parton- & 23.6 & 23.6 & 22.4 & 20.3 & 17.0 & 13.9\\
rate \IN{\%} & $\pm 0.9$ & $\pm 0.9$ & $\pm 0.9$ & $\pm 0.8$ & $\pm 0.8$ & $\pm 0.7$\\
\hline
Detektor- & 21.8 & 21.8 & 21.3 & 18.3 & 14.0 & 10.9\\
rate \IN{\%} & $\pm 1.0$ & $\pm 1.0$ & $\pm 1.0$ & $\pm 0.9$ & $\pm 0.8$ & $\pm 0.8$\\
\hline
Korrektur- & 1.09 & 1.09 & 1.06 & 1.12 & 1.22 & 1.28\\
faktor & $\pm0.09$ & $\pm0.09$ & $\pm0.09$ & $\pm0.10$ & $\pm0.12$ & $\pm0.15$\\
\hline
$\tjet$ \IN{^{\circ}}&12&14&16&18&20&\\
\hline
Parton- & 11.5 &  9.4 &  8.0 &  7.1 &  6.6 &\\
rate \IN{\%} & $\pm 0.7$ & $\pm 0.6$ & $\pm 0.6$ & $\pm 0.6$ & $\pm 0.6$ &\\
\hline
Detektor- &  8.2 &  6.7 &  5.4 &  4.5 &  4.2&\\
rate \IN{\%} & $\pm 0.7$ & $\pm 0.6$ & $\pm 0.6$ & $\pm 0.6$ & $\pm 0.6$ &\\
\hline
Korrektur- & 1.40 & 1.41 & 1.50 & 1.57 & 1.57 &\\
faktor & $\pm0.19$ & $\pm0.22$ & $\pm0.26$ & $\pm0.31$ & $\pm0.33$ &\\
\hline
\end{tabular}
\end{center}
\caption[Raten- und Korrekturfaktorbetrachtung bei $\tjet$ Schnitt]{{\bf
Raten- und Korrekturfaktorbetrachtung bei $\tjet$ Schnitt,
in $Q^2$ Bins}}
\label{tabtjcut}
\end{table}

\begin{figure}[tbp]                
\begin{center}
\epsfig{file=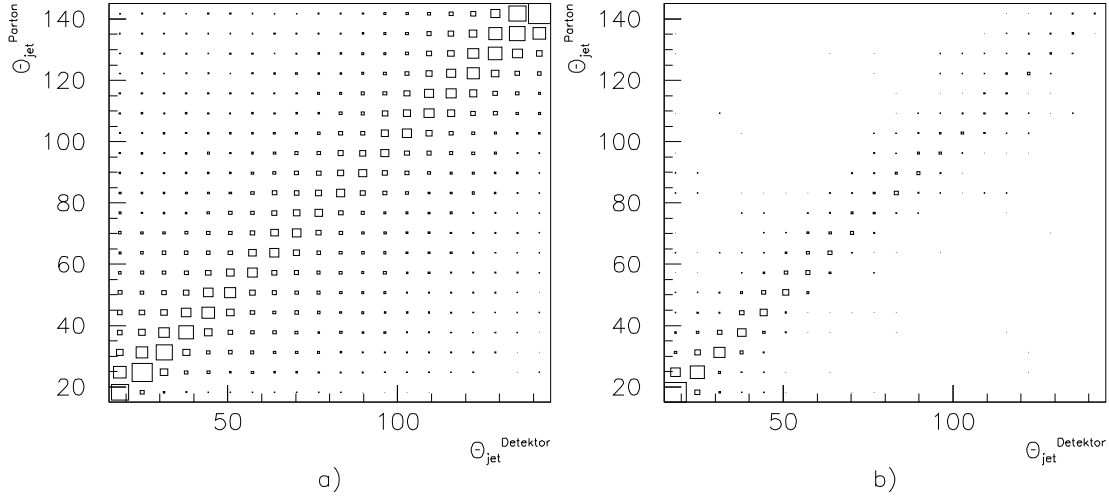,width=\hsize}
\end{center}
\caption[Jetwinkelkorrelation im mittleren $\tjet$ Bereich]{{\bf
Vergleich der Jetwinkel zwischen Parton- und Detektorniveau im Bereich
$20^{\circ} < \tjet < 145^{\circ}.$}{\it\ a)~alle Ereignisse,
b)~2+1 Ereignisse.}}
\label{abbtjetkorr}
\end{figure}

Betrachten wir nun den mittleren Jetwinkelbereich. In Abbildung
\ref{abbtjetkorr} ist die Korrelation des Jetwinkels auf Parton-
gegen den auf Detektorniveau gezeigt. In Bild a sind alle Ereignisse
verzeichnet. Die "Ubereinstimmung ist gut zu sehen, es erfolgen jedoch
auch viele Fehl-Erkennungen. Bei den 2+1 Ereignissen, die in Bild b
gezeigt sind, ist die Korrelation ebenfalls gut, die Statistik ist
jedoch begrenzt.

In Anbetracht der schlechten Beschreibung der Jetwinkel im Bereich
kleiner und gro"ser Winkel, ist ein Schnitt in dieser Gr"o"se sinnvoll.
Wir erreichen dadurch eine gute Beschreibung im mittleren Winkelbereich.
Dies ist aufgrund der guten Korrelation dort erlaubt. Der obere Schnitt
sollte einen Wert von 145 bis 150 Grad haben. Bei der unteren Grenze ist
die Angabe eines Wertes komplizierter. Ein zu tiefer Wert beh"alt einen
zu gro"sen Teil der Problemzone bei. Um eine gute "Ubereinstimmung der
beiden Monte-Carlo-Niveaus zu erreichen, w"are ein Schnitt von
mindestens 20 Grad angebracht. Dabei verliert man jedoch viel Statistik,
so da"s sich der entsprechende Fehler vergr"o"sert, und der
Korrekturfaktor steigt, was eine Ratenmessung problematisch werden
l"a"st. Der Schnitt sollte deshalb im Bereich zwischen zehn und 20 Grad
je nach Anforderungen an die Messung liegen. Ist eine hohe Statistik
erw"unscht, so sollten schw"achere Grenzen gew"ahlt werden, wobei sich
der systematische Fehler dann erh"oht. Wir w"ahlen die Grenzen 15 und
145 Grad.

\subsection{Skalenvariable $z_p$}

Eine weitere Gr"o"se, die ich untersucht habe, ist die Verteilung
in der Skalenvariable $z_p.$ Sie ist definiert f"ur 2+1 Jet Ereignisse
durch den Viererimpuls des einlaufenden Protons $\vv{P}$ und die 
Viererimpulse der beiden harten Jets $\vv{p}_i$
\begin{eqnarray}
z_{p,i} &=& {\vv{P} \cdot \vv{p}_i \over \vv{P} \cdot (\vv{p}_1 +
\vv{p}_2)} \\
&=& {1\over 2} \left( 1 - \cos(\theta_i^{*}) \right)
\end{eqnarray}
Der Winkel $\theta_i^{*}$ ist der Winkel zwischen den beiden harten
Jets im hadronischen Schwerpunktsystem. Es gelten die beiden
Beziehungen
\begin{eqnarray}
z_{p,1} &=& 1-z_{p,2} \\
\theta_1^{*} &=& 180 - \theta_2^{*}
\end{eqnarray}

\begin{figure}[tbp]                
\begin{center}
\epsfig{file=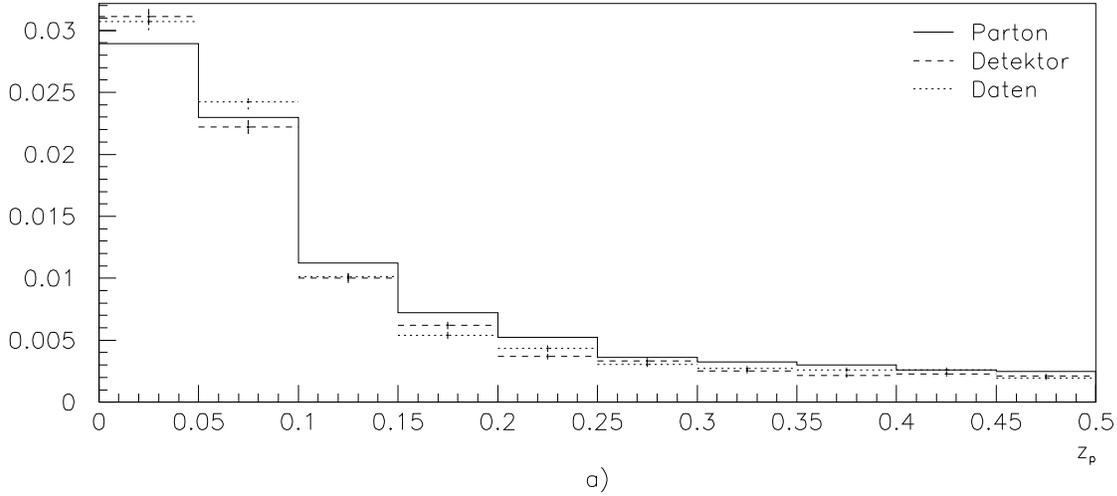,width=\hsize}
\end{center}
\caption[Vergleich der $z_p$ Verteilungen]{{\bf
Vergleich der $z_p$ Verteilungen auf Parton- und Detektorniveau
und bei den Daten nach Standardschnitten.}
{\it\ Dargestellt sind Parton- (durchgezogen) und Detektorniveau (gestrichelt)
der Lepto-Monte-Carlos und die Daten (gepunktet). Die Verteilungen sind auf
gleiche Fl"achen normiert.}}
\label{abbzp}
\end{figure}

Abbildung \ref{abbzp} zeigt die Verteilung des jeweils minimalen
$z_p = \min (z_{p,1},z_{p,2})$ f"ur Parton- und Detektorniveau
und f"ur die Daten. Die Verteilungen sind auf gleiche Fl"achen normiert.
Wir erkennen eine gute "Ubereinstimmung. Ein Schnitt in dieser Gr"o"se
wird keine "Anderung des Korrekturfaktors hervorrufen, da Parton- und 
Detektorniveau in gleicher Weise getroffen werden. 

Um jedoch einen Einflu"s des Schnittes auf die Migrationen und die
Reinheit zu untersuchen, berechnen wir die Raten in Abh"angigkeit
eines minimalen $z_p$ Schnittes. Tabelle \ref{tabzpmigmeps} zeigt das
Ergebnis f"ur die $Q^2$ Bins drei, sechs und acht f"ur drei 
unterschiedliche Schnittwerte. Diese Werte k"onnen mit denen in Tabelle
\ref{tabstdmigmeps} verglichen werden. F"ur die einzelnen Schnitte
gilt, da"s die Gesamtereigniszahl gleich bleibt, wobei jedoch einige
der 2+1 Ereignisse durch die Schnitte nicht mehr als 2+1 Ereignisse
gez"ahlt werden. Auch hier ist dieses Verfahren sinnvoller als ein
Verwerfen der Ereignisse, da Ereignisse, deren zweiter Jet nicht
gefunden wurde, ebenfalls zur Gesamtanzahl gerechnet werden.
Dadurch werden die Raten kleiner.
Wir erkennen, da"s der Einflu"s des Schnittes auf den Korrekturfaktor
im Rahmen der Fehler wie erwartet keine "Anderung bewirkt, wobei das
niedrige $Q^2$ Bin etwas empfindlicher reagiert. Die Reinheit
zeigt auch keine erkennbare Ver"anderung. 

\begin{table}[tbp]
\begin{center}
\begin{tabular}{@{\extracolsep{\fill}}c|c||c|c}
\multicolumn{2}{c||}{Bin 3} & \multicolumn{2}{c}{Parton} \\
\multicolumn{2}{c||}{Schnitt 1} & 1+1 & 2+1 \\
\hline
\hline
Clus- & 1+1 & 9330 & 188 \\
\cline{2-4}
ter & 2+1 & 168 & 198 \\
\end{tabular}
\hfill
\begin{tabular}{@{\extracolsep{\fill}}c|c||c|c}
\multicolumn{2}{c||}{Bin 3} & \multicolumn{2}{c}{Parton} \\
\multicolumn{2}{c||}{Schnitt 2} & 1+1 & 2+1 \\
\hline
\hline
Clus- & 1+1 & 9563 & 123 \\
\cline{2-4}
ter & 2+1 & 91 & 107 \\
\end{tabular}
\hfill
\begin{tabular}{@{\extracolsep{\fill}}c|c||c|c}
\multicolumn{2}{c||}{Bin 3} & \multicolumn{2}{c}{Parton} \\
\multicolumn{2}{c||}{Schnitt 3} & 1+1 & 2+1 \\
\hline
\hline
Clus- & 1+1 & 9668 & 85 \\
\cline{2-4}
ter & 2+1 & 60 & 71 \\
\end{tabular}
\end{center}
\vspace {0.25cm}
\begin{center}
\begin{tabular}{@{\extracolsep{\fill}}c|c||c|c}
\multicolumn{2}{c||}{Bin 6} & \multicolumn{2}{c}{Parton} \\
\multicolumn{2}{c||}{Schnitt 1} & 1+1 & 2+1 \\
\hline
\hline
Clus- & 1+1 & 6367 & 306 \\
\cline{2-4}
ter & 2+1 & 177 & 392 \\
\end{tabular}
\hfill
\begin{tabular}{@{\extracolsep{\fill}}c|c||c|c}
\multicolumn{2}{c||}{Bin 6} & \multicolumn{2}{c}{Parton} \\
\multicolumn{2}{c||}{Schnitt 2} & 1+1 & 2+1 \\
\hline
\hline
Clus- & 1+1 & 6678 & 195 \\
\cline{2-4}
ter & 2+1 & 118 & 251 \\
\end{tabular}
\hfill
\begin{tabular}{@{\extracolsep{\fill}}c|c||c|c}
\multicolumn{2}{c||}{Bin 6} & \multicolumn{2}{c}{Parton} \\
\multicolumn{2}{c||}{Schnitt 3} & 1+1 & 2+1 \\
\hline
\hline
Clus- & 1+1 & 6852 & 140 \\
\cline{2-4}
ter & 2+1 & 67 & 183 \\
\end{tabular}
\end{center}
\vspace {0.25cm}
\begin{center}
\begin{tabular}{@{\extracolsep{\fill}}c|c||c|c}
\multicolumn{2}{c||}{Bin 8} & \multicolumn{2}{c}{Parton} \\
\multicolumn{2}{c||}{Schnitt 1} & 1+1 & 2+1 \\
\hline
\hline
Clus- & 1+1 & 1431 & 95 \\
\cline{2-4}
ter & 2+1 & 71 & 229 \\
\end{tabular}
\hfill
\begin{tabular}{@{\extracolsep{\fill}}c|c||c|c}
\multicolumn{2}{c||}{Bin 8} & \multicolumn{2}{c}{Parton} \\
\multicolumn{2}{c||}{Schnitt 2} & 1+1 & 2+1 \\
\hline
\hline
Clus- & 1+1 & 1536 & 82 \\
\cline{2-4}
ter & 2+1 & 52 & 156 \\
\end{tabular}
\hfill
\begin{tabular}{@{\extracolsep{\fill}}c|c||c|c}
\multicolumn{2}{c||}{Bin 8} & \multicolumn{2}{c}{Parton} \\
\multicolumn{2}{c||}{Schnitt 3} & 1+1 & 2+1 \\
\hline
\hline
Clus- & 1+1 & 1596 & 69 \\
\cline{2-4}
ter & 2+1 & 44 & 117 \\
\end{tabular}
\end{center}
\vspace {0.25cm}
\begin{center}
\begin{tabular}{|c||c|c|c|c|c|c|c|c|c|}
\hline
Bin & \multicolumn{3}{|c|}{3} & \multicolumn{3}{|c|}{6} & \multicolumn{3}{|c|}{8}\\
\hline
Schnitt & 1 & 2 & 3 & 1 & 2 & 3 & 1 & 2 & 3 \\
\hline
Parton&  3.9 &  2.3 &  1.6 &  9.6 &  6.2 &  4.5 & 17.7 & 13.0 & 10.2\\
in \% & $\pm 0.2$&$\pm 0.2$&$\pm 0.2$&$\pm 0.4$&$\pm 0.3$&$\pm 0.3$&$\pm 0.9$&$\pm 0.8$&$\pm 0.8$\\
\hline
Detektor&  3.7 &  2.0 &  1.3 &  7.9 &  5.1 &  3.5 & 16.4 & 11.4 &  8.8\\
in \%&$\pm 0.2$&$\pm 0.2$&$\pm 0.2$&$\pm 0.4$&$\pm 0.3$&$\pm 0.3$&$\pm 0.9$&$\pm 0.8$&$\pm 0.7$\\
\hline
Korrek- & 1.05 & 1.16 & 1.19 & 1.23 & 1.21 & 1.29 & 1.08 & 1.14 & 1.16\\
turfa. &$\pm0.11$&$\pm0.16$&$\pm0.20$&$\pm0.10$&$\pm0.12$&$\pm0.16$&$\pm0.12$&$\pm0.15$&$\pm0.17$\\
\hline
Reinheit & 54.1 & 54.0 & 54.2 & 68.9 & 68.0 & 73.2 & 76.3 & 75.0 & 72.7\\
in \%&&&&&&&&&\\
\hline
\end{tabular}
\end{center}
\caption[Migrationen und Korrekturfaktoren mit verschiedenen $z_p$ Schnitten]
{{\bf Migrationen und Korrekturfaktoren mit verschiedenen $z_p$ Schnitten}
{\it\ Exemplarisch werden drei $Q^2$ Bins mit drei verschiedenen
$z_p$ Schnitten gezeigt. Schnitt 1 ist hier $z_p>0.05$, Schnitt 2 $z_p>0.10$
und Schnitt 3 $z_p>0.15.$}}
\label{tabzpmigmeps}
\end{table}

Ein Schnitt in $z_p$ erscheint aus diesen Gr"unden nicht n"otig,
aber auch nicht verboten.

\subsection{Transversaler Impuls $p_t^*$}
\label{kapupt}

Der transversale Impuls der Jets im hadronischen Schwerpunktsystem ist
ebenfalls untersucht worden. Wir erwarten hier nur geringe Impulswerte
f"ur 1+1 Jet Ereignisse, da der Protonrest den Detektor weiterhin in 
$z$-Richtung verl"a"st. Bei 2+1 Jet Ereignissen sollten die transversalen
Impulse der beiden harten Jets die gleiche Gr"o"se haben, da die 
transversale Impulssumme der harten Jets ebenfalls klein sein sollte.

\begin{figure}[tbp]                
\begin{center}
\epsfig{file=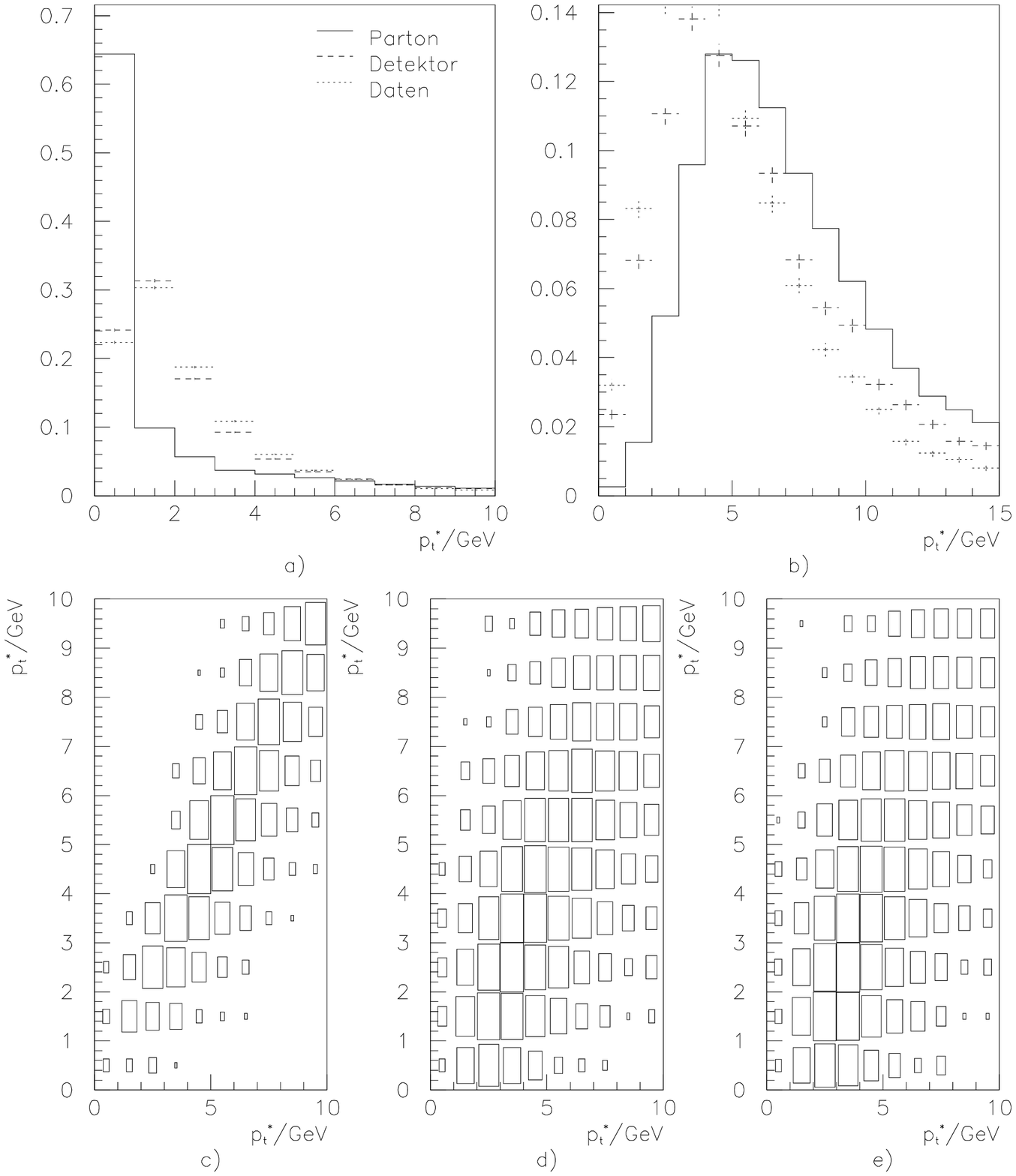,width=0.9\hsize}
\end{center}
\caption[Vergleich der $p_t^{*}$ Verteilungen]{{\bf
Vergleich der $p_t^{*}$ Verteilungen auf Parton- und Detektorniveau
und bei den Daten nach Standardschnitten.}
{\it\ a)~Verteilung f"ur alle Ereignisse, dargestellt sind Parton-
(durchgezogen) und Detektorniveau (gestrichelt) und die Daten
(gepunktet), b)~Verteilung f"ur 2+1 Jet Ereignisse, c,d,e)~Korrelation
des transversalen Impulses f"ur beide harte Jets mit logarithmischer
Boxengr"o"se bei 2+1 Ereignissen f"ur Parton- (c), Detektorniveau (d)
und Daten (e)}}
\label{abbpt}
\end{figure}

Die zugeh"origen Verteilungen sind in Abbildung \ref{abbpt} dargestellt.
Im Bild links oben ist die $p_t^{*}$ Verteilung f"ur alle Ereignisse
in den drei Niveaus, daneben f"ur die 2+1 Ereignisse gezeigt. 
Wir erkennen, da"s der mittlere $p_t^{*}$ Wert f"ur die
2+1 Verteilungen klar von Null abweicht, w"ahrend die Verteilung
f"ur alle Ereignisse bei Null einen Peak zeigt. Im
unteren Teil ist die Korrelation der beiden $p_t^{*}$ Werte der
harten Jets f"ur 2+1 Ereignisse abgebildet. Die Gr"o"se der Boxen ist
wieder logarithmisch.

Die Korrelationen auf Partonniveau sind gut. Auf Detektorniveau und bei
den Daten zeigt die Gegen"uberstellung schlechtere Wert. Dies liegt an
der Ungenauigkeit des transversalen Impulses im Laborsystem, die sich auch
hier zeigen, da die Lo"-rentz"-transformation fast ausschlie"slich in
$z$ Richtung stattfindet. Im Rahmen der in Kapitel \ref{kaprekon}
betrachteten Impulsverteilung ist die Korrelation auch hier gut.

Vergleichen wir die Niveaus in Abbildung \ref{abbpt}a, so f"allt auf, 
da"s auf Detektorniveau und bei den Daten weniger Jets mit einem transversalen
Impuls kleiner als $1\;\GeV$ gefunden werden. Dies ist aufgrund der begrenzten
Energieaufl"osung erkl"arbar. Insbesondere ist die Berechnung der
Lorentztransformation auf die Genauigkeit der Rekonstruktion des
hadronischen Endzustandes angewiesen. Dies verursacht wie erw"ahnt
auch die Verbreiterung der Korrelationen in den Bildern c bis e, die in 
der gleichen Gr"o"senordnung liegen. Bei gr"o"seren transversalen Impulsen 
stimmen jedoch alle drei Verteilungen unter Beachtung der ge"anderten
Normierung einigerma"sen "uberein. Dies gilt ebenso f"ur die
2+1 Ereignisse, da deren transversale Impulse, wie bereits erw"ahnt,
gr"o"ser sind. Die Beschreibung der Daten gelingt auch hier durch
das Detektorniveau sehr gut.

Bei einem Schnitt in den transversalen Impuls in der Gr"o"se von
$2\;\GeV$ erwarten wir also den Wegfall vieler 1+1 Jet Ereignisse,
jedoch nur weniger mit zwei harten Jets. Dies wird in Tabelle
\ref{tabpt2migmeps} an den Migrationsmatrizen deutlich. Die Raten
sind stark angestiegen. Beim Vergleich mit Tabelle \ref{tabstdmigmeps}
stellen wir einen Anstieg des Korrekturfaktorwertes fest. Die Variation
ist mit 1.26 bis 1.40 kleiner und alle Werte sind im Rahmen ihrer
Fehler mit dem Mittelwert 1.33 vertr"aglich. Die Reinheit der 2+1 Jetklasse
ist in allen Bins deutlich gestiegen. Dieser Schnitt verbessert auch
aufgrund der ge"anderten Normierung die Verteilungen der transversalen
Impulse.

\begin{table}[tbp]
\begin{center}
\begin{tabular}{@{\extracolsep{\fill}}c|c||c|c}
\multicolumn{2}{c||}{Bin} & \multicolumn{2}{c}{Parton} \\
\multicolumn{2}{c||}{1} & 1+1 & 2+1 \\
\hline
\hline
Clus- & 1+1 & 607 & 221 \\
\cline{2-4}
ter & 2+1 & 108 & 326 \\
\end{tabular}
\hfill
\begin{tabular}{@{\extracolsep{\fill}}c|c||c|c}
\multicolumn{2}{c||}{Bin} & \multicolumn{2}{c}{Parton} \\
\multicolumn{2}{c||}{2} & 1+1 & 2+1 \\
\hline
\hline
Clus- & 1+1 & 419 & 172 \\
\cline{2-4}
ter & 2+1 & 72 & 247 \\
\end{tabular}
\hfill
\begin{tabular}{@{\extracolsep{\fill}}c|c||c|c}
\multicolumn{2}{c||}{Bin} & \multicolumn{2}{c}{Parton} \\
\multicolumn{2}{c||}{3} & 1+1 & 2+1 \\
\hline
\hline
Clus- & 1+1 & 453 & 204 \\
\cline{2-4}
ter & 2+1 & 78 & 269 \\
\end{tabular}
\end{center}
\vspace {0.25cm}
\begin{center}
\begin{tabular}{@{\extracolsep{\fill}}c|c||c|c}
\multicolumn{2}{c||}{Bin} & \multicolumn{2}{c}{Parton} \\
\multicolumn{2}{c||}{4} & 1+1 & 2+1 \\
\hline
\hline
Clus- & 1+1 & 404 & 230 \\
\cline{2-4}
ter & 2+1 & 87 & 330 \\
\end{tabular}
\hfill
\begin{tabular}{@{\extracolsep{\fill}}c|c||c|c}
\multicolumn{2}{c||}{Bin} & \multicolumn{2}{c}{Parton} \\
\multicolumn{2}{c||}{5} & 1+1 & 2+1 \\
\hline
\hline
Clus- & 1+1 & 320 & 176 \\
\cline{2-4}
ter & 2+1 & 61 & 274 \\
\end{tabular}
\hfill
\begin{tabular}{@{\extracolsep{\fill}}c|c||c|c}
\multicolumn{2}{c||}{Bin} & \multicolumn{2}{c}{Parton} \\
\multicolumn{2}{c||}{6} & 1+1 & 2+1 \\
\hline
\hline
Clus- & 1+1 & 811 & 369 \\
\cline{2-4}
ter & 2+1 & 105 & 556 \\
\end{tabular}
\end{center}
\vspace {0.25cm}
\begin{center}
\begin{tabular}{@{\extracolsep{\fill}}c|c||c|c}
\multicolumn{2}{c||}{Bin} & \multicolumn{2}{c}{Parton} \\
\multicolumn{2}{c||}{7} & 1+1 & 2+1 \\
\hline
\hline
Clus- & 1+1 & 543 & 283 \\
\cline{2-4}
ter & 2+1 & 97 & 560 \\
\end{tabular}
\hfill
\begin{tabular}{@{\extracolsep{\fill}}c|c||c|c}
\multicolumn{2}{c||}{Bin} & \multicolumn{2}{c}{Parton} \\
\multicolumn{2}{c||}{8} & 1+1 & 2+1 \\
\hline
\hline
Clus- & 1+1 & 239 & 127 \\
\cline{2-4}
ter & 2+1 & 33 & 274 \\
\end{tabular}
\hfill
\hphantom{
\begin{tabular}{@{\extracolsep{\fill}}c|c||c|c}
\multicolumn{2}{c||}{} & \multicolumn{2}{c}{Parton} \\
\multicolumn{2}{c||}{} & 1+1 & 2+1 \\
\hline
\hline
Clus- & 1+1 & & \\
\cline{2-4}
ter & 2+1 & & \\
\end{tabular}
}
\end{center}
\vspace {0.25cm}
\begin{center}
\begin{tabular}{@{\extracolsep{\fill}}|c||c|c|c|c|c|c|c|c|}
\hline
Bin & 1 & 2 & 3 & 4 & 5 & 6 & 7 & 8 \\
\hline
Rate Parton& 43.3 & 46.0 & 47.1 & 53.3 & 54.2 & 50.2 & 56.8 & 59.6\\
in \% & $\pm 1.4$&$\pm 1.7$&$\pm 1.6$&$\pm 1.6$&$\pm 1.8$&$\pm 1.2$&$\pm 1.3$&$\pm 1.9$\\
\hline
Rate Detektor& 34.4 & 35.1 & 34.6 & 39.7 & 40.3 & 35.9 & 44.3 & 45.6\\
in \%&$\pm 1.4$&$\pm 1.6$&$\pm 1.6$&$\pm 1.6$&$\pm 1.8$&$\pm 1.2$&$\pm 1.3$&$\pm 2.0$\\
\hline
Korrektur- & 1.26 & 1.31 & 1.36 & 1.34 & 1.34 & 1.40 & 1.28 & 1.31\\
faktor &$\pm0.09$&$\pm0.11$&$\pm0.11$&$\pm0.09$&$\pm0.10$&$\pm0.08$&$\pm0.07$&$\pm0.10$\\
\hline
Reinheit in \%& 75.1 & 77.4 & 77.5 & 79.1 & 81.8 & 84.1 & 85.2 & 89.3\\
\hline
\end{tabular}
\end{center}
\caption[Migrationen und Korrekturfaktoren bei Lepto, $p_t^{*}$ Schnitt]{{\bf
Migrationen und Korrekturfaktoren bei Lepto mit Standardschnitten und 
zus"atzlichem $p_t^{*} > 2\;\GeV$ Schnitt,
in $Q^2$ Bins}}
\label{tabpt2migmeps}
\end{table}

Die Qualit"at des Schnittes tr"ugt jedoch. Dies sieht man an der
Raten"anderung in Abh"angigkeit der Gr"o"se des $p_t^{*}$ Schnittes.
Wir erkennen hier sehr gro"se Korrekturfaktoren. Der Unterschied
zwischen den Detektorraten bei den Migrationsmatrizen und bei der Berechnung 
in der Tabelle \ref{tabptcut} liegt darin begr"undet, da"s bei den
Migrationsmatrizen nur die Ereignisse gezeigt werden, die die Schnitte
auf beiden Niveaus erf"ullen. Viele Ereignisse haben aber, wie wir in
Abbildung \ref{abbpt}a gesehen haben, einen verschwindenden transversalen
Impuls auf Partonniveau, jedoch einen h"oheren auf Detektorniveau. Eine 
Stabilisierung auf hohem Niveau tritt erst ein, wenn die Verteilung
der transversalen Impulse auf beiden Niveaus "ubereinstimmt.

\begin{table}[tbp]
\begin{center}
\begin{tabular}{@{\extracolsep{\fill}}|c|c|c|c|c|c|c|}
\hline
\multicolumn{7}{|c|}{Bin 6}\\
\hline
$p_{t,\mb{min}}^{*}$ \IN{\GeV}&0&1&2&3&4&5\\
\hline
Parton- & 14.6 & 32.2 & 46.9 & 60.6 & 68.9 & 73.3\\
rate \IN{\%} & $\pm 0.4$ & $\pm 0.7$ & $\pm 1.0$ & $\pm 1.1$ & $\pm 1.2$ & $\pm 1.2$\\
\hline
Detektor- & 12.4 & 14.2 & 18.5 & 23.7 & 27.6 & 31.4\\
rate \IN{\%} & $\pm 0.4$ & $\pm 0.5$ & $\pm 0.6$ & $\pm 0.8$ & $\pm 1.0$ & $\pm 1.2$\\
\hline
Korrektur- & 1.19 & 2.28 & 2.55 & 2.56 & 2.50 & 2.34\\
faktor & $\pm0.07$ & $\pm0.12$ & $\pm0.13$ & $\pm0.13$ & $\pm0.13$ & $\pm0.12$\\
\hline
$p_{t,\mb{min}}^{*}$ \IN{\GeV}&6&7&8&9&10&\\
\hline
Parton- & 75.9 & 77.3 & 77.9 & 75.5 & 74.3 &\\
rate \IN{\%} & $\pm 1.3$ & $\pm 1.4$ & $\pm 1.5$ & $\pm 1.7$ & $\pm 1.9$ &\\
\hline
Detektor- & 34.5 & 38.3 & 41.0 & 42.6 & 45.1&\\
rate \IN{\%} & $\pm 1.4$ & $\pm 1.7$ & $\pm 1.9$ & $\pm 2.2$ & $\pm 2.6$ &\\
\hline
Korrektur- & 2.21 & 2.02 & 1.90 & 1.77 & 1.65 &\\
faktor & $\pm0.12$ & $\pm0.12$ & $\pm0.12$ & $\pm0.13$ & $\pm0.13$ &\\
\hline
\hline
\multicolumn{7}{|c|}{Bin 7}\\
\hline
$p_{t,\mb{min}}^{*}$ \IN{\GeV}&0&1&2&3&4&5\\
\hline
Parton- & 21.2 & 37.6 & 53.6 & 65.5 & 72.3 & 77.4\\
rate \IN{\%} & $\pm 0.6$ & $\pm 0.9$ & $\pm 1.1$ & $\pm 1.2$ & $\pm 1.3$ & $\pm 1.3$\\
\hline
Detektor- & 19.1 & 20.3 & 23.5 & 26.8 & 29.4 & 30.7\\
rate \IN{\%} & $\pm 0.6$ & $\pm 0.7$ & $\pm 0.8$ & $\pm 0.9$ & $\pm 1.1$ & $\pm 1.2$\\
\hline
Korrektur- & 1.11 & 1.86 & 2.29 & 2.45 & 2.46 & 2.52\\
faktor & $\pm0.06$ & $\pm0.10$ & $\pm0.12$ & $\pm0.12$ & $\pm0.13$ & $\pm0.14$\\
\hline
$p_{t,\mb{min}}^{*}$ \IN{\GeV}&6&7&8&9&10&\\
\hline
Parton- & 78.1 & 78.6 & 77.7 & 77.9 & 76.9 &\\
rate \IN{\%} & $\pm 1.4$ & $\pm 1.5$ & $\pm 1.7$ & $\pm 1.8$ & $\pm 2.0$ &\\
\hline
Detektor- & 32.0 & 31.6 & 31.7 & 32.0 & 33.1&\\
rate \IN{\%} & $\pm 1.4$ & $\pm 1.6$ & $\pm 1.8$ & $\pm 2.0$ & $\pm 2.3$ &\\
\hline
Korrektur- & 2.45 & 2.49 & 2.46 & 2.44 & 2.33 &\\
faktor & $\pm0.15$ & $\pm0.17$ & $\pm0.19$ & $\pm0.21$ & $\pm0.22$ &\\
\hline
\end{tabular}
\end{center}
\caption[Raten- und Korrekturfaktorbetrachtung bei $p_t^{*}$ Schnitt]{{\bf
Raten- und Korrekturfaktorbetrachtung bei $p_t^{*}$ Schnitt,
in $Q^2$ Bins}}
\label{tabptcuta}
\end{table}

\begin{table}[tbp]
\begin{center}
\begin{tabular}{@{\extracolsep{\fill}}|c|c|c|c|c|c|c|}
\hline
\multicolumn{7}{|c|}{Bin 8}\\
\hline
$p_{t,\mb{min}}^{*}$ \IN{\GeV}&0&1&2&3&4&5\\
\hline
Parton- & 23.6 & 39.4 & 54.7 & 66.1 & 73.9 & 77.8\\
rate \IN{\%} & $\pm 0.9$ & $\pm 1.3$ & $\pm 1.6$ & $\pm 1.7$ & $\pm 1.8$ & $\pm 1.8$\\
\hline
Detektor- & 21.8 & 22.7 & 25.2 & 27.4 & 29.2 & 31.2\\
rate \IN{\%} & $\pm 1.0$ & $\pm 1.0$ & $\pm 1.2$ & $\pm 1.3$ & $\pm 1.5$ & $\pm 1.7$\\
\hline
Korrektur- & 1.09 & 1.74 & 2.17 & 2.42 & 2.54 & 2.50\\
faktor & $\pm0.09$ & $\pm0.13$ & $\pm0.16$ & $\pm0.18$ & $\pm0.18$ & $\pm0.19$\\
\hline
$p_{t,\mb{min}}^{*}$ \IN{\GeV}&6&7&8&9&10&\\
\hline
Parton- & 81.1 & 81.9 & 79.0 & 79.4 & 79.5 &\\
rate \IN{\%} & $\pm 1.8$ & $\pm 1.9$ & $\pm 2.1$ & $\pm 2.3$ & $\pm 2.5$ &\\
\hline
Detektor- & 32.3 & 34.0 & 32.4 & 31.4 & 32.0&\\
rate \IN{\%} & $\pm 1.9$ & $\pm 2.1$ & $\pm 2.2$ & $\pm 2.4$ & $\pm 2.6$ &\\
\hline
Korrektur- & 2.51 & 2.41 & 2.44 & 2.53 & 2.49 &\\
faktor & $\pm0.20$ & $\pm0.20$ & $\pm0.23$ & $\pm0.26$ & $\pm0.28$ &\\
\hline
\end{tabular}
\end{center}
\caption[Raten- und Korrekturfaktorbetrachtung bei $p_t^{*}$ Schnitt]{{\bf
Raten- und Korrekturfaktorbetrachtung bei $p_t^{*}$ Schnitt,
in $Q^2$ Bins}}
\label{tabptcut}
\end{table}

Ein Schnitt in dieser Gr"o"se ist gef"ahrlich, da die Verteilung in keinem
Bereich gut beschrieben wird.

\subsection{Hadronische Schwerpunktenergie $\hat s$}
\label{kapushat}

Die bereits in Kapitel \ref{kapkine} in Gleichung \ref{eqnshat}
eingef"uhrte Variable $\hat s$ kann in der Praxis f"ur 2+1 Jet
Ereignisse durch zwei unterschiedliche Methoden berechnet werden. Die
eine berechnet die invariante Masse des harten hadronischen Endzustandes
und die andere benutzt die Jetwinkel im hadronischen Schwerpunktsystem.
W"ahrend die zweite Berechnung eindeutig vom Jetalgorithmus abh"angig
ist, trifft dies f"ur die erste Methode nur in abgeschw"achter Form zu.
Das Problem bei der Berechnung "uber den hadronischen Endzustand ist,
da"s nur die Cluster verwendet werden d"urfen, die zu den harten Jets
geh"oren, nicht jedoch die es Protonrestes. Dazu ist jedoch ein
Jetalgorithmus n"otig.

Die Berechnung kann dann durch die Formeln
\begin{eqnarray}
\hat s^{\had} &=& \left(\displaystyle\sum_{\mb{hadr.Partikel}} \vv{p}_i 
\right)^2 \\
\hat s^{\mb{jet}} &=& W^2 e^{\displaystyle (\eta_1^{*}+\eta_2^{*})}
\end{eqnarray}                                    
mit den Pseudorapidit"aten $\eta_i^{*}$ der beiden harten Jets im 
hadronischen Schwerpunktsystem erfolgen.

\begin{figure}[tbp]                
\begin{center}
\epsfig{file=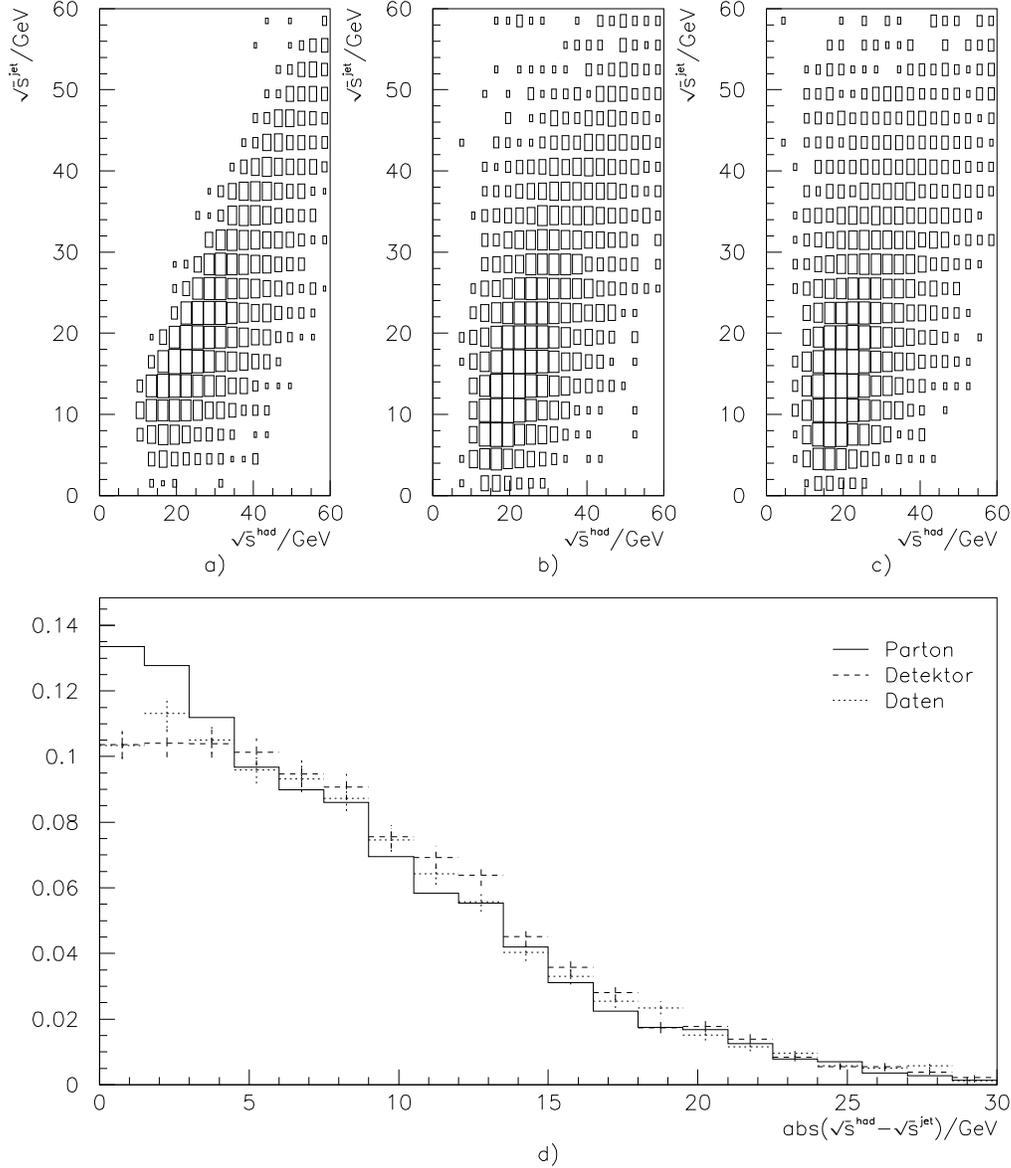,width=0.9\hsize}
\end{center}
\caption[Vergleich der $\hat s$ Verteilungen]{{\bf
Vergleich der $\sqrt{\hat s}$ Verteilungen auf Parton- und Detektorniveau
und bei den Daten nach Standardschnitten}
{\it\ a,b,c)~Korrelation der Jet- und der Summations-Methode mit
logarithmischer Boxenskalierung f"ur
Parton- (a), Detektorniveau (b) und Daten (c). d)~Absolutbetrag der
Differenz der Methoden. Dargestellt sind Parton- (durchgezogen) und
Detektorniveau (gestrichelt) und die Daten (gepunktet).}}
\label{abbshat}
\end{figure}

In Abbildung \ref{abbshat} sind die Werte gegen"ubergestellt, die
mit den unterschiedlichen Methoden berechnet wurden. Aufgetragen wurde
die Wurzel von $\hat s.$ Dabei ist oben f"ur Daten, Parton- und
Detektorniveau die Korrelation dargestellt und in Bild d der
Absolutbetrag der Differenz aufgetragen. Wir erkennen, da"s die
Verteilung in allen Niveaus "ahnlich ist, nur im Bereich kleiner
Unerschiede macht sich die zus"atzliche Begrenzung durch die
Energieaufl"osung bemerkbar. Die Korrelationen zwischen den beiden
Berechnungen ist jedoch nicht sehr gut. Die Skalierung der Boxen ist
logarithmisch.

Die zugeh"origen Migrationen wurden berechnet und in Tabelle
\ref{tabshatmigmeps} exemplarisch f"ur die $Q^2$ Bins drei,
sechs und acht mit drei verschiedenen Schnitten aufgetragen. Der
h"ochste Schnitt 3 ist hier der loseste und gibt deshalb die h"ochste
Statistik. Der Korrekturfaktor "andert sich nur im Bin acht, dort
vergr"o"sert er sich jedoch stark. Die Reinheit verschlechtert sich
in allen Bins mit st"arkerem Schnitt. 

\begin{table}[tbp]
\begin{center}
\begin{tabular}{@{\extracolsep{\fill}}c|c||c|c}
\multicolumn{2}{c||}{Bin 3} & \multicolumn{2}{c}{Parton} \\
\multicolumn{2}{c||}{Schnitt 1} & 1+1 & 2+1 \\
\hline
\hline
Clus- & 1+1 & 9507 & 147 \\
\cline{2-4}
ter & 2+1 & 157 & 73 \\
\end{tabular}
\hfill
\begin{tabular}{@{\extracolsep{\fill}}c|c||c|c}
\multicolumn{2}{c||}{Bin 3} & \multicolumn{2}{c}{Parton} \\
\multicolumn{2}{c||}{Schnitt 2} & 1+1 & 2+1 \\
\hline
\hline
Clus- & 1+1 & 9233 & 235 \\
\cline{2-4}
ter & 2+1 & 251 & 165 \\
\end{tabular}
\hfill
\begin{tabular}{@{\extracolsep{\fill}}c|c||c|c}
\multicolumn{2}{c||}{Bin 3} & \multicolumn{2}{c}{Parton} \\
\multicolumn{2}{c||}{Schnitt 3} & 1+1 & 2+1 \\
\hline
\hline
Clus- & 1+1 & 9043 & 294 \\
\cline{2-4}
ter & 2+1 & 314 & 233 \\
\end{tabular}
\end{center}
\vspace {0.25cm}
\begin{center}
\begin{tabular}{@{\extracolsep{\fill}}c|c||c|c}
\multicolumn{2}{c||}{Bin 6} & \multicolumn{2}{c}{Parton} \\
\multicolumn{2}{c||}{Schnitt 1} & 1+1 & 2+1 \\
\hline
\hline
Clus- & 1+1 & 6674 & 261 \\
\cline{2-4}
ter & 2+1 & 184 & 123 \\
\end{tabular}
\hfill
\begin{tabular}{@{\extracolsep{\fill}}c|c||c|c}
\multicolumn{2}{c||}{Bin 6} & \multicolumn{2}{c}{Parton} \\
\multicolumn{2}{c||}{Schnitt 2} & 1+1 & 2+1 \\
\hline
\hline
Clus- & 1+1 & 6352 & 342 \\
\cline{2-4}
ter & 2+1 & 261 & 287 \\
\end{tabular}
\hfill
\begin{tabular}{@{\extracolsep{\fill}}c|c||c|c}
\multicolumn{2}{c||}{Bin 6} & \multicolumn{2}{c}{Parton} \\
\multicolumn{2}{c||}{Schnitt 3} & 1+1 & 2+1 \\
\hline
\hline
Clus- & 1+1 & 6125 & 404 \\
\cline{2-4}
ter & 2+1 & 288 & 425 \\
\end{tabular}
\end{center}
\vspace {0.25cm}
\begin{center}
\begin{tabular}{@{\extracolsep{\fill}}c|c||c|c}
\multicolumn{2}{c||}{Bin 8} & \multicolumn{2}{c}{Parton} \\
\multicolumn{2}{c||}{Schnitt 1} & 1+1 & 2+1 \\
\hline
\hline
Clus- & 1+1 & 1540 & 149 \\
\cline{2-4}
ter & 2+1 & 56 & 81 \\
\end{tabular}
\hfill
\begin{tabular}{@{\extracolsep{\fill}}c|c||c|c}
\multicolumn{2}{c||}{Bin 8} & \multicolumn{2}{c}{Parton} \\
\multicolumn{2}{c||}{Schnitt 2} & 1+1 & 2+1 \\
\hline
\hline
Clus- & 1+1 & 1412 & 171 \\
\cline{2-4}
ter & 2+1 & 77 & 166 \\
\end{tabular}
\hfill
\begin{tabular}{@{\extracolsep{\fill}}c|c||c|c}
\multicolumn{2}{c||}{Bin 8} & \multicolumn{2}{c}{Parton} \\
\multicolumn{2}{c||}{Schnitt 3} & 1+1 & 2+1 \\
\hline
\hline
Clus- & 1+1 & 1361 & 153 \\
\cline{2-4}
ter & 2+1 & 85 & 227 \\
\end{tabular}
\end{center}
\vspace {0.25cm}
\begin{center}
\begin{tabular}{|c||c|c|c|c|c|c|c|c|c|}
\hline
Bin & \multicolumn{3}{|c|}{3} & \multicolumn{3}{|c|}{6} & \multicolumn{3}{|c|}{8}\\
\hline
Schnitt & 1 & 2 & 3 & 1 & 2 & 3 & 1 & 2 & 3 \\
\hline
Parton&  2.2 &  4.0 &  5.3 &  5.3 &  8.7 & 11.4 & 12.6 & 18.5 & 20.8\\
in \% & $\pm 0.2$&$\pm 0.2$&$\pm 0.3$&$\pm 0.3$&$\pm 0.4$&$\pm 0.4$&$\pm 0.8$&$\pm 1.0$&$\pm 1.0$\\
\hline
Detektor&  2.3 &  4.2 &  5.5 &  4.2 &  7.6 &  9.8 &  7.5 & 13.3 & 17.1\\
in \%&$\pm 0.2$&$\pm 0.3$&$\pm 0.3$&$\pm 0.3$&$\pm 0.4$&$\pm 0.4$&$\pm 0.7$&$\pm 0.8$&$\pm 0.9$\\
\hline
Korrek- & 0.96 & 0.96 & 0.96 & 1.25 & 1.15 & 1.16 & 1.68 & 1.39 & 1.22\\
turfa. &$\pm0.13$&$\pm0.10$&$\pm0.09$&$\pm0.14$&$\pm0.10$&$\pm0.08$&$\pm0.25$&$\pm0.16$&$\pm0.12$\\
\hline
Reinheit & 31.7 & 39.7 & 42.6 & 40.1 & 52.4 & 59.6 & 59.1 & 68.3 & 72.8\\
in \%&&&&&&&&&\\
\hline
\end{tabular}
\end{center}
\caption[Migrationen und Korrekturfaktoren mit verschiedenen $\Delta \hat s$ 
Schnitten]
{{\bf Migrationen und Korrekturfaktoren mit verschiedenen $\Delta\sqrt{\hat s}$ 
Schnitten}
{\it\ Exemplarisch werden drei $Q^2$ Bins mit drei verschiedenen
$\Delta\sqrt{\hat s}$ Schnitten gezeigt. Schnitt 1 ist hier
$\Delta\sqrt{\hat s}<5\;\GeV$, Schnitt 2 $\Delta\sqrt{\hat s}<10\;\GeV$
und Schnitt 3 $\Delta\sqrt{\hat s}<15\;\GeV.$}}
\label{tabshatmigmeps}
\end{table}

Ein Schnitt in dieser Variable ist nicht empfehlenswert.
 
\subsection{Rapidit"atsdifferenz $\Delta \eta$}

\begin{figure}[tb]                
\begin{center}
\epsfig{file=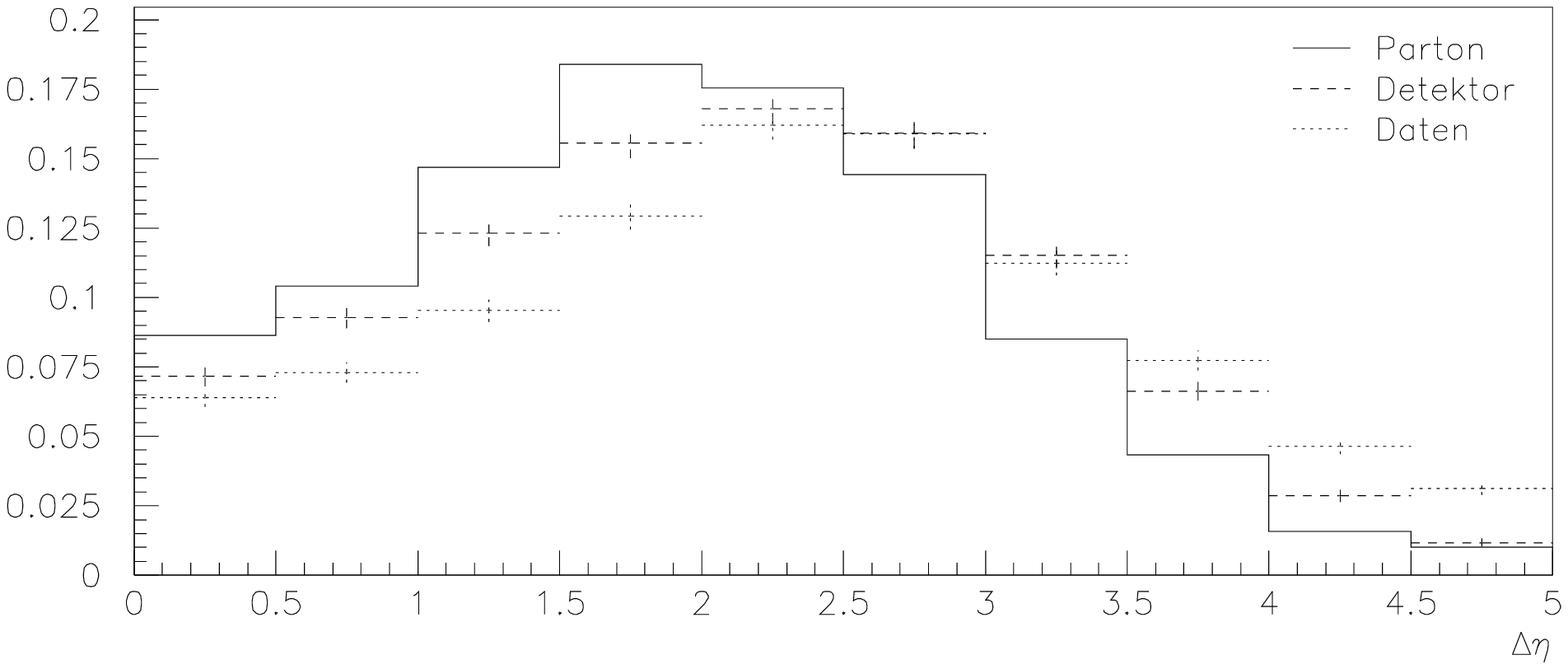,width=\hsize}
\end{center}
\caption[Vergleich der normierten $\Delta \eta$ Verteilungen]{{\bf
Vergleich der $\Delta \eta$ Verteilungen} {\it\ Die Verteilungen sind
auf gleiche Fl"achen normiert. Dargestellt sind Parton- (durchgezogen)
und Detektorniveau (gestrichelt) und die Daten (gepunktet).}}
\label{abbdeta}
\end{figure}

Ein weiterer Schnitt, den es zu untersuchen lohnt, bezieht sich auf
die r"aumliche Verteilung der Jets. Schon anschaulich ist klar, da"s
Verwechslungen verst"arkt auftreten, wenn die Jets nahe beieinander
liegen und sich so nach der Hadronisierung teilweise durchdringen.
Um diese Ereignisse zu unterdr"ucken ist ein Schnitt sinnvoll, der
einen minimalen Rapidit"atsunterschied zwischen den beiden harten
Jets verlangt. In Abbildung \ref{abbdeta} ist deshalb die Verteilung
des Rapidit"atsunterschiedes auf den beiden Monte-Carlo-Niveaus und 
bei den Daten aufgetragen. Wir erkennen eine gute
Beschreibung der Daten durch den Detektorlevel des Monte-Carlos. Die
Partonjets sind jedoch im Durchschnitt etwas dichter beieinander und die
Verteilung ist leicht verschoben.

In Tabelle \ref{tabdeta2migmeps} sind die Migrationen f"ur einen 
Mindestabstand von $\Delta \eta > 2$ eingetragen.  Im Vergleich zu
Tabelle \ref{tabstdmigmeps} erkennen wir eine Erniedrigung des
Korrekturfaktors und eine Verschlechterung der Reinheit. Auff"allig
ist weiterhin die "Anderung der $Q^2$ Abh"angigkeit. Die Raten
schwanken zwar auf beiden Niveaus und steigen leicht an, jedoch ist
die Steigung minimal und im achten Bin ist eine starke Reduktion zu
sehen.

\begin{table}[tbp]
\begin{center}
\begin{tabular}{@{\extracolsep{\fill}}c|c||c|c}
\multicolumn{2}{c||}{Bin} & \multicolumn{2}{c}{Parton} \\
\multicolumn{2}{c||}{1} & 1+1 & 2+1 \\
\hline
\hline
Clus- & 1+1 & 12705 & 260 \\
\cline{2-4}
ter & 2+1 & 420 & 176 \\
\end{tabular}
\hfill
\begin{tabular}{@{\extracolsep{\fill}}c|c||c|c}
\multicolumn{2}{c||}{Bin} & \multicolumn{2}{c}{Parton} \\
\multicolumn{2}{c||}{2} & 1+1 & 2+1 \\
\hline
\hline
Clus- & 1+1 & 9593 & 220 \\
\cline{2-4}
ter & 2+1 & 327 & 129 \\
\end{tabular}
\hfill
\begin{tabular}{@{\extracolsep{\fill}}c|c||c|c}
\multicolumn{2}{c||}{Bin} & \multicolumn{2}{c}{Parton} \\
\multicolumn{2}{c||}{3} & 1+1 & 2+1 \\
\hline
\hline
Clus- & 1+1 & 9208 & 231 \\
\cline{2-4}
ter & 2+1 & 323 & 122 \\
\end{tabular}
\end{center}
\vspace {0.25cm}
\begin{center}
\begin{tabular}{@{\extracolsep{\fill}}c|c||c|c}
\multicolumn{2}{c||}{Bin} & \multicolumn{2}{c}{Parton} \\
\multicolumn{2}{c||}{4} & 1+1 & 2+1 \\
\hline
\hline
Clus- & 1+1 & 8400 & 276 \\
\cline{2-4}
ter & 2+1 & 294 & 133 \\
\end{tabular}
\hfill
\begin{tabular}{@{\extracolsep{\fill}}c|c||c|c}
\multicolumn{2}{c||}{Bin} & \multicolumn{2}{c}{Parton} \\
\multicolumn{2}{c||}{5} & 1+1 & 2+1 \\
\hline
\hline
Clus- & 1+1 & 5755 & 169 \\
\cline{2-4}
ter & 2+1 & 229 & 125 \\
\end{tabular}
\hfill
\begin{tabular}{@{\extracolsep{\fill}}c|c||c|c}
\multicolumn{2}{c||}{Bin} & \multicolumn{2}{c}{Parton} \\
\multicolumn{2}{c||}{6} & 1+1 & 2+1 \\
\hline
\hline
Clus- & 1+1 & 6555 & 276 \\
\cline{2-4}
ter & 2+1 & 230 & 181 \\
\end{tabular}
\end{center}
\vspace {0.25cm}
\begin{center}
\begin{tabular}{@{\extracolsep{\fill}}c|c||c|c}
\multicolumn{2}{c||}{Bin} & \multicolumn{2}{c}{Parton} \\
\multicolumn{2}{c||}{7} & 1+1 & 2+1 \\
\hline
\hline
Clus- & 1+1 & 4083 & 152 \\
\cline{2-4}
ter & 2+1 & 154 & 115 \\
\end{tabular}
\hfill
\begin{tabular}{@{\extracolsep{\fill}}c|c||c|c}
\multicolumn{2}{c||}{Bin} & \multicolumn{2}{c}{Parton} \\
\multicolumn{2}{c||}{8} & 1+1 & 2+1 \\
\hline
\hline
Clus- & 1+1 & 1731 & 35 \\
\cline{2-4}
ter & 2+1 & 32 & 28 \\
\end{tabular}
\hfill
\hphantom{
\begin{tabular}{@{\extracolsep{\fill}}c|c||c|c}
\multicolumn{2}{c||}{} & \multicolumn{2}{c}{Parton} \\
\multicolumn{2}{c||}{} & 1+1 & 2+1 \\
\hline
\hline
Clus- & 1+1 & & \\
\cline{2-4}
ter & 2+1 & & \\
\end{tabular}
}
\end{center}
\vspace {0.25cm}
\begin{center}
\begin{tabular}{@{\extracolsep{\fill}}|c||c|c|c|c|c|c|c|c|}
\hline
Bin & 1 & 2 & 3 & 4 & 5 & 6 & 7 & 8 \\
\hline
Rate Parton&  3.2 &  3.4 &  3.6 &  4.5 &  4.7 &  6.3 &  5.9 &  3.5\\
in \% & $\pm 0.2$&$\pm 0.2$&$\pm 0.2$&$\pm 0.3$&$\pm 0.3$&$\pm 0.3$&$\pm 0.4$&$\pm 0.5$\\
\hline
Rate Detektor&  4.4 &  4.4 &  4.5 &  4.7 &  5.6 &  5.7 &  6.0 &  3.3\\
in \%&$\pm 0.2$&$\pm 0.3$&$\pm 0.3$&$\pm 0.3$&$\pm 0.3$&$\pm 0.3$&$\pm 0.4$&$\pm 0.5$\\
\hline
Korrektur- & 0.73 & 0.77 & 0.79 & 0.96 & 0.83 & 1.11 & 0.99 & 1.05\\
faktor &$\pm0.07$&$\pm0.08$&$\pm0.08$&$\pm0.10$&$\pm0.10$&$\pm0.11$&$\pm0.12$&$\pm0.27$\\
\hline
Reinheit in \%& 29.5 & 28.3 & 27.4 & 31.1 & 35.3 & 44.0 & 42.8 & 46.7\\
\hline
\end{tabular}
\end{center}
\caption[Migrationen und Korrekturfaktoren bei Lepto, $\Delta\eta$ Schnitt]{{\bf
Migrationen und Korrekturfaktoren bei Lepto mit Standardschnitten und 
zus"atzlichem $\Delta\eta > 2$ Schnitt,
in $Q^2$ Bins}}
\label{tabdeta2migmeps}
\end{table}

Ein Schnitt in dieser Variable ist f"ur Jetanalysen ungeeignet.

\subsection{Quarkwinkel $\theta_q$}

Wie wir schon in Kapitel \ref{kaputhejet} an Abbildung \ref{abbthejet}a 
und b gesehen haben, treten gro"se Jetwinkel haupts"achlich bei 1+1 
Ereignissen auf, wohingegen nahezu alle 2+1 Ereignisse einen Jet unter
kleinen Winkeln haben. Durch einen Schnitt lie"se sich somit das
Verh"altnis von 2+1 zu 1+1 Jet Ereignissen und damit die Rate
ver"andern. Besser als der Jetwinkel eignet sich hierf"ur der Winkel des
nach dem harten Sto"s auslaufenden Quarks im Quark Parton Modell. Dieser
Winkel l"a"st sich wegen der relativ einfachen Reaktionskinematik des
Born Graphen aus der Messung des gestreuten Elektrons durch die folgende
Gleichung herleiten~:
\begin{eqnarray}
\tquark &=& \acos\left( { (1-y_B) Q^2 - 4 E_e^2 y_B^2 \over
(1-y_B) Q^2 + 4 E_e^2 y_B^2 }\right)
\end{eqnarray}
Die Werte f"ur $Q^2$ und $y_B$ werden dabei wie in Kapitel \ref{kapkinvar}
erl"autert berechnet.

\begin{figure}[tbp]                
\begin{center}
\epsfig{file=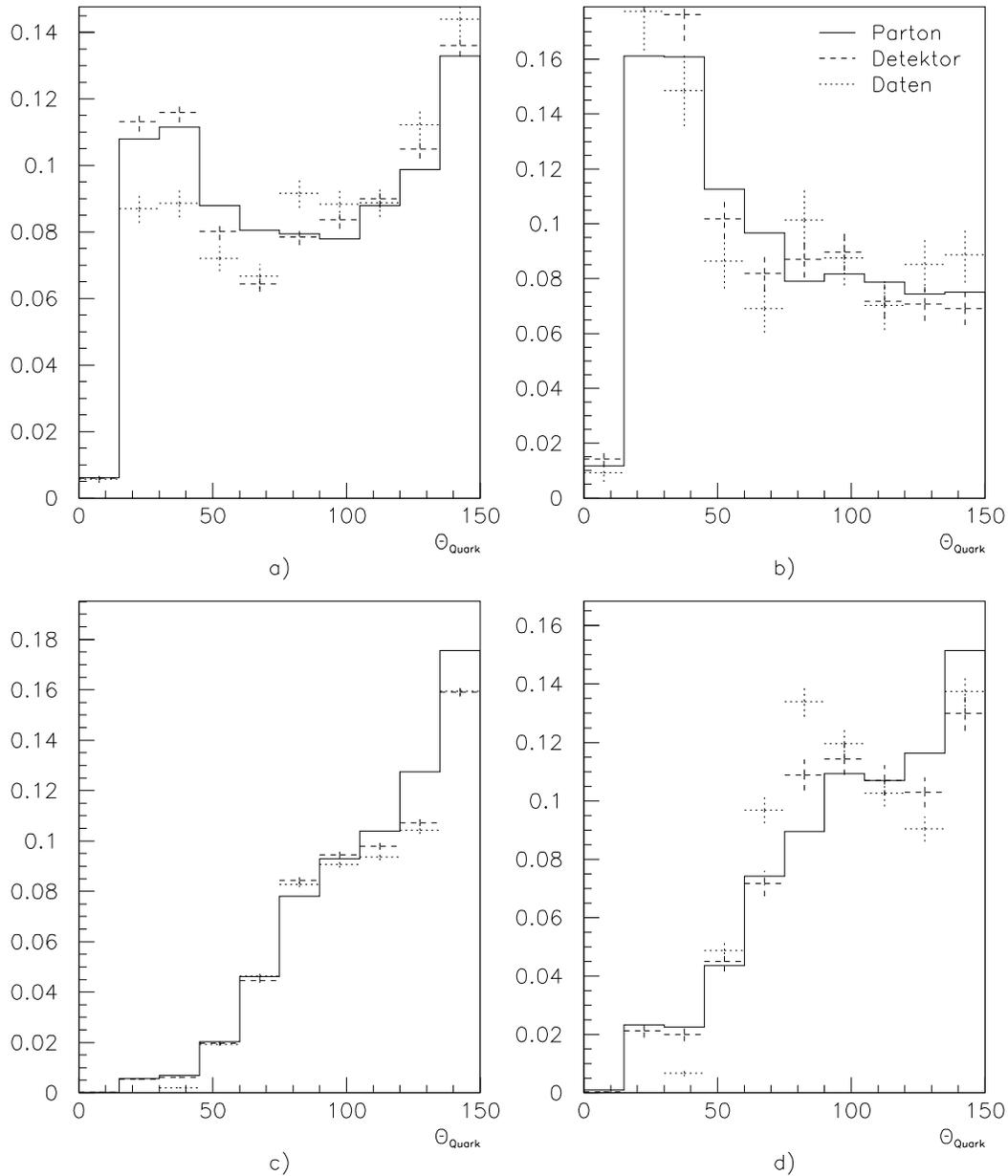,width=0.95\hsize}
\end{center}
\caption[Vergleich der normierten $\tquark$ Verteilungen]{{\bf
Vergleich der $\tquark$ Verteilungen}{\it\ a) zeigt alle Ereignisse mit
$Q^2>100\;\GeV,$ b) die 2+1 Ereignisse. c) und d) sind die analogen
Graphen f"ur die Bins mit $Q^2<100\;\GeV$} {\it\ Dargestellt sind
Parton-(durchgezogen) und Hadronniveau (gestrichelt) und die Daten
(gepunktet). Die Verteilungen sind jeweils auf gleiche Fl"achen
normiert.}}
\label{abbtquark}
\end{figure}

Abbildung \ref{abbtquark} zeigt die Verteilung des Quarkwinkels f"ur
alle und f"ur die 2+1 Ereignisse in beiden $Q^2$ Bereiche getrennt. Die
Verteilungen zeigen eine unbefriedigende Beschreibung der Daten. Nur im
Bereich $Q^2 < 100\;\GeV$ ist die Beschreibung gut. Das gleiche gilt
f"ur den Unterschied zwischen den Monte-Carlo-Niveaus. Ein Schnitt auf
einen maximalen Quarkwinkel verwirft im hohen $Q^2$ Bereich wesentlich
mehr 1+1 Ereignisse als solche mit 2+1 Jets. Wir erwarten daher eine
Ratensteigerung. Im $Q^2<100\;\GeV$ Bereich gilt die gleiche Aussage.
Die Verteilungen sind jedoch steiler, so da"s ein Schnitt hier die
Statistik stark reduziert.

Ein Schnitt auf einen maximalen Quarkwinkel verwirft mehr 1+1 Ereignisse
als solche mit 2+1 Jets. Wir erwarten somit eine Ratensteigerung. Der
Effekt ist bei einer Grenze von 100 Grad am deutlichsten zu sehen. Die
Migrationen, die sich aus der Anwendung dieses Schnittes ergeben, sond
in Tabelle \ref{tabtq100migmeps} gezeigt. Ein Vergleich mit Tabelle
\ref{tabstdmigmeps} zeigt die Best"atigung des oben gesagten. Die Raten
sind auf beiden Niveaus gestiegen, wobei es in den verschiedenen Bins
jedoch starke Unterschiede gibt. Die Reinheit ist in etwa konstant
geblieben und der Korrekturfaktor hat sich nur in den drei unteren Bins
ver"andert. In diesen Bins macht sich die Reduktion der Statistik stark
bemerkbar. Der relative Fehler ist von ca.\ 5\% ohne $\tquark$ Schnitt
auf ca.\ 20\% mit Schnitt angestiegen.

\begin{table}[tbp]
\begin{center}
\begin{tabular}{@{\extracolsep{\fill}}c|c||c|c}
\multicolumn{2}{c||}{Bin} & \multicolumn{2}{c}{Parton} \\
\multicolumn{2}{c||}{1} & 1+1 & 2+1 \\
\hline
\hline
Clus- & 1+1 & 198 & 5 \\
\cline{2-4}
ter & 2+1 & 14 & 10 \\
\end{tabular}
\hfill
\begin{tabular}{@{\extracolsep{\fill}}c|c||c|c}
\multicolumn{2}{c||}{Bin} & \multicolumn{2}{c}{Parton} \\
\multicolumn{2}{c||}{2} & 1+1 & 2+1 \\
\hline
\hline
Clus- & 1+1 & 668 & 17 \\
\cline{2-4}
ter & 2+1 & 29 & 20 \\
\end{tabular}
\hfill
\begin{tabular}{@{\extracolsep{\fill}}c|c||c|c}
\multicolumn{2}{c||}{Bin} & \multicolumn{2}{c}{Parton} \\
\multicolumn{2}{c||}{3} & 1+1 & 2+1 \\
\hline
\hline
Clus- & 1+1 & 1183 & 27 \\
\cline{2-4}
ter & 2+1 & 55 & 44 \\
\end{tabular}
\end{center}
\vspace {0.25cm}
\begin{center}
\begin{tabular}{@{\extracolsep{\fill}}c|c||c|c}
\multicolumn{2}{c||}{Bin} & \multicolumn{2}{c}{Parton} \\
\multicolumn{2}{c||}{4} & 1+1 & 2+1 \\
\hline
\hline
Clus- & 1+1 & 1726 & 89 \\
\cline{2-4}
ter & 2+1 & 83 & 100 \\
\end{tabular}
\hfill
\begin{tabular}{@{\extracolsep{\fill}}c|c||c|c}
\multicolumn{2}{c||}{Bin} & \multicolumn{2}{c}{Parton} \\
\multicolumn{2}{c||}{5} & 1+1 & 2+1 \\
\hline
\hline
Clus- & 1+1 & 2165 & 121 \\
\cline{2-4}
ter & 2+1 & 133 & 148 \\
\end{tabular}
\hfill
\begin{tabular}{@{\extracolsep{\fill}}c|c||c|c}
\multicolumn{2}{c||}{Bin} & \multicolumn{2}{c}{Parton} \\
\multicolumn{2}{c||}{6} & 1+1 & 2+1 \\
\hline
\hline
Clus- & 1+1 & 1526 & 187 \\
\cline{2-4}
ter & 2+1 & 136 & 261 \\
\end{tabular}
\end{center}
\vspace {0.25cm}
\begin{center}
\begin{tabular}{@{\extracolsep{\fill}}c|c||c|c}
\multicolumn{2}{c||}{Bin} & \multicolumn{2}{c}{Parton} \\
\multicolumn{2}{c||}{7} & 1+1 & 2+1 \\
\hline
\hline
Clus- & 1+1 & 2187 & 253 \\
\cline{2-4}
ter & 2+1 & 178 & 490 \\
\end{tabular}
\hfill
\begin{tabular}{@{\extracolsep{\fill}}c|c||c|c}
\multicolumn{2}{c||}{Bin} & \multicolumn{2}{c}{Parton} \\
\multicolumn{2}{c||}{8} & 1+1 & 2+1 \\
\hline
\hline
Clus- & 1+1 & 1069 & 137 \\
\cline{2-4}
ter & 2+1 & 92 & 260 \\
\end{tabular}
\hfill
\hphantom{
\begin{tabular}{@{\extracolsep{\fill}}c|c||c|c}
\multicolumn{2}{c||}{} & \multicolumn{2}{c}{Parton} \\
\multicolumn{2}{c||}{} & 1+1 & 2+1 \\
\hline
\hline
Clus- & 1+1 & & \\
\cline{2-4}
ter & 2+1 & & \\
\end{tabular}
}
\end{center}
\vspace {0.25cm}
\begin{center}
\begin{tabular}{@{\extracolsep{\fill}}|c||c|c|c|c|c|c|c|c|}
\hline
Bin & 1 & 2 & 3 & 4 & 5 & 6 & 7 & 8 \\
\hline
Rate Parton&  6.6 &  5.0 &  5.4 &  9.5 & 10.5 & 21.2 & 23.9 & 25.5\\
in \% & $\pm 1.7$&$\pm 0.9$&$\pm 0.7$&$\pm 0.7$&$\pm 0.7$&$\pm 0.9$&$\pm 0.8$&$\pm 1.2$\\
\hline
Rate Detektor& 10.6 &  6.7 &  7.6 &  9.2 & 10.9 & 18.8 & 21.5 & 22.6\\
in \%&$\pm 2.1$&$\pm 1.0$&$\pm 0.8$&$\pm 0.7$&$\pm 0.7$&$\pm 0.9$&$\pm 0.8$&$\pm 1.1$\\
\hline
Korrektur- & 0.62 & 0.76 & 0.72 & 1.03 & 0.96 & 1.13 & 1.11 & 1.13\\
faktor &$\pm0.28$&$\pm0.23$&$\pm0.16$&$\pm0.15$&$\pm0.11$&$\pm0.10$&$\pm0.08$&$\pm0.11$\\
\hline
Reinheit in \%& 41.7 & 40.8 & 44.4 & 54.6 & 52.7 & 65.7 & 73.4 & 73.9\\
\hline
\end{tabular}
\end{center}
\caption[Migrationen und Korrekturfaktoren bei Lepto, $\tquark$ Schnitt]{{\bf
Migrationen und Korrekturfaktoren bei Lepto mit Standardschnitten und 
zus"atzlichem $\tquark<100^{\circ}$ Schnitt,
in $Q^2$ Bins}}
\label{tabtq100migmeps}
\end{table}

Unter Ber"ucksichtigung dieser Ergebnisse ist ein Schnitt im Quarkwinkel
nur in den hohen $Q^2$ Bins sinnvoll. Er verbessert die Migrationen
jedoch nicht.

\section {Alternative Ratenberechnung}

\begin{figure}[tbp]                
\begin{center}
\epsfig{file=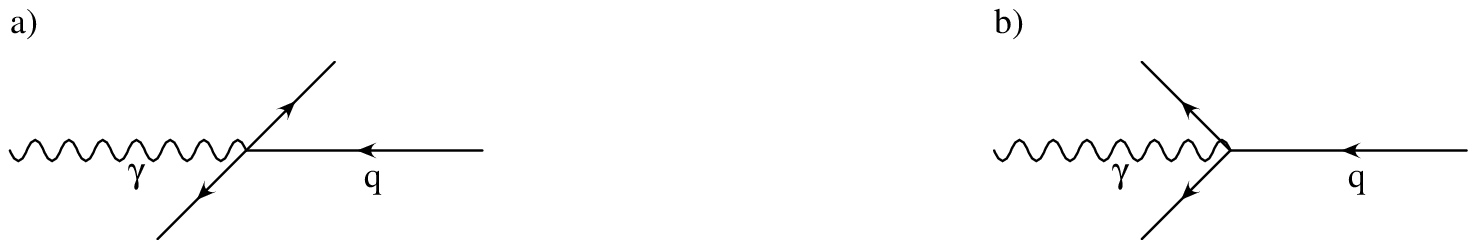,width=\hsize}
\end{center}
\caption[2+1 Jetereignisse im Breitsystem]{{\bf
2+1 Jetereignisse im Breitsystem}{\it\ a) bei $\myxi = 2 x_B,$ b) bei
$\myxi >> 2 x_B.$}} 
\label{abbbreit2}
\end{figure}

Im letzten Abschnitt dieses Kapitels untersuchen wir einen alternativen
Ansatz die Jetrate zu bestimmen. Dieser verzichtet auf einen
Jetalgorithmus und macht sich die Tatsache zu nutze, da"s die
Energieverteilung im Breitsystem je nach Anzahl der Jets unterschiedlich
ist\cite{hoeger:95}. Hierzu machen wir uns klar, wie sich 2+1 Ereignisse
im Breitsystem verhalten. Dabei tritt das Problem auf, da"s $\hat s$
ungleich Null ist (siehe Kapitel \ref{kapushat}) und somit nach Gleichung
\ref{eqnxishat} die Bjorken Skalenvariable $x_B$ nicht dem
Partonimpulsbruchteil $\myxi$ entspricht. Deshalb wird das
einlaufende Parton nach der Transformation nicht mehr durch das
Austauschteilchen wie an einer harten Wand reflektiert (siehe Kapitel
\ref{kapkt} und Abbildung \ref{abbbreit}). Es verbleibt statt dessen ein
Restimpuls. Im Spezialfall $\myxi = 2 x_B$ ist der Impuls des Photons
entgegengesetzt gleich gro"s wie der Impuls des Partons. Die Jets laufen
in entgegengesetzte Richtungen aus. Ist $\myxi$ noch gr"o"ser, werden
sogar beide Jets in die Photon Hemisph"are gelenkt. Dies wird in
Abbildung \ref{abbbreit2} verdeutlicht.

\begin{figure}[tbp]                
\begin{center}
\epsfig{file=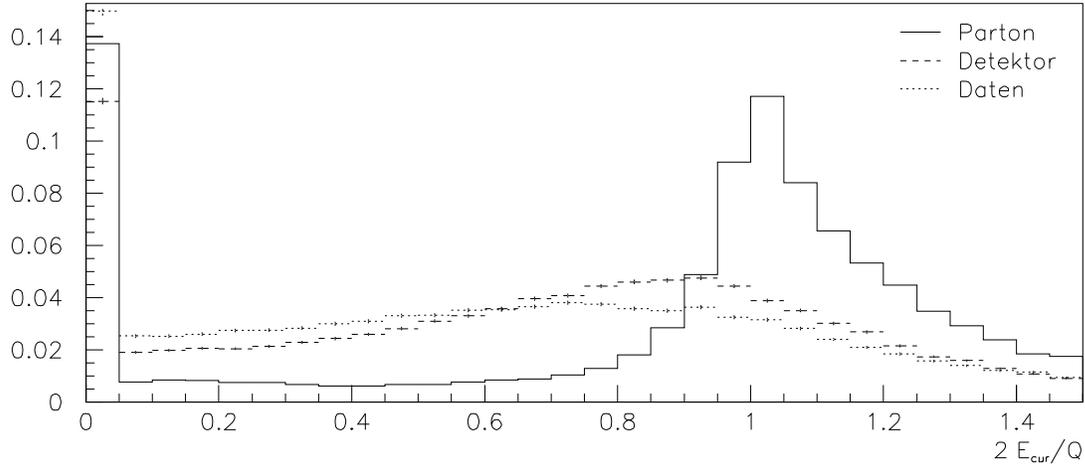,width=\hsize}
\end{center}
\caption[Vergleich der Energieverteilung in der Current Hemisph"are im Breitsystem]
{{\bf Vergleich der Energieverteilung in der Current Hemisph"are im Breitsystem}
{\it\ Dargestellt sind Parton- (durchgezogen) und Detektorniveau (gestrichelt)
des Lepto-Monte-Carlos und die Daten (gepunktet). Die Verteilungen sind auf
gleiche Fl"achen normiert.}}
\label{abbalter}
\end{figure}

Legen wir nun zwei Bereiche im Breitsystem fest. Die Grenze wird durch
einen Winkel $\theta^B$ relativ zur einlaufenden Partonrichtung
festgelegt. Wir benutzen im folgenden 90 Grad zur Trennung der
sogenannten {\it Current Hemisph"are,} d.i.\ die Seite mit dem
einlaufenden Parton, und der {\it Target Hemisph"are,} d.i.\ die Seite
mit einlaufendem Photon. Wir erwarten in der Current Hemisph"are die
gesamte Schwerpunktenergie ${1\over2}Q$ im Falle eines 1+1 Ereignisses
und bei 2+1 Jet Ereignissen mit $x_B \approx \myxi.$ Bei $\myxi >> 2
x_B$ erwarten wir keine Energie in der Current Hemisph"are. Abbildung
\ref{abbalter} zeigt die Verteilung auf den beiden
Monte-Carlo-Niveaus und bei den Daten. Auf Partonniveau entspricht die 
Verteilung den Erwartungen. Das Detektorniveau beschreibt die Daten 
einigerma"sen, die Verteilungen sind aber sehr breit. Wir benutzen nun einen
Abschneideparameter $\epsilon =0.25$ und definieren Ereignisse
mit $E_{\mb{cur}} < {\epsilon\over 2} Q$ als 2+1 Jet Ereignisse
und die mit gr"o"serer Current Energie als 1+1 Jet Ereignisse. 

Die Migrationen in Tabelle \ref{tabstdmigalt} zeigen ein merkw"urdiges
$Q^2$ Verhalten. Die Raten fallen mit steigendem $Q^2.$ Ob dies durch
die Theorie beschrieben wird, ist unklar, da die vorhandenen
Theorieprogramme die Partonjets erzeugen, nicht jedoch die Clus"-ter. Ein
Vergleich in n"achst-zu-f"uhrender Ordnung ist somit nicht m"oglich.
Auch die Reinheit der Jetklassen ist in den unteren vier $Q^2$ Bins
besser als beim JADE Algorithmus, w"ahrend die oberen Bins eine
schlechtere Reinheit zeigen. Der Korrekturfaktor zeigt ein
uneinheitliches Verhalten.
         
\begin{table}[tbp]
\begin{center}
\begin{tabular}{@{\extracolsep{\fill}}c|c||c|c}
\multicolumn{2}{c||}{Bin} & \multicolumn{2}{c}{Parton} \\
\multicolumn{2}{c||}{1} & 1+1 & 2+1 \\
\hline
\hline
Clus- & 1+1 & 8569 & 1297 \\
\cline{2-4}
ter & 2+1 & 1474 & 2221 \\
\end{tabular}
\hfill
\begin{tabular}{@{\extracolsep{\fill}}c|c||c|c}
\multicolumn{2}{c||}{Bin} & \multicolumn{2}{c}{Parton} \\
\multicolumn{2}{c||}{2} & 1+1 & 2+1 \\
\hline
\hline
Clus- & 1+1 & 6709 & 894 \\
\cline{2-4}
ter & 2+1 & 1187 & 1479 \\
\end{tabular}
\hfill
\begin{tabular}{@{\extracolsep{\fill}}c|c||c|c}
\multicolumn{2}{c||}{Bin} & \multicolumn{2}{c}{Parton} \\
\multicolumn{2}{c||}{3} & 1+1 & 2+1 \\
\hline
\hline
Clus- & 1+1 & 6844 & 739 \\
\cline{2-4}
ter & 2+1 & 1053 & 1248 \\
\end{tabular}
\end{center}
\vspace {0.25cm}
\begin{center}
\begin{tabular}{@{\extracolsep{\fill}}c|c||c|c}
\multicolumn{2}{c||}{Bin} & \multicolumn{2}{c}{Parton} \\
\multicolumn{2}{c||}{4} & 1+1 & 2+1 \\
\hline
\hline
Clus- & 1+1 & 6691 & 535 \\
\cline{2-4}
ter & 2+1 & 932 & 945 \\
\end{tabular}
\hfill
\begin{tabular}{@{\extracolsep{\fill}}c|c||c|c}
\multicolumn{2}{c||}{Bin} & \multicolumn{2}{c}{Parton} \\
\multicolumn{2}{c||}{5} & 1+1 & 2+1 \\
\hline
\hline
Clus- & 1+1 & 5117 & 252 \\
\cline{2-4}
ter & 2+1 & 499 & 410 \\
\end{tabular}
\hfill
\begin{tabular}{@{\extracolsep{\fill}}c|c||c|c}
\multicolumn{2}{c||}{Bin} & \multicolumn{2}{c}{Parton} \\
\multicolumn{2}{c||}{6} & 1+1 & 2+1 \\
\hline
\hline
Clus- & 1+1 & 6648 & 165 \\
\cline{2-4}
ter & 2+1 & 165 & 264 \\
\end{tabular}
\end{center}
\vspace {0.25cm}
\begin{center}
\begin{tabular}{@{\extracolsep{\fill}}c|c||c|c}
\multicolumn{2}{c||}{Bin} & \multicolumn{2}{c}{Parton} \\
\multicolumn{2}{c||}{7} & 1+1 & 2+1 \\
\hline
\hline
Clus- & 1+1 & 4308 & 38 \\
\cline{2-4}
ter & 2+1 & 77 & 81 \\
\end{tabular}
\hfill
\begin{tabular}{@{\extracolsep{\fill}}c|c||c|c}
\multicolumn{2}{c||}{Bin} & \multicolumn{2}{c}{Parton} \\
\multicolumn{2}{c||}{8} & 1+1 & 2+1 \\
\hline
\hline
Clus- & 1+1 & 1803 & 5 \\
\cline{2-4}
ter & 2+1 & 11 & 7 \\
\end{tabular}
\hfill
\hphantom{
\begin{tabular}{@{\extracolsep{\fill}}c|c||c|c}
\multicolumn{2}{c||}{} & \multicolumn{2}{c}{Parton} \\
\multicolumn{2}{c||}{} & 1+1 & 2+1 \\
\hline
\hline
Clus- & 1+1 & & \\
\cline{2-4}
ter & 2+1 & & \\
\end{tabular}
}
\end{center}
\vspace {0.25cm}
\begin{center}
\begin{tabular}{@{\extracolsep{\fill}}|c||c|c|c|c|c|c|c|c|}
\hline
Bin & 1 & 2 & 3 & 4 & 5 & 6 & 7 & 8 \\
\hline
Rate Parton& 25.9 & 23.1 & 20.1 & 16.3 & 10.5 &  5.9 &  2.6 &  0.7\\
in \% & $\pm 0.4$&$\pm 0.5$&$\pm 0.5$&$\pm 0.4$&$\pm 0.4$&$\pm 0.3$&$\pm 0.3$&$\pm 0.2$\\
\hline
Rate Detektor& 27.2 & 26.0 & 23.3 & 20.6 & 14.5 &  5.9 &  3.5 &  1.0\\
in \%&$\pm 0.4$&$\pm 0.5$&$\pm 0.5$&$\pm 0.5$&$\pm 0.5$&$\pm 0.3$&$\pm 0.3$&$\pm 0.3$\\
\hline
Korrektur- & 0.95 & 0.89 & 0.86 & 0.79 & 0.73 & 1.00 & 0.75 & 0.67\\
faktor &$\pm0.03$&$\pm0.04$&$\pm0.04$&$\pm0.04$&$\pm0.05$&$\pm0.10$&$\pm0.13$&$\pm0.35$\\
\hline
Reinheit in \%& 60.1 & 55.5 & 54.2 & 50.3 & 45.1 & 61.5 & 51.3 & 38.9\\
\hline
\end{tabular}
\end{center}
\caption[Migrationen und Korrekturfaktoren bei Lepto bei alternativer
Jetbestimmung]{{\bf Migrationen und Korrekturfaktoren bei Lepto in $Q^2$
Bins mit Standardschnitten bei Anzahlbestimmung durch die Energie in der
Current Hemisph"are}}
\label{tabstdmigalt}
\end{table}

Ein Vergleich zwischen der Jetklassifizierung "uber die Current Energie
mit dem JADE Algorithmus ist in Tabelle \ref{tabalter} dargestellt.
Wie wir schon aufgrund des v"ollig unterschiedlichen $Q^2$ Verhaltens
erwarten konnten, ist die Korrelation sehr schlecht.

\begin{table}[tbp]
\begin{center}
\begin{tabular}{@{\extracolsep{\fill}}c|c||c|c}
\multicolumn{2}{c||}{Parton-} & \multicolumn{2}{c}{alternativ} \\
\multicolumn{2}{c||}{niveau} & 1+1 & 2+1 \\
\hline
\hline
JADE & 1+1 & 56980 & 9442 \\
\cline{2-4}
& 2+1 & 3766 & 2969 \\
\end{tabular}
\hfill
\begin{tabular}{@{\extracolsep{\fill}}c|c||c|c}
\multicolumn{2}{c||}{Detektor-} & \multicolumn{2}{c}{alternativ} \\
\multicolumn{2}{c||}{niveau} & 1+1 & 2+1 \\
\hline
\hline
JADE & 1+1 & 49991 & 10659 \\
\cline{2-4}
& 2+1 & 3379 & 2529 \\
\end{tabular}
\end{center}
\caption[Vergleich von JADE und alternativem Algorithmus]{{\bf
Vergleich vom JADE Algorithmus mit der Ratenbestimmung durch die
Energie in der Current Hemisph"are}}
\label{tabalter}
\end{table}

Diese Jetanzahlbestimmung hat den Vorteil, da"s sie die Klassifizierung
nicht "uber einen Algorithmus durchf"uhrt, sondern durch Berechnung nur
eines Wertes. Der Nachteil dieses Verfahrens, neben den oben
angesprochenden, ist, da"s die Jets selber nicht bestimmt werden. Es
sind somit keine weiteren Schnitte auf z.B.\ Jetwinkel m"oglich.

\chapter{Zusammenfassung und Ausblick}
\label {kap_zusammen}

\section {Zusammenfassung}

Aus den im vorigen Kapitel gezeigten Untersuchungen werden die unten
genannten Schnitte gefolgert~:
\begin{itemize}
\item die Standardschnitte f"ur die Kinematik (Kapitel \ref{kapkinschn}), 
die Rekonstruktionsqualit"at (Kapitel \ref{kaprekon}) und die Elektronfinder
(Kapitel \ref{kapelfin}).
\item und der Jetwinkelschnitt $15^{\circ} < \tjet < 145^{\circ}$
\end{itemize}

Mit diesen Schnitten ergeben sich die in Tabelle \ref{tabt15migmeps}
aufgetragenen Migrationen. Im Vergleich zur Tabelle \ref{tabstdmigmeps}
zeigen sich hier die kleineren Raten und der erh"ohte Korrekturfaktor.
Der ungewichtete Mittelwert des Korrekturfaktors ist $1.683\pm0.191.$ Im
Rahmen ihrer Fehler sind alle Werte mit dem mittleren Korrekturfaktor
vertr"aglich. Auch die Schwankungen in der Reinheit sind kleiner. Die
Reinheit selber hat sich durch den Schnitt erh"oht.

\begin{table}[tbp]
\begin{center}
\begin{tabular}{@{\extracolsep{\fill}}c|c||c|c}
\multicolumn{2}{c||}{Bin} & \multicolumn{2}{c}{Parton} \\
\multicolumn{2}{c||}{1} & 1+1 & 2+1 \\
\hline
\hline
Clus- & 1+1 & 6716 & 128 \\
\cline{2-4}
ter & 2+1 & 61 & 81 \\
\end{tabular}
\hfill
\begin{tabular}{@{\extracolsep{\fill}}c|c||c|c}
\multicolumn{2}{c||}{Bin} & \multicolumn{2}{c}{Parton} \\
\multicolumn{2}{c||}{2} & 1+1 & 2+1 \\
\hline
\hline
Clus- & 1+1 & 6343 & 74 \\
\cline{2-4}
ter & 2+1 & 30 & 52 \\
\end{tabular}
\hfill
\begin{tabular}{@{\extracolsep{\fill}}c|c||c|c}
\multicolumn{2}{c||}{Bin} & \multicolumn{2}{c}{Parton} \\
\multicolumn{2}{c||}{3} & 1+1 & 2+1 \\
\hline
\hline
Clus- & 1+1 & 6584 & 114 \\
\cline{2-4}
ter & 2+1 & 34 & 52 \\
\end{tabular}
\end{center}
\vspace {0.25cm}
\begin{center}
\begin{tabular}{@{\extracolsep{\fill}}c|c||c|c}
\multicolumn{2}{c||}{Bin} & \multicolumn{2}{c}{Parton} \\
\multicolumn{2}{c||}{4} & 1+1 & 2+1 \\
\hline
\hline
Clus- & 1+1 & 6647 & 121 \\
\cline{2-4}
ter & 2+1 & 40 & 56 \\
\end{tabular}
\hfill
\begin{tabular}{@{\extracolsep{\fill}}c|c||c|c}
\multicolumn{2}{c||}{Bin} & \multicolumn{2}{c}{Parton} \\
\multicolumn{2}{c||}{5} & 1+1 & 2+1 \\
\hline
\hline
Clus- & 1+1 & 5351 & 91 \\
\cline{2-4}
ter & 2+1 & 26 & 43 \\
\end{tabular}
\hfill
\begin{tabular}{@{\extracolsep{\fill}}c|c||c|c}
\multicolumn{2}{c||}{Bin} & \multicolumn{2}{c}{Parton} \\
\multicolumn{2}{c||}{6} & 1+1 & 2+1 \\
\hline
\hline
Clus- & 1+1 & 5413 & 240 \\
\cline{2-4}
ter & 2+1 & 66 & 176 \\
\end{tabular}
\end{center}
\vspace {0.25cm}
\begin{center}
\begin{tabular}{@{\extracolsep{\fill}}c|c||c|c}
\multicolumn{2}{c||}{Bin} & \multicolumn{2}{c}{Parton} \\
\multicolumn{2}{c||}{7} & 1+1 & 2+1 \\
\hline
\hline
Clus- & 1+1 & 3954 & 172 \\
\cline{2-4}
ter & 2+1 & 57 & 137 \\
\end{tabular}
\hfill
\begin{tabular}{@{\extracolsep{\fill}}c|c||c|c}
\multicolumn{2}{c||}{Bin} & \multicolumn{2}{c}{Parton} \\
\multicolumn{2}{c||}{8} & 1+1 & 2+1 \\
\hline
\hline
Clus- & 1+1 & 1508 & 62 \\
\cline{2-4}
ter & 2+1 & 21 & 74 \\
\end{tabular}
\hfill
\hphantom{
\begin{tabular}{@{\extracolsep{\fill}}c|c||c|c}
\multicolumn{2}{c||}{} & \multicolumn{2}{c}{Parton} \\
\multicolumn{2}{c||}{} & 1+1 & 2+1 \\
\hline
\hline
Clus- & 1+1 & & \\
\cline{2-4}
ter & 2+1 & & \\
\end{tabular}
}
\end{center}
\vspace {0.25cm}
\begin{center}
\begin{tabular}{@{\extracolsep{\fill}}|c||c|c|c|c|c|c|c|c|}
\hline
Bin & 1 & 2 & 3 & 4 & 5 & 6 & 7 & 8 \\
\hline
Rate Parton&  3.0 &  1.9 &  2.4 &  2.6 &  2.4 &  7.1 &  7.2 &  8.2\\
in \% & $\pm 0.3$&$\pm 0.2$&$\pm 0.2$&$\pm 0.2$&$\pm 0.3$&$\pm 0.4$&$\pm 0.4$&$\pm 0.7$\\
\hline
Rate Detektor&  2.0 &  1.3 &  1.3 &  1.4 &  1.3 &  4.1 &  4.5 &  5.7\\
in \%&$\pm 0.2$&$\pm 0.2$&$\pm 0.2$&$\pm 0.2$&$\pm 0.2$&$\pm 0.3$&$\pm 0.4$&$\pm 0.6$\\
\hline
Korrektur- & 1.47 & 1.54 & 1.93 & 1.84 & 1.94 & 1.72 & 1.59 & 1.43\\
faktor &$\pm0.23$&$\pm0.31$&$\pm0.36$&$\pm0.33$&$\pm0.40$&$\pm0.19$&$\pm0.20$&$\pm0.27$\\
\hline
Reinheit in \%& 57.0 & 63.4 & 60.5 & 58.3 & 62.3 & 72.7 & 70.6 & 77.9\\
\hline
\end{tabular}
\end{center}
\caption[Migrationen und Korrekturfaktoren bei Lepto, $\tjet$]{{\bf
Migrationen und Korrekturfaktoren bei Lepto mit Standardschnitten und
zus"atzlichem $15^{\circ} < \tjet < 145^{\circ}$ Schnitt,
in $Q^2$ Bins}}
\label{tabt15migmeps}
\end{table}

\begin{figure}[tbp]                
\begin{center}
\epsfig{file=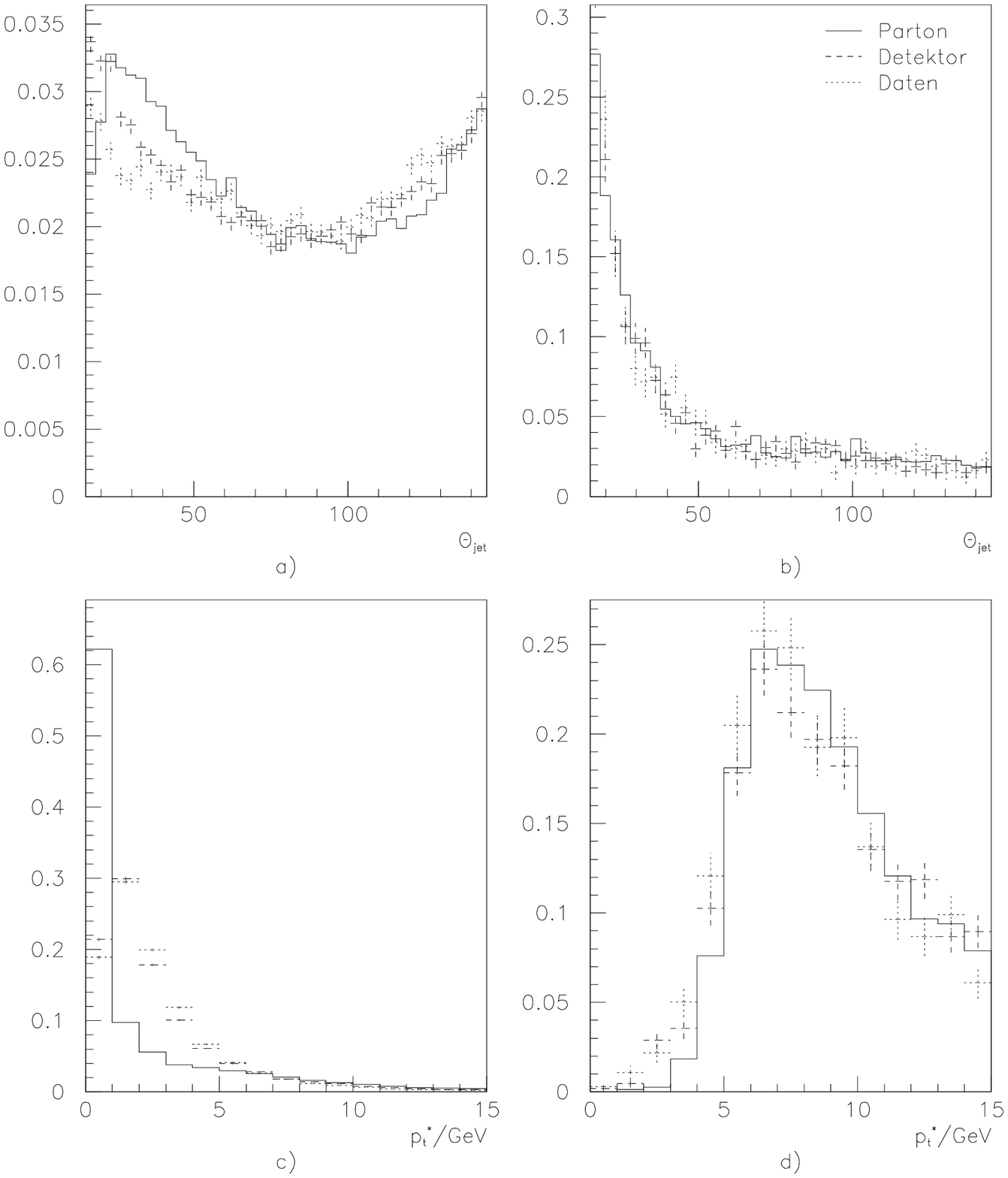,width=0.95\hsize}
\end{center}
\caption[Vergleich der wichtigsten Verteilungen mit abschlie"senden Schnitten]
{{\bf Vergleich der wichtigsten Verteilungen mit abschlie"senden Schnitten}
{\it\ a) Jetwinkelverteilung f"ur alle Ereignisse, b) inklusive 
Jetwinkelverteilung f"ur 2+1 Jet Ereignisse, c) und d) transversale
Impulsverteilung f"ur das hadronische Schwerpunktsystem f"ur alle bzw.\ nur
f"ur 2+1 Jet Ereignisse. Dargestellt sind Parton- (durchgezogen),
Detektorniveau (gestrichelt) und Daten (gepunktet)}}
\label{abbvert1}
\end{figure}

\begin{figure}[tbp]                
\begin{center}
\epsfig{file=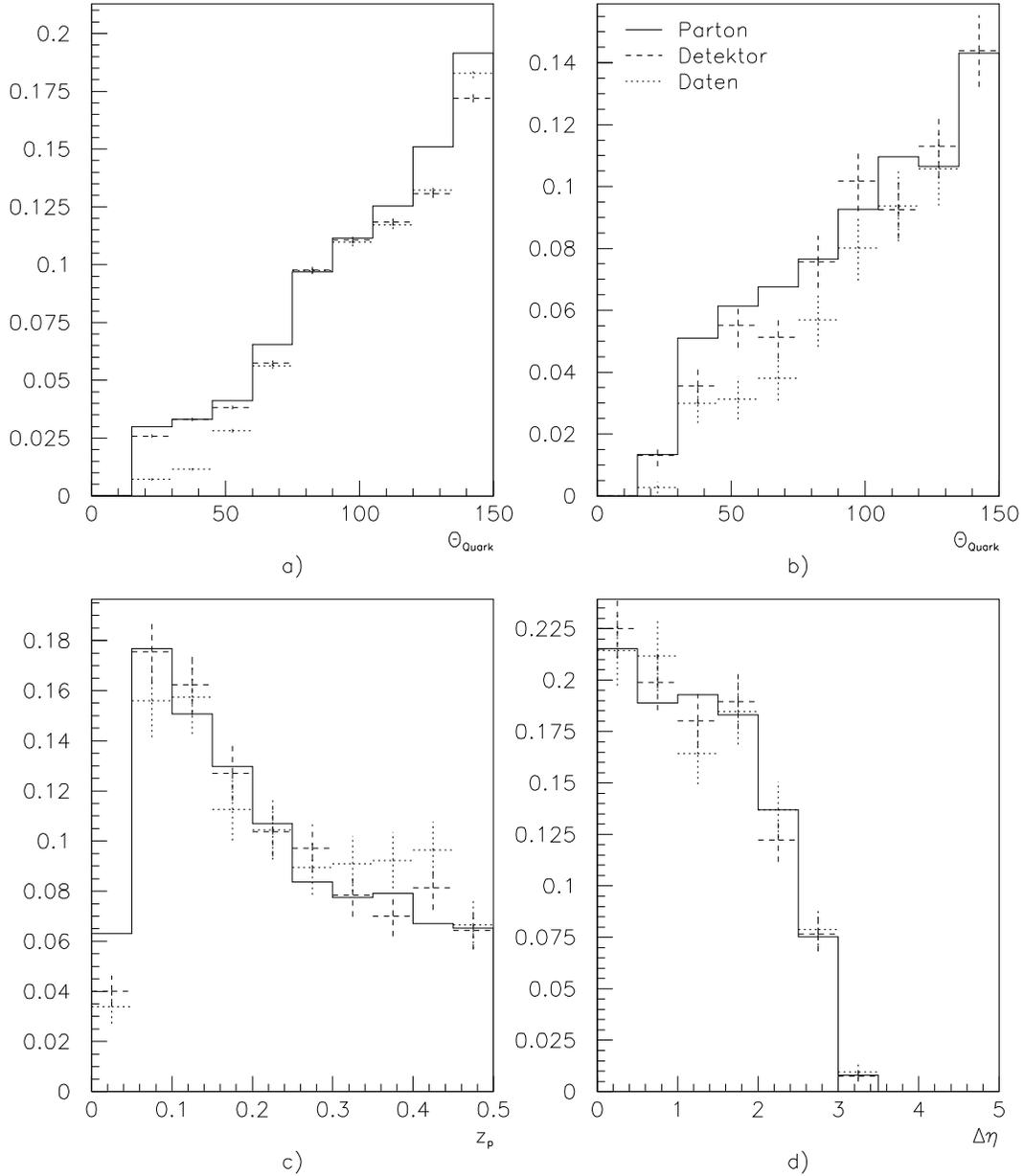,width=0.95\hsize}
\end{center}
\caption[Vergleich der wichtigsten Verteilungen mit abschlie"senden Schnitten]
{{\bf Vergleich der wichtigsten Verteilungen mit abschlie"senden Schnitten}
{\it\ a) und b) Quarkwinkelverteilung f"ur alle bzw.\ 2+1 Jet Ereignisse,
c) $z_p$ Verteilung und d) Rapidit"atsdifferenz der beiden harten Jets.
Dargestellt sind Parton- (durchgezogen),
Detektorniveau (gestrichelt) und Daten (gepunktet)}}
\label{abbvert2}
\end{figure}

Vergleichen wir nun die in Kapitel \ref{kap_unter} untersuchten
Verteilungen noch einmal unter Ber"ucksichtigung des Jetwinkelschnittes
in den Abbildungen \ref{abbvert1} und \ref{abbvert2}. Bei der $\tjet$
Verteilung haben wir im Vorw"artsbereich nicht den gesamten ungen"ugend
beschriebenen Bereich durch unseren Schnitt beseitigt. Die Gr"unde
f"ur die Wahl dieser Grenzen sind schon in Kapitel \ref{kaputhejet}
besprochen worden. Die Verteilung f"ur die 2+1 Raten stimmt nun jedoch
auf allen drei Niveau hervorragend "uberein. (vergleiche Abbildung
\ref{abbthejet}). F"ur die Verteilung des transversalen Impulses im
hadronischen Schwerpunktsystem gilt die gleiche Aussage (vergleiche
Abbildung \ref{abbpt}). Die Gesamtverteilung stimmt aus Gr"unden der
Energiekalibration nur f"ur den Vergleich zwischen
Monte-Carlo-Detektorniveau und Daten. Auch hier hat sich die Verteilung
der 2+1 Jet Ereignisse wesentlich verbessert. Die Verschiebung zwischen
Parton- und Detektorniveau wurde also durch die schlecht beschriebene
Jetwinkelverteilung ausgel"ost.

Die Quarkwinkelverteilungen waren vor dem Schnitt unbefriedigend
beschrieben (siehe Abbildung \ref{abbtquark}). Dies hat sich auch mit
der Einschr"ankung des Jetwinkelbereiches nicht ge"andert (Abbildung
\ref{abbvert2}a und b). Das Aussehen der $z_p$ Verteilung hat sich
ge"andert. Im untersten Bin sind nun wesentlich weniger Eintr"age. In
diesem Bin ist die "Ubereinstimmung zwischen Parton- und Detektorniveau
nicht mehr gegeben. Dies kann in Anbetracht der St"arke der Steigung
jedoch ein reiner Migrationseffekt sein (vergleiche Abbildung
\ref{abbzp}). Die Verteilung des Rapidit"atsunterschiedes hat sich am
st"arksten ge"andert. Dies ist auch verst"andlich, da durch den
Jetwinkelschnitt die Jets unter extremen Winkeln wegfallen. Daher finden
wir nur noch solche Ereignisse, deren Jets wesentlich dichter
zusammen liegen.

\begin{figure}[tbp]                
\begin{center}
\epsfig{file=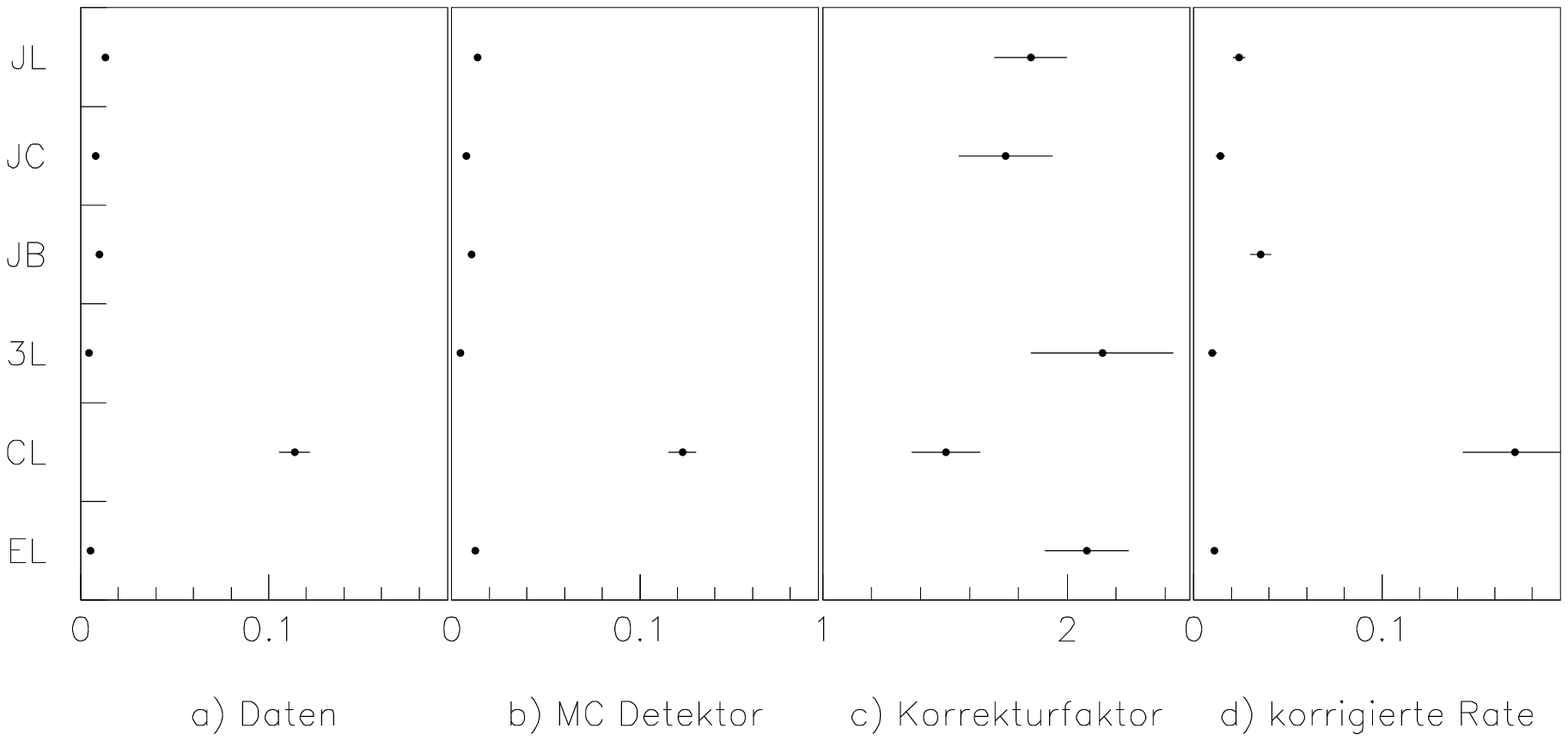,width=\hsize}
\end{center}
\caption[Vergleich verschiedener Jetalgorithmen, Bin 1-4]{{\bf
Vergleich verschiedener Jetalgorithmen, Bin 1-4.}{\it\ Dargestellt sind auf
der Ordinate die Jetalgorithmen JADE im Laborsystem (JL), JADE im hadronischen
Schwerpunktsystem (JC), JADE im Breitsystem (JB), JADE mit $\ycut = 0.03$ (3L),
Cone Algorithmus (CL) und der E Algorithmus (EL) (siehe auch Tabelle
\ref{tabualgo}). In den vier Graphen
sind die Raten f"ur die Daten (a) und das Monte-Carlo-Detektorniveau (b). 
Der Korrekturfaktor ist in c gezeigt und die sich ergebende
korrigierte Rate in d.}}
\label{abbt15meps4}
\end{figure}

\begin{figure}[tbp]                
\begin{center}
\epsfig{file=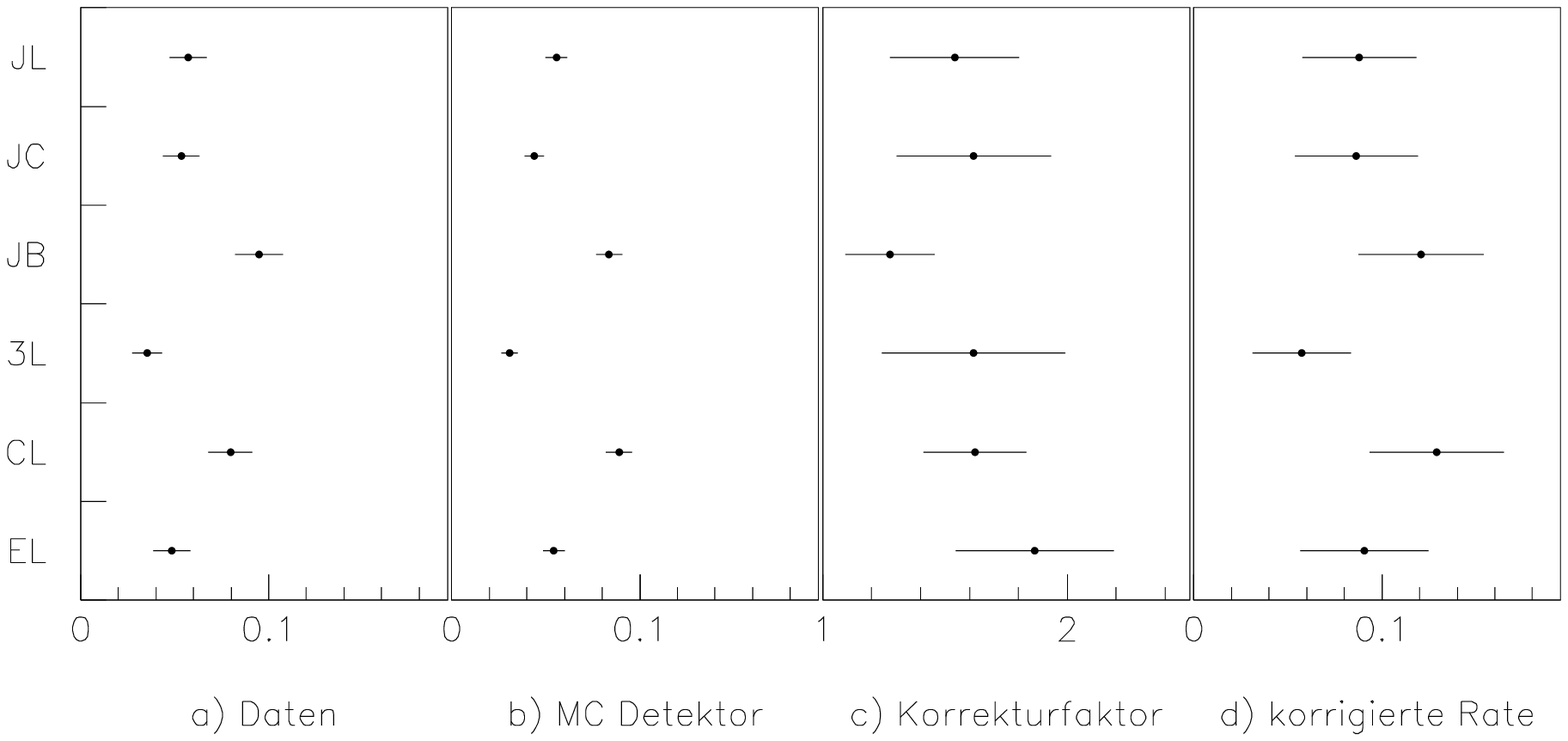,width=\hsize}
\end{center}
\caption[Vergleich verschiedener Jetalgorithmen, Bin 8]{{\bf
Vergleich verschiedener Jetalgorithmen, Bin 8.}{\it\ F"ur n"ahere
Erl"auterungen siehe Abbildung \ref{abbt15meps4} und Tabelle
\ref{tabualgo} }}
\label{abbt15meps8}
\end{figure}

Betrachten wir nun noch einmal die Abh"angigkeit des Ergebnisses vom
Jet"-algorithmus. Der Jetwinkelschnitt wurde dabei immer im Laborsystem
angewandt, d.h.\ die Partikel wurden erst in das entsprechende System
transformiert, danach der Jetalgorithmus angewandt und danach wurden die
Jets zur"ucktransformiert. Anschlie"send wurde der Jetwinkelschnitt
durchgef"uhrt. Das Ergebnis ist in den Abbildungen \ref{abbt15meps4} bis
\ref{abbt15meps8} und den Tabellen \ref{tabjett151} bis \ref{tabjett153}
gezeigt. Ein Vergleich mit den entsprechenden Abbildungen und Tabellen
im Kapitel \ref{kapmotiv} verdeutlicht, da"s wir durch den Schnitt im
Jetwinkel wesentlich kleinere Raten und einen erh"ohten Korrekturfaktor
finden. Die Werte bei den verschiedenen Jetalgorithmen zeigen nun jedoch
kleinere Abweichungen. Die Abh"angigkeit vom Jetalgorithmus und vom
Lorentzsystem ist deutlich kleiner geworden. Sicherlich kann noch nicht
von einer "Ubereinstimmung der verschiedenen Algorithmen gesprochen
werden.

\begin{figure}[tbp]                
\begin{center}
\epsfig{file=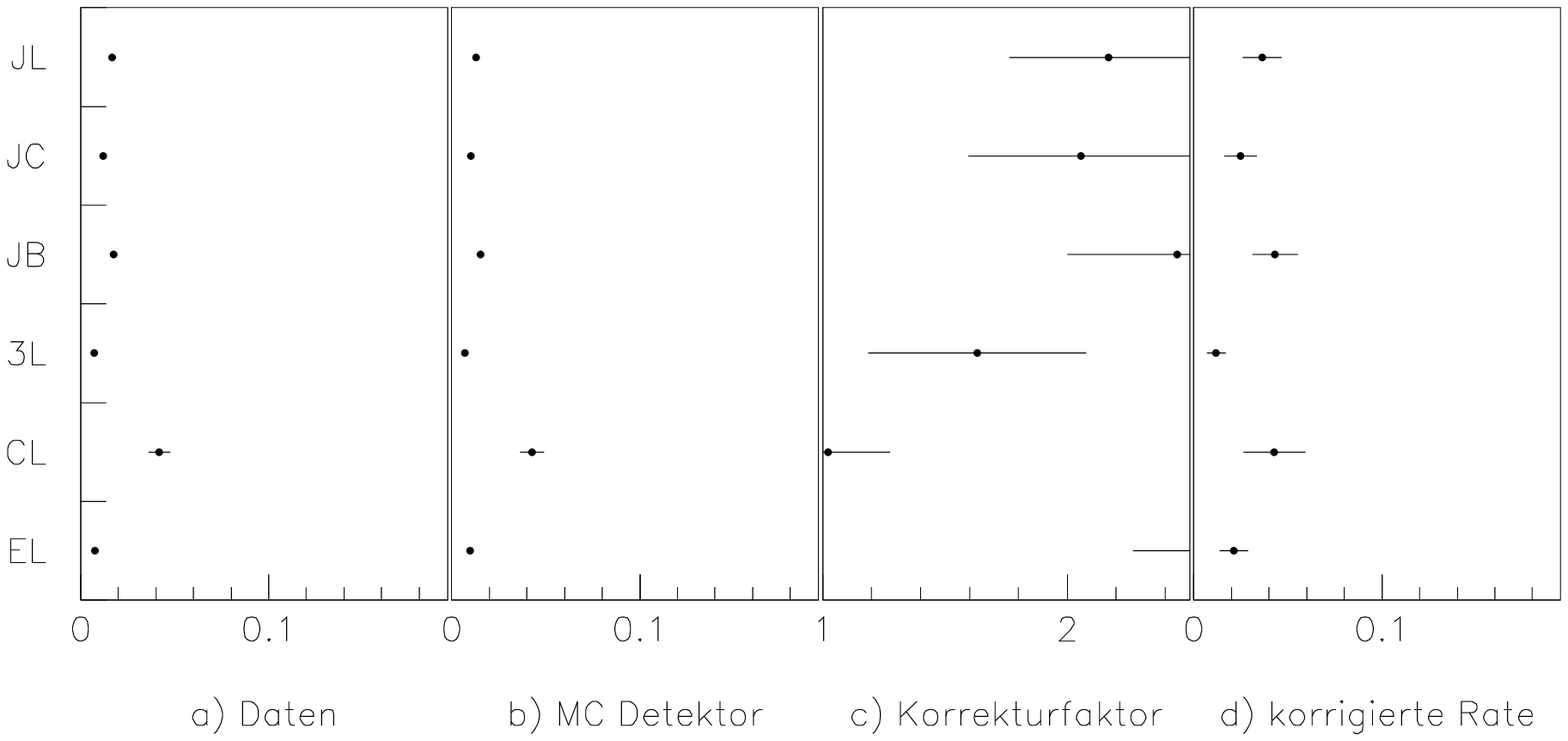,width=\hsize}
\end{center}
\caption[Vergleich verschiedener Jetalgorithmen, Bin 5]{{\bf
Vergleich verschiedener Jetalgorithmen, Bin 5.}{\it\ F"ur n"ahere
Erl"auterungen siehe Abbildung \ref{abbt15meps4} und Tabelle
\ref{tabualgo} }}
\label{abbt15meps5}
\end{figure}

\begin{figure}[tbp]                
\begin{center}
\epsfig{file=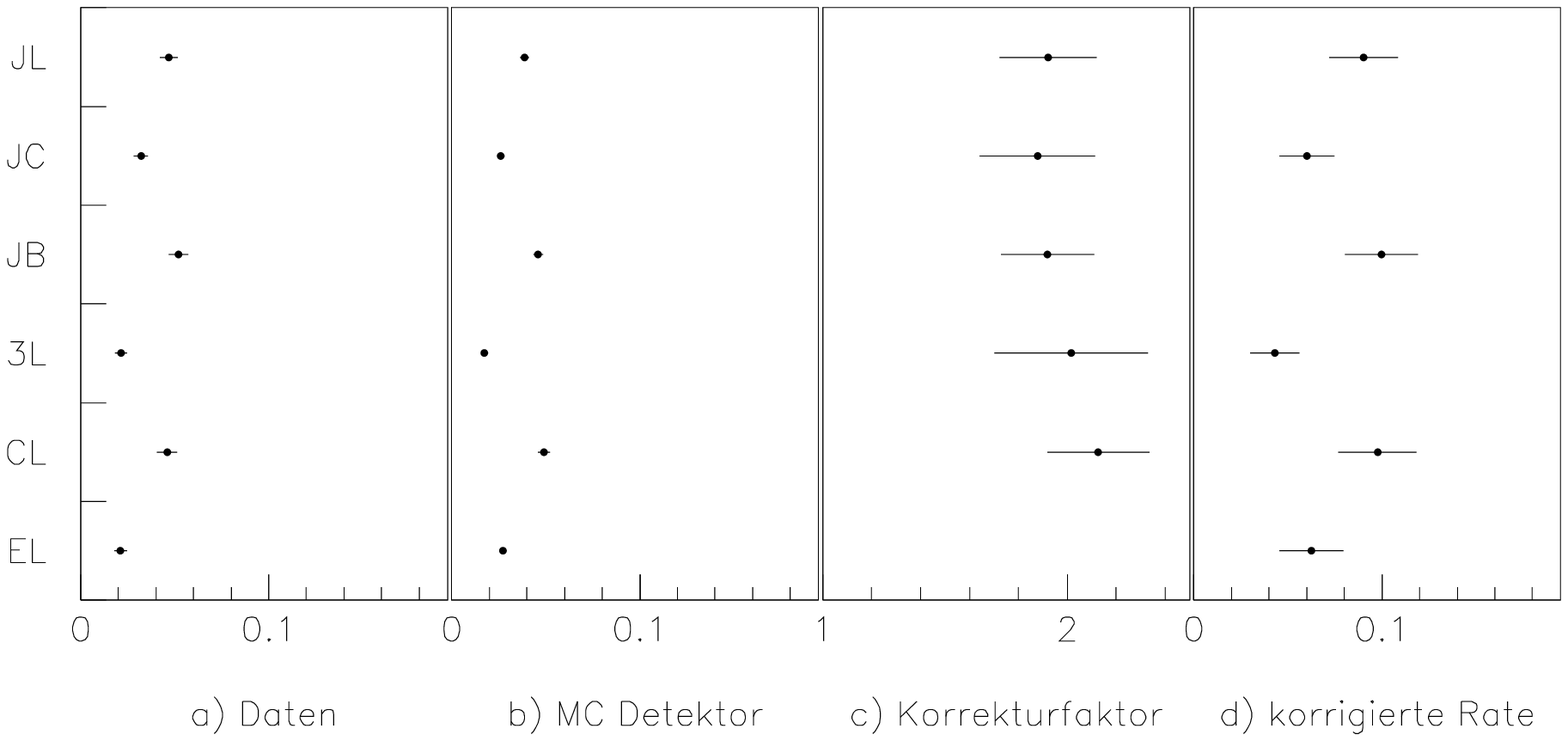,width=\hsize}
\end{center}
\caption[Vergleich verschiedener Jetalgorithmen, Bin 6]{{\bf
Vergleich verschiedener Jetalgorithmen, Bin 6.}{\it\ F"ur n"ahere
Erl"auterungen siehe Abbildung \ref{abbt15meps4} und Tabelle
\ref{tabualgo} }}
\label{abbt15meps6}
\end{figure}

\begin{figure}[tbp]                
\begin{center}
\epsfig{file=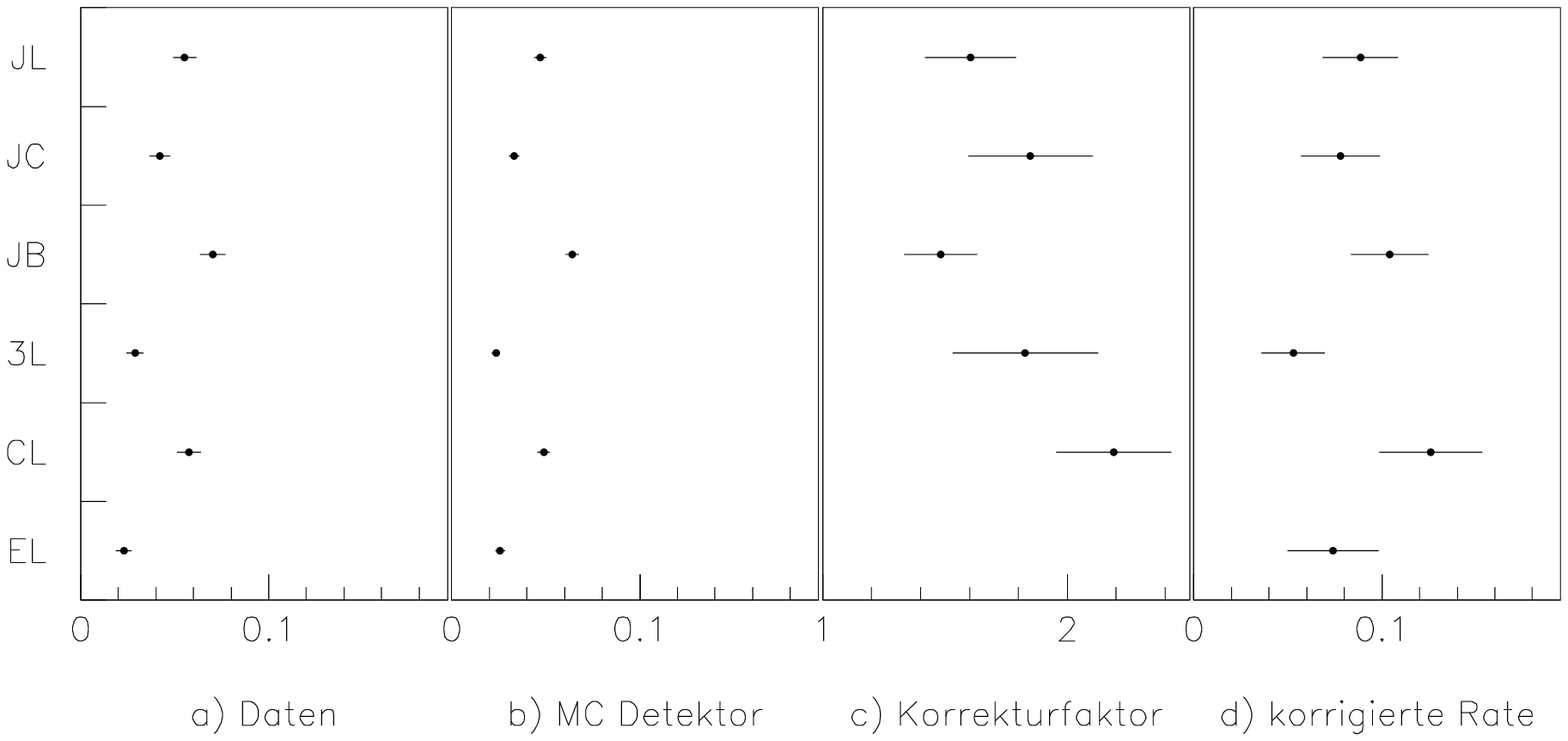,width=\hsize}
\end{center}
\caption[Vergleich verschiedener Jetalgorithmen, Bin 7]{{\bf
Vergleich verschiedener Jetalgorithmen, Bin 7.}{\it\ F"ur n"ahere
Erl"auterungen siehe Abbildung \ref{abbt15meps4} und Tabelle
\ref{tabualgo} }}
\label{abbt15meps7}
\end{figure}

\begin{table}[tbp]
\begin{center}
\begin{tabular}{@{\extracolsep{\fill}}|c||c|c|c|c|c|c|c|c|}
\hline
& 1 & 2 & 3 & 4 & 5 & 6 & 7 & 8\\
\hline
\hline
\multicolumn{9}{|c|}{Datenrate $R_{2+1,\mb{Daten}}$ in \%}\\
\hline
\hline
JL &  1.2 &  1.3 &  1.2 &  1.5 &  1.7 &  4.7 &  5.5 &  5.7\\
   & $\pm 0.6$ & $\pm 0.6$ & $\pm 0.6$ & $\pm 0.6$ & $\pm 0.7$ & $\pm 1.0$ & $\pm 1.1$ & $\pm 1.5$\\
\hline
JC &  0.8 &  0.8 &  0.6 &  1.0 &  1.2 &  3.2 &  4.2 &  5.3\\
   & $\pm 0.6$ & $\pm 0.6$ & $\pm 0.6$ & $\pm 0.6$ & $\pm 0.7$ & $\pm 0.9$ & $\pm 1.1$ & $\pm 1.5$\\
\hline
JB &  0.8 &  1.0 &  1.0 &  1.3 &  1.8 &  5.2 &  7.0 &  9.5\\
   & $\pm 0.6$ & $\pm 0.6$ & $\pm 0.6$ & $\pm 0.6$ & $\pm 0.7$ & $\pm 1.0$ & $\pm 1.2$ & $\pm 1.8$\\
\hline
3L &  0.5 &  0.5 &  0.3 &  0.6 &  0.7 &  2.1 &  2.9 &  3.5\\
   & $\pm 0.6$ & $\pm 0.6$ & $\pm 0.6$ & $\pm 0.6$ & $\pm 0.6$ & $\pm 0.8$ & $\pm 1.0$ & $\pm 1.3$\\
\hline
CL & 17.0 & 14.8 &  9.2 &  8.8 &  4.2 &  4.6 &  5.7 &  8.0\\
   & $\pm 2.8$ & $\pm 2.7$ & $\pm 2.0$ & $\pm 1.6$ & $\pm 1.1$ & $\pm 1.0$ & $\pm 1.1$ & $\pm 1.7$\\
\hline
EL &  0.5 &  0.4 &  0.5 &  0.8 &  0.7 &  2.1 &  2.3 &  4.9\\
   & $\pm 0.6$ & $\pm 0.6$ & $\pm 0.6$ & $\pm 0.6$ & $\pm 0.6$ & $\pm 0.8$ & $\pm 0.9$ & $\pm 1.5$\\
\hline
\end{tabular}
\end{center}
\caption[Zusammenfassung des Vergleichs verschiedener Jetalgorithmen
mit $\tjet$ Schnitt]{{\bf
Zusammenfassung des Vergleichs verschiedener Jetalgorithmen mit
$15^{\circ} < \tjet < 145^{\circ}$ Schnitt}}
\label{tabjett151}
\end{table}

\begin{table}[tbp]
\begin{center}
\begin{tabular}{@{\extracolsep{\fill}}|c||c|c|c|c|c|c|c|c|}
\hline
& 1 & 2 & 3 & 4 & 5 & 6 & 7 & 8\\
\hline
\hline
\multicolumn{9}{|c|}{Monte-Carlo-Detektorrate $R_{2+1,\mb{Cluster}}$ in \%}\\
\hline
\hline
JL &  1.8 &  1.1 &  1.2 &  1.3 &  1.3 &  3.9 &  4.7 &  5.6\\
   & $\pm 0.6$ & $\pm 0.6$ & $\pm 0.6$ & $\pm 0.6$ & $\pm 0.6$ & $\pm 0.7$ & $\pm 0.8$ & $\pm 1.1$\\
\hline
JC &  0.8 &  0.7 &  0.8 &  0.9 &  1.0 &  2.6 &  3.3 &  4.4\\
   & $\pm 0.6$ & $\pm 0.6$ & $\pm 0.6$ & $\pm 0.6$ & $\pm 0.6$ & $\pm 0.7$ & $\pm 0.8$ & $\pm 1.0$\\
\hline
JB &  1.3 &  0.8 &  0.9 &  1.2 &  1.5 &  4.6 &  6.4 &  8.4\\
   & $\pm 0.6$ & $\pm 0.6$ & $\pm 0.6$ & $\pm 0.6$ & $\pm 0.7$ & $\pm 0.8$ & $\pm 0.9$ & $\pm 1.2$\\
\hline
3L &  0.6 &  0.4 &  0.3 &  0.5 &  0.7 &  1.7 &  2.4 &  3.1\\
   & $\pm 0.6$ & $\pm 0.6$ & $\pm 0.6$ & $\pm 0.6$ & $\pm 0.6$ & $\pm 0.7$ & $\pm 0.7$ & $\pm 0.9$\\
\hline
CL & 16.5 & 11.3 &  9.6 & 11.2 &  4.3 &  4.9 &  4.9 &  8.9\\
   & $\pm 2.1$ & $\pm 2.2$ & $\pm 1.8$ & $\pm 1.8$ & $\pm 1.1$ & $\pm 0.8$ & $\pm 0.8$ & $\pm 1.2$\\
\hline
EL &  1.6 &  1.2 &  1.0 &  1.1 &  1.0 &  2.7 &  2.6 &  5.4\\
   & $\pm 0.6$ & $\pm 0.6$ & $\pm 0.6$ & $\pm 0.6$ & $\pm 0.6$ & $\pm 0.7$ & $\pm 0.7$ & $\pm 1.1$\\
\hline
\hline
\multicolumn{9}{|c|}{Monte-Carlo-Partonrate $R_{2+1,\mb{Parton}}$ in \%}\\
\hline
\hline
JL &  2.9 &  1.9 &  2.5 &  2.8 &  2.8 &  7.4 &  7.5 &  8.6\\
   & $\pm 0.7$ & $\pm 0.6$ & $\pm 0.7$ & $\pm 0.7$ & $\pm 0.7$ & $\pm 0.8$ & $\pm 0.9$ & $\pm 1.1$\\
\hline
JC &  1.5 &  1.2 &  1.3 &  1.6 &  2.1 &  4.9 &  6.1 &  7.1\\
   & $\pm 0.6$ & $\pm 0.6$ & $\pm 0.6$ & $\pm 0.6$ & $\pm 0.7$ & $\pm 0.8$ & $\pm 0.8$ & $\pm 1.1$\\
\hline
JB &  5.1 &  3.0 &  3.6 &  3.5 &  3.7 &  8.8 &  9.5 & 10.7\\
   & $\pm 0.9$ & $\pm 0.8$ & $\pm 0.8$ & $\pm 0.8$ & $\pm 0.8$ & $\pm 0.9$ & $\pm 0.9$ & $\pm 1.2$\\
\hline
3L &  1.1 &  0.8 &  0.9 &  1.1 &  1.1 &  3.5 &  4.3 &  5.0\\
   & $\pm 0.6$ & $\pm 0.6$ & $\pm 0.6$ & $\pm 0.6$ & $\pm 0.6$ & $\pm 0.7$ & $\pm 0.8$ & $\pm 1.0$\\
\hline
CL & 24.2 & 19.5 & 16.4 & 14.2 &  4.4 & 10.4 & 10.7 & 14.4\\
   & $\pm 1.8$ & $\pm 1.9$ & $\pm 1.7$ & $\pm 1.5$ & $\pm 0.9$ & $\pm 0.9$ & $\pm 0.9$ & $\pm 1.2$\\
\hline
EL &  3.1 &  1.9 &  2.4 &  2.8 &  2.8 &  8.0 &  8.3 & 10.1\\
   & $\pm 0.7$ & $\pm 0.6$ & $\pm 0.7$ & $\pm 0.7$ & $\pm 0.7$ & $\pm 0.8$ & $\pm 0.9$ & $\pm 1.2$\\
\hline
\end{tabular}
\end{center}
\caption[Zusammenfassung des Vergleichs verschiedener Jetalgorithmen
mit $\tjet$ Schnitt]{{\bf
Zusammenfassung des Vergleichs verschiedener Jetalgorithmen mit
$15^{\circ} < \tjet < 145^{\circ}$ Schnitt}}
\label{tabjett152}
\end{table}

\begin{table}[tbp]
\begin{center}
\begin{tabular}{@{\extracolsep{\fill}}|c||c|c|c|c|c|c|c|c|}
\hline
& 1 & 2 & 3 & 4 & 5 & 6 & 7 & 8\\
\hline
\hline
\multicolumn{9}{|c|}{Korrekturfaktor $R$}\\
\hline
\hline
JL & 1.63 & 1.77 & 2.06 & 2.11 & 2.17 & 1.92 & 1.60 & 1.54\\
   & $\pm0.22$ & $\pm0.33$ & $\pm0.36$ & $\pm0.35$ & $\pm0.41$ & $\pm0.20$ & $\pm0.19$ & $\pm0.27$\\
\hline
JC & 1.94 & 1.70 & 1.63 & 1.71 & 2.06 & 1.88 & 1.85 & 1.62\\
   & $\pm0.39$ & $\pm0.43$ & $\pm0.38$ & $\pm0.36$ & $\pm0.47$ & $\pm0.24$ & $\pm0.26$ & $\pm0.32$\\
\hline
JB & 3.80 & 3.62 & 4.10 & 2.88 & 2.45 & 1.92 & 1.48 & 1.27\\
   & $\pm0.66$ & $\pm0.83$ & $\pm0.86$ & $\pm0.54$ & $\pm0.45$ & $\pm0.20$ & $\pm0.15$ & $\pm0.19$\\
\hline
3L & 1.82 & 2.24 & 2.82 & 2.09 & 1.63 & 2.02 & 1.83 & 1.62\\
   & $\pm0.43$ & $\pm0.70$ & $\pm0.88$ & $\pm0.55$ & $\pm0.45$ & $\pm0.32$ & $\pm0.30$ & $\pm0.38$\\
\hline
CL & 1.47 & 1.73 & 1.70 & 1.27 & 1.02 & 2.13 & 2.19 & 1.62\\
   & $\pm0.22$ & $\pm0.39$ & $\pm0.37$ & $\pm0.23$ & $\pm0.26$ & $\pm0.21$ & $\pm0.24$ & $\pm0.22$\\
\hline
EL & 1.98 & 1.60 & 2.34 & 2.58 & 2.86 & 2.94 & 3.22 & 1.87\\
   & $\pm0.28$ & $\pm0.29$ & $\pm0.43$ & $\pm0.45$ & $\pm0.59$ & $\pm0.35$ & $\pm0.47$ & $\pm0.33$\\
\hline
\hline
\multicolumn{9}{|c|}{korrigierte Rate $R_{2+1,\mb{korr.}}$ in \%}\\
\hline
\hline
JL &  2.0 &  2.3 &  2.5 &  3.1 &  3.6 &  9.0 &  8.9 &  8.8\\
   & $\pm 1.0$ & $\pm 1.2$ & $\pm 1.2$ & $\pm 1.3$ & $\pm 1.5$ & $\pm 2.3$ & $\pm 2.5$ & $\pm 3.5$\\
\hline
JC &  1.5 &  1.3 &  1.0 &  1.7 &  2.5 &  6.0 &  7.8 &  8.6\\
   & $\pm 1.0$ & $\pm 1.0$ & $\pm 0.9$ & $\pm 1.1$ & $\pm 1.4$ & $\pm 2.0$ & $\pm 2.6$ & $\pm 3.8$\\
\hline
JB &  3.0 &  3.6 &  4.1 &  3.6 &  4.3 & 10.0 & 10.4 & 12.1\\
   & $\pm 1.4$ & $\pm 1.8$ & $\pm 1.9$ & $\pm 1.5$ & $\pm 1.7$ & $\pm 2.5$ & $\pm 2.6$ & $\pm 3.8$\\
\hline
3L &  0.9 &  1.1 &  0.8 &  1.2 &  1.2 &  4.3 &  5.3 &  5.7\\
   & $\pm 0.8$ & $\pm 1.0$ & $\pm 0.9$ & $\pm 1.0$ & $\pm 1.0$ & $\pm 1.8$ & $\pm 2.2$ & $\pm 3.1$\\
\hline
CL & 25.0 & 25.6 & 15.7 & 11.2 &  4.3 &  9.8 & 12.6 & 12.9\\
   & $\pm 7.6$ & $\pm 9.9$ & $\pm 6.4$ & $\pm 3.9$ & $\pm 2.1$ & $\pm 2.6$ & $\pm 3.2$ & $\pm 4.1$\\
\hline
EL &  0.9 &  0.6 &  1.2 &  1.9 &  2.1 &  6.3 &  7.4 &  9.1\\
   & $\pm 0.8$ & $\pm 0.7$ & $\pm 0.9$ & $\pm 1.1$ & $\pm 1.2$ & $\pm 2.2$ & $\pm 2.9$ & $\pm 3.9$\\
\hline
\end{tabular}
\end{center}
\caption[Zusammenfassung des Vergleichs verschiedener Jetalgorithmen
mit $\tjet$ Schnitt]{{\bf
Zusammenfassung des Vergleichs verschiedener Jetalgorithmen mit
$15^{\circ} < \tjet < 145^{\circ}$ Schnitt}}
\label{tabjett153}
\end{table}

\vfill        
\section {Ausblick}

Zusammenfassend konnte deutlich gemacht werden, da"s die Betrachtung von
Jet--Parton--Korrelationen einem Problem in einem multi-dimensionalen
Phasenraum entspricht. Es war nicht m"oglich einen Schnitt in einer
einzelnen kinematischen Gr"o"se zu finden, der eine Beseitigung
der Migrationen bewirkt, d.h.\ der eine Verkleinerung des
Korrekturfaktors und gleichzeitig eine Verbesserung der Reinheit der 2+1
Jetereignisklasse erreicht. Immerhin konnte die Reduktion der
Abh"angigkeit von Rekombinationsschema, Jetalgorithmus und Bezugssystem
durch Einf"uhrung eines Jetwinkelschnittes gezeigt werden.

Weitere Arbeit ist n"otig um neue Verteilungen zu "uberpr"ufen. Zu den
dringendsten Problemen geh"oren sicherlich die Erweiterung der
Untersuchungen auf das Theorieniveau. Diese wird besonders interessant,
wenn es gilt, verschiedene Programme zu vergleichen. Neben dem von uns
benutzten Programm PROJET \cite{Projet} sind zur Zeit zwei weitere
Programme in der Test- (MEPJET, \cite{mepjet}) bzw.\ Entwicklungsphase.
Diese neuen Programme werden eine h"ohere Flexibilit"at in Bezug auf die
Auswahl des Jetalgorithmus haben.

Daneben sind jedoch auch neue M"oglichkeiten entstanden durch den Austausch 
des r"uckw"artigen Kalorimeters BEMC durch das SPACAL \cite{DESY:95-067}, 
welches nun auch einen hadronischen Teil besitzt. Sobald die 
Energiekalibration dort verstanden ist, kann der obere Winkelschnitt 
vielleicht fallen gelassen oder zumindest ver"andert werden. Wie wir
aber an der Jetwinkelverteilung in Abbildung \ref{abbthejet} gesehen
haben, sind dort nur wenige Jets von 2+1 Jetereignissen. Au"serdem hat
sich die Luminosit"at bei der Datennahme 1995 wieder erh"oht, so da"s
wir mehr Ereignisse und damit geringere statistische Fehler erwarten
d"urfen.

Das immer besser werdende Verst"andnis von Monte-Carlo-Simulationen und
die Untersuchung der Hadronisierungseffekte werden sicher auch weiterhin
neue Aspekte zum Verst"andnis der Jet--Parton--Korrelation beitragen.
W"unschenswert w"are jedoch ein Monte-Carlo-Ereignisgenerator, der auf
den Partonschaueransatz zur Beschreibung der n"achst-zu-f"uhrenden
Ordnung Korrekturen verzichtet und sie statt dessen bereits in den
Matrixelementen beinhaltet. Dem stehen jedoch nicht nur software-technische
Probleme entgegen.

Die in Kapitel \ref{kapgluon} erw"ahnte Messung der Gluondichte des
Protons durch eine Jetanalyse stellt durch die zus"atzliche $x_B$
Einteilung neue Anforderungen an das Verst"andnis der
Jet--Parton--Korrelationen. Hier ist eine Untersuchung der Gr"o"sen auf
eine $x_B$ Abh"angigkeit von N"oten.

Wie wir aus dieser sicher unvollst"andigen Aufz"ahlung erkennen k"onnen,
bleiben noch viele Fragen offen, die die Neugier des Physikers anregen.

\cleardoublepage 

\addcontentsline{toc}{chapter}{{\bf Abbildungsverzeichnis}}
\listoffigures

\cleardoublepage 

\addcontentsline{toc}{chapter}{{\bf Tabellenverzeichnis}}
\listoftables

\cleardoublepage 

\addcontentsline{toc}{chapter}{{\bf Literaturverzeichnis}}
\bibliographystyle{alpha}
\bibliography{diplom}

\begin{thebibliography}{MKSW93}

\bibitem[BB94]{DESY:94-231}
U.~Bassler und G.~Bernardi.
\newblock On the kinematic reconstruction of deep inelastic scattering at
  {HERA}: the ${\Sigma}$ method.
\newblock DESY,  94-231, Dezember 1994.

\bibitem[Ber92]{Berger:92}
Ch. Berger.
\newblock {\em Teilchenphysik}.
\newblock Springer ISBN 3-540-54218-3, 1992.

\bibitem[Bes93]{H1-12:93-336}
M.~Besan\c{c}on.
\newblock A proposal for the use of the $k_t$ clustering algorithm for jets in
  {DIS}.
\newblock DAPNIA, Saclay, Frankreich,  93-336, 1993.

\bibitem[BM94]{Disjet}
T.~Brodkorb und E.~Mirkes.
\newblock {\em {DISJET} manual}, 1994.
\newblock MAD/PH/821.

\bibitem[BW95]{DESY:95-067}
K.~Borras und M.~Weber.
\newblock The {H}1 calorimetry: {P}erformance and {U}pgrade {P}rogram.
\newblock DESY,  95-067, April 1995.

\bibitem[CDB92]{Webber:92}
S.~Catani, Yu.L. Dokshitzer, und B.R.Webber.
\newblock The $k_{\perp}-$ clustering algorithms for jets in deep inelastic
  scattering and hadron collisions.
\newblock {\em Physics Letters B}, 285:291--299, 1992.

\bibitem[{CDF}95]{Fermi:95-022-E}
{CDF Collaboration}.
\newblock Observation of {T}op {Q}uark {P}roduction in $p\bar p$ {C}ollisions.
\newblock Fermilab,  95/022-E, Februar 1995.

\bibitem[CSS93]{Django}
K.~Charchula, G.~A. Schuler, und H.~Spiesberger.
\newblock {\em Combined {QED} and {QCD} Radiative Effects in Deep Inelastic
  Lepton Proton Scattering the Monte Carlo Generator DJANGO6}, 1993.

\bibitem[{DZe}95]{Fermi:95-028-E}
{DZero Collaboration}.
\newblock Observation of the {T}op {Q}uark.
\newblock Fermilab,  95/028-E, Februar 1995.

\bibitem[Eis95]{Eisele:95}
F.~Eisele.
\newblock Deep {I}nelastic {S}cattering.
\newblock In {\em Invited Talk at the International Europhysics Conference On
  High Energy Physics, Brussels, 27 July - 2 August, 1995}, 1995.
\newblock Available on WWW http://dice2.desy.de/psfiles/proceedings/list.html.

\bibitem[{GEA}93]{Geant:321}
{GEANT Team}.
\newblock Geant -- detector description and simulation tool.
\newblock CERN Program Library,  W5053, 1993.

\bibitem[GHVB95]{DESY:95-107}
D.~Graudenz, M.~Hampel, A.~Vogt, und Ch. Berger.
\newblock The {M}ellin transform technique for the extraction of the gluon
  density.
\newblock DESY, CERN, RWTH-Aachen,  95-107, Juni 1995.
\newblock auch : CERN-TH/95-149.

\bibitem[GM91]{Graudenz:91}
D.~Graudenz und M.~Magnussen.
\newblock Jet cross section in deeply inelastic scattering at {HERA}.
\newblock In {\em Physics at HERA, Proceedings of the workshop}, Band~1, Seiten
  261--273. DESY, Oktober 1991.

\bibitem[Gra93]{Projet}
D.~Graudenz.
\newblock {\em PROJET jet cross sections in deeply inelastic electron proton
  scattering --- version 3.6}, 1993.

\bibitem[{H1 }93a]{DESY:93-078}
The {H1 Calorimeter Group}.
\newblock The {H}1 {L}iquid {A}rgon {C}alorimeter {S}ystem.
\newblock DESY,  93-078, Juni 1993.

\bibitem[{H1 }93b]{DESY:93-103}
{H1 Collaboration}.
\newblock The {H}1 {D}etector at {HERA}.
\newblock DESY,  93-103, Juli 1993.

\bibitem[{H1 }95a]{H1sim}
{H1 Collaboration}.
\newblock {\em Guide to simulation program {H1SIM}}, 1995.
\newblock H1 Intern.

\bibitem[{H1 }95b]{H1phan}
{H1 Collaboration}.
\newblock {\em {H1PHAN} - {F}unktionsbibliothek von {H1}}, 1995.
\newblock H1 Intern, Version 2.3.4.

\bibitem[{H1 }95c]{H1rec}
{H1 Collaboration}.
\newblock {\em {H1REC} - {R}ekonstruktionsprogramm von {H1}}, 1995.
\newblock H1 Intern.

\bibitem[Ham93]{Hampel:93}
M.~Hampel.
\newblock Monte-{C}arlo {U}ntersuchungen zur tief inelastischen
  {E}lektronen-{S}treuung.
\newblock Diplomarbeit, RWTH Aachen, Februar 1993.

\bibitem[HM84]{Halzen:84}
F.~Halzen und A.~Martin.
\newblock {\em Quarks and leptons}.
\newblock John Wiley \& Sons, ISBN 0-471-81187-4, 1984.

\bibitem[Hoe95]{hoeger:95}
Ch. Hoeger.
\newblock Extracting $xg(x,{Q}^2)$ with {E}nergy {D}istributions in the {B}reit
  {F}rame.
\newblock In {\em Talk at internal jet meeting on 3th March}, 1995.

\bibitem[Ing92]{Lepto}
G.~Ingelmann.
\newblock {\em LEPTO --- The Lund Monte Carlo for Deep Inelastic Lepton-Nucleon
  Scattering}, 1992.
\newblock Version 6.01.

\bibitem[Itt93]{Itterbeck:93}
H.~Itterbeck.
\newblock Untersuchungen am {M}yonsystem des {H}1-{D}etektors.
\newblock Diplomarbeit, RWTH Aachen, Februar 1993.

\bibitem[IZ85]{Itzykson:85}
C.~Itzykson und J.-B. Zuber.
\newblock {\em Quantum field theory}.
\newblock McGraw-Hill ISBN 0-07-066353-X, 1985.

\bibitem[{JAD}86]{JADE:86}
{JADE Collaboration}.
\newblock Experimental studies on multi--jet production in $\ep-\emm$
  annihilation at {PETRA} energies.
\newblock {\em Z.Phys.}, C33:23f, 1986.

\bibitem[KM95]{H1-12:95-466}
S.~Kermiche und J.~Marks.
\newblock {LA}r electromagnetic energy scale studies using {NC DIS} (1+1) jet
  and {QED} {C}ompton events for the {$e^+p$} 1994 {H1} data.
\newblock Centre de physique des particules de Marseille, Frankreich,  95-466,
  1995.

\bibitem[K"o95]{Koehler:95}
T.~K"ohler.
\newblock {\em Suche nach angeregten {L}eptonen mit dem {H}1-{D}etektor}.
\newblock Promotionsarbeit, RWTH Aachen, Februar 1995.

\bibitem[L"o94]{Ariadne}
Leif L"onnblad.
\newblock {\em {ARIADNE} --- a program for simulation of {QCD}-cascades
  implementing the colour dipole model}.
\newblock CERN, 1994.
\newblock Revision 5.

\bibitem[MKSW93]{Herwig}
G.~Marchesini, I.G. Knowles, M.H. Seymour, und B.R. Webber.
\newblock {\em {HERWIG} --- a {M}onte {C}arlo event generator for simulating
  {H}adron {E}mission {R}eactions {W}ith {I}nterfering {G}luons}, 1993.
\newblock Version 5.8, Computer Phys. Commun. 67 (1992) 465.

\bibitem[MZ95]{mepjet}
E.~Mirkes und D.~Zeppenfeld.
\newblock {\em {MEPJET} 1.0 : {A} next-to-leading order event generator for
  $ep\rightarrow n$ jets}, 1995.
\newblock Preliminary Version.

\bibitem[Nac92]{Nachtmann:92}
O.~Nachtmann.
\newblock {\em Elementarteilchenphysik}.
\newblock Vieweg ISBN 3-528-08926-1, 1992.

\bibitem[Nis94a]{NisQCD:94}
R.~Nisius.
\newblock Jet {S}tudies in {D}eep {I}nelastic ep {S}cattering at {HERA}.
\newblock In {\em QCD 94, Montpellier, France, July 1994}, 1994.
\newblock Available on WWW http://dice2.desy.de/psfiles/proceedings/list.html,
  published in Nucl. Phys. B (Proc. Suppl.) 39B,C (1995) March 1995.

\bibitem[Nis94b]{Nisius:94}
R.~Nisius.
\newblock {\em Measurement of the strong coupling constant $\alpha_s$ from jet
  rates in deep inelastic scattering}.
\newblock Promotionsarbeit, RWTH Aachen, Juni 1994.
\newblock Available on WWW
  http://www.physik.rwth-aachen.de/group/IIIphys/H1/papers/nisius\_dr.ps.Z.

\bibitem[Nis95a]{Nisius:95}
R.~Nisius.
\newblock The alpha\_s {M}easurement from {J}et {R}ates in {D}eep {I}nelastic
  ep {S}cattering at {HERA}.
\newblock In {\em Talk at the XXXth Rencontre de Moriond, Les Arcs, France,
  March 1995}, 1995.
\newblock Available on WWW http://dice2.desy.de/psfiles/proceedings/list.html.

\bibitem[Nis95b]{NisBrue:95}
R.~Nisius.
\newblock Determination of the {S}trong {C}oupling {C}onstant alpha\_s in
  {D}eep {I}nelastic $ep$ {S}cattering at {HERA}.
\newblock In {\em Talk at the International Europhysics Conference On High
  Energy Physics, Brussels, 27 July - 2 August, 1995}, 1995.
\newblock Available on WWW
  http://dice2.desy.de/psfiles/confpap/pap\_list.html\#EPS95.

\bibitem[{Par}94]{PDG:94}
{Particle Data Group}.
\newblock Particle {P}hysics {B}ooklet.
\newblock American Institut of Physics, Juli 1994.
\newblock Auszug aus Review of Partice Properties, Physical Review, D50,
  1173(1194).

\bibitem[PB95]{PDF:606}
H.~Plothow-Besch.
\newblock {PDFLIB}.
\newblock CERN Program Library,  W5051, 1995.
\newblock Version 6/06/02.

\bibitem[Ros95]{Rosenbauer:95}
K.~Rosenbauer.
\newblock {\em Suche nach {L}eptoquarks und {L}eptogluonen im {H1}-{E}xperiment
  bei {HERA}}.
\newblock Promotionsarbeit, RWTH Aachen, Juli 1995.

\bibitem[Sey95]{Seymour:95}
M.~Seymour.
\newblock Monte {C}arlo models and hadronization.
\newblock In {\em Folienkopien vom 4.-5.Oktober}, 1995.
\newblock Graduate School of Partice Physics, 25.September - 6.Oktober 1995,
  Cul-des-Sarts, Belgien.

\bibitem[{ZEU}95]{DESY:95-182}
{ZEUS Collaboration}.
\newblock Measurement of $\alpha_s$ from jet rates in deep inelastic scattering
  at {HERA}.
\newblock DESY,  95-182, September 1995.

\end{thebibliography}

\cleardoublepage 

\chapter*{Danksagung\markboth{DANKSAGUNG}{DANKSAGUNG}}
\addcontentsline{toc}{chapter}{{\bf Danksagung}}

Am Ende dieser Arbeit m"ochte ich nun all denen danken, ohne die meine
Diplomarbeit nicht m"oglich gewesen w"are.

Als Erstes bin ich Herrn Prof.~Ch.~Berger zu Dank verpflichtet. Nicht nur die
interessante Themenstellung ist sein Verdienst, sondern auch die hervorragende
Betreuung. Ebenso hat er mir durch Reisen zum DESY und zur {\it graduate
school of particle physics } in Belgien erm"oglicht, die Vielfalt der
physikalischen Grundlagenforschung kennenzulernen.

Prof.~G.~Fl"ugge danke ich f"ur die M"uhe bei der Erstellung des
Zweitgutachtens.

F"ur die Einarbeitung in das Thema habe ich Dr.~Richard Nisius zu danken. Er
hat trotz der eigenen Arbeit immer wieder die Zeit gefunden, mir bei Problemen
zu helfen und mich in die wissenschaftliche Arbeitsweise einzuf"uhren.
Au"serdem hat er mir durch neue Ideen immer wieder die M"oglichkeit gegeben an
der aktuellen Analyse teil zu haben.

Der inzwischen recht gro"s gewordenen Aachener Jetgruppe, namentlich
Martin Hampel, Christian Niedzballa, Klaus Rabbertz, Dr.~Konrad
Rosenbauer und J"urgen Scheins, danke ich f"ur den engen
Informationsaustausch und die gute Zusammenarbeit. Das freundliche
Arbeitsklima, sowohl in Aachen als auch in Hamburg, ist sicherlich nicht
selbstverst"andlich. Eingeschlossen sind hier auch diejenigen, die sich
nicht direkt mit Jetanalysen besch"aftigen, aber trotzdem immer ein
offenes Ohr f"ur Fragen und Probleme hatten. Erw"ahnt seien noch
Dr.~Torsten K"ohler, Oliver Mang, Claus Keuker und Heiko Itterbeck,
letzterer hat mich auf das faszinierende Arbeitsfeld des
I.\ Physikalischen Instituts hingewiesen.

Sicher haben alle Mitarbeiter der H1 Kollaboration durch das Erstellen von
Softwarepaketen, die Instandhaltung des Detektors und die Messung der Daten zu
dieser Arbeit beigetragen.

Mein besonderer Dank gilt Martin Hampel, Patrick Kandel, Richard Nisius,
Klaus Rabbertz und Konrad Rosenbauer, die das Manuskript durchgesehen
haben und denen ich viele Tips und Anregungen verdanke.

Ein besonderer Dank gilt meinen Eltern, die mir mein Studium nicht nur
finanziell erm"oglicht haben, sondern mir auch immer wieder viel Verst"andnis
entgegenbrachten.

\cleardoublepage 

\chapter*{Lebenslauf\markboth{Lebenslauf}{Lebenslauf}}
\addcontentsline{toc}{chapter}{{\bf Lebenslauf}}

\begin{description}
\item[24.\ Juni 1969] geboren in Grevenbroich als einziger Sohn der
N"aherin Marlies Hadig, geb.\ Roumen, und des Starkstromelektrikers
Wolfgang Hadig.
\item[1975 - 1979] Besuch der Gemeinschaftsgrundschule Hochneukirch 
in Otzenrath
\item[ab 1979] Besuch des Franz--Meyers--Gymnasium in 
Giesenkirchen/M"onchenglad"-bach (fr"uher Gymnasium i.E. Giesenkirchen (bis 
1980), dann Franz-Meyers-Gym"-na"-sium i.E. (1980-1982)).
\item[1988] Erlangung der Allgemeinen Hochschulreife
\item[1988-1989] Ableistung eines 15-monatigen Wehrdienstes bei der Bundeswehr
\item[Oktober 1989] Beginn des Studiums an der RWTH Aachen im Studiengang
Physik
\item[Fr"uhjahr 1992] Vordiplom in Physik
\item[Oktober 1990 - M"arz 1993] Studentische Hilfskraft am
Rogowski--Institut f"ur\linebreak[4] Elek"-tro"-tech"-nik. Beaufsichtigung
von Computer-unterst"utztem Unterricht und administrative Aufgaben
auf Sun-Workstations.
\item[Juli 1993 - Juni 1994] Studentische Hilfskraft am Lehrstuhl f"ur
Informatik I. Betreuung eines CIP-Pools mit UNIX-Workstation und
PC-Rechnern.
\item[Juli 1994] Beginn der Mitarbeit in der H1 Kollaboration
\item[Februar 1995] Beginn der Diplomarbeit bei Prof.\ Dr.\ Berger am 
I.\ Physikalischen Institut der RWTH Aachen im Rahmen der H1 Kollaboration 
\end{description}

\end{document}